\documentclass[11pt]{article}
\usepackage[english]{babel}
\usepackage{graphicx,rotating}
\usepackage{a4p,here}
\usepackage{cite}

\renewcommand{\Huge}{\huge}
\newcommand{\updates}[1]%
 {\fbox{\parbox{\linewidth}{\textbf{Updates with respect to last summer:}\\#1}}}
\def\pz{\phantom{0}}
\def\pzz{\phantom{00}}
\def\enw     { \mathrm{e} \nu \mathrm{W} }
\def\nng     { \nu \bar{\nu} \gamma }
\def\WWg     { {\mathrm{WW}\gamma} }
\def\ifmath#1{\relax\ifmmode #1\else $#1$\fi}%
\def\GeV{\ifmmode {\mathrm{ Ge\kern -0.1em V}}\else
                   \textrm{Ge\kern -0.1em V}\fi}%
\def\MeV{\ifmmode {\mathrm{ Me\kern -0.1em V}}\else
                   \textrm{Me\kern -0.1em V}\fi}%
\def\keV{\ifmmode {\mathrm{ ke\kern -0.1em V}}\else
                   \textrm{ke\kern -0.1em V}\fi}%
\def\eV{\ifmmode  {\mathrm{ e\kern -0.1em V}}\else
                   \textrm{e\kern -0.1em V}\fi}%

\newcommand{\LEPI}{\ifmath{\mathrm{\hbox{LEP-I}}}}
\newcommand{\LEPII}{\ifmath{\mathrm{\hbox{LEP-II}}}}
\newcommand{\pb}{\ifmath{\mathrm{pb^{-1}}}}

\def\mca#1#2 {\multicolumn{#1}{|c|}{#2}}

\newcommand{\Afbpol}{A^{0,\,\ell}_{\rm {FB}}}
\newcommand{\Afbzl}{A^{0,\,\ell}_{\rm {FB}}}
\newcommand{\Afbze}{A^{0,\,{\rm e}}_{\rm {FB}}}
\newcommand{\Afbzf}{A^{0,\,{\rm f}}_{\rm {FB}}}
\newcommand{\Afbzm}{A^{0,\,\mu}_{\rm {FB}}}
\newcommand{\Afbzt}{A^{0,\,\tau}_{\rm {FB}}}
\newcommand{\Afbzq}{A^{0,\,{\rm q}}_{\rm {FB}}}
\newcommand{\Afbzb}{A^{0,\,{\rm b}}_{\rm {FB}}}
\newcommand{\Afbzc}{A^{0,\,{\rm c}}_{\rm {FB}}}

\renewcommand{\rm}{\mathrm}
\renewcommand{\thefootnote}{\fnsymbol{footnote}}

\newcommand{\alfas}{\alpha_{\rm{s}}}
\newcommand{\alfmz}{\alfas(\MZ^2)}

\newcommand {\spr}     {{s^{\prime}}} 
\newcommand{\Zzero}{\mbox{${\mathrm{Z}}$}}

\newcommand{\SM}{\mbox{Standard Model}}
\newcommand{\MZ}{m_{\mathrm{Z}}}
\newcommand{\MW}{m_{\mathrm{W}}}
\newcommand{\GF}{G_{\mathrm{F}}}

\newcommand{\MH}{m_{\mathrm{H}}}
\newcommand{\Mt}{m_{\mathrm{t}}}

\newcommand{\GZ}{\Gamma_{\mathrm{Z}}}
\newcommand{\Afb}{A_{\mathrm{FB}}}

\newcommand{\Nnu}{N_\nu}
\newcommand{\RZ}{R_{\ell}}
\newcommand{\Ree}{R_{\mathrm{e}}}
\newcommand{\Rmu}{R_{\mu}}
\newcommand{\Rtau}{R_{\tau}}
\newcommand{\Rl}{R_{\ell}}

\newcommand{\thw}{\theta_{\mathrm{W}}}

\newcommand{\swsq}{1-\MW^2/\MZ^2}

\newcommand{\swsqeffl}{\sin^2\!\theta_{\rm{eff}}^{\rm {lept}}}

\newcommand{\ff}{{\rm f}\overline{\rm f}}
\newcommand{\lept}{\ell^+\ell^-}
\newcommand{\ee}{\mathrm{e}^+\mathrm{e}^-}
\newcommand{\bb}{{\mathrm b\overline{\mathrm b}}}
\newcommand{\cc}{{\rm c\overline{\rm c}}}

\newcommand{\mumu}{\mu^+\mu^-}

\newcommand{\tautau}{\tau^+\tau^-}
\newcommand{\eeff}{\ee\rightarrow\ff}
\newcommand{\eecc}{\ee\rightarrow\cc}
\newcommand{\eebb}{\ee\rightarrow\bb}

\newcommand{\eeee}{\ee\rightarrow\ee}

\newcommand{\eemumu}{\ee\rightarrow \mu^+\mu^-}
\newcommand{\eetautau}{\ee\rightarrow \tau^+\tau^-}

\newcommand{\eeqq}{\ee\rightarrow {\rm q}\overline{\rm q}}

\newcommand{\eell}{\ee\rightarrow \lept}

\newcommand{\Ztomumu}{\mathrm{Z}\rightarrow \mu^+\mu^-}
\newcommand{\Ztotautau}{\mathrm{Z}\rightarrow \tau^+\tau^-}

\newcommand{\Gee}{\Gamma_{\rm {ee}}}

\newcommand{\Gmumu}{\Gamma_{\mu\mu}}
\newcommand{\Gtautau}{\Gamma_{\tau\tau}}
\newcommand{\Ginv}{\Gamma_{\mathrm{inv}}}
\newcommand{\Ghad}{\Gamma_{\mathrm{had}}}
\newcommand{\Gnu}{\Gamma_{\nu\nu}}

\newcommand{\Gll}{\Gamma_{\ell\ell}}

\newcommand{\roots}{\ifmath{\sqrt{s}}}

\newcommand{\ra}{\rightarrow}

\newcommand{\shad}{\sigma_{\mathrm{h}}^{0}}

\newcommand{\sll}{\sigma_{\ell}^{\mathrm{0}}}

\newcommand{\ptau}{\mbox{$\cal P_{\tau}$}}
\newcommand{\ALR}{\mbox{$A_{\rm {LR}}$}}
\newcommand{\ALRz}{\mbox{$A^0_{\rm{LR}}$}}

\newcommand{\cAe}{\mbox{$\cal A_{\rm e}$}}
\newcommand{\cAm}{\mbox{$\cal A_{\mu}$}}
\newcommand{\cAt}{\mbox{$\cal A_{\tau}$}}
\newcommand{\cAf}{\mbox{$\cal A_{\rm f}$}}

\newcommand{\cAl}{\mbox{$\cal A_{\ell}$}}
\newcommand{\cAb}{\mbox{$\cal A_{\rm b}$}}
\newcommand{\cAc}{\mbox{$\cal A_{\rm c}$}}

\newcommand{\Abb}{\ifmath{A_{\mathrm{FB}}^{\mathrm{b\bar{b}}}}}

\newcommand{\Acc}{\ifmath{A_{\mathrm{FB}}^{\mathrm{c\bar{c}}}}}

\newcommand{\Rb}{\ifmath{R_{\mathrm{b}}}}

\newcommand{\Rc}{\ifmath{R_{\mathrm{c}}}}
\newcommand{\Rbz}{\ifmath{R_{\mathrm{b}}^0}}
\newcommand{\Rcz}{\ifmath{R_{\mathrm{c}}^0}}
\newcommand{\Gbb}{\ifmath{\Gamma_{\mathrm{b\bar{b}}}}}
\newcommand{\Gcc}{\ifmath{\Gamma_{\mathrm{c\bar{c}}}}}

\newcommand{\bl}{\rm b \rightarrow \ell }

\newcommand{\cl}{\rm c \rightarrow \ell }

\newcommand{\pp}{\mbox{$\mathrm{p}\overline{\mathrm{p}}$}}

\newcommand{\gahate}{g_{{A}{\rm e}}}
\newcommand{\gvhate}{g_{{V}{\rm e}}}
\newcommand{\gahatf}{g_{{A}{\rm f}}}
\newcommand{\gvhatf}{g_{{V}{\rm f}}}

\newcommand{\gahatmu}{g_{{A}{\mu}}}
\newcommand{\gvhatmu}{g_{{V}{\mu}}}
\newcommand{\gahattau}{g_{{A}{\tau}}}
\newcommand{\gvhattau}{g_{{V}{\tau}}}
\newcommand{\gahatl}{g_{{A}{\ell}}}
\newcommand{\gvhatl}{g_{{V}{\ell}}}
\newcommand{\effb}{\varepsilon_{\mathrm{b}}}
\newcommand{\effc}{\varepsilon_{\mathrm{c}}}
\newcommand{\effuds}{\varepsilon_{\mathrm{uds}}}

\newcommand{\Dstarp}{{\rm D}^{*+}}

\newcommand{\Dstarpm}{\mbox{${\mathrm D}^{*\pm}$}}

\newcommand{\Dzero}{{\rm D}^0}
\newcommand{\Dplus}{{\rm D}^+}

\newcommand{\Bzero}{{\rm B}^0}
\newcommand{\Bzerob}{{\overline{{\rm B}^0}}}

\newcommand{\Brcl}{\mathrm{BR(c \rightarrow \ell)}}
\newcommand{\Brbclp}{\ifmath{\mathrm{BR(b \ra c \ra \bar\ell)}}}

\newcommand{\Ds}{{\rm{D_s}}}
\newcommand{\Lc}{{\rm{\Lambda_c}}}
\def\Ups4s{\mbox{$\Upsilon(4S)$}}
\newcommand{\Cq}{{\cal C}_{\rm q}}

\newcommand{\Cc}{{\cal C}_{\rm c}}

\newcommand{\Cb}{{\cal C}_{\rm b}}
\newcommand{\HI}{\mbox{$100~\GeV \le \MH \le 1000~\GeV$}}

\newcommand{\chiM}{\ifmath{\overline{\chi}}}
\newcommand{\Abh}{\ifmath{\Abb(+2)}}
\newcommand{\Ach}{\ifmath{\Acc(+2)}}
\newcommand{\Abp}{\ifmath{\Abb(\mathrm{pk})}}
\newcommand{\Acp}{\ifmath{\Acc(\mathrm{pk})}}
\newcommand{\Abl}{\ifmath{\Abb(-2)}}
\newcommand{\Acl}{\ifmath{\Acc(-2)}}
\newcommand{\gahatn}{g_{{A}{\nu}}}
\newcommand{\gvhatn}{g_{{V}{\nu}}}
\newcommand{\ghatn}{g_{\nu}}
\newcommand{\mco}{\multicolumn {1}{|c|}}

\newcommand{\mcc}[1]{\multicolumn{1}{c|}{#1}}

\newcommand{\tmcite}[1]{\multicolumn{1}{c|}{\mbox{\cite{#1}}}}

\newcommand{\apm}[2] {^{+#1}_{-#2}}
\newcommand{\gz}{\mbox{$g_1^{\mathrm{z}}$}}
\newcommand{\kg}{\mbox{$\kappa_\gamma$}}
\newcommand{\kz}{\mbox{$\kappa_{\mathrm{z}}$}}
\renewcommand{\lg}{\mbox{$\lambda_\gamma$}}
\newcommand{\lz}{\mbox{$\lambda_{\mathrm{z}}$}}
\newcommand{\dgz}{\mbox{$\Delta g_1^{\mathrm{z}}$}}

\newcommand{\dkg}{\mbox{$\Delta \kappa_\gamma$}}
\newcommand{\dkz}{\mbox{$\Delta \kappa_{\mathrm{z}}$}}

\newcommand{\twsq}{\mbox{$\tan^2\thw$}}

\newcommand{\LL}{-\Delta \log{\cal L}}

\newcommand{\avQfb}{{\langle Q_{\mathrm{FB}} \rangle}}
\newcommand{\Brbl}{\mbox{$\mathrm{BR(b\rightarrow \ell)}$}}

\newcommand{\fDp}{\mbox{$f(\Dplus)$}}
\newcommand{\fDs}{\mbox{$f(\Ds)$}}
\newcommand{\fLc}{\mbox{$f(\Lc)$}}
\newcommand{\fcb}{\mbox{$f({\mathrm{c_{baryon}}})$}}
\newcommand{\PcDst}{\ensuremath{\mathrm{P( c \rightarrow D^{*+})}}
      \ensuremath{\times}
      \ensuremath{\mathrm{BR( D^{*+} \rightarrow \pi^+ D^0 )}}}
\newcommand{\RcfDs}{\mbox{${\rm{R_c f_{D_s}}}$}}
\newcommand{\RcfLc}{\mbox{${\rm{R_c  f_{\Lambda_c}}}$}}

\newcommand{\RcfDp}{\mbox{${\rm{R_c f_{D^+}}}$}}
\newcommand{\RcfDz}{\mbox{${\rm{R_c f_{D^0}}}$}}
\newcommand{\RcPcDst}{\mbox{$\mathrm{R_c P( c \rightarrow D^{*+}) \times
      BR( D^{*+} \rightarrow \pi^+ D^0 )}$}}

\newcommand{\LepII}{\ifmath{\mathrm{\hbox{LEP-II}}}}
\def\WWZg    { {\mathrm{WWZ}\gamma} }
\def\WWgg    { {\mathrm{WW}\gamma \gamma} }
\def\nngg    { \nu \bar{\nu} \gamma \gamma }

\def\ZZgg    { {\mathrm{ZZ}\gamma \gamma} }

\def\WWZg    { {\mathrm{WWZ}\gamma} }

\def\Zgg     { {\mathrm{Z} \gamma \gamma} }
\newcommand{\gvz}{\mbox{$g_5^{\mathrm{Z}}$}}
\newcommand{\azl}{\mbox{$a_0/\Lambda^2$}}
\newcommand{\acl}{\mbox{$a_c/\Lambda^2$}}
\newcommand{\anl}{\mbox{$a_n/\Lambda^2$}}
\renewcommand{\lg}{\mbox{$\lambda_\gamma$}}

\newcommand{\CoM}        {centre-of-mass}
\newcommand{\epem}       {\mbox{$\mathrm{e^+e^-}$}}
\newcommand{\MC}         {Monte Carlo}
\newcommand{\RacoonWW}   {\mbox{R{\sc acoon}WW}}
\newcommand{\Gentle}     {\mbox{G{\sc entle}}}
\newcommand{\YFSWW}      {\mbox{YFSWW3}}
\newcommand{\Grace}      {\mbox{\tt grc4f}}
\newcommand{\YFSZZ}      {\mbox{YFSZZ}}
\newcommand{\ZZTO}       {\mbox{ZZTO}}
\newcommand{\WTO}        {\mbox{WTO}}
\newcommand{\WPHACT}     {\mbox{WPHACT}}
\newcommand{\SWAP}       {\mbox{SWAP}}
\newcommand{\phz}        {\phantom{0}}

\newcommand{\Wtolnu}     {\mbox{$\mathrm{W}\rightarrow\ell\overline{\nu}_{\ell}$}}
\newcommand{\Wtoenu}     {\mbox{$\mathrm{W}\rightarrow\mathrm{e\overline{\nu}_{e}}$}}
\newcommand{\Wtomnu}     {\mbox{$\mathrm{W}\rightarrow\mu\overline{\nu}_{\mu}$}}
\newcommand{\Wtotnu}     {\mbox{$\mathrm{W}\rightarrow\tau\overline{\nu}_{\tau}$}}

\newcommand{\Mw}         {\mbox{$m_{\mathrm{W}}$}}
\newcommand{\Gw}         {\mbox{$\Gamma_{\mathrm{W}}$}}

\newcommand{\qqqq}{\mbox{\qq\qq}}
\newcommand{\qqln}{\mbox{\qq\lnu}}
\newcommand{\lnu}{\mbox{$\ell\overline{\nu}_{\ell}$}}
\newcommand{\WW} {\mbox{$\mathrm{W^+W^-}$}}
\newcommand{\qq}{\mbox{$\mathrm{q\overline{q}}$}}
\newcommand{\WWqqqq}{\mbox{\WW$\rightarrow$\qq\qq}}

\newcommand{\WWqqln}{\mbox{\WW$\rightarrow$\qq\lnu}}
\newcommand{\qqlv}{\mbox{\qq\lnu}}

\newcommand{\WWlnln}     {\mbox{$\mathrm{W^+W^-}\rightarrow\ell^{-}\overline{\nu}_
                                 {\ell}\ell^{\prime +} \nu_{\ell^{\prime}}$}}

\newcommand {\dsdc}    {\frac{{\mathrm{d}}\sigma}{{\mathrm{d}}\cos\theta}}
\newcommand {\eemm}       {\ee \rightarrow \mumu}
\newcommand {\eett}       {\ee \rightarrow \tautau}
\newcommand {\Zprime} {\mathrm{Z^{\prime}}}
\newcommand {\thtzzp} {\mbox{$\Theta_{\mathrm{Z} \mathrm{Z}^{'}}$}}
\newcommand {\MZp}    {\mbox{$\mathrm{M}_{\mathrm{Z}^{'}}$}}
\newcommand {\MZplim} {\mbox{$\mathrm{M}_{\mathrm{Z}^{'}}^{limit}$}}
\newcommand {\jbctot} {\mbox{$j_{\mathrm{b,c}}^{\mathrm{tot}}$}}
\newcommand {\jbcfb}  {\mbox{$j_{\mathrm{b,c}}^{\mathrm{fb}}$}}
\newcommand {\qqb}        {\mathrm{q\bar{q}}}
\newcommand {\Rq}    {\mbox{$\mathrm{R_{q}}$}}
\newcommand {\porm}   {\mbox{$\pm$}}
\newcommand {\rhadtot}{\mbox{$r_{\mathrm{had}}^{\mathrm{tot}}$}}
\newcommand {\jhadtot}{\mbox{$j_{\mathrm{had}}^{\mathrm{tot}}$}}
\newcommand {\rbctot} {\mbox{$r_{\mathrm{b,c}}^{\mathrm{tot}}$}}
\newcommand {\rbcfb}  {\mbox{$r_{\mathrm{b,c}}^{\mathrm{fb}}$}}
\newcommand {\rltot}  {\mbox{$r_{\mathrm{l}}^{\mathrm{tot}}$}}
\newcommand {\jltot}  {\mbox{$j_{\mathrm{l}}^{\mathrm{tot}}$}}
\newcommand {\rlfb}   {\mbox{$r_{\mathrm{l}}^{\mathrm{fb}}$}}
\newcommand {\jlfb}   {\mbox{$j_{\mathrm{l}}^{\mathrm{fb}}$}}

\input{rotate}
\begin{document}
\flushbottom
\begin{titlepage}
\begin{center}
\Large {EUROPEAN ORGANIZATION FOR NUCLEAR RESEARCH\\}
\end{center}
\vspace*{0.2cm}
\begin{flushright}
       CERN-EP/2001-021\\
       February 28, 2001 \\
\end{flushright}

\vspace*{1cm}
\begin{center}
\boldmath
\Huge {\bf A Combination of Preliminary\\
               Electroweak Measurements and \\
            Constraints on the Standard Model\\[.5cm]

}
\unboldmath



\vspace*{1.0cm}
\Large {\bf
The LEP Collaborations\footnote{The LEP Collaborations each take
responsibility for the preliminary results of their own.}
 ALEPH, DELPHI, L3, OPAL,\\
    the LEP Electroweak Working Group\footnote{%
The members of the 
LEP Electroweak Working Group 
who contributed significantly to this
note are :
D.~Abbaneo,         
J.~Alcaraz,         
P.~Antilogus,       
S.~Arcelli,         
P.~Bambade,         
E.~Barberio,        
G.~Bella,           
A.~Blondel,         
S.~Blyth,           
D.~Bourilkov,       
R.~Chierici,        
R.~Clare,           
P.~de Jong,         
G.~Duckeck,         
A.~Ealet,           
M.~Elsing,          
F.~Fiedler,         
P.~Garcia-Abia,     
A.~Gurtu,           
M.W.~Gr\"unewald,   
J.B.~Hansen,        
R.~Hawkings,        
J.~Holt,            
S.~Jezequel,        
R.W.L.~Jones,       
N.~Kjaer,           
M.~Kobel,           
E.~Lan{\c c}on,     
W.~Lohmann,         
C.~Mariotti,        
M.~Martinez,        
F.~Matorras,        
C.~Matteuzzi,       
S.~Mele,            
E.~Migliore,        
M.N.~Minard,        
K.~M\"onig,         
A.~Olshevski,       
C.~Parkes,          
U.~Parzefall,       
Ch.~Paus,           
M.~Pepe-Altarelli,  
B.~Pietrzyk,        
G.~Quast,           
P.~Renton,          
H.~Rick,            
S.~Riemann,         
J.M.~Roney,         
K.~Sachs,           
C.~Sbarra,          
A.~Schmidt-Kaerst,  
S.~Spagnolo,        
A.~Straessner,      
R.~Str\"ohmer       
D.~Strom,           
R.~Tenchini,        
F.~Terranova,       
F.~Teubert,         
M.A.~Thomson,       
E.~Tournefier,      
M.~Verzocchi,       
H.~Voss,            
C.P.~Ward,          
St.~Wynhoff.         
}\\
and the SLD Heavy Flavour and Electroweak Groups\footnote{%
T.~Abe,             
N.~de Groot,        
M.~Iwasaki,         
P.C.~Rowson,        
D.~Su,              
M.~Swartz.          
}\\
}
\vskip 1cm
\large\textbf{Prepared from Contributions of the LEP and SLD
  experiments \\
to the 2000 Summer conferences.}\\
\end{center}
\vfill
\begin{abstract}
  This note presents a combination of published and preliminary
  electroweak results from the four LEP collaborations and the SLD
  collaboration which were prepared for the 2000 summer conferences.
  Averages from  $\Zzero$ resonance results
  are derived for hadronic and leptonic cross sections, the
  leptonic forward-backward asymmetries, the $\tau$ polarisation
  asymmetries, the $\bb$ and $\cc$ partial widths and forward-backward
  asymmetries and the $\qq$ charge asymmetry.  
  Above the  $\Zzero$ resonance, averages are derived for
  di--fermion cross sections
  and asymmetries, 
  W--pair, Z--pair and single--W production cross section,
  electroweak gauge boson couplings
  and W mass and decay branching ratios.
  The major changes with
  respect to results presented in summer 1999 are 
  final $\Zzero$  lineshape results from LEP, updates to the
  W mass and gauge-boson couplings from LEP, 
  and $\ALR$ from SLD.
  The
  results are compared with precise electroweak measurements from
  other experiments.  
  The parameters of the Standard Model are
  evaluated, first using the combined LEP electroweak measurements,
  and then using the full set of electroweak results.
\end{abstract}
\end{titlepage}
\setcounter{page}{2}
\renewcommand{\thefootnote}{\arabic{footnote}}
\setcounter{footnote}{0}

\section{Introduction}
\label{sec-Intro}

The four LEP experiments and SLD have previously presented
\cite{bib-EWEP-99} parameters derived from the $\Zzero$ resonance
using published and preliminary results based on data recorded until
the end of 1995 for the LEP experiments and 1998 for SLD.  
Since 1996 LEP has run at energies 
above the W-pair production threshold.  In 1999 the delivered
luminosity was significantly higher than in previous years, and thus
the knowledge of the properties of the W boson
has been significantly improved.
To allow a quick assessment, a box highlighting the updates is given
at the beginning of each section.  

$\LEPI$  (1990-1995) $\Zzero$-pole measurements 
are the hadronic and leptonic
cross sections, the leptonic forward-backward asymmetries, the $\tau$
polarisation asymmetries, the $\bb$ and $\cc$ partial widths and
forward-backward asymmetries and the $\qq$ charge asymmetry.  The
measurements of 
the left-right cross section asymmetry, 
the $\bb$ and $\cc$ partial widths and
left-right-forward-backward asymmetries for b and c quarks from SLD
are treated consistently with the LEP data. Many technical aspects of
their combination are described in
References~\citen{LEPLS}, \citen{ref:lephf} and references
therein. 

The $\LEPII$ (1996-2000) measurements are di--fermion cross sections
and asymmetries; W--pair, Z--pair and single--W production cross sections, 
electroweak gauge boson couplings. W boson properties, like mass, width
and decay branching ratios are also measured.

Several measurements included in the
combinations are still preliminary.

This note is organised as follows:
\begin{description}
\item [Section~\ref{sec-LS}] $\Zzero$ Line Shape and Leptonic
  Forward-Backward Asymmetries;
\item [Section~\ref{sec-TP}] $\tau$ Polarisation;
\item [Section~\ref{sec-ALR}] $\ALR$ Measurement at SLD;
\item [Section~\ref{sec-HF}] Heavy Flavour Analyses;
\item [Section~\ref{sec-QFB}] Inclusive Hadronic Charge Asymmetry;
\item [Section~\ref{sec-FF}] $\ff$ Production at Energies above the Z;
\item [Section~\ref{sec-MW}] W Boson Properties, including $\MW$,
  Branching Ratios, W--pair Production Cross Section;
\item [Section~\ref{sec-SGW}] Single--W Production Cross Section;
\item [Section~\ref{sec-ZZ}] ZZ Production Cross Section;
\item [Section~\ref{sec-GC}] Electroweak Gauge Boson Couplings;
\item [Section~\ref{sec-IofR}] Interpretation of the Results,
  Including the Combination of Results from LEP, SLD, Neutrino
  Interaction Experiments and from CDF and D\O;
\item [Section~\ref{sec-Future}] Prospects for the Future.
\end{description}

\clearpage

\boldmath
\section{Results from the $\Zzero$ Peak Data}
\label{sec-LS}
\unboldmath

\updates{
All experiments have updated their results, and all are now final.
Recent theoretical developments have been included in the results 
and fits.
}

\boldmath
\subsection{$\Zzero$ Lineshape and Lepton Forward-Backward Asymmetries}\label{sec-LS-SM}
\unboldmath
\label{sec-LSpole}

The results presented here are based on the full \LEPI{} data set.
This includes the data taken during the
energy scans in 1990 and 1991 in the range\footnote{In this note
  $\hbar=c=1$.}  $|\roots-\MZ|<3$~\GeV{}, the data collected at the
$\Zzero$ peak in 1992 and 1994 and the precise energy scans in
1993 and 1995 ($|\roots-\MZ|<1.8$~\GeV{}).  
The total event statistics are given in Table~\ref{tab-LSstat}.
Details of the individual analyses can be found in 
References~\citen{ALEPHLS,DELPHILS,L3LS,OPALLS}. 

\begin{table}[hbtp]
\begin{center}\begin{tabular}{lr} 
\begin{minipage}[b]{0.49\textwidth}
\begin{center}\begin{tabular}{r||rrrr||r}
  \multicolumn{6}{c}{$\qq$}  \\
\hline
   year & A &   D  &   L  &  O  & all \\
\hline
'90/91  & 433 &  357 &  416 &  454 &  1660\\
'92     & 633 &  697 &  678 &  733 &  2741\\
'93     & 630 &  682 &  646 &  649 &  2607\\
'94     &1640 & 1310 & 1359 & 1601 &  5910\\
'95     & 735 &  659 &  526 &  659 &  2579\\
\hline
 total  & 4071 & 3705 & 3625 & 4096 & 15497\\
\end{tabular}\end{center}
\end{minipage}
   &
\begin{minipage}[b]{0.49\textwidth}
\begin{center}\begin{tabular}{r||rrrr||r}
  \multicolumn{6}{c}{$\lept$} \\
\hline
   year & A &   D  &   L  &  O  & all \\
\hline
'90/91  &  53 &  36 &  39  &  58  &  186 \\
'92     &  77 &  70 &  59  &  88  &  294 \\
'93     &  78 &  75 &  64  &  79  &  296 \\
'94     & 202 & 137 & 127  & 191  &  657 \\
'95     &  90 &  66 &  54  &  81  &  291 \\
\hline
total   & 500 & 384 & 343  & 497  & 1724 \\
\end{tabular}\end{center}
\end{minipage} \\
\end{tabular} \end{center}
\caption[Recorded event statistics]{
The $\qq$ and $\lept$ event statistics, in units of $10^3$, used
for the analysis of the $\Zzero$ line shape and lepton forward-backward
asymmetries by the experiments ALEPH (A), DELPHI (D), L3
(L) and OPAL (O).
}
\label{tab-LSstat}
\end{table}

For the averaging of results the LEP experiments provide a standard
set of 9 parameters describing the information contained in hadronic
and leptonic cross sections and leptonic forward-backward
asymmetries.  These parameters are
convenient for fitting and averaging since they have small
correlations. They are:
\begin{itemize}
\item The mass and total width of the Z boson, where the definition is
  based on the Breit-Wigner denominator $(s-\MZ^2+is\GZ/\MZ)$
  with $s$-dependent width~\cite{ref:QEDCONV}.
\item The hadronic pole cross section of Z exchange:
\begin{equation}
\shad\equiv{12\pi\over\MZ^2}{\Gee\Ghad\over\GZ^2}\,.
\end{equation}
Here $\Gee$ and $\Ghad$ are the partial widths of the $\Zzero$ for
decays into electrons and hadrons.

\item The ratios:
\begin{equation}\label{eqn-sighad}
 \Ree\equiv\Ghad/\Gee, \;\; \Rmu\equiv\Ghad/\Gmumu \;\mbox{and}\;
\Rtau\equiv\Ghad/\Gtautau.
\end{equation}
Here $\Gmumu$ and $\Gtautau$ are the partial widths of the $\Zzero$
for the decays $\Ztomumu$ and $\Ztotautau$.  Due to the mass of
the $\tau$ lepton, a difference of 0.2\% is expected between the
values for $\Ree$ and $\Rmu$, and the value for $\Rtau$, even under
the assumption of lepton universality~\cite{ref:consoli}.
\item The pole asymmetries, $\Afbze$, $\Afbzm$ and $\Afbzt$, for the
  processes $\eeee$, $\eemumu$ and $\eetautau$. In terms of the real
  parts of the effective vector and axial-vector neutral current
  couplings of fermions, $\gvhatf$ and $\gahatf$, the pole asymmetries
  are expressed as
\begin{equation}
\label{eqn-apol}
\Afbzf \equiv {3\over 4} \cAe\cAf
\end{equation}
with
\begin{equation}
\label{eqn-cAf}
\cAf\equiv\frac{2\gvhatf \gahatf} {\gvhatf^{2}+\gahatf^{2}}\ = 2 \frac{\gvhatf/\gahatf} {1+(\gvhatf/\gahatf)^{2}}\,.
\end{equation}
\end{itemize}
The imaginary parts of the vector and axial-vector coupling constants
as well as real and imaginary parts of the photon vacuum polarisation
are taken into account explicitly in the fitting formulae and are fixed to
their Standard Model values.
The fitting procedure takes into account the effects of initial-state
radiation~\cite{ref:QEDCONV} to 
${\cal O}(\alpha^3)$~\cite{ref:Jadach91,ref:Skrzypek92,ref:Montagna96}, as
well as the $t$-channel and  the $s$-$t$ interference contributions in the case 
of $\ee$ final states.

The set of 9 parameters does not describe hadron and
lepton-pair production completely, because it does not include the
interference of the $s$-channel $\Zzero$ exchange with the $s$-channel
$\gamma$ exchange.  For the results presented in this section and used
in the rest of the note, the $\gamma$-exchange contributions and the
hadronic $\gamma\Zzero$ interference terms are fixed to their Standard
Model values.  The leptonic $\gamma\Zzero$ interference terms are
expressed in terms of the effective couplings.

\begin{table}[tp] \begin{center}{\small
\begin {tabular} {lr|@{\,}r@{\,}r@{\,}r@{\,}r@{\,}r@{\,}r@{\,}r@{\,}r@{\,}r}
\hline 
\multicolumn{2}{c|}{~}& \multicolumn{9}{c}{correlations} \\
\multicolumn{2}{c|}{~} & $\MZ$ & $\GZ$ & $\shad$ &
     $\Ree$ &$\Rmu$ & $\Rtau$ & $\Afbze$ & $\Afbzm$ & $\Afbzt$ \\
\hline 
\multicolumn{2}{l}{ $\pzz \chi^2/N_{\rm df}\,=\, 169/176$}& 
                                      \multicolumn{9}{c}{ALEPH} \\
\hline 
 $\MZ$\,[\GeV{}]\hspace*{-.5pc} & 91.1891 $\pm$ 0.0031     &   
  1.00 \\
 $\GZ$\,[\GeV]\hspace*{-2pc}  &  2.4959 $\pm$ 0.0043     & 
  .038 & ~1.00 \\ 
 $\shad$\,[nb]\hspace*{-2pc}  &  41.558 $\pm$ 0.057$\pz$ &   
 $-$.091 & $-$.383 & ~1.00 \\
 $\Ree$        &  20.690 $\pm$ 0.075$\pz$ &   
  .102  &~.004 & ~.134 & ~1.00 \\
 $\Rmu$        &  20.801 $\pm$ 0.056$\pz$ &   
 $-$.003 & ~.012 & ~.167 & ~.083 & ~1.00 \\ 
 $\Rtau$       &  20.708 $\pm$ 0.062$\pz$ &   
 $-$.003 & ~.004 & ~.152 & ~.067 & ~.093 & ~1.00 \\ 
 $\Afbze$      &  0.0184 $\pm$ 0.0034     &   
 $-$.047 & ~.000 & $-$.003 & $-$.388 & ~.000 & ~.000 & ~1.00 \\ 
 $\Afbzm$      &  0.0172 $\pm$ 0.0024     &   
 .072 & ~.002 & ~.002 & ~.019 & ~.013 & ~.000 & $-$.008 & ~1.00 \\ 
 $\Afbzt$      &  0.0170 $\pm$ 0.0028     &   
 .061 & ~.002 & ~.002 & ~.017 & ~.000 & ~.011 & $-$.007 & ~.016 & ~1.00 \\
    ~          & \multicolumn{2}{c}{~}                \\[-0.5pc]
\hline 
\multicolumn{2}{l}{$\pzz \chi^2/N_{\rm df}\,=\, 177/168$} & 
                                        \multicolumn{9}{c}{DELPHI} \\
\hline 
 $\MZ$\,[\GeV{}]\hspace*{-.5pc}   &  91.1864 $\pm$ 0.0028    &
 ~1.00 \\ 
 $\GZ$\,[\GeV]\hspace*{-2pc}   &  2.4876 $\pm$ 0.0041     &
 ~.047 & ~1.00 \\ 
 $\shad$\,[nb]\hspace*{-2pc}   &  41.578 $\pm$ 0.069$\pz$ &
 $-$.070 & $-$.270 & ~1.00 \\ 
 $\Ree$        &  20.88  $\pm$ 0.12$\pzz$ &
 ~.063 & ~.000 & ~.120 & ~1.00 \\ 
 $\Rmu$        &  20.650 $\pm$ 0.076$\pz$ &
 $-$.003 & $-$.007 & ~.191 & ~.054 & ~1.00 \\ 
 $\Rtau$       &  20.84  $\pm$ 0.13$\pzz$ &
 ~.001 & $-$.001 & ~.113 & ~.033 & ~.051 & ~1.00 \\ 
 $\Afbze$      &  0.0171 $\pm$ 0.0049     &
 ~.057 & ~.001 & $-$.006 & $-$.106 & ~.000 & $-$.001 & ~1.00 \\ 
 $\Afbzm$      &  0.0165 $\pm$ 0.0025    &
 ~.064 & ~.006 & $-$.002 & ~.025 & ~.008 & ~.000 & $-$.016 & ~1.00 \\ 
 $\Afbzt$      &  0.0241 $\pm$ 0.0037     & 
 ~.043 & ~.003 & $-$.002 & ~.015 & ~.000 & ~.012 & $-$.015 & ~.014 & ~1.00 \\
    ~          & \multicolumn{2}{c}{~}                \\[-0.5pc]
\hline 
\multicolumn{2}{l}{$\pzz \chi^2/N_{\rm df}\,=\, 158/166 $}  & 
                                    \multicolumn{9}{c}{L3} \\
\hline %
 $\MZ$\,[\GeV{}]\hspace*{-.5pc}   &  91.1897 $\pm$ 0.0030      & 
 ~1.00 \\ 
 $\GZ$\,[\GeV]\hspace*{-2pc}   &   2.5025 $\pm$ 0.0041      & 
 ~.065 & ~1.00 \\ 
 $\shad$\,[nb]\hspace*{-2pc}   &   41.535 $\pm$ 0.054$\pz$  & 
 ~.009 & $-$.343 & ~1.00 \\ 
 $\Ree$        &   20.815  $\pm$ 0.089$\pz$ & 
 ~.108 & $-$.007 & ~.075 & ~1.00 \\ 
 $\Rmu$        &   20.861  $\pm$ 0.097$\pz$ & 
 $-$.001 & ~.002 & ~.077 & ~.030 & ~1.00 \\ 
 $\Rtau$       &   20.79 $\pz\pm$ 0.13$\pzz$& 
 ~.002 & ~.005 & ~.053 & ~.024 & ~.020 & ~1.00 \\ 
 $\Afbze$      &   0.0107 $\pm$ 0.0058      & 
 $-$.045 & ~.055 & $-$.006 & $-$.146 & $-$.001 & $-$.003 & ~1.00 \\ 
 $\Afbzm$      &   0.0188 $\pm$ 0.0033      & 
 ~.052 & ~.004 & ~.005 & ~.017 & ~.005 & ~.000 & ~.011 & ~1.00 \\ 
 $\Afbzt$      &   0.0260 $\pm$ 0.0047      & 
 ~.034 & ~.004 & ~.003 & ~.012 & ~.000 & ~.007 & $-$.008 & ~.006 & ~1.00 \\
    ~          & \multicolumn{2}{c}{~}                \\[-0.5pc]
\hline 
\multicolumn{2}{l}{$\pzz \chi^2/N_{\rm df}\,=\, 155/194 $} & 
                                      \multicolumn{9}{c}{OPAL} \\
\hline %
 $\MZ$\,[\GeV{}]\hspace*{-.5pc}   & 91.1858 $\pm$ 0.0030 &
 ~1.00 \\ 
 $\GZ$\,[\GeV]\hspace*{-2pc}   & 2.4948  $\pm$ 0.0041 & 
 ~.049 & ~1.00 \\ 
 $\shad$\,[nb]\hspace*{-2pc}   & 41.501 $\pm$ 0.055$\pz$ & 
 ~.031 & $-$.352 & ~1.00 \\ 
 $\Ree$        & 20.901 $\pm$ 0.084$\pz$ &
 ~.108 & ~.011 & ~.155 & ~1.00 \\ 
 $\Rmu$        & 20.811 $\pm$ 0.058$\pz$ &
 ~.001 & ~.020 & ~.222 & ~.093 & ~1.00 \\ 
 $\Rtau$       & 20.832 $\pm$ 0.091$\pz$ &
 ~.001 & ~.013 & ~.137 & ~.039 & ~.051 & ~1.00 \\ 
 $\Afbze$      & 0.0089 $\pm$ 0.0045 &
 $-$.053 & $-$.005 & ~.011 & $-$.222 & $-$.001 & ~.005 & ~1.00 \\ 
 $\Afbzm$     & 0.0159 $\pm$ 0.0023 &
 ~.077 & $-$.002 & ~.011 & ~.031 & ~.018 & ~.004 & $-$.012 & ~1.00 \\ 
 $\Afbzt$      & 0.0145 $\pm$ 0.0030 & 
 ~.059 & $-$.003 & ~.003 & ~.015 & $-$.010 & ~.007 & $-$.010 & ~.013 & ~1.00 \\
\hline 
\end{tabular}}
\caption[Nine parameter results]{\label{tab-ninepar}
Line Shape and asymmetry parameters from fits to the data of the four
LEP experiments and their correlation coefficients. }
\end{center}
\end{table} 

The four sets of 9 parameters provided by the LEP experiments are
presented in Table~\ref{tab-ninepar}. 
For performing the average over these four sets of nine parameters, 
the overall covariance matrix is constructed from the covariance
matrices of the individual LEP experiments and common systematic
errors~\cite{LEPLS}.   The common systematic errors include
theoretical errors as well as errors arising from the uncertainty in
the LEP beam energy.  
The beam energy uncertainty contributes an uncertainty of
$\pm1.7~\MeV$ to $\MZ$ and $\pm1.2~\MeV$ 
to $\GZ$. In addition, the uncertainty in the \CoM\ energy spread of
about $\pm1~\MeV$ contributes $\pm0.2~\MeV$ to $\GZ$.
The theoretical error on calculations
of the small-angle Bhabha cross section is $\pm$0.054\,\%\cite{bib-lumthopal} 
for OPAL and $\pm$0.061\,\%\cite{bib-lumth99} for all other experiments, 
and results in the largest common systematic uncertainty on $\shad$.
QED radiation, dominated by photon radiation from the initial state 
electrons, contributes a common uncertainty of $\pm$0.02\,\% 
on $\shad$, of $\pm0.3$~\MeV{} on $\MZ$ and of $\pm0.2$~\MeV{} on $\GZ$.
The contribution of $t$-channel diagrams
and the $s$-$t$ interference in $\Zzero\ra\ee$ leads to an additional 
theoretical
uncertainty estimated to be $\pm0.24$ on $\Ree$ and $\pm0.0014$
on $\Afbze$, which are fully anti--correlated.
Uncertainties from the model-independent parameterisation
of the energy dependence of the cross section are almost negligible,
if the definitions of Reference\,\cite{bib-PCP99} are applied. Through
unavoidable Standard Model remnants, dominated by the need to
fix the $\gamma$-$\Zzero$ interference contribution in the $\qq$
channel, there is some small dependence of $\pm 0.2$ \MeV{} of $\MZ$
on the Higgs mass, $\MH$ (in the range 100 \GeV{} to 1000 \GeV{}) and
the value of the electromagnetic 
coupling constant. Such ``parametric'' errors are negligible for
the other pseudo-observables. The combined parameter set and its 
correlation matrix are given in Table~\ref{tab-zparavg}.

\begin{figure}[p]
\vspace*{-0.6cm}
\begin{center}
  \mbox{\includegraphics[width=0.9\linewidth]{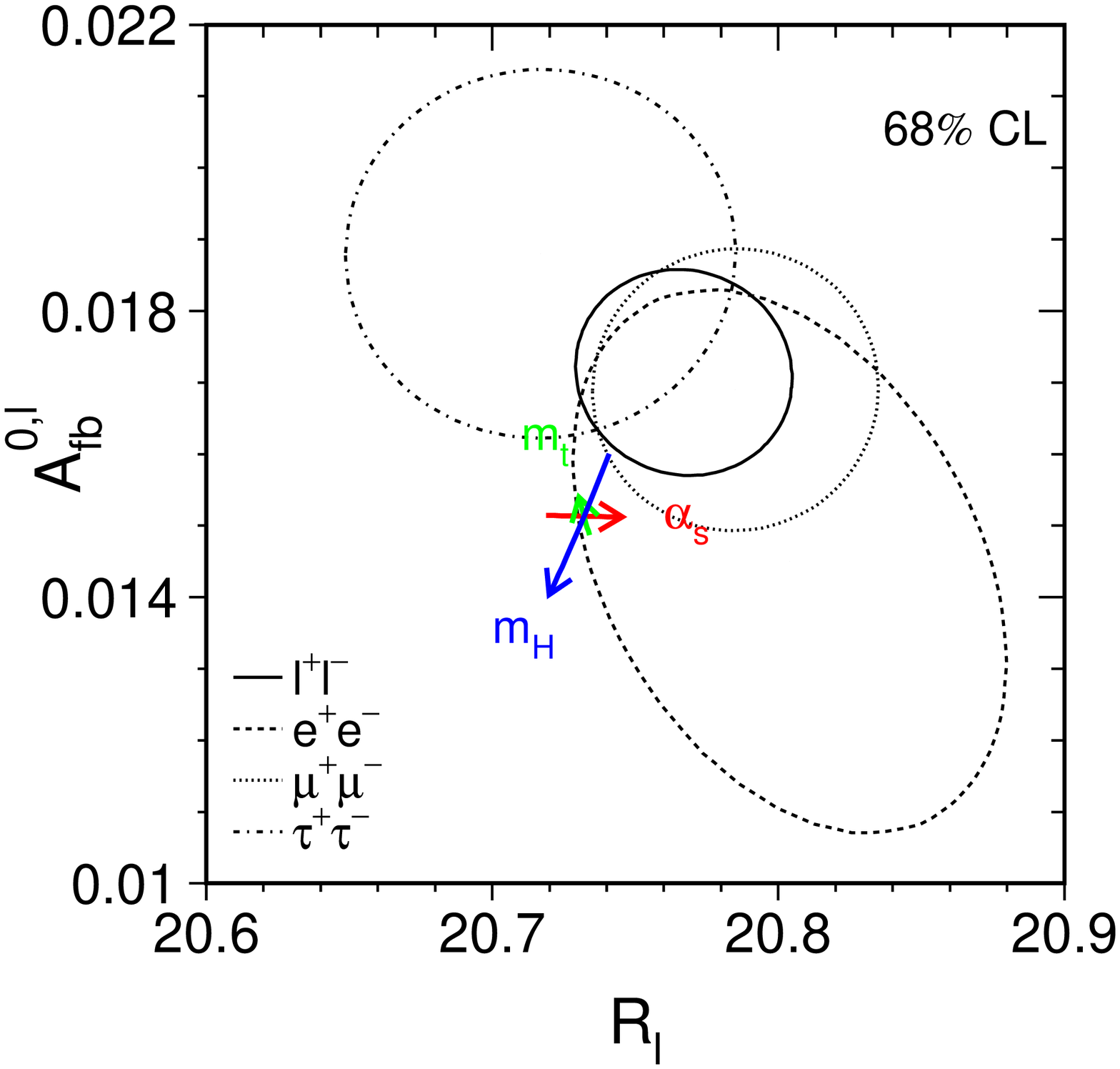}}
\end{center}
\caption[]{
  Contours of 68\% probability in the $\RZ$-$\Afbpol$ plane. For
  better comparison the results for the $\tau$ lepton are corrected to
  correspond to the massless case.  The $\SM$ prediction for
  $\MZ=91.1875$~\GeV{}, $\Mt=174.3$~\GeV{}, $\MH=300$~\GeV{}, and
  $\alfmz=0.119$ is also shown. The lines with arrows correspond to
  the variation of the $\SM$ prediction when $\Mt$, $\MH$ and $\alfmz$
  are varied in the intervals $\Mt=174.3\pm5.1$~\GeV{},
  $\MH=300^{+700}_{-187}~\GeV$, and $\alfmz=0.119\pm0.002$,
  respectively. The arrows point in the direction of increasing values
  of $\Mt$, $\MH$ and $\alfas$.  }
\label{fig-LU} 
\end{figure}

If lepton universality is assumed, the set of 9 parameters  
is reduced to a set of 5 parameters. $\RZ$ is defined as 
$\RZ\equiv\Ghad/\Gll$, where $\Gll$ refers to the partial 
$\Zzero$ width for the decay into a pair of massless charged 
leptons.  The data of each of the four LEP experiments are 
consistent with lepton universality (the difference in $\chi^2$ 
over the difference in d.o.f.{} with and without the assumption 
of lepton universality is 3/4, 6/4, 5/4 and 3/4 for ALEPH, DELPHI, 
L3 and OPAL, respectively). The lower part of Table~\ref{tab-zparavg} 
gives the combined result and the corresponding correlation matrix.  
Figure~\ref{fig-LU} shows, for each lepton species and for the 
combination assuming lepton universality, the resulting 68\% 
probability contours in the $\RZ$-$\Afbpol$ plane. Good agreement
is observed. 

For completeness the partial decay widths of the $\Zzero$ boson 
are listed in Table~\ref{tab-widths}, although 
they are more correlated than the ratios given in 
Table~\ref{tab-zparavg}. The leptonic pole cross-section, 
$\sll$, defined as
\[ \sll\equiv{12\pi\over\MZ^2}{\Gll^2\over\GZ^2} \, , \]
in analogy to $\shad$, is shown in the last line of the Table.
Because QCD final state corrections appear quadratically in the 
denominator via $\GZ$, $\sll$ has a higher sensitivity to $\alpha_s$ 
than $\shad$ or $\Rl$, where the dependence on QCD corrections is 
only linear.

\begin{table}[htb]\begin{center}
\begin {tabular} {lr|r@{\,}r@{\,}r@{\,}r@{\,}r@{\,}r@{\,}r@{\,}r@{\,}r}
\hline 
\multicolumn{2}{c|} {without lepton universality} & 
                                    \multicolumn{9}{l}{~~~correlations} \\
\hline 
\multicolumn{2}{c|}{$\pzz \chi^2/N_{\rm df}\,=\,32.6/27 $} &
   $\MZ$ & $\GZ$ & $\shad$ &
     $\Ree$ &$\Rmu$ & $\Rtau$ & $\Afbze$ & $\Afbzm$ & $\Afbzt$ \\
\hline 
 $\MZ$ [\GeV{}]  & 91.1876$\pm$ 0.0021 &
 ~1.00 \\
 $\GZ$ [\GeV]  & 2.4952 $\pm$ 0.0023 &
 $-$.024 & ~1.00 \\ 
 $\shad$ [nb]  & 41.541 $\pm$ 0.037$\pz$ &
 $-$.044 & $-$.297 & ~1.00 \\ 
 $\Ree$        & 20.804 $\pm$ 0.050$\pz$ &
 ~.078 & $-$.011 & ~.105 & ~1.00 \\ 
 $\Rmu$        & 20.785 $\pm$ 0.033$\pz$ & 
 ~.000 & ~.008 & ~.131 & ~.069 & ~1.00 \\ 
 $\Rtau$       & 20.764 $\pm$ 0.045$\pz$ &  
 ~.002 & ~.006 & ~.092 & ~.046 & ~.069 & ~1.00 \\ 
 $\Afbze$      & 0.0145 $\pm$ 0.0025 &
 $-$.014 & ~.007 & ~.001 & $-$.371 & ~.001 & ~.003 & ~1.00 \\ 
 $\Afbzm$      & 0.0169 $\pm$ 0.0013 &
 ~.046 & ~.002 & ~.003 & ~.020 & ~.012 & ~.001 & $-$.024 & ~1.00 \\ 
 $\Afbzt$      & 0.0188 $\pm$ 0.0017 &
 ~.035 & ~.001 & ~.002 & ~.013 & $-$.003 & ~.009 & $-$.020 & ~.046 & ~1.00 \\ 
\multicolumn{3}{c}{~}\\[-0.5pc]
\multicolumn{2}{c} {with lepton universality} \\
\hline 
\multicolumn{2}{c|}{$\pzz \chi^2/N_{\rm df}\,=\,36.5/31 $}  & 
   $\MZ$ & $\GZ$ & $\shad$ & $\Rl$ &$\Afbpol$ \\
\hline 
 $\MZ$ [\GeV{}]  & 91.1875$\pm$ 0.0021$\pz$    &
 ~1.00 \\ 
 $\GZ$ [\GeV]  & 2.4952 $\pm$ 0.0023$\pz$    &
 $-$.023  & ~1.00 \\ 
 $\shad$ [nb]  & 41.540 $\pm$ 0.037$\pzz$ &
 $-$.045 & $-$.297 &  ~1.00 \\ 
 $\Rl$         & 20.767 $\pm$ 0.025$\pzz$ &
 ~.033 & ~.004 & ~.183 & ~1.00 \\ 
 $\Afbpol$     & 0.0171 $\pm$ 0.0010   & 
 ~.055 & ~.003 & ~.006 & $-$.056 &  ~1.00 \\ 
\hline 
\end{tabular} 
\caption[]{
  Average line shape and asymmetry parameters from the data of the
  four LEP experiments,  without and with the
  assumption of lepton universality. }
\label{tab-zparavg}
\end{center}
\end{table}

\begin{table}[hbtp] \begin{center} 
\begin{tabular} {lr|r@{\,}r@{\,}r@{\,}r}
\hline 
 \multicolumn{2}{c|}{without lepton universality} &
                            \multicolumn{4}{l}{~~~correlations} \\
\hline 
$\Ghad$ [\MeV]     & 1745.8$\pz\pm$2.7$\pz\pzz$ & ~1.00 \\
$\Gee$ [\MeV]      & 83.92$\pm$0.12$\pzz$& $-$0.29 & ~1.00 \\ 
$\Gmumu$ [\MeV]    & 83.99$\pm$0.18$\pzz$& ~0.66 & $-$0.20 & ~1.00 \\
$\Gtautau$ [\MeV]  & 84.08$\pm$0.22$\pzz$&  0.54 & $-$0.17 & ~0.39 & ~1.00 \\  
\hline 
    \multicolumn{6}{c}{~} \\[-0.5pc]
    \multicolumn{2}{c}{with lepton universality} \\
\hline 
$\Ginv$ [\MeV]     & $\pz$499.0$\pzz\pm$1.5$\pz\pzz$    & ~1.00 \\
$\Ghad$ [\MeV]     & 1744.4$\pzz\pm$2.0$\pz\pzz$   & $-$0.29 & ~1.00 \\
$\Gll$ [\MeV]       & 83.984$\pm$0.086$\pz$  & ~0.49 & ~0.39 & ~1.00 \\
\hline 
$\Ginv/\Gll$        & {$\pz$5.942\pz$\pm$0.016$\pz$} &    \\
\hline 
$\sll$ [nb]      & {2.0003$\pm$0.0027} &  \\
\hline 
\end{tabular} 
\caption[]{
  Partial decay widths of the $\Zzero$ boson, derived from the results of the
  9-parameter averages in Table~\ref{tab-zparavg}. In the
  case of lepton universality, $\Gll$ refers to the partial $\Zzero$ width for
  the decay into a pair of massless charged leptons.  }
\label{tab-widths}
\end{center}
\end{table}

\clearpage
\boldmath
\section{The $\tau$ Polarisation}
\label{sec-TP}
\unboldmath

\updates{%
DELPHI have finalised their results.

OPAL have updated their results.
}

\noindent
The longitudinal $\tau$ polarisation $\cal {P}_{\tau}$ of $\tau$
pairs produced 
in $\Zzero$ decays is defined as
\begin{eqnarray}
{\cal P}_{\tau} & \equiv & 
\frac{\sigma_{\mathrm{R}} - \sigma_{\mathrm{L}}}
{\sigma_{\mathrm{R}} + \sigma_{\mathrm{L}}} \, ,
\end{eqnarray}
where $\sigma_{\mathrm{R}}$ and $\sigma_{\mathrm{L}}$ are the $\tau$-pair cross sections for
the production of a right-handed and left-handed $\tau^-$,
respectively. The distribution of $\ptau$ as a function of the polar
scattering angle $\theta$ between the $\mathrm{e}^-$ and the $\tau^-$,
at $\roots = \MZ$, is given by
\begin{eqnarray}
\label{eqn-taupol}
{\cal P}_{\tau}(\cos\theta) & = &
 - \frac{\cAt(1+\cos^2\theta) + 2    \cAe\cos\theta}
             {1+\cos^2\theta  + 2\cAt\cAe\cos\theta} \, ,
\end{eqnarray}
with $\cAe$ and $\cAt$ as defined in Equation~(\ref{eqn-cAf}).
Equation~(\ref{eqn-taupol}) is valid for pure Z exchange.  The effects of
$\gamma$ exchange, $\gamma$-$\Zzero$ interference and electromagnetic
radiative corrections in the initial and final states
are taken into account in the experimental analyses.  In
particular, these corrections account for the $\sqrt{s}$ dependence of
the $\tau$ polarisation, which is important because
the off-peak data are included in the event samples for all
experiments.
When averaged over all production angles $\cal {P}_{\tau}$ is a
measurement of $\cAt$. As a function of $\cos\theta$, $\cal
{P}_{\tau}(\cos\theta)$ provides nearly independent determinations of
both $\cAt$ and $\cAe$, thus allowing a test of the universality of
the couplings of the $\Zzero$ to $\mathrm{e}$ and $\tau$.

Each experiment makes separate $\ptau$ measurements using the five
$\tau$ decay modes e$\nu \overline{\nu}$, $\mu\nu \overline{\nu}$,
$\pi\nu$, $\rho\nu$ and
$a_{1}\nu$\cite{bib-ALEPHTAU,bib-DELPHITAUnew,bib-L3TAUfin,bib-OPALTAU}.
The $\rho\nu$ and $\pi\nu$ are the most sensitive channels,
contributing weights of about $40\%$ each in the average.  DELPHI and L3 have
also used an inclusive hadronic analysis. The combination is made
using the results from each experiment already averaged over the
$\tau$ decay modes.


\subsection{Results}

Tables~\ref{tab-tau1} and~\ref{tab-tau2} show the most recent results
for $\cAt$ and $\cAe$ obtained by the four LEP
collaborations\cite{bib-ALEPHTAU,bib-DELPHITAUnew,bib-L3TAUfin,bib-OPALTAU}
and their combination.  
Common systematic errors arise from uncertainties in the decay
radiation in the $\pi\nu$ and $\rho\nu$ channels, and in the modelling
of the $a_{1}$ decays\cite{bib-EWPPE187}. 
These errors need further
investigation and might need to be taken into account for the final
results (see Reference~\citen{bib-L3TAUfin}).
For the current combination the systematic errors on $\cAt$ and $\cAe$
are treated as uncorrelated between the experiments.
The statistical
correlation between the extracted values of $\cAt$ and $\cAe$ is small
($\le$ 5\%), and is neglected.

The average values for $\cAt$ and $\cAe$:
\begin{eqnarray}
  \cAt & = & 0.1439 \pm 0.0042 \\
  \cAe & = & 0.1498 \pm 0.0048 \,,
\end{eqnarray}
are compatible, in agreement with lepton universality. Assuming
$\mathrm{e}$-$\tau$ universality, the values for $\cAt$ and $\cAe$ can
be combined. This combination is performed neglecting any possible
common systematic error between $\cAt$ and $\cAe$ within a given
experiment, as these errors are also estimated to be small.  The
combined result of $\cAt$ and $\cAe$ is:
\begin{eqnarray}
  \cAl & = & 0.1464 \pm 0.0032 \,.
\end{eqnarray}

\begin{table}[htbp]
\renewcommand{\arraystretch}{1.15}
\begin{center}
\begin{tabular}{|ll||c|}
\hline
Experiment & & $\cAt$ \\
\hline
\hline
ALEPH  &(90 - 95), prel.       & $0.1452\pm0.0052\pm0.0032$  \\
DELPHI &(90 - 95), final       & $0.1359\pm0.0079\pm0.0055$  \\
L3     &(90 - 95), final       & $0.1476\pm0.0088\pm0.0062$  \\
OPAL   &(90 - 95), prel.       & $0.1456\pm0.0075\pm0.0057$  \\
\hline
\hline
LEP Average &                    & $0.1439\pm0.0042$           \\
\hline
\end{tabular}
\end{center}
\caption[]{
  LEP results for $\cAt$. The $\chi^2$/d.o.f.{} for the average is 0.9/3. The
  first error is statistical and the second systematic. In the LEP average,
  statistical and systematic errors are combined in quadrature. The systematic
  component of the error 
  is $\pm0.0023$.}
\label{tab-tau1}
\end{table}

\begin{table}[htbp]
\renewcommand{\arraystretch}{1.15}
\begin{center}
\begin{tabular}{|ll||c|}
\hline
Experiment & & $\cAe$ \\
\hline
\hline
ALEPH   &(90 - 95), prel.       & $0.1505\pm0.0069\pm0.0010$  \\
DELPHI  &(90 - 95), final       & $0.1382\pm0.0116\pm0.0005$  \\
L3      &(90 - 95), final       & $0.1678\pm0.0127\pm0.0030$  \\
OPAL    &(90 - 95), prel.       & $0.1456\pm0.0103\pm0.0034$  \\
\hline
\hline
LEP Average &                     & $0.1498\pm0.0048$  \\
\hline
\end{tabular}
\end{center}
\caption[]{
  LEP results for $\cAe$. The $\chi^2$/d.o.f.{} for the average is 3.1/3. The
  first error is statistical and the second systematic. In the LEP average,
  statistical and systematic errors are combined in quadrature. The systematic
  component of the error 
  is $\pm 0.0010$.}
\label{tab-tau2}
\end{table}


\clearpage
\boldmath
\section{Measurement of $\ALR$ at SLC}
\unboldmath
\label{sec-ALR}

\updates{SLD have final results for \ALR{} and the leptonic left-right forward-backward asymmetries.}                             

The measurement of the left-right cross section asymmetry ($\ALR$) by
SLD\cite{ref:sld-s00} at the SLC provides a systematically precise,
statistics-dominated determination of the coupling $\cAe$, and is
presently the most precise single measurement, with the smallest
systematic error, 
of this quantity.  In principle the analysis is
straightforward: one counts the numbers of Z bosons produced by left
and right longitudinally polarised electrons, forms an asymmetry, and
then divides by the luminosity-weighted e$^-$ beam polarisation
magnitude (the e$^+$ beam is not polarised):
\begin{equation}
  \label{eq:ALR}
  \ALR = \frac{N_{\mathrm{L}} - N_{\mathrm{R}}}%
              {N_{\mathrm{L}} + N_{\mathrm{R}}}%
         \frac{1}{P_{\mathrm{e}}}.
\end{equation}
Since the advent of high polarisation ``strained lattice'' GaAs
photocathodes (1994), the average electron
polarisation at the interaction point has been in the range 73\% to 77\%.
The method requires no
detailed final state event identification ($\ee$ final state events
are removed, as are non-Z backgrounds) and is insensitive to all
acceptance and efficiency effects.  The small total systematic error
of  0.64\% 
is
dominated by the 0.50\%
systematic error in the determination of the e$^-$ polarisation.
The 
statistical error on $\ALR$ is about 1.3\%.

The precision Compton polarimeter detects beam electrons
that have been scattered by photons from a circularly polarised laser.
Two additional polarimeters that are sensitive to the Compton-scattered
photons and which are operated in the absence of positron beam, 
have verified the precision polarimeter result and are used to set a
calibration uncertainty of 0.4\%.
In 1998, a dedicated experiment was performed in order to directly test
the expectation that accidental polarisation of the positron beam was
negligible; the e$^+$ polarisation was found to be consistent with 
zero ($-0.02\pm 0.07$)\%.

The $\ALR$ analysis includes several very small corrections. The
polarimeter result is corrected for higher order QED and
accelerator related effects, a total of
($-0.22\pm0.15$)\% for 1997/98 data. 
The event asymmetry is
corrected for backgrounds and accelerator asymmetries, a total of
($+0.15\pm0.07$)\%, for 1997/98 data.

The translation of the $\ALR$ result to a ``pole'' value is a ($-2.5\pm0.4$)\%
effect, where the uncertainty arises from the precision of the
centre-of-mass
energy
determination.  
This small error due to the beam energy measurement is slightly larger
than seen previously (it was closer to 0.3\%) and reflects the results
of a scan of the Z peak used to calibrate the energy spectrometers 
to $\MZ$ from LEP data, 
which was performed for the first time during the most
recent SLC run.
The pole value, $\ALRz$, is equivalent to a measurement
of $\cAe$.

The 2000  result is included in a running average of all
of the SLD $\ALR$ measurements (1992, 1993, 1994/1995, 1996, 1997 and 1998).
This updated result for $\ALRz$ ($\cAe$)
is $0.1514 \pm 0.0022$.
In addition, the left-right forward-backward asymmetries for leptonic
final states are measured\cite{ref:sld-asym}.  From these, the parameters $\cAe$,
$\cAm$ and $\cAt$ can be determined.  The results are
$\cAe = 0.1544 \pm 0.0060$, $\cAm = 0.142 \pm 0.015$ and $\cAt =
0.136 \pm 0.015$. 
The lepton-based result for $\cAe$ can be combined 
with the $\ALRz$ result to yield
$\cAe = 0.1516 \pm 0.0021$, 
including small correlations in the systematic errors.
The correlation of this measurement with  
$\cAm$ and $\cAt$ is 
indicated in Table~\ref{tab:corr-As}.

Assuming lepton universality, the $\ALR$ result and the results on the
leptonic left-right forward-backward asymmetries 
can be combined, while accounting
for small
correlated systematic errors, yielding
\begin{equation}
 \cAl = 0.1513 \pm 0.0021.
\end{equation}

\begin{table}[h]
\centering
\begin{tabular}{c|ccc} 
       & $\cAe$ & $\cAm$ & $\cAt$ \\
\hline
$\cAe$ &  1.000  \\
$\cAm$ &  0.038 & 1.000 \\
$\cAt$ &  0.033 & 0.007  & 1.000 \\
\end{tabular}
\caption{Correlation coefficients  between $\cAe$,
$\cAm$ and $\cAt$}
\label{tab:corr-As}
\end{table}


\clearpage
\boldmath
\section{Results from b and c Quarks}
\label{sec-HF}
\unboldmath

\updates{
DELPHI has presented new measurements of $\Abb$ and $\Acc$.

SLD has presented updated measurements of \Rb, \Rc, $\cAb$ with leptons
and vertex charge and $\cAc$ with leptons and D-mesons.

ALEPH has presented a new measurement of $\Brbl$ and $\Brbclp$ 
and L3 has published their measurement of \Rb{} and $\Brbl$.
}

\noindent
The relevant quantities in the heavy quark sector at LEP/SLD which are
currently determined by the combination procedure are:
\begin{itemize}
\item The ratios
  of the b and c quark partial widths of the Z to its total hadronic
  partial width: $\Rbz \equiv \Gbb / \Ghad$ and $\Rcz \equiv \Gcc /
  \Ghad$.
\item The forward-backward asymmetries, \Abb{} and \Acc.
\item The final state coupling parameters $\cAb,\,\cAc$ obtained from the
  left-right-forward-backward asymmetry at SLD.
\item The semileptonic branching ratios, $\Brbl$, $\Brbclp$ and $\Brcl$, and
  the average time-integrated $\Bzero\Bzerob$ mixing parameter, $\chiM$. 
  These are often determined at the same time or with similar methods
  as the asymmetries.
  Including them in the combination greatly reduces the errors.
  For example the measurements of $\chiM$ act as an effective measurement of 
  the   charge tagging efficiency, so that all errors coming from the mixture of
  different lepton sources in $\bb$ events cancel in the asymmetries.

\item The probability that a c quark produces a $\Dplus$, $\Ds$, 
 $\Dstarp$ meson\footnote{%
   Actually the product $\PcDst$ is fitted because this quantity is
   needed and measured by the LEP experiments.}  or a charmed baryon.
 The probability that a c quark fragments into a $\Dzero$ is
 calculated from the constraint that the probabilities for the weakly
 decaying charmed hadrons add up to one.  
\end{itemize}
A full description of the averaging procedure is published in \cite{ref:lephf};
the main motivations for the procedure are outlined here.  
Several analyses measure
more than one parameter simultaneously, for example the asymmetry measurements
with leptons or D mesons.
Some of the measurements of electroweak parameters depend explicitly
on the values of other parameters, for example \Rb{} depends on \Rc.
The common tagging and analysis techniques lead to common sources of
systematic uncertainty, in particular for the double-tag measurements
of \Rb.  The starting point for the combination is to ensure that all
the analyses use a common set of assumptions for input parameters
which give rise to systematic uncertainties.  
The input parameters are updated
and extended \cite{ref:lephfnew,bib-EWEP-99} to accommodate new analyses
and more recent measurements.  The correlations and interdependences
of the input measurements are then taken into account in a $\chi^2$
minimisation which results in the combined electroweak parameters and
their correlation matrix.

In a first fit the asymmetry measurements on peak, above peak and
below peak are corrected to three common centre-of-mass energies and
are then  combined at each energy point. The results of
this fit, including the SLD results, are given in
Appendix~\ref{app-HF}.  The dependence of the average asymmetries on
centre-of-mass energy agrees with the prediction of the Standard
Model. 
A second fit is made to derive the pole asymmetries $\Afbzq$ from the
measured quark asymmetries, in which all the off-peak asymmetry
measurements are corrected to the peak energy before combining. This fit
determines a total of 14 parameters:
the two partial widths, two LEP asymmetries, 
two coupling parameters from SLD,
three semileptonic branching ratios, the average mixing parameter and the
probabilities for c quark to fragment into a $\Dplus$, a $\Ds$, a
$\Dstarp$, or a charmed baryon.
If the SLD measurements are excluded from the fit there are 12 parameters to
be determined.

\subsection{Summary of Measurements and Averaging Procedure}

All measurements are presented by the LEP and SLD collaborations in
a consistent manner for the purpose of combination.
The tables prepared by the experiments include a detailed breakdown of
the systematic error of each measurement and its dependence on other
electroweak parameters. Where necessary, the experiments apply small
corrections to their results in order to use agreed values and ranges
for the input parameters to calculate systematic errors.  The
measurements, corrected where necessary, are summarised in
Appendix~\ref{app-HF} in Tables~\ref{tab:Rbinp}--\ref{tab:RcPcDstinp},
where the statistical and systematic errors are quoted separately.
The correlated systematic entries are from physics sources shared with one or
more other results in the table and are derived from the full
breakdown of common systematic uncertainties. The uncorrelated
systematic entries come from the remaining sources.


\subsubsection{Averaging Procedure}

A $\chi^2$ minimisation procedure is used to derive the values of the
heavy-flavour electroweak parameters as published in
Reference~\citen{ref:lephf}.  The full statistical and systematic
covariance matrix for all measurements is calculated.  This
correlation matrix takes into account correlations between different measurements
of one experiment and between different experiments.  The
explicit dependence of each measurement on the other parameters is
also accounted for.  

Since c-quark events form the main background in the \Rb{} analyses,
in the lifetime \Rb{} analyses,
the value of \Rb{} depends on the value of \Rc. If \Rb{} and \Rc{} are
measured in the same analysis, this is reflected in the correlation
matrix for the results.  However the analyses do not determine \Rb\ 
and \Rc\ simultaneously but instead measure \Rb\ for an assumed value
of \Rc. In this case the dependence is parameterised as
\begin{eqnarray}
 \Rb & = & 
 \Rb^{\rm{meas}} + a(\Rc) \frac {(\Rc - \Rc^{\rm{used}} )} {\Rc}.
\label{eq:rbrc}
\end{eqnarray}
In this expression, $\Rb^{\rm{meas}}$ is the result of the analysis
assuming a value of $\Rc = \Rc^{\rm{used}}$. The values of
$\Rc^{\rm{used}}$ and the coefficients $a(\Rc)$ are given in
Table~\ref{tab:Rbinp} where appropriate. The dependence of all other
measurements on other electroweak parameters is treated in the same
way, with coefficients $a(x)$ describing the dependence on parameter
$x$.

\subsubsection{Partial Width Measurements}

The measurements of \Rb{} and \Rc{} fall into two categories. In the
first, called a single-tag measurement, a method to select b or c
events is devised, and the number of tagged events is counted. This
number must then be corrected for backgrounds from other flavours and
for the tagging efficiency to calculate the true fraction of hadronic
\Zzero{} decays of that flavour. The dominant systematic errors come
from understanding the branching ratios and detection efficiencies
which give the overall tagging efficiency. For the second technique,
called a double-tag measurement, each event is divided into two
hemispheres.  With $N_t$ being the number of tagged hemispheres,
$N_{tt}$ the number of events with both hemispheres tagged and
$N_{\rm{had}}$ the total number of hadronic \Zzero{} decays one has
\begin{eqnarray}
   \frac{N_t}{2N_{\rm{had}}} &=& \effb \Rb
                        + \effc  \Rc +
                        \effuds ( 1 - \Rb - \Rc ) ,\\
   \frac{N_{tt}}{N_{\rm{had}}} &=& \Cb \effb^2 \Rb
                +    \Cc \effc^2 \Rc +
                          {\cal C}_{\mathrm{uds}} \effuds^2 ( 1 - \Rb - \Rc ) ,
\end{eqnarray}
where $\effb$, $\effc$ and $\effuds$ are the tagging efficiencies per
hemisphere for b, c and light-quark events, and $\Cq \ne 1$ accounts
for the fact that the tagging efficiencies between the hemispheres may
be correlated.  In the case of \Rb{} one has $\effb\gg\effc\gg\effuds$,
$\Cb \approx 1$. The correlations for the other flavours can be
neglected. These equations can be solved to give \Rb{} and $\effb$.
Neglecting the c and uds backgrounds and the correlations they are
approximately given by
\begin{eqnarray}
\effb &\approx& 2 N_{tt} / N_t  , \\
\Rb &\approx& N_t^2 / (4N_{tt}N_{\rm{had}}).
\end{eqnarray}
The double-tagging method has the advantage that the b tagging
efficiency is derived 
from the data, reducing the systematic
error. The residual background of other flavours in the sample, and
the evaluation of the correlation between the tagging efficiencies in
the two hemispheres of the event are the main sources of systematic
uncertainty in such an analysis.

This method can be enhanced by including more tags. All additional 
efficiencies can be determined from the data, reducing the statistical 
uncertainties without adding new systematic uncertainties.

Small corrections must be applied to the results to obtain the partial
width ratios \Rbz{} and \Rcz{} from the cross section ratios \Rb{} and \Rc{}.
These corrections depend slightly on the 
invariant mass cutoff of the simulations used by the experiments, so that
they are applied by the collaborations before the combination.

The partial width measurements included are:
\begin{itemize}
\item Lifetime (and lepton) double tag measurements for \Rb{} from
  ALEPH\cite{ref:alife}, DELPHI\cite{ref:drb}, L3\cite{ref:lrbmixed},
  OPAL\cite{ref:omixed} and SLD\cite{ref:SLD_RB_RC}.  These are the most
  precise determinations of \Rb.
  Since they completely dominate the combined result, no other \Rb{}
  measurements are used at present.
  The basic features of the double-tag technique are discussed above.
  In the ALEPH, DELPHI, OPAL and SLD measurements the charm rejection is
  enhanced by using the invariant mass information. DELPHI, OPAL and SLD
  also add kinematic information from the particles at the
  secondary vertex.
  The ALEPH and DELPHI measurements make use of several different
  tags; this improves the statistical accuracy and reduces the
  systematic errors due to hemisphere correlations and charm
  contamination, compared with the simple single/double tag.
\item Analyses with D/$\Dstarpm$ mesons to measure \Rc{} from
  ALEPH, DELPHI and OPAL.
  All measurements are constructed in such a way that no assumptions on the
  energy dependence of charm fragmentation are necessary.  The
  available measurements can be divided into four groups:
\begin{itemize}
\item inclusive/exclusive double tag (ALEPH\cite{ref:arcd}, 
  DELPHI\cite{ref:drcd,ref:drcc}, OPAL\cite{ref:orcd}): In a first
  step $\Dstarpm$ mesons are reconstructed in several decay channels
  and their production rate is measured, which depends on the product
  $\Rc \times \PcDst$.  This sample of $\cc$ (and $\bb$) events is
  then used to measure $\PcDst$ using a slow pion tag in the opposite
  hemisphere.  In the ALEPH measurement \Rc{} is unfolded internally
  in the analysis so that no explicit $\PcDst$ is available. 
\item exclusive double tag (ALEPH\cite{ref:arcd}): 
  This analysis uses exclusively
  reconstructed $\Dstarp$, $\Dzero$ and $\Dplus$ mesons in different
  decay channels. It has lower statistics but better purity than the
  inclusive analyses.
\item reconstruction of all weakly decaying charmed states
  (ALEPH\cite{ref:arcc},  DELPHI\cite{ref:drcc}, OPAL\cite{ref:orcc}): 
  These analyses make the assumption that the production rates
  of $\Dzero$, $\Dplus$, $\Ds$ and $\Lc$ 
  in $\cc$ events add up to one with small corrections
  due to unmeasured charms strange baryons.
  This is a single tag measurement, relying only on knowing
  the decay branching ratios of the charm hadrons.  
  These analyses are also used to measure the c hadron production
  ratios which are needed for the \Rb{} analyses.  
\end{itemize}
\item A lifetime plus mass double tag from SLD to measure
  \Rc\cite{ref:SLD_RB_RC}.  This analysis uses the same tagging
  algorithm as the SLD \Rb{} analysis, but with the neural net tuned to
  tag charm. Although the
  charm tag has a purity of about 84\%, most of the background is from
  b which can be measured with high precision from the b/c mixed tag rate.
\item A measurement of \Rc{} using single leptons assuming $\Brcl$ from
  ALEPH \cite{ref:arcd}.
\end{itemize}
To avoid effects from non linearities in the fit, for the inclusive/exclusive
single/double tag and for the charm-counting analyses, the products
\RcPcDst, \RcfDz, \RcfDp, \RcfDs{} and \RcfLc that are actually
measured in the analyses are directly used as inputs to the fit.
The measurements of the production rates of weakly decaying charmed
hadrons, especially \RcfDs{} and \RcfLc{} have a substantial error due
to the branching ratio of the decay mode used. Since this error is a
relative one there is a potential bias towards lower measurements.
To avoid this bias, for the production rates of weakly decaying charmed 
hadrons the logarithm of the production rates instead of the rates themselves
are input to the fit. For \RcfDz{} and \RcfDp{} the difference between
the results using the logarithm or the value itself is negligible. For
\RcfDs{} and \RcfLc{} the difference in the \Rc-result is about one
tenth of a standard deviation.

\subsubsection{Asymmetry Measurements}
\label{sec:asycorrections}
All b and c asymmetries given by the experiments are corrected to full
acceptance.

The QCD corrections to the forward-backward asymmetries depend
strongly on the experimental analyses.  For this reason the numbers
given by the collaborations are also corrected for QCD effects. A
detailed description of the procedure can be found
in \cite{ref:afbqcd} with updates reported in \cite{bib-EWEP-99}

For the 12- and 14-parameter fits described above, the LEP peak and
off-peak asymmetries are corrected to $ \sqrt {s} = 91.26$ \GeV{}
using the predicted dependence from ZFITTER\cite{ref:ZFITTER}. The
slope of the asymmetry around $\MZ$ depends only on the axial coupling
and the charge of the initial and final state fermions and is thus
independent of the value of the asymmetry itself.

After calculating the overall averages, the quark pole asymmetries,
$\Afbzq$, are derived by applying the corrections described below.
To relate
the pole asymmetries to the measured ones a few corrections that are
summarised in Table~\ref{tab:aqqcor} have to be applied. These
corrections are due to the energy shift from 91.26 \GeV{} to
$\MZ$, initial state radiation, $\gamma$ exchange and $\gamma$-$\Zzero$
interference.  A very small correction due to the nonzero value of the
b quark mass is included in the correction called $\gamma$-$\Zzero$
interference.  All corrections are calculated using ZFITTER.

\begin{table}[bhtb]
\begin{center}
\begin{tabular}{|l||l|l|}
\hline
Source   & $\delta A_{\mathrm{FB}}^{\mathrm{b}}$
         & $\delta A_{\mathrm{FB}}^{\mathrm{c}}$ \\
\hline
\hline
$\sqrt{s} = \MZ $       & $ -0.0013 $  & $ -0.0034$  \\
QED corrections         & $ +0.0041 $  & $ +0.0104$  \\
$\gamma$, $\gamma$-$\Zzero$, mass & $ -0.0003 $  & $ -0.0008$  \\
\hline
\hline
Total                   & $ +0.0025 $  & $ +0.0062$  \\
\hline
\end{tabular}
\end{center}
\caption[]{%
  Corrections to be applied to the quark asymmetries as 
   $A_{\mathrm{FB}}^0 = A_{\mathrm{FB}}^{\mathrm{meas}}
  + \delta A_{\mathrm{FB}}$.}
\label{tab:aqqcor}
\end{table}

The SLD left-right-forward-backward asymmetries are also corrected for all
radiative effects and are directly presented in terms of $\cAb$ and $\cAc$.

The measurements used are:
\begin{itemize}
\item Measurements of \Abb{} and \Acc{} using leptons from 
  ALEPH\cite{ref:alasy}, DELPHI\cite{ref:dlasy}, L3\cite{ref:llasy} and
  OPAL\cite{ref:olasy}.  
  These analyses measure either \Abb{} only from a high $p_t$ lepton
  sample or they obtain \Abb{} and \Acc{} from a fit to the lepton
  spectra. In the case of OPAL the lepton information is combined
  with hadronic variables in a neural net. DELPHI uses in addition lifetime
  information and jet-charge in the hemisphere opposite to the lepton to
  separate the different lepton sources.
  Some asymmetry analyses also measure $\chiM$.
\item Measurements of \Abb{} based on lifetime tagged events with a
  hemisphere charge measurement from ALEPH\cite{ref:ajet}, 
  DELPHI\cite{ref:djasy}, L3\cite{ref:ljet} and OPAL\cite{ref:ojet}.
  These measurements contribute roughly the same weight to the
  combined result as the lepton fits.  
\item Analyses with D mesons to measure \Acc{} from
  ALEPH\cite{ref:adsac} or \Acc{} and \Abb{} from
  DELPHI\cite{ref:ddasy} and OPAL\cite{ref:odsac}.
\item Measurements of \cAb{} and \cAc{} from SLD.
  These results include measurements using 
  lepton \cite{ref:SLD_ABL,ref:SLD_ACL}, 
  D meson \cite{ref:SLD_ACD} and 
  vertex mass plus hemisphere charge \cite{ref:SLD_ABJ} 
  tags, which have similar sources of
  systematic errors as the LEP asymmetry measurements. 
  SLD also uses vertex mass for bottom or charm tag in conjunction
  with a kaon tag or a vertex charge tag for both $\cAb$ and $\cAc$ 
  measurements \cite{ref:SLD_ABK,ref:SLD_vtxasy,ref:SLD_ACV}.
\end{itemize}

\subsubsection{Other Measurements}

The measurements of the charmed hadron fractions $\PcDst$, $\fDp$, $\fDs$
and $\fcb$ are included in the \Rc{} measurements and are described there.

ALEPH\cite{ref:abl}, DELPHI\cite{ref:dbl}, L3\cite{ref:lbl,ref:lrbmixed} and 
OPAL\cite{ref:obl} measure $\Brbl$, $\Brbclp$ and $\chiM$ or a subset of them
from a sample of leptons opposite to a b-tagged hemisphere and from a
double lepton sample. 
DELPHI\cite{ref:drcd} and OPAL\cite{ref:ocl} measure $\Brcl$ from a sample
opposite to a high energy $\Dstarpm$.
%
\subsection{Results}\label{sec-HFSUM}


\subsubsection{Results of the 12-Parameter Fit to the LEP Data}
\label{sec-HFSUM-LEP}

Using the full averaging procedure gives the following combined
results for the electroweak parameters:
\begin{eqnarray}
  \label{eqn-hf4}
  \Rbz    &=& 0.21648  \pm 0.00075   \\
  \Rcz    &=& 0.1674   \pm 0.0047    \nonumber \\
  \Afbzb  &=& 0.0989   \pm 0.0020    \nonumber \\
  \Afbzc  &=& 0.0688   \pm 0.0035    \,,\nonumber
\end{eqnarray}
where all corrections to the asymmetries and partial widths are
applied.  The $\chi^2/$d.o.f.{} is $49/(89-12)$. The corresponding
correlation matrix is given in Table~\ref{tab:12parcor}.

\begin{table}[htbp]
\begin{center}
\begin{tabular}{|l||rrrr|}
\hline
&\makebox[1.2cm]{\Rbz}
&\makebox[1.2cm]{\Rcz}
&\makebox[1.2cm]{$\Afbzb$}
&\makebox[1.2cm]{$\Afbzc$}\\
\hline
\hline
\Rbz      &$  1.00$&$ -0.15$&$ -0.03$&$  0.01$  \\
\Rcz      &$ -0.15$&$  1.00$&$  0.07$&$ -0.01$  \\
$\Afbzb$  &$ -0.03$&$  0.07$&$  1.00$&$  0.11$  \\
$\Afbzc$  &$  0.01$&$ -0.01$&$  0.11$&$  1.00$  \\
\hline
\end{tabular}
\end{center}
\caption[]{
  The correlation matrix for the four electroweak parameters from the
  12-parameter fit.}
\label{tab:12parcor}
\end{table}

\subsubsection{Results of the 14-Parameter Fit to LEP and SLD Data}
\label{sec-HFSUM-LEP-SLD}

Including the SLD results for \Rb, \Rc, \cAb{} and \cAc{} into the fit the
following results are obtained:
\begin{eqnarray}
  \label{eqn-hf6}
  \Rbz    &=& 0.21653  \pm  0.00069  \\
  \Rcz    &=& 0.1709   \pm  0.0034  \nonumber \\
  \Afbzb  &=& 0.0990   \pm  0.0020  \nonumber \\
  \Afbzc  &=& 0.0689   \pm  0.0035  \nonumber \\
  \cAb    &=& 0.922    \pm  0.023   \nonumber \\
  \cAc    &=& 0.631    \pm  0.026   \, , \nonumber
\end{eqnarray}
with a $\chi^2/$d.o.f.{} of $54/(98-14)$. The corresponding
correlation matrix is given in Table~\ref{tab:14parcor}
and the largest errors for the electroweak parameters are listed in Table
\ref{tab:hferrbk}.

\begin{figure}[htbp]
\begin{center}
  \mbox{\includegraphics[width=0.62\linewidth,bb=24 36 565 565]{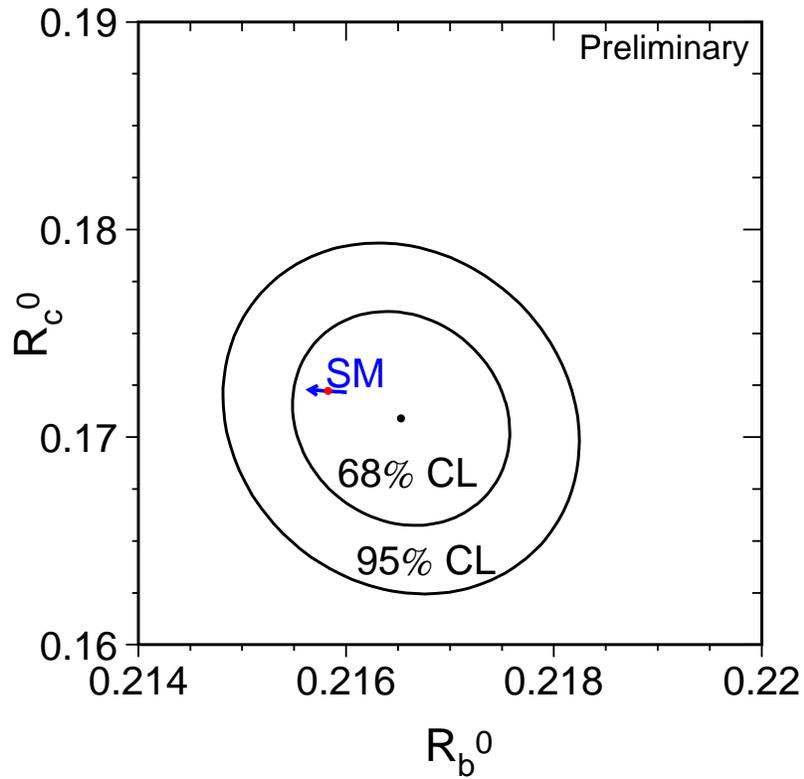}}
\end{center}
\caption[]{%
  Contours in the $\Rbz$-$\Rcz$ plane derived from the LEP+SLD
  data, corresponding to 68\% and 95\% confidence levels assuming
  Gaussian systematic errors. The Standard Model prediction for
  $\Mt=174.3 \pm 5.1$~\GeV{} is also shown. The arrow points in the
  direction of increasing values of $\Mt$.  }
\label{fig-RbRc}
\end{figure}

In deriving
these results the parameters $\cAb$ and $\cAc$ are treated as
independent of the forward-backward asymmetries $\Afbzb$ and $\Afbzc$.  In
Figure~\ref{fig-RbRc} the results for $\Rbz$ and $\Rcz$ are shown
compared with the Standard Model expectation.

\begin{table}[htbp]
\begin{center}
\begin{tabular}{|l||rrrrrr|}
\hline
&\makebox[1.2cm]{\Rbz}
&\makebox[1.2cm]{\Rcz}
&\makebox[1.2cm]{$\Afbzb$}
&\makebox[1.2cm]{$\Afbzc$}
&\makebox[0.9cm]{\cAb}
&\makebox[0.9cm]{\cAc}\\
\hline
\hline
\Rbz     & $  1.00$&$ -0.13$&$ -0.02$&$  0.01$&$ -0.04$&$  0.02$   \\
\Rcz     & $ -0.13$&$  1.00$&$  0.05$&$ -0.01$&$  0.02$&$ -0.02$   \\  
$\Afbzb$ & $ -0.02$&$  0.05$&$  1.00$&$  0.10$&$  0.02$&$  0.00$   \\
$\Afbzc$ & $  0.01$&$ -0.01$&$  0.10$&$  1.00$&$  0.00$&$  0.01$   \\
\cAb     & $ -0.04$&$  0.02$&$  0.02$&$  0.00$&$  1.00$&$  0.14$   \\
\cAc     & $  0.02$&$ -0.02$&$  0.00$&$  0.01$&$  0.14$&$  1.00$   \\
\hline
\end{tabular}
\end{center}
\caption[]{
  The correlation matrix for the six electroweak parameters from the
  14-parameter fit.  }
\label{tab:14parcor}
\end{table}

\begin{table}[htbp]
\begin{center}
\begin{tabular}{|c|c|c|c|c|c|c|}
\hline
&\makebox[1.2cm]{\Rbz}
&\makebox[1.2cm]{\Rcz}
&\makebox[1.2cm]{$\Afbzb$}
&\makebox[1.2cm]{$\Afbzc$}
&\makebox[0.9cm]{\cAb}
&\makebox[0.9cm]{\cAc}\\
 & $(10^{-3})$ & $(10^{-3})$ & $(10^{-3})$ & $(10^{-3})$ 
 & $(10^{-2})$ & $(10^{-2})$ \\
\hline
statistics & 
$0.43$ & $2.6$ & $1.7$ & $3.0$ & $1.6$ & $2.0$ \\
internal systematics &
$0.31$ & $1.8$ & $0.7$ & $1.4$ & $1.6$ & $1.6$ \\
QCD effects &
$0.19$ & $0.1$ & $0.2$ & $0.1$ & $0.6$ & $0.3$ \\
BR(D $\rightarrow$ neut.)&
$0.14$ & $0.1$ & $0$   & $0$ & $0$ & $0$ \\
D decay multiplicity &
$0.12$ & $0.2$ & $0  $ & $0  $ & $0$ & $0$ \\
BR(D$^+ \rightarrow$ K$^- \pi^+ \pi^+) $&
$0.10$ & $0.3$ & $0.1$ & $0$ & $0$ & $0  $ \\
BR($\Ds \rightarrow \phi \pi^+) $&
$0.02$ & $0.7$ & $0.1$ & $0$ & $0$ & $0$ \\
BR($\Lambda_{\mathrm{c}} \rightarrow $p K$^- \pi^+) $&
$0.06$ & $0.6$ & $0$ & $0.1$ & $0$ & $0  $ \\
D lifetimes&
$0.06$    & $0.1$ & $0  $ & $0.1$ & $0$ & $0$ \\
gluon splitting &
$0.26$ & $0.6$ & $0$ & $0.2$ & $0.1$ & $0.1$ \\
c fragmentation &
$0.10$ & $0.3$ & $0.1$ & $0.2$ & $0.1$ & $0.1$ \\
light quarks&
$0.07$ & $0.3$ & $0.5$ & $0.1$ & $0$ & $0  $ \\
\hline
total &
$0.69$ & $3.4$ & $2.0$ & $3.5$ & $2.3$ & $2.6$ \\
\hline
\end{tabular}
\end{center}
\caption[]{
The dominant error sources for the electroweak parameters from the 14-parameter
fit.
}
\label{tab:hferrbk}
\end{table}

The 14 parameter fit yields the $\bl$ branching ratio:
\begin{equation}
\Brbl \, = \, 0.1057 \pm 0.0019.
\end{equation}
The largest error sources on this quantity are the dependences on the 
semileptonic decay models $\bl$, $\cl$ with 
\begin{eqnarray*}
\Delta \Brbl |_{(\bl) \rm{model}} &  = & 0.0008,\\
\Delta \Brbl |_{(\cl) \rm{model}} &  = & 0.0005.
\end{eqnarray*}
Extensive studies are made to understand the size of these errors.
If all the asymmetry measurements are excluded from the fit
a consistent result is
obtained with modelling errors of 0.0010 and 0.0006. 
The reduction of the modelling uncertainty is due to the inclusion of
asymmetry measurements using different methods. Those using leptons depend
on the semileptonic decay models while those using a lifetime tag and
jet charge or D mesons do not. The mutual consistency of the asymmetry
measurements effectively constrains the semileptonic decay models, and
reduces the uncertainty in the semileptonic branching ratio.

The result of the full fit to the LEP+SLC results including the
off-peak asymmetries and the non-electroweak parameters can be
found in Appendix~\ref{app-HF}.  
Results
for
the non-electroweak parameters are independent of the
treatment of the off-peak asymmetries and the SLD data.

\clearpage
\boldmath
\section{The Hadronic Charge Asymmetry $\avQfb$}
\label{sec-QFB}
\unboldmath

\updates{L3 have published their result.}

\noindent
The LEP experiments
ALEPH\cite{ALEPHcharge,ALEPHcharge1992,ALEPHcharge1996},
DELPHI\cite{DELPHIcharge,DELPHIcharge1996}, L3\cite{ref:ljet} and
OPAL\cite{OPALcharge, OPALcharge1995} have provided measurements of
the hadronic charge asymmetry based on the mean difference in jet
charges measured in the forward and backward event hemispheres,
$\avQfb$. DELPHI have also provided a related measurement of the total
charge asymmetry by making a charge assignment on an event-by-event
basis and performing a likelihood fit\cite{DELPHIcharge}.  The
experimental values quoted for the average forward-backward charge
difference, $\avQfb$, cannot be directly compared as some of them
include detector dependent effects such as acceptances and
efficiencies.  Therefore the effective electroweak mixing angle,
$\swsqeffl$, as defined in Section~\ref{sec-SW}, is used as a means of
combining the experimental results summarised in Table~\ref{partab}.

\begin{table}[htb]
\begin{center}
\renewcommand{\arraystretch}{1.1}
\begin{tabular}{|ll||c|}
\hline
Experiment & & $\swsqeffl$ \\
\hline
\hline
ALEPH & (90-94), final & $0.2322\pm0.0008\pm0.0011$ \\
DELPHI& (91-94), prel. & $0.2311\pm0.0010\pm0.0014$ \\
L3    & (91-95), final & $0.2327\pm0.0012\pm0.0013$ \\
OPAL  & (91-94), prel. & $0.2326\pm0.0012\pm0.0013$ \\
\hline
\hline
LEP Average  &           & $0.2321\pm0.0010$ \\
\hline
\end{tabular}
\caption[]{
  Summary of the determination of $\swsqeffl$ from inclusive hadronic
  charge asymmetries at LEP. For each experiment, the first error is
  statistical and the second systematic. The latter is dominated by
  fragmentation and decay modelling uncertainties.  }
\label{partab}
\end{center}
\end{table}
%

The dominant source of systematic error arises from the modelling of
the charge flow in the fragmentation process for each flavour. All
experiments measure the required charge properties for $\Zzero\ra\bb$
events from the data. ALEPH also determines the charm charge
properties from the data. The fragmentation model implemented in the
JETSET Monte Carlo program\cite{JETSET} is used by all experiments as
reference; the one of the HERWIG Monte Carlo program\cite{HERWIG} is
used for comparison. The JETSET fragmentation parameters are varied to
estimate the systematic errors. The central values chosen by the
experiments for these parameters are, however, not the same. 
The smaller of
the two fragmentation errors in any pair of results is treated as
common to both.  The present average of $\swsqeffl$ from $\avQfb$ and
its associated error are not very sensitive to the treatment of common
uncertainties.
The ambiguities due to QCD corrections may cause changes in the
derived value of $\swsqeffl$. These are, however, well below the
fragmentation uncertainties and experimental errors. The effect of
fully correlating the estimated systematic uncertainties from this
source between the experiments has a negligible effect upon the
average and its error.

There is also some correlation between these results and those for
$\Abb$ using jet charges. The dominant source of correlation is again
through uncertainties in the fragmentation and decay models used. The
typical correlation between the derived values of $\swsqeffl$ from
the $\avQfb$ and the $\Abb$ jet charge measurements is estimated
to be about 20\% to 25\%. This leads to only a small change in the
relative weights for the $\Abb$ and $\avQfb$ results when averaging
their $\swsqeffl$ values (Section~\ref{sec-SW}). Furthermore, the jet
charge method contributes at most half of the weight of the $\Abb$
measurement.  Thus, the correlation between $\avQfb$ and $\Abb$ from
jet charge will have little impact on the overall Standard Model fit,
and is neglected at present.



\clearpage
\boldmath
\section{Averages for ${\mathbf{f\overline{f}}}$ Production at \LEPII}
\label{sec-FF}
\unboldmath

\updates{Results are updated with data taken in 1999 and 2000.}

Since the start of the \LEPII ~program LEP has delivered collisions
at energies from $\sim 130$ GeV to $\sim 209$ GeV. The four LEP experiments
have made measurements on the $\eeff$ process over this range of energies,
and preliminary combinations of these data are discussed in this note.
 
For the combination presented here, only data taken up to the end of
June 2000 are considered. 
The nominal and actual \CoM\ energies to which the LEP data are
averaged for each year are given in Table~\ref{tab:lep2-ff-ecms}.

A number of measurements on the process $\eeff$ exist and are 
combined.
\begin{list}{$\bullet$}{\setlength{\itemsep}{0ex}
                        \setlength{\parsep}{0ex}
                        \setlength{\topsep}{0ex}}
 \item preliminary averages of cross section and forward-backward asymmetry
       measurements
 \item a preliminary average of the differential cross section measurements, $\dsdc$,  for the channels $\eemm$ and $\eett$

 \item  heavy flavour results $\Rb$, $\Rc$, $\Abb$ and $\Acc$ 
\end{list}
Complete results of the combinations are available on the web 
page~\cite{ref:lep2-ff-ffbar_web} and are discussed 
in~\cite{ref:lep2-ff-all-osaka}.

The combined results are interpreted
in terms of contact interactions, the exchange of $\Zprime$ bosons, 
and contours of the S-Matrix parameters $\jbctot$ and 
$\jbcfb$, that describe $\gamma$-$\Zzero$ interference, are derived.
The interpretations are discussed fully in~\cite{ref:lep2-ff-all-osaka}.

\begin{table}[htp]
 \begin{center}
 \begin{tabular}{|c|c|c|c|}
  \hline
   Year & Nominal Energy & Actual Energy & Luminosity \\
        &     $\GeV$     &    $\GeV$     &  pb$^{-1}$ \\
  \hline
  \hline
   1995 &      130       &    130.2      & $\sim 3  $ \\
        &      136       &    136.2      & $\sim 3  $ \\
  \cline{2-4}
        &  $133^{\ast}$ &     133.2      & $\sim 6  $ \\
  \hline
   1996 &      161       &    161.3      & $\sim 10 $ \\
        &      172       &    172.1      & $\sim 10 $ \\
  \cline{2-4}
        &  $167^{\ast}$ &     166.6      & $\sim 20 $ \\
  \hline
   1997 &      130       &    130.2      & $\sim 2  $ \\
        &      136       &    136.2      & $\sim 2  $ \\
        &      183       &    182.7      & $\sim 50 $ \\
  \hline
   1998 &      189       &    188.6      & $\sim 170$ \\
  \hline
   1999 &      192       &    191.6      & $\sim 30 $ \\
        &      196       &    195.5      & $\sim 80 $ \\
        &      200       &    199.5      & $\sim 80 $ \\
        &      202       &    201.6      & $\sim 40 $ \\
  \hline
   2000 &      205       &    204.9      & $\sim 60 $ \\
        &      207       &    206.7      & $\sim 30 $ \\
  \cline{2-4}
        &  $206^{\ast}$ &     205.5      & $\sim 90 $ \\
  \hline
 \end{tabular}
 \caption{The nominal and actual \CoM\ energies for data
          collected during $\LEPII$ operation in each year. The approximate
          average luminosity analysed per experiment at each energy is also
          shown. Values marked with 
          a $^{\ast}$ are average energies for 1995, 1996 and 2000 used 
          for heavy flavour results. The data taken at nominal energies of
          130 and 136 in 1995 and 1997 are combined by most experiments.}
 \label{tab:lep2-ff-ecms}
 \end{center}
\end{table}

\subsection{Cross Sections and Asymmetry Measurements}
\label{sec-ave-xsc-afb}

Cross section results are combined for the $\eeqq$, $\eemm$ and $\eett$ 
channels, forward-backward asymmetry measurements are combined for
the $\mumu$ and $\tautau$ final states. The averages are made for the
samples of events with high $\sqrt{\spr}$. 

As before~\cite{bib-EWEP-18}, the averaged results are given 
for two signal definitions:
\begin{itemize}
 \item {\bf Definition 1:} $\sqrt{\spr}$ is taken to be the mass of the
 $s$-channel propagator, with the $\ff$ signal being defined by the cut 
 $\sqrt{\spr/s} > 0.85$. ISR-FSR photon interference is subtracted to 
 render the propagator mass unambiguous. 
 \item {\bf Definition 2:} For dilepton events, $\sqrt{\spr}$ is taken to be 
 the bare invariant mass of the outgoing difermion pair. For hadronic events,
 it is taken to be the mass of the $s$-channel propagator. In both cases,
 ISR-FSR photon interference is included and the signal is defined by
 the cut $\sqrt{\spr/s} > 0.85$. When calculating the contribution to
 the hadronic cross section due to ISR-FSR interference, since the propagator 
 mass is ill-defined, it is replaced by the bare $\qq$ mass.
\end{itemize}
Events containing additional fermion pairs from radiative 
processes are considered to be signal, providing that the primary 
pair passes the cut on $\sqrt{\spr/s}$ and that the secondary pair
has a mass below 70~GeV/c$^2$.

The data are split into 3 sets: data taken at energies from 130--189 GeV,
data taken during 1999, and data taken in 2000. Averages are performed 
separately for each of these data sets. Within each subset correlations between
experiments and energies and channels are considered.

\begin{figure}[htp]
 \begin{center}
   \mbox{
     \includegraphics[width=0.48\linewidth]{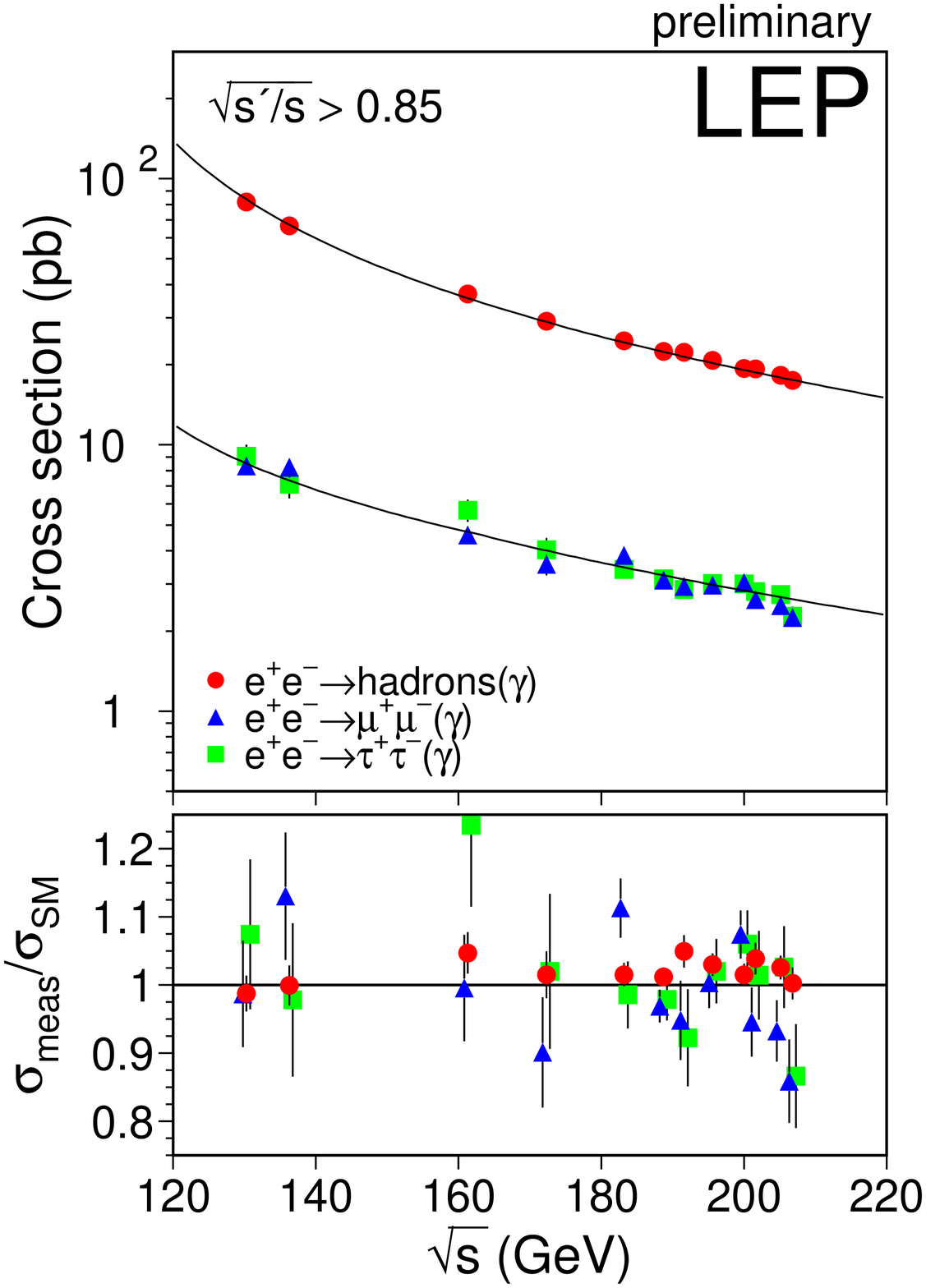}
     \includegraphics[width=0.48\linewidth]{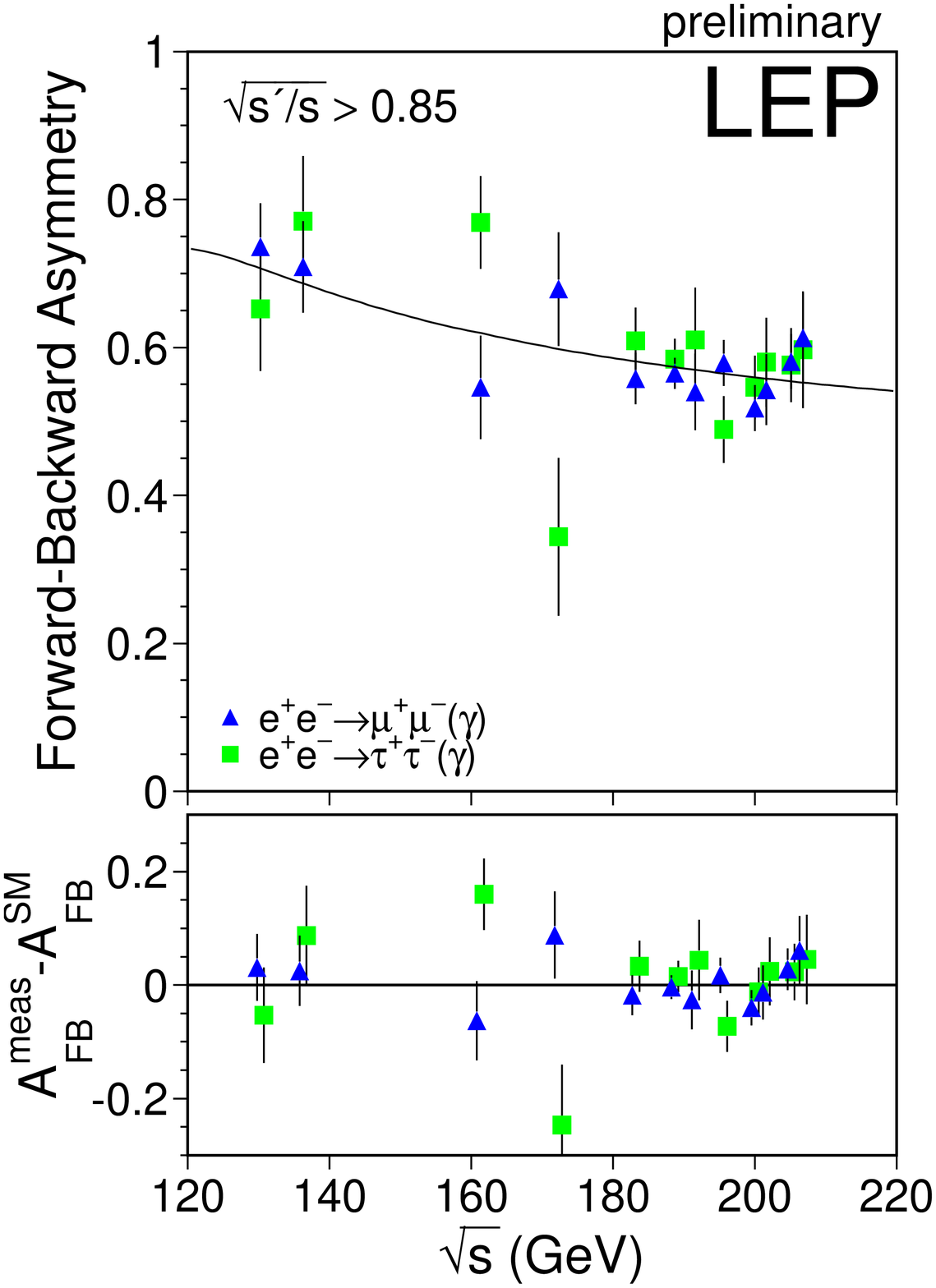}
     }
 \caption{Preliminary combined LEP results on the cross sections for 
 $\qqb$, $\mumu$ and $\tautau$ final states, and forward-backward asymmetries 
 for $\mumu$ and $\tautau$ final states as a function of \CoM\
 energy. The values at 130--189~GeV are taken 
 from~\cite{ref:lep2-ff-moriond2000}. The prediction of the Standard Model, 
 computed with ZFITTER~\cite{ref:lep2-ff-zfitter}, are shown as curves.
 The lower plots shows the ratio of the data divided by the predictions 
 for the cross sections, and the difference between the measurements and 
 the predictions for the asymmetries. For better visibility, the
 $\mumu$ and $\tautau$ data points are slightly shifted apart from
 their nominal \CoM\ value. 
}
 \label{fig:lep2-ff-xs-afb-lep}
 \end{center}
\end{figure}

Tables~\ref{tab:lep2-ff-res-xsc-afb-a}
and~\ref{tab:lep2-ff-res-xsc-afb-b} show the preliminary combined results 
for the 1995--1999 data corresponding to the signal definition~1 and the 
difference in the results if definition~2 is used. The results for the 
averages of the 130--189 GeV data are identical to those given 
in~\cite{ref:lep2-ff-moriond2000}.
Results for the more preliminary data taken during 2000 are not given in 
numerical form but are shown in Figure~\ref{fig:lep2-ff-xs-afb-lep}
which show the LEP averaged cross sections and asymmetries 
(based on definition~1), respectively, as a function of the 
\CoM\ energy, together with the SM predictions. 

The $\chi^2$ per degree of freedom for the average of the 1999 data is 52.5/60.
The correlations are rather small, with the largest components
at any given pair of energies being between the hadronic cross sections.

There is good agreement between the SM expectations and the measurements of the
individual experiments and the combined averages.
The cross sections for hadronic final states at most of the energy points 
are somewhat above the SM expectations. Taking into account the correlations
between the data points and also assigning a theory error of 
$\pm 0.2\%$~\cite{ref:lep2-ff-lepffwrkshp} to the SM predictions, the 
difference of the cross section from the SM expectations averaged over 
all energies is approximately a 2.5 standard deviation excess.
It is concluded that there is no significant evidence in the results of the
combinations for physics beyond the SM in the process $\eeff$.

\begin{table}[htp]
 \begin{center}
 \begin{tabular}{|c|c|r@{$\pm$}l|r|r|}
 \hline
$\sqrt{s}$ (GeV) &Quantity &\multicolumn{2}{|c|}{Value}   &
\multicolumn{1}{c|}{SM} & \multicolumn{1}{|c|}{$\Delta$}  \\
\hline\hline
  130 & $\sigma(q\overline{q})$ \hfill [pb] & 81.938 &  2.220 & 82.803 &
-0.251 \\
      & $\sigma(\mu^{+}\mu^{-})$ [pb]       &  8.592 &  0.682 &  8.439 &
-0.331 \\
      & $\sigma(\tau^{+}\tau^{-})$ [pb]     &  9.082 &  0.931 &  8.435 &
-0.108 \\
      & $\mathrm{A_{fb}}(\mu^{+}\mu^{-})$   &  0.692 &  0.060 &  0.705 &
0.012 \\
      & $\mathrm{A_{fb}}(\tau^{+}\tau^{-})$ &  0.663 &  0.076 &  0.704 &
0.012 \\
\hline
  136 & $\sigma(q\overline{q})$ \hfill [pb] & 66.570 &  1.967 & 66.596 &
-0.224 \\
      & $\sigma(\mu^{+}\mu^{-})$ [pb]       &  8.231 &  0.678 &  7.281 &
-0.280 \\
      & $\sigma(\tau^{+}\tau^{-})$ [pb]     &  7.123 &  0.821 &  7.279 &
-0.091 \\
      & $\mathrm{A_{fb}}(\mu^{+}\mu^{-})$   &  0.704 &  0.060 &  0.684 &
0.013 \\
      & $\mathrm{A_{fb}}(\tau^{+}\tau^{-})$ &  0.752 &  0.088 &  0.683 &
0.014 \\
\hline
 \end{tabular}
 \caption{Preliminary combined LEP results for $\eeff$
 for \CoM\ energies below the W--pair production threshold. 
 All the results correspond to the signal definition~1. The Standard Model
 predictions are from ZFITTER v6.10~\cite{ref:lep2-ff-zfitter}.
 The difference, $\Delta$, in the averages for the measurements for 
 definition~2 relative to definition~1 are also indicated.
 The quoted uncertainties do not include the theoretical 
 uncertainties on the corrections discussed in ~\cite{ref:lep2-ff-all-osaka}
}
 \label{tab:lep2-ff-res-xsc-afb-a}
 \end{center}
\end{table}

\begin{table}[htp]
 \begin{center}
 \begin{tabular}{|c|c|r@{$\pm$}l|r|r|}
 \hline
$\sqrt{s}$ (GeV) &Quantity &\multicolumn{2}{|c|}{Value}   &
\multicolumn{1}{c|}{SM} & \multicolumn{1}{|c|}{$\Delta$}  \\
\hline\hline
  161 & $\sigma(q\overline{q})$ \hfill [pb] & 36.909 &  1.071 & 35.247 &
-0.143 \\
      & $\sigma(\mu^{+}\mu^{-})$ [pb]       &  4.586 &  0.364 &  4.613 &
-0.178 \\
      & $\sigma(\tau^{+}\tau^{-})$ [pb]     &  5.692 &  0.545 &  4.613 &
-0.061 \\
      & $\mathrm{A_{fb}}(\mu^{+}\mu^{-})$   &  0.535 &  0.067 &  0.609 &
0.017 \\
      & $\mathrm{A_{fb}}(\tau^{+}\tau^{-})$ &  0.646 &  0.077 &  0.609 &
0.016 \\
\hline
  172 & $\sigma(q\overline{q})$ \hfill [pb] & 29.172 &  0.987 & 28.738 &
-0.124 \\
      & $\sigma(\mu^{+}\mu^{-})$ [pb]       &  3.556 &  0.317 &  3.952 &
-0.157 \\
      & $\sigma(\tau^{+}\tau^{-})$ [pb]     &  4.026 &  0.450 &  3.951 &
-0.054 \\
      & $\mathrm{A_{fb}}(\mu^{+}\mu^{-})$   &  0.672 &  0.077 &  0.591 &
0.018 \\
      & $\mathrm{A_{fb}}(\tau^{+}\tau^{-})$ &  0.342 &  0.094 &  0.591 &
0.017 \\
\hline
  183 & $\sigma(q\overline{q})$ \hfill [pb] & 24.567 &  0.421 & 24.200 &
-0.109 \\
      & $\sigma(\mu^{+}\mu^{-})$ [pb]       &  3.484 &  0.147 &  3.446 &
-0.139 \\
      & $\sigma(\tau^{+}\tau^{-})$ [pb]     &  3.398 &  0.174 &  3.446 &
-0.050 \\
      & $\mathrm{A_{fb}}(\mu^{+}\mu^{-})$   &  0.558 &  0.035 &  0.576 &
0.018 \\
      & $\mathrm{A_{fb}}(\tau^{+}\tau^{-})$ &  0.608 &  0.045 &  0.576 &
0.018 \\
\hline
  189 & $\sigma(q\overline{q})$ \hfill [pb] & 22.420 &  0.248 & 22.156 &
-0.101 \\
      & $\sigma(\mu^{+}\mu^{-})$ [pb]       &  3.109 &  0.077 &  3.207 &
-0.131 \\
      & $\sigma(\tau^{+}\tau^{-})$ [pb]     &  3.140 &  0.100 &  3.207 &
-0.048 \\
      & $\mathrm{A_{fb}}(\mu^{+}\mu^{-})$   &  0.565 &  0.021 &  0.569 &
0.019 \\
      & $\mathrm{A_{fb}}(\tau^{+}\tau^{-})$ &  0.584 &  0.028 &  0.569 &
0.018 \\
\hline
 192 & \hfill $\sigma(q\overline{q})$ [pb] & 22.292 & 0.514 & 21.237 &
--0.098 \\
     & $\sigma(\mu^{+}\mu^{-})$ [pb]       &  2.941 & 0.175 &  3.097 &
--0.127 \\
     & $\sigma(\tau^{+}\tau^{-})$ [pb]     &  2.863 & 0.216 &  3.097 &
--0.047 \\
     & $\Afb(\mu^{+}\mu^{-})$              &  0.540 & 0.052 &  0.566 &
0.019 \\
     & $\Afb(\tau^{+}\tau^{-})$            &  0.610 & 0.071 &  0.566 &
0.019 \\
 \hline
 196 & \hfill $\sigma(q\overline{q})$ [pb] & 20.730 & 0.330 & 20.127 &
--0.094 \\
     & $\sigma(\mu^{+}\mu^{-})$ [pb]       &  2.965 & 0.106 &  2.962 &
--0.123 \\
     & $\sigma(\tau^{+}\tau^{-})$ [pb]     &  3.015 & 0.139 &  2.962 &
--0.045 \\
     & $\Afb(\mu^{+}\mu^{-})$              &  0.579 & 0.031 &  0.562 &
0.019 \\
     & $\Afb(\tau^{+}\tau^{-})$            &  0.489 & 0.045 &  0.562 &
0.019 \\
 \hline
 200 & \hfill $\sigma(q\overline{q})$ [pb] & 19.376 & 0.306 & 19.085 &
--0.090 \\
     & $\sigma(\mu^{+}\mu^{-})$ [pb]       &  3.038 & 0.104 &  2.834 &
--0.118 \\
     & $\sigma(\tau^{+}\tau^{-})$ [pb]     &  2.995 & 0.135 &  2.833 &
--0.044 \\
     & $\Afb(\mu^{+}\mu^{-})$              &  0.518 & 0.031 &  0.558 &
0.019 \\
     & $\Afb(\tau^{+}\tau^{-})$            &  0.546 & 0.043 &  0.558 &
0.019 \\
 \hline
 202 & \hfill $\sigma(q\overline{q})$ [pb] & 19.291 & 0.425 & 18.572 &
--0.088 \\
     & $\sigma(\mu^{+}\mu^{-})$ [pb]       &  2.621 & 0.139 &  2.770 &
--0.116 \\
     & $\sigma(\tau^{+}\tau^{-})$ [pb]     &  2.806 & 0.183 &  2.769 &
--0.043 \\
     & $\Afb(\mu^{+}\mu^{-})$              &  0.543 & 0.048 &  0.556 &
0.020 \\
     & $\Afb(\tau^{+}\tau^{-})$            &  0.580 & 0.060 &  0.556 &
0.019 \\
 \hline
 \end{tabular}
 \caption{Preliminary combined LEP results for $\eeff$
 for \CoM\ energies above the W--pair production threshold. 
 All the results correspond to the signal definition~1. The Standard Model
 predictions are from ZFITTER v6.10~\cite{ref:lep2-ff-zfitter}.
 The difference, $\Delta$, in the averages for the measurements for 
 definition~2 relative to definition~1 are also indicated.
 The quoted uncertainties do not include the theoretical 
 uncertainties on the corrections discussed in ~\cite{ref:lep2-ff-all-osaka}
}
 \label{tab:lep2-ff-res-xsc-afb-b}
 \end{center}
 \vskip 5cm
\end{table}

\subsection{Differential cross sections}
\label{sec-dsdc}

The LEP experiments have measured the differential cross section, $\dsdc$, 
for the $\eemm$ and $\eett$ channels. This section discusses
a procedure to combine these measurements and presents preliminary results.

Using a Monte Carlo simulation 
it is found that a $\chi^{2}$ fit to the measured differential cross sections,
using the expected error on the differential cross sections, computed from 
the expected cross sections and the expected numbers of events in each 
experiment, provided a very good approximation to the exact likelihood method
based on Poisson statistics. Further details are given 
in~\cite{ref:lep2-ff-all-osaka}.

Data are binned in 10 bins of $\cos\theta$. 
The scattering angle, $\theta$, is the angle of the
negative lepton with respect to the incoming electron direction
in the lab coordinate system. The outer acceptances of the most forward 
and most backward bins for which the four experiments have presented 
their data are different. This is accounted for as a correction to a common 
signal definition. The signal definition used corresponds to definition 1
of Section~\ref{sec-ave-xsc-afb}.  

Correlated small systematic errors between different experiments, channels and 
energies, arising from uncertainties on the overall normalisation are 
considered.

\begin{figure}[]
 \begin{center}
   \mbox{
    \includegraphics[width=0.75\linewidth]{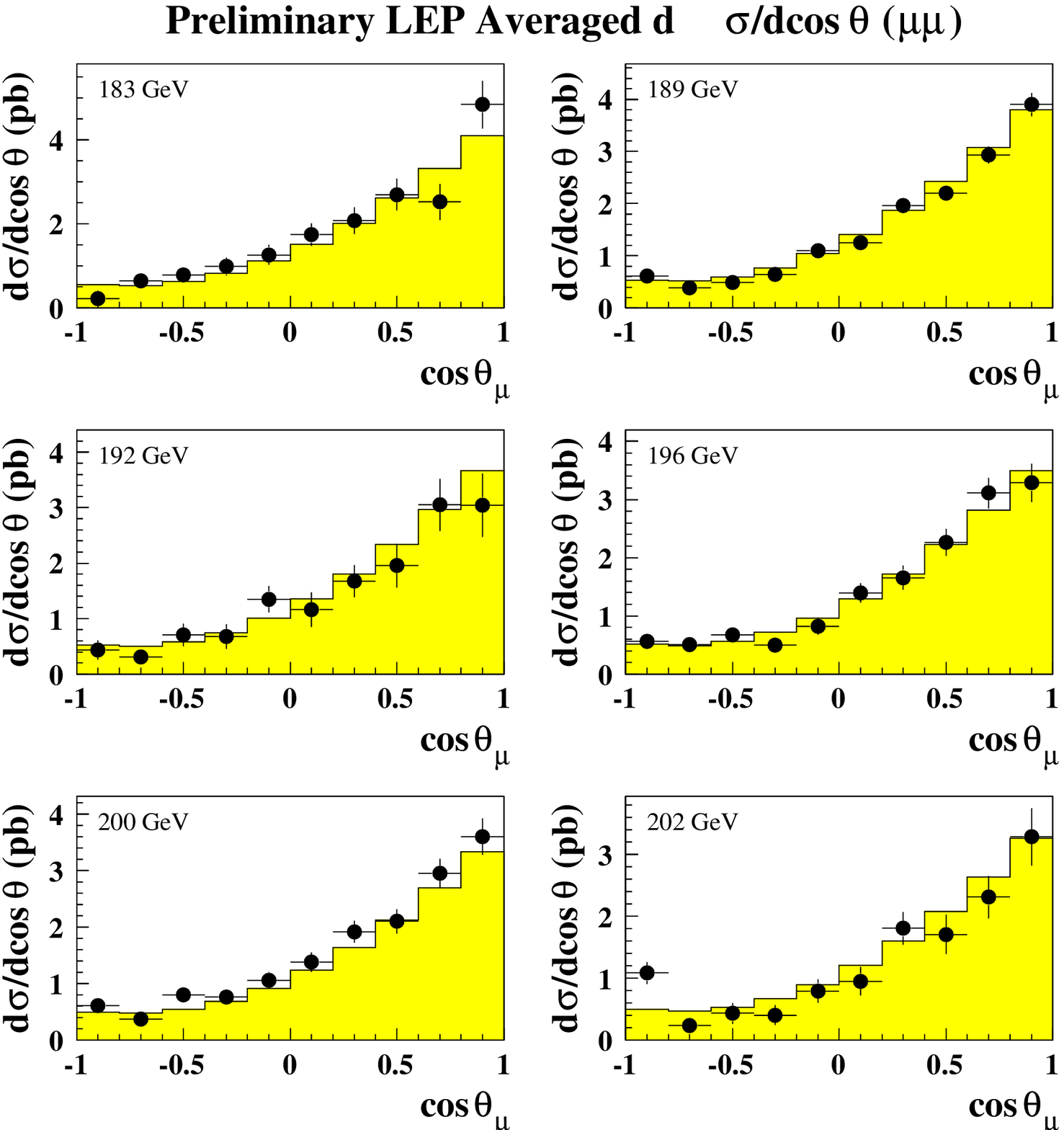}
   }
 \caption{LEP averaged differential cross sections for $\eemm$
          at energies of 183-202 GeV. The Standard Model Predictions shown as
          solid histograms are computed with 
          ZFITTER~\cite{ref:lep2-ff-zfitter}}
 \label{fig:lep2-ff-dsdc-mm-res}
 \end{center}
\end{figure}

\begin{figure}[]
 \begin{center}
   \mbox{
    \includegraphics[width=0.75\linewidth]{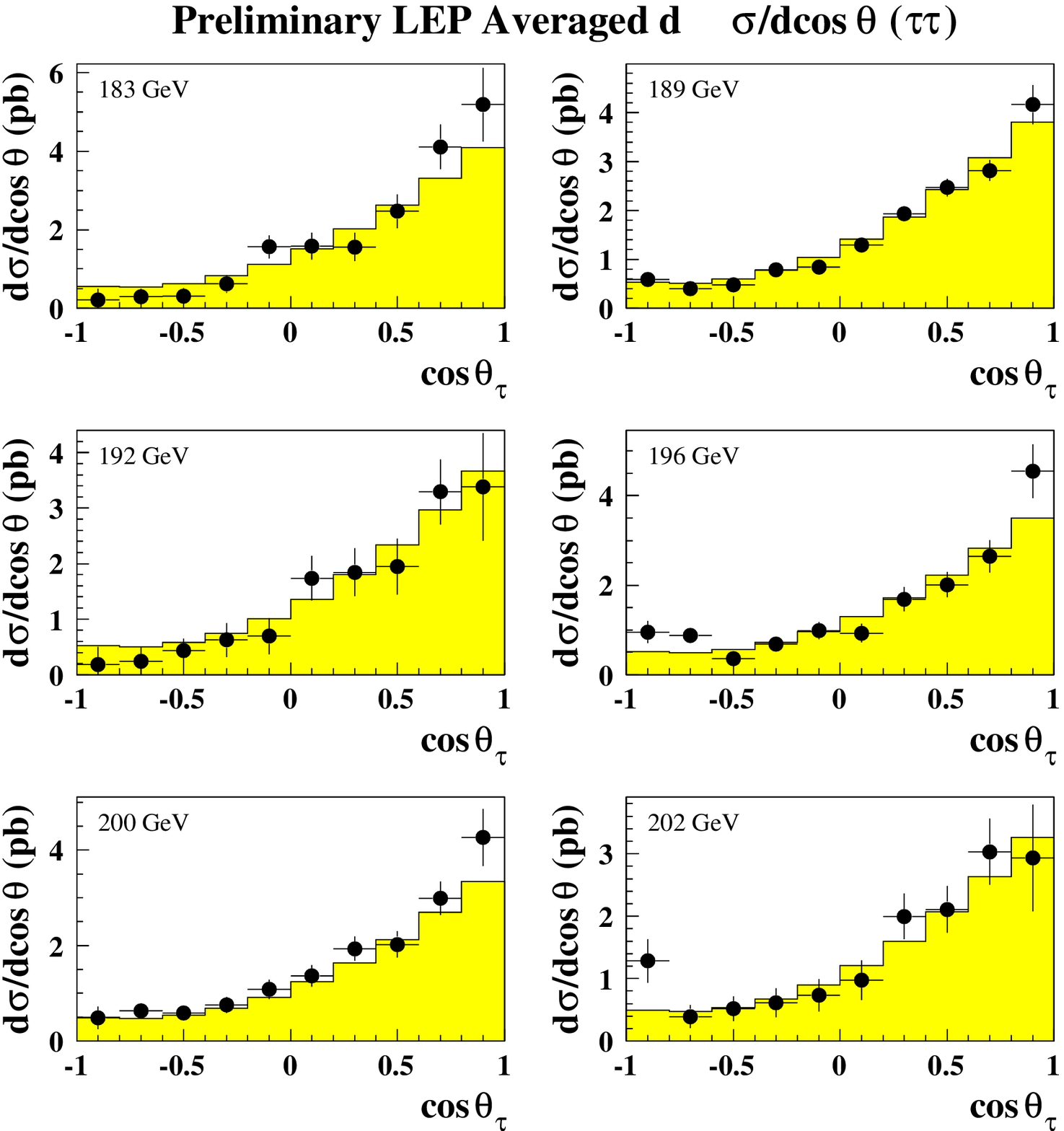}
   }
 \caption{LEP averaged differential cross sections for $\eett$
          at energies of 183-202 GeV. The Standard Model Predictions shown as
          solid histograms are computed with 
          ZFITTER~\cite{ref:lep2-ff-zfitter}}
 \label{fig:lep2-ff-dsdc-tt-res}
 \end{center}
\end{figure}

The data are subdivided into two energy ranges, 183 and 189 GeV and
   192--202 GeV, and averages are made for each energy point within each
   of these subsets.
The results of the averages are shown in 
Figures~\ref{fig:lep2-ff-dsdc-mm-res} and~\ref{fig:lep2-ff-dsdc-tt-res}.

The correlations between bins in the average are less that 
$2\%$ of the total error on the averages in each bin.
The overall agreement between the averaged data and the predictions
is good, with a $\chi^{2}$ of 114 for 120 degrees of freedom. 
At 202 GeV the cross section in the most backward bin,
$-1.00 < \cos\theta < -0.8$, for both muon and tau final states is 
above the predictions. For the muons the excess in data corresponds to 3.3
standard deviations. For the taus the excess is 2.3 standard deviations,
however, for this measurement the individual experiments are somewhat
inconsistent, having a chi-squared with respect to the average of
10.5 for 2 degrees of freedom.

\subsection{Heavy Flavour Measurements}
\label{sec-hvflv}

A combination of measurements of the ratios\footnote{Unlike at $\LEPI$, 
$\Rq$ is defined as 
$\mathrm{\frac{\sigma_{q \overline{q}}}{\sigma_{had}}}$.},
$\Rb$ and $\Rc$
 and the forward-backward asymmetries, $\Abb$ and 
$\Acc$, from the LEP collaborations at \CoM\ 
energies in the range of 130 to 209 GeV is performed. 

A common signal definition is defined for all the 
measurements, requiring:
\begin{list}{$\bullet$}{\setlength{\itemsep}{0ex}
                        \setlength{\parsep}{0ex}
                        \setlength{\topsep}{0ex}}
 \item{an effective \CoM\ energy $\sqrt{s^{\prime}} > 0.85 \sqrt{s}$}
 \item{the inclusion of ISR and FSR photon interference contribution and}
 \item{extrapolation to full angular acceptance.}
\end{list}

Systematic errors are divided into three  categories: uncorrelated errors, 
errors correlated between the measurements of each experiment, and 
errors common to all experiments.

\begin{figure}[]
\begin{center}
   \mbox{
    \includegraphics[width=0.48\linewidth]{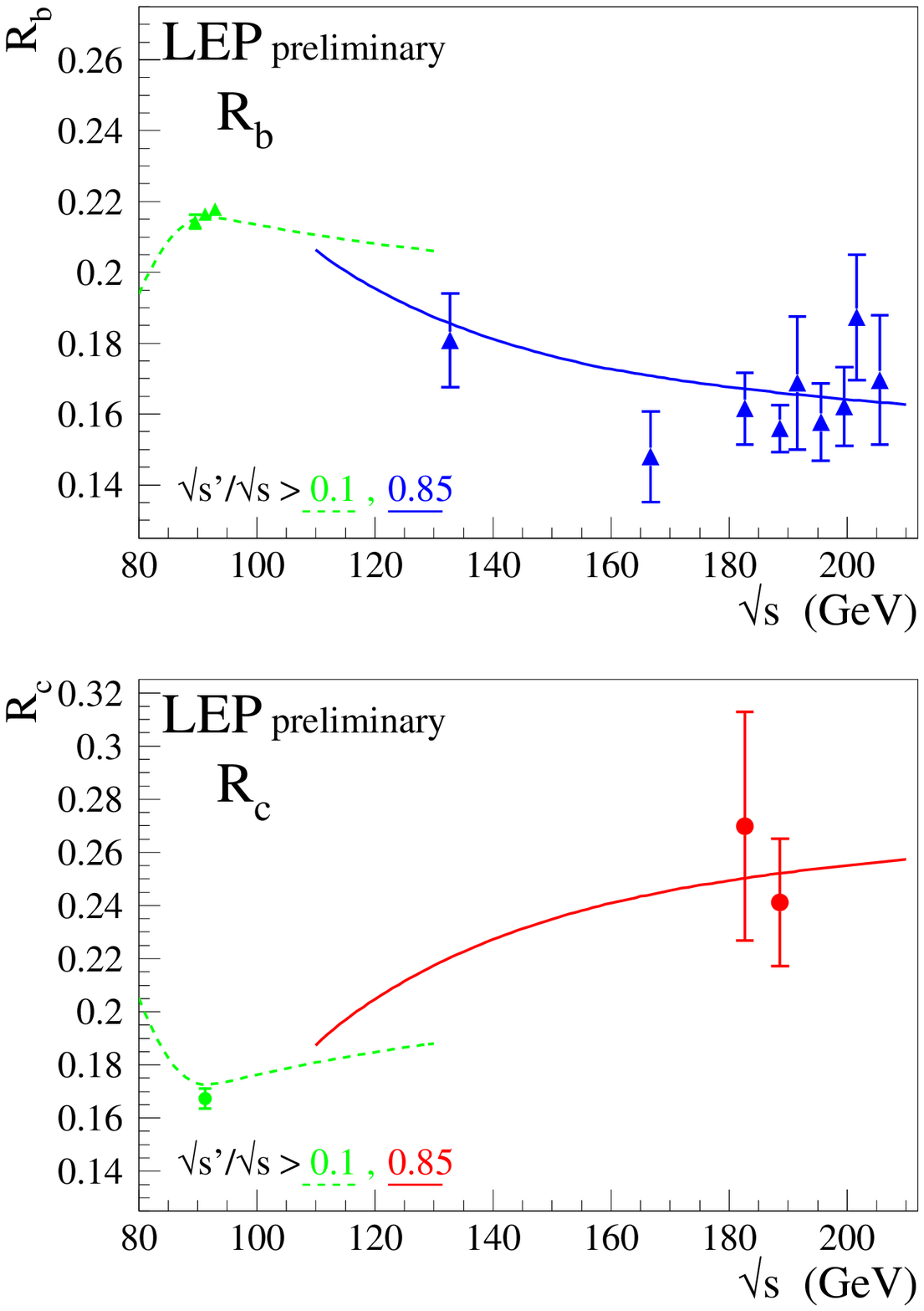}
    \includegraphics[width=0.48\linewidth]{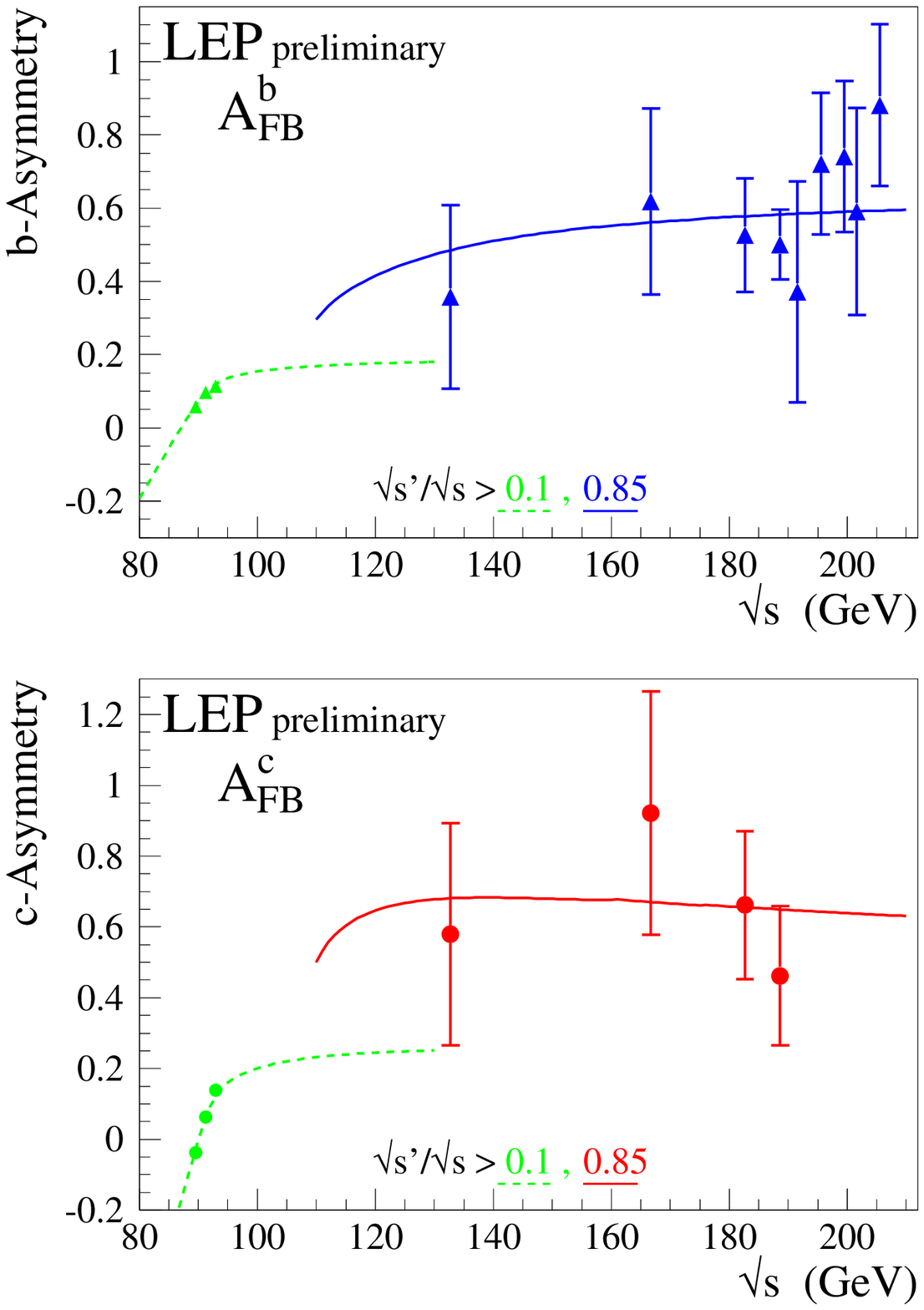}
  }
\caption{ Preliminary combined LEP measurements of 
$\Rb$, $\Rc$ and $\Abb$ and $\Acc$. Solid lines represent the 
Standard Model prediction for the signal definition and dotted lines 
the inclusive prediction. The theoretical predictions are computed 
with ZFITTER~\cite{ref:lep2-ff-hfzfit}. 
The $\LEPI$ measurements are taken from \cite{ref:lep2-ff-hflep1-99,bib-EWEP-99}.
\label{fig:lep2-ff-hfres} }
\end{center}
\end{figure}

The results of the combination are presented in 
Table~\ref{tab:lep2-ff-hfresults} and Figure~\ref{fig:lep2-ff-hfres}.
Because of the large correlation (-0.36)
with $\Rc$ at 183~GeV and 189~GeV,
the errors on the corresponding measurements of $\Rb$ receive an 
additional contribution which is absent at the other energy points.
For other energies where there is no measurement of $\Rc$, the
Standard Model value of $\Rc$ is used in extracting $\Rb$. The
error that this introduces on $\Rb$ is assumed to be negligible.
The results are consistent with the Standard Model predictions of ZFITTER.

\begin{table}[]
\begin{center}
\begin{tabular}{|l||c|c|c|c|}
\hline 
$\sqrt{s}$ (GeV) & $\Rb$
                 & $\Rc$ 
                 & $\Abb$
                 & $\Acc$ \\
\hline\hline
133      & 0.1809 \porm 0.0133    & -   & 0.357 \porm 0.251 &  0.580 \porm 0.314   \\
         & (0.1853)  & -  & (0.487)  &  (0.681)  \\
\hline
167      & 0.1479 \porm 0.0127    &  -  & 0.618 \porm 0.254    &  0.921 \porm 0.344  \\ 
         & (0.1708)  & -  & (0.561)  &  (0.671) \\
\hline 
183      & 0.1616 \porm 0.0101    & 0.270 \porm 0.043  & 0.527 \porm 0.155 & 0.662 \porm 0.209 \\
         & (0.1671)  & (0.250) & (0.578)  &  (0.656) \\
\hline 
189      & 0.1559 \porm 0.0066    & 0.241 \porm 0.024  & 0.500 \porm 0.096 & 0.462 \porm 0.197 \\ 
         & (0.1660)  & (0.252) &  (0.583) & (0.649) \\
\hline
192      & 0.1688 \porm 0.0187    &  -  & 0.371 \porm 0.302 &  - \\
         & (0.1655)  &   - & (0.585)  & - \\ 
\hline 
196      & 0.1577 \porm 0.0109    &  -  & 0.721 \porm 0.194   &  - \\ 
         & (0.1648)  &   - & (0.587)  & - \\ 
\hline
200      & 0.1621 \porm 0.0111    &  -  & 0.741 \porm 0.206    & - \\
         & (0.1642)  &   - & (0.590)  & - \\ 
\hline
202      & 0.1873 \porm 0.0177    &  -  & 0.591 \porm 0.284  & - \\
         & (0.1638)  &   - & (0.591)  & - \\ 
\hline
206      & 0.1696 \porm 0.0182   &  -  & 0.881 \porm 0.221  & - \\
         & (0.1633)  &   - & (0.593)  & - \\ 
\hline
\end{tabular}
\end{center}
\caption{\label{tab:lep2-ff-hfresults}
Results of the global fit, compared to the Standard Model predictions for the 
signal definition, computed with ZFITTER~\cite{ref:lep2-ff-hfzfit}
in parentheses. Quoted errors 
represent the statistical and systematic errors added in quadrature. 
}
\end{table}

\subsection{Interpretation}
\label{sec-interp}

The combined cross sections, asymmetries and results on heavy flavour 
production are interpreted in a variety of models.
The cross section and asymmetry results are used to place limits 
on the mass of a possible additional heavy neutral boson, $\Zprime$.
Limits on contact interactions between leptons and on contact interaction 
between electrons and $b$ and $c$ quarks are obtained.
Heavy flavour results are also used 
within the S-Matrix formalism to give information on the $\gamma$-$\Zzero$
interference for heavy quarks.

\subsection{Models with $\mathbf{\Zprime}$ Bosons}

The combined hadronic and leptonic cross sections and the leptonic 
forward-backward asymmetries 
are used to fit the data to models including an additional heavy
neutral boson, $\Zprime$ within a variety of models~\cite{ref:zprime-thry}.

Fits are made to the mass of a $\Zprime$, $\MZp$, for 4 different models 
referred to as $\chi$, $\psi$, $\eta$ and L-R and
for the Sequential Standard Model \cite{ref:sqsm}, which proposes the 
existence of a $\Zprime$ with exactly the same coupling to fermions as 
the standard Z. $\LEPII$ data alone does not significantly constrain
the mixing angle between the Z and $\Zprime$ fields, $\thtzzp$.
However results from a single experiment where $\LEPI$ data is used in the 
fit show that the mixing is consistent with zero (see for 
example~\cite{ref:lep1zprime}). So for these fits $\thtzzp$ is fixed to 
zero. 

No evidence is found for the existence of a $\Zprime$ boson
in any of the models.
$95\%$ confidence level lower limits on $\MZp$ are obtained, by
integrating the likelihood function\footnote{To be able to obtain 
confidence limits from the likelihood function it is necessary to convert 
the likelihood to a probability density function; this is done 
in a Bayesian approach by 
multiplying by a prior probability function. Simply integrating the 
likelihood is equivalent to multiplying by a uniform prior probability 
function.}. The lower limits on the ${\Zprime}$ mass are shown in 
Table~\ref{tab:zprime_mass_lim}.

\begin{table}[htp]
\begin{center}
\renewcommand{\arraystretch}{1.5}
\begin{tabular}{|cc|c|c|c|c|c|}
\hline
 \multicolumn{2}{|c|}{Model}          & $\chi$  & $\psi$ & $\eta$ & L-R  & SSM   \\
\hline \hline
 $\MZplim$           & ($\GeV/c^{2}$) & 630     & 510    & 400    &  950 & 2260  \\
\hline
\end{tabular}
\end{center}
\caption{
           {95\% confidence level lower limits on the $\Zprime$ mass and
           $\chi$, $\psi$, $\eta$, L-R and SSM models.
                }}
  \label{tab:zprime_mass_lim}
\end{table}

\subsection{Contact Interactions between Leptons}
\label{sec-cntc}
Following reference~\cite{ELPthr}, contact interactions are parameterised 
by an effective Lagrangian, $\cal{L}_{\mathrm{eff}}$, which is added to the 
Standard Model Lagrangian and has the form:
\begin{displaymath}
 \mbox{$\cal{L}$}_{\mathrm{eff}} = 
                        \frac{g^{2}}{(1+\delta)\Lambda^{2}} 
                          \sum_{i,j=L,R} \eta_{ij} 
                           \overline{e}_{i} \gamma_{\mu} e_{i}
                            \overline{f}_{j} \gamma^{\mu} f_{j},
\end{displaymath}
where $g^{2}/{4\pi}$ is taken to be 1 by convention, $\delta=1 (0)$ for 
$f=e ~(f \neq e)$, $\eta_{ij}=\pm 1$ or $0$,
$\Lambda$ is the scale of the contact interactions,
$e_{i}$ and $f_{j}$ are left or right-handed spinors. 
By assuming different helicity coupling between the initial 
state and final state currents, a set of different models can be defined
from this Lagrangian~\cite{Kroha}, with either
constructive ($+$) or destructive ($-$) interference between the 
Standard Model process and the contact interactions. The models and 
corresponding choices of $\eta_{ij}$ are given in Table~\ref{tab:cntcmoddef}.
The models LL, RR, VV, AA, LR, RL, V0, A0 are considered here since 
these models lead to large deviations in the $\eemm$ and $\eett$ channels.
The total hadronic cross section on its own is not particularly
sensitive to contact interactions involving quarks.
For the purpose of fitting contact interaction models to the data, 
a new parameter 
$\epsilon=1/\Lambda^{2}$ is defined; $\epsilon=0$ in the limit that
there are no contact interactions. This parameter is allowed to 
take both positive and negative values in the fits. 

The averaged measurements of the cross sections and forward-backward 
asymmetries for $\eemm$ and $\eett$ from all energies from 130 to 207 GeV
are used. Theoretical uncertainties on the SM predictions of 
$\pm 0.5\%$~\cite{ref:lepffwrkshp} on the cross sections and $\pm 0.005$
on the forward-backward asymmetries, fully correlated between all energies, 
are assumed. 

\begin{figure}[]
 \begin{center}
   \mbox{
    \includegraphics[width=0.55\linewidth]{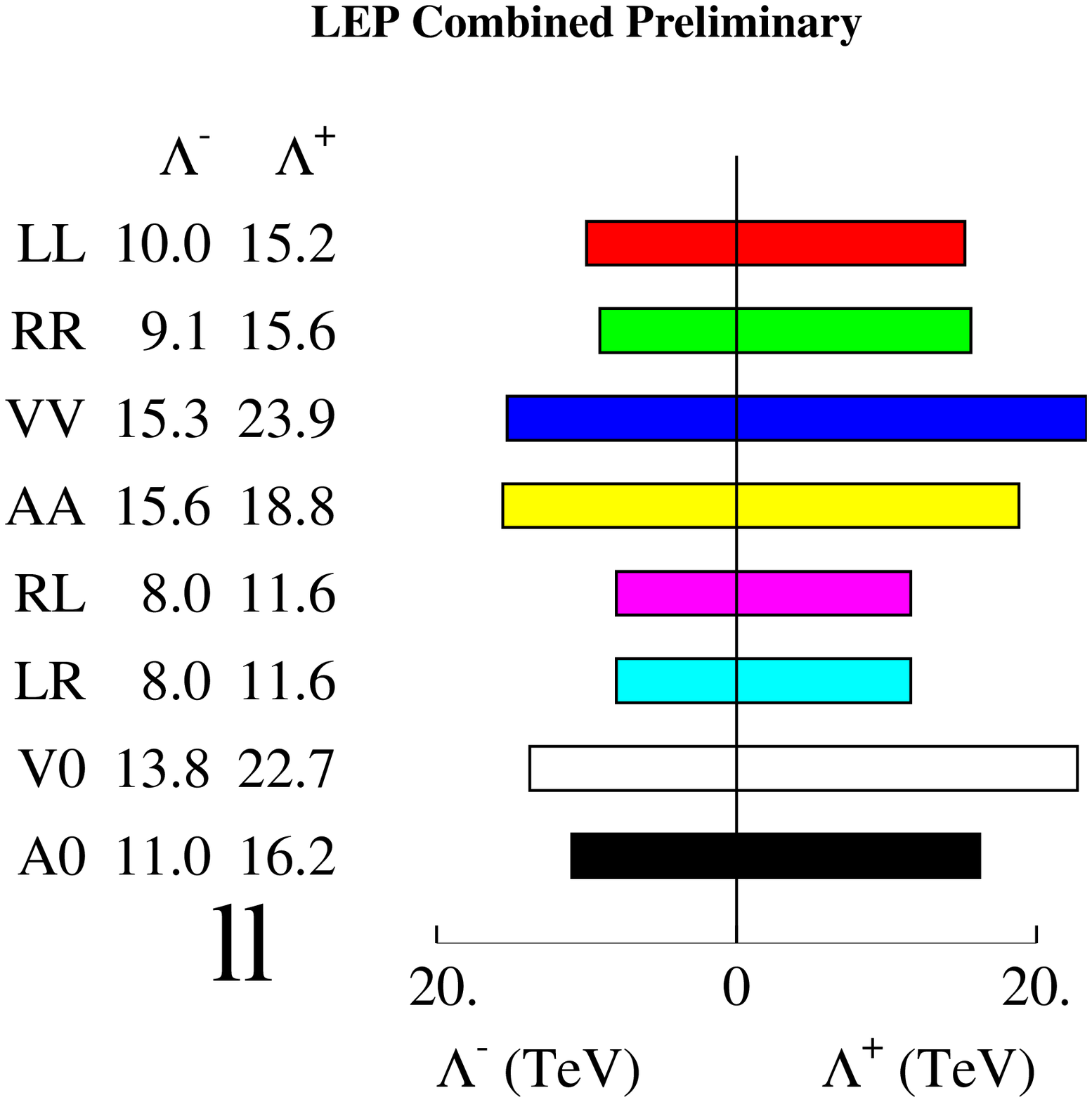}
   }
 \end{center}
 \caption[]{The limits on $\Lambda$ for $\eell$ assuming
            universality in the contact interactions between 
            $\eemm$ and $\eett$.}
 \label{fig:cntc}
\end{figure}

The values of $\epsilon$ extracted for each model are all compatible 
with the Standard Model expectation $\epsilon=0$, at the two standard 
deviation level. These errors on $\epsilon$ are typically a factor of two 
smaller than those obtained from a single LEP experiment with the same data
set. The fitted values of $\epsilon$ are converted into  
$95\%$ confidence level lower limits on $\Lambda$. 
The limits are obtained
by integrating the likelihood function over the physically allowed values, 
$\epsilon \ge 0$ for each $\Lambda^{+}$ limit and $\epsilon \le 0$ for 
$\Lambda^{-}$ limits. The fitted values of $\epsilon$ and the extracted 
limits are shown in Table~\ref{tab:cntc-leptons}. 
Figure~\ref{fig:cntc} shows the limits obtained on the scale $\Lambda$ for
the different models assuming universality between contact interactions
for $\eemm$ and $\eett$.

\begin{table}[]
 \begin{center}
  \begin{tabular}{|c|c|c|c|c|}
   \hline
   Model      & $\eta_{LL}$ & $\eta_{RR}$ & $\eta_{LR}$ & $\eta_{RL}$ \\
   \hline\hline
   LL$^{\pm}$ &   $\pm 1$   &      0      &      0      &      0      \\
   \hline
   RR$^{\pm}$ &      0      &   $\pm 1$   &      0      &      0      \\
   \hline
   VV$^{\pm}$ &   $\pm 1$   &   $\pm 1$   &   $\pm 1$   &   $\pm 1$   \\
   \hline
   AA$^{\pm}$ &   $\pm 1$   &   $\pm 1$   &   $\mp 1$   &   $\mp 1$   \\
   \hline
   LR$^{\pm}$ &      0      &      0      &   $\pm 1$   &      0      \\
   \hline
   RL$^{\pm}$ &      0      &      0      &      0      &   $\pm 1$   \\
   \hline
   V0$^{\pm}$ &   $\pm 1$   &   $\pm 1$   &      0      &      0      \\
   \hline
   A0$^{\pm}$ &      0      &      0      &  $\pm 1$    &   $\pm 1$   \\
   \hline
  \end{tabular}
 \end{center}
 \caption{Choices of $\eta_{ij}$ for different contact interaction models}
 \label{tab:cntcmoddef}.
\end{table}

\begin{table}[]
 \begin{center}
 \newcommand{\TeV}{\mathrm{TeV}}
 \renewcommand{\arraystretch}{1.1}
  \begin{tabular}{|c|r|c|c|}
   \hline
   \multicolumn{4}{|c|}{\boldmath $e^{+}e^{-} \rightarrow \mu^{+}\mu^{-}$\unboldmath} \\
   \hline
   Model  & $\epsilon$ (TeV$^{-2}$)     &  $\Lambda^{-} (\TeV)$ & $\Lambda^{+} (\TeV)$ \\
   \hline
   \hline
   LL & $-0.0066^{+ 0.0039}_{- 0.0042}$ &    8.2 &   14.3 \\
   \hline
   RR & $-0.0069^{+ 0.0045}_{- 0.0054}$ &    8.0 &   13.4 \\
   \hline
   VV & $-0.0023^{+ 0.0017}_{- 0.0018}$ &   13.7 &   21.6 \\
   \hline
   AA & $-0.0033^{+ 0.0032}_{- 0.0012}$ &   13.1 &   19.2 \\
   \hline
   RL & $-0.0052^{+ 0.0067}_{- 0.0074}$ &    7.2 &   10.1 \\
   \hline
   LR & $-0.0052^{+ 0.0067}_{- 0.0074}$ &    7.2 &   10.1 \\
   \hline
   V0 & $-0.0036^{+ 0.0024}_{- 0.0022}$ &   11.9 &   20.6 \\
   \hline
   A0 & $-0.0027^{+ 0.0035}_{- 0.0033}$ &   10.8 &   14.4 \\
   \hline
  \end{tabular}
  \vskip 0.5cm
  \begin{tabular}{|c|r|c|c|}
   \hline
   \multicolumn{4}{|c|}{\boldmath $e^{+}e^{-} \rightarrow \tau^{+} \tau^{-}$\unboldmath} \\
   \hline
   Model  & $\epsilon$ (TeV$^{-2})$     & $\Lambda^{-} (\TeV)$ & $\Lambda^{+} (\TeV)$ \\
   \hline
   \hline
   LL & $-0.0005^{+ 0.0057}_{- 0.0055}$ &    9.5 &    9.8 \\
   \hline
   RR & $-0.0005^{+ 0.0060}_{- 0.0063}$ &    8.7 &    9.5 \\
   \hline
   VV & $-0.0008^{+ 0.0023}_{- 0.0036}$ &   14.4 &   16.1 \\
   \hline
   AA & $-0.0008^{+ 0.0033}_{- 0.0016}$ &   13.4 &   12.2 \\
   \hline
   RL & $-0.0052^{+ 0.0093}_{- 0.0102}$ &    6.5 &    8.8 \\
   \hline
   LR & $-0.0052^{+ 0.0093}_{- 0.0102}$ &    6.5 &    8.8 \\
   \hline
   V0 & $-0.0003^{+ 0.0029}_{- 0.0029}$ &   12.9 &   13.7 \\
   \hline
   A0 & $-0.0026^{+ 0.0049}_{- 0.0050}$ &    9.5 &   12.4 \\
   \hline
  \end{tabular}
  \vskip 0.5cm
  \begin{tabular}{|c|r|c|c|}
   \hline
   \multicolumn{4}{|c|}{\boldmath $e^{+}e^{-} \rightarrow l^{+} l^{-}$\unboldmath} \\
   \hline
   Model  & $\epsilon$ (TeV$^{-2})$     & $\Lambda^{-} (\TeV)$ & $\Lambda^{+} (\TeV)$ \\
   \hline
   \hline
   LL & $-0.0046^{+ 0.0038}_{- 0.0036}$ &   10.0 &   15.2 \\
   \hline
   RR & $-0.0046^{+ 0.0038}_{- 0.0044}$ &    9.1 &   15.6 \\
   \hline
   VV & $-0.0019^{+ 0.0024}_{- 0.0012}$ &   15.3 &   23.9 \\
   \hline
   AA & $-0.0013^{+ 0.0018}_{- 0.0015}$ &   15.6 &   18.8 \\
   \hline
   RL & $-0.0052^{+ 0.0054}_{- 0.0060}$ &    8.0 &   11.6 \\
   \hline
   LR & $-0.0052^{+ 0.0054}_{- 0.0060}$ &    8.0 &   11.6 \\
   \hline
   V0 & $-0.0023^{+ 0.0018}_{- 0.0020}$ &   13.8 &   22.7 \\
   \hline
   A0 & $-0.0027^{+ 0.0028}_{- 0.0028}$ &   11.0 &   16.2 \\
   \hline
  \end{tabular}
 \end{center}
 \caption[]{Fitted values of $\epsilon$
            and $95\%$ confidence limits on the scale,
            $\Lambda$, for constructive ($+$) and destructive interference
            ($-$) with the Standard Model, for the contact interaction models
            discussed in the text. Results are given for $\eemm$, $\eett$
            and $\eell$, assuming universality in the contact interactions 
            between $\eemm$ and $\eett$.}            
 \label{tab:cntc-leptons}
\end{table}

\subsection{Contact Interactions from Heavy Flavour Averages}

Limits on contact interactions between electrons and b and c quarks
are obtained. These results are of particular 
interest since they are inaccessible to ${\mathrm{p\bar{p}}}$ or ep colliders. 
The formalism for describing contact interactions including heavy flavours 
is identical to that described above for leptons.

All heavy flavour $\LEPII$ combined results from 
133 to 205 GeV listed in Table~\ref{tab:lep2-ff-hfresults} are used as inputs. 
For the purpose of fitting contact interaction models to the data, 
$\Rb$ and $\Rc$ are converted to cross sections 
$\sigma_{\bb}$ and $\sigma_{\cc}$ using the averaged ${\qq}$ cross section of 
section \ref{sec-ave-xsc-afb} corresponding to signal definition 2. 
In the calculation of errors, the correlations between $\Rb$, $\Rc$ and 
$\sigma_{\qq}$ are assumed to be negligible. 

\begin{figure}[]
 \begin{center}
  \begin{tabular}{c}
    \includegraphics[width=0.48\linewidth]{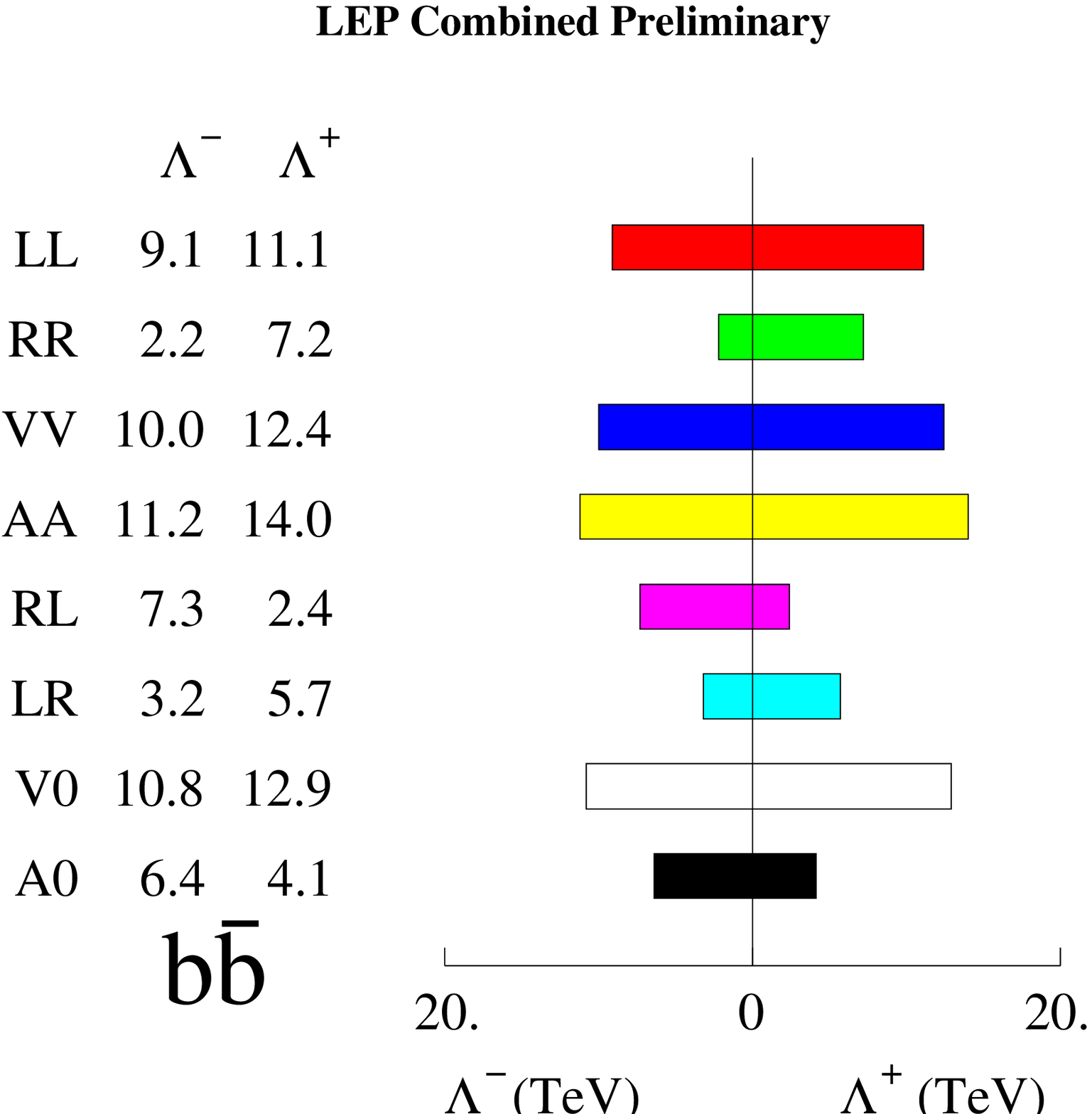}
    \includegraphics[width=0.48\linewidth]{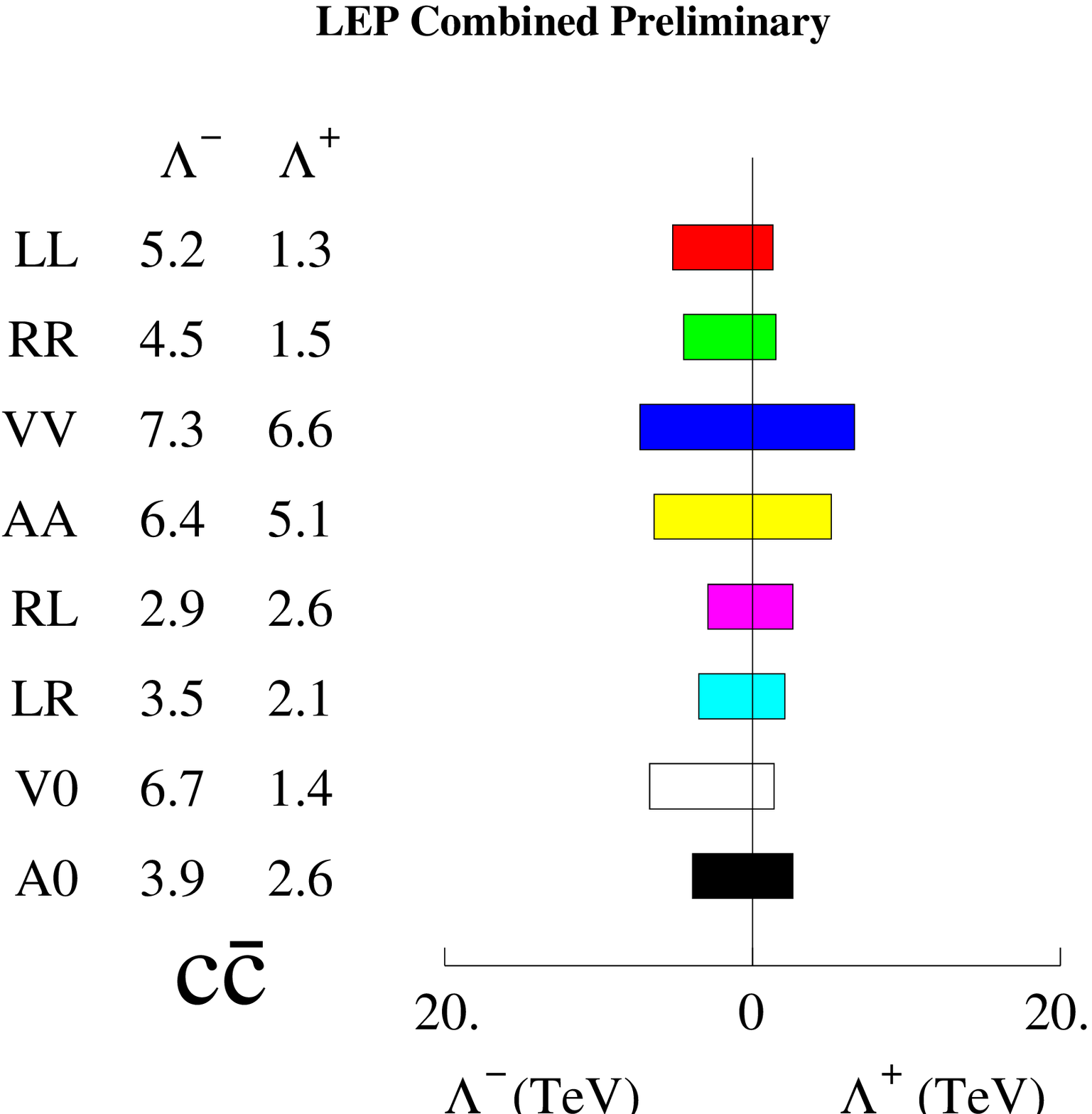}
  \end{tabular}
 \end{center}
\caption{95\% CL limits on the scale of contact interactions in $\eebb$
and $\eecc$ using Heavy Flavour LEP combined results from 133 to 205 GeV.} 
 \label{fig:cibc}
\end{figure}

The fitted values of ${\mathrm \epsilon \ = \ \frac{1}{\Lambda^2}}$ and their
68$\%$ confidence level uncertainties together with the 95$\%$ confidence 
level lower limits 
on ${\mathrm{\Lambda}}$ are shown in Table~\ref{tab:cibc}.
Figure~\ref{fig:cibc} shows the limits obtained on the scale,
${\mathrm \Lambda}$, of models with different helicity combinations 
involved in the interactions.

\begin{table}[]
\begin{center}
\begin{tabular}{|c|r|c|c|}
\hline
\multicolumn{4}{|c|}{\boldmath $e^{+}e^{-} \rightarrow b\overline{b}$\unboldmath} \\
\hline 
Model & $\epsilon$ (TeV$^{-2}$)        & $\Lambda^{-}$ (TeV) & $\Lambda^{+}$ (TeV) \\
\hline 
\hline 
LL    & $-0.0025^{+0.0049}_{-0.0052}$  &        9.1          &   11.1  \\
\hline 
RR    & $-0.1890^{+0.1290}_{-0.0151}$  &        2.2          &  7.2  \\
\hline 
VV    & $-0.0020^{+0.0041}_{-0.0043}$  &       10.0          & 12.4  \\
\hline 
AA    & $-0.0018^{+0.0032}_{-0.0034}$  &       11.2          & 14.0  \\
\hline 
RL    & $ 0.0190^{+0.1299}_{-0.0201}$  &        7.3          &  2.4  \\
\hline 
LR    & $-0.0428^{+0.0408}_{-0.0367}$  &        3.2          &  5.7  \\
\hline 
V0    & $-0.0018^{+0.0035}_{-0.0037}$  &       10.8          & 12.9  \\
\hline 
A0    & $ 0.0266^{+0.0234}_{-0.0255}$  &        6.4          &  4.1  \\
\hline 
\end{tabular}
\end{center}
\vskip 0.5cm
\begin{center}
\begin{tabular}{|c|r|c|c|}
\hline
\multicolumn{4}{|c|}{\boldmath $e^{+}e^{-} \rightarrow c\overline{c}$\unboldmath} \\
\hline 
Model & $\epsilon$ (TeV$^{-2}$)       & $\Lambda^{-}$ (TeV) & $\Lambda^{+}$ (TeV) \\
\hline 
\hline
LL    & $ 0.0127^{+0.5957}_{-0.0264}$ &       5.2     &  1.6   \\
\hline
RR    & $ 0.0466^{+0.3781}_{-0.0576}$ &       4.5     &  1.5   \\
\hline
VV    & $-0.0008^{+0.0109}_{-0.0103}$ &       7.3     &  6.6   \\
\hline
AA    & $ 0.0046^{+0.0168}_{-0.0151}$ &       6.4     &  5.1   \\
\hline
RL    & $ 0.0127^{+0.0845}_{-0.0845}$ &       2.8     &  2.6   \\
\hline
LR    & $ 0.0874^{+0.1049}_{-0.1127}$ &       3.5     &  2.1   \\
\hline
V0    & $ 0.0036^{+0.0181}_{-0.0135}$ &       6.7     &  1.4   \\
\hline
A0    & $ 0.0499^{+0.0691}_{-0.0691}$ &       3.9     &  2.6   \\
\hline 
\end{tabular}
\end{center}
 \caption[]{Fitted values of $\epsilon$
            and $95\%$ confidence limits on the scale,
            $\Lambda$, for constructive ($+$) and destructive interference
            ($-$) with the Standard Model, for the contact interaction models
            discussed in the text. From
            combined $\mathrm{b \bar{b}}$ and $\mathrm{c \bar{c}}$ results with \CoM\ 
            energies from 133 to 205 GeV.}
 \label{tab:cibc}
\end{table}

\subsection{S-Matrix Parameters for Heavy Flavour Production}
\label{sec-hvflv-smat}

The S-Matrix formalism~\cite{ref:SMat} parameterises the cross sections and 
forward-backward asymmetries for two-fermion production in terms of the 
exchange of a massless ($\gamma$) and a massive vector boson (Z):

\begin{displaymath}
 \sigma_{\mathrm{tot}}^{0,f}(s) = \frac{4}{3} \pi \alpha^{2} +
      \left[ 
        \frac{g_{f}^{\mathrm{tot}}}{s} + 
              \frac{j_{f}^{\mathrm{tot}} (s-{\overline{m}}_{\mathrm{Z}}^{2}) + r_{f}^{\mathrm{tot}} s}
                   {(s-{\overline{m}}_{\mathrm{Z}}^{2})^{2} +
                         {\overline{m}}_{\mathrm{Z}}^{2} {\overline{\Gamma}}_{\mathrm{Z}}^{2}}
      \right]
\end{displaymath}
\begin{displaymath}
  \sigma_{\mathrm{fb}}^{0,f}(s) = \frac{4}{3} \pi \alpha^{2} +
      \left[ 
        \frac{g_{f}^{\mathrm{fb}}}{s} + 
              \frac{j_{f}^{\mathrm{fb}} (s-{\overline{m}}_{\mathrm{Z}}^{2}) + r_{f}^{\mathrm{fb}} s}
                   {(s-{\overline{m}}_{\mathrm{Z}}^{2})^{2} +
                         {\overline{m}}_{\mathrm{Z}}^{2} {\overline{\Gamma}}_{\mathrm{Z}}^{2}}
      \right]
\end{displaymath}

\begin{displaymath}
  A_{\mathrm{fb}}^{0,f}(s) = \frac{3}{4}\frac{\sigma_{\mathrm{fb}}^{0,f}(s)}
                                  {\sigma_{\mathrm{tot}}^{0,f}(s)}
\end{displaymath}
The mass, ${\overline{m}}_{\mathrm{Z}}$, and width, ${\overline{\Gamma}}_{\mathrm{Z}}$, used in
the S-Matrix fits are slightly different from the usual mass $\MZ$, and width
$\GZ$ which are defined using an $s$-dependent width term in the
Breit-Wigner resonance of the Z:
\begin{eqnarray}
  {\overline{m}}_{\mathrm{Z}}      & \sim & \MZ - 34.1 \mathrm{Mev} \nonumber \\ 
  {\overline{\Gamma}}_{\mathrm{Z}} & \sim & \GZ - 0.9  \mathrm{Mev} \nonumber
\end{eqnarray}
The parameters $g$, $r$ and $j$ parameterise the
cross sections and forward-backward asymmetries arising from the exchange
of the $\gamma$ ($g$) and the Z ($r$) and the interference of the two ($j$).
Values for these parameters can be obtained for each fermion species
and also for a combination of all $\qq$ final states, $f=\mathrm{had}$.
Values of the  $g$, $r$ and $j$ parameters and ${\overline{\Gamma}}_{\mathrm{Z}}$
are computed in the Standard Model for comparison with the results of 
the fits.

S-Matrix fits have already been performed using the hadronic and leptonic
cross section and leptonic forward-backward asymmetry
data from $\LEPI$ and $\LEPII$ for energies up to 
172 GeV~\cite{bib-EWEP-18},
providing constraints on $\rhadtot$ and $\jhadtot$, the parameters describing
charged lepton production and ${\overline{m}}_{\mathrm{Z}}$ and 
${\overline{\Gamma}}_{\mathrm{Z}}$.
The results of these existing fits and the full error matrix are used to 
constrain these parameters in the fit performed here.

$\LEPI$ and SLD heavy flavour averages~\cite{ref:hflep1-00}, $\Rb$, $\Rc$, $\Abb$ and 
$\Acc$ are used to constrain the S-Matrix parameters $\rbctot$ and 
$\rbcfb$ that describe the heavy quark couplings to the Z.
By including the combined $\LEPII$ heavy flavour measurements from 
133 to 205 GeV listed in Table~\ref{tab:lep2-ff-hfresults} values of the parameters 
$\jbctot$ and $\jbcfb$ that describe $\gamma$-$\Zzero$ interference in 
heavy quark production are obtained. Contours for these parameters
are given at the end of this section.

The $\LEPII$ average values of $\Rb$ and $\Rc$ from Section~\ref{sec-hvflv}
for \CoM\ energies from 130 up to 166 GeV are used 
directly in the fit. These are largely uncorrelated with the
measured total cross sections, therefore, the correlations between
the existing S-Matrix fits parameters and these measurements are neglected. 
The measurements are fitted using the ratio of the predicted
b and c cross sections and the total hadronic cross section.

For energies of 183 GeV and above, the flavour tagged measurements are 
more precise than those from 133--166 GeV. The uncertainties on 
$\jhadtot$ from the existing fits introduces a sizeable uncertainty on
the prediction of $\Rb$ and $\Rc$. For these energies the measurements
of $\Rb$ and $\Rc$ are first converted into cross sections for $\bb$ and 
$\cc$ production by multiplying by the the total hadronic cross sections
from section~\ref{sec-ave-xsc-afb}. The full error matrix of the quantities
and the correlations with the lower energy $\Rb$ and $\Rc$ values 
are computed. Correlations between the existing S-Matrix fit results and
the total hadronic cross sections from 183 GeV and above are neglected. 
For this reason the highest energy hadronic cross sections 
are not used to improve the fits to the inclusive hadronic S-Matrix 
parameters.

The forward-backward asymmetries for $b$ and $c$ quark production are fitted
directly to predictions of the asymmetries at all energies.

Predictions of the S-Matrix formalism are made using the
SMATASY~\cite{ref:SMATASY} program. The couplings to the photon are 
fixed to the expectation from the Standard Model.

In summary, the following parameters are obtained from the fit:
\begin{list}{$\bullet$}{\setlength{\itemsep}{0ex}
                        \setlength{\parsep}{0ex}
                        \setlength{\topsep}{0ex}}

 \item
 The mass and total width of the $\mathrm{Z}$ boson, the S-Matrix parameters 
 $\rhadtot$ and $\jhadtot$ for the total 
 hadronic cross section, and the parameters $\rltot$, 
 $\jltot$, $\rlfb$, and $\jlfb$ for 
 leptons\footnote{Lepton universality is assumed.}.

 \item
 The S-Matrix parameters $\rbctot$ and $\rbcfb$ that describe 
 the total $\bb$ and $\cc$ cross sections and asymmetries due to Z boson 
 exchange.

 \item
 The four parameters $\jbctot$ and $\jbcfb$ that describe the 
 the effect of 
 $\gamma$-$\Zzero$ interference on the energy dependence 
 of the bottom and charm 
 cross sections and asymmetries.

\end{list}

The following data are used:

\begin{list}{$\bullet$}{\setlength{\itemsep}{0ex}
                        \setlength{\parsep}{0ex}
                        \setlength{\topsep}{0ex}}

 \item
 The results of the existing S-Matrix fits~\cite{bib-EWEP-18}, 
 derived from $\LEPI$ data and including total hadronic cross sections 
 up to 172 GeV.

 \item 
 $\LEPI$ and SLD heavy flavour averages~\cite{ref:hflep1-00}.

 \item 
 $\LEPII$ heavy flavour averages $\Rb$ and $\Rc$ for 
 energies up to and including 172 GeV from Table~\ref{tab:lep2-ff-hfresults}.

 \item 
 Values of the flavour tagged b and c cross sections $\sigma_{b}$ and 
 $\sigma_{c}$ derived from measurements of $\Rb$ and $\Rc$ and the total 
 hadronic cross section for energies of 183 GeV and above, 
 from section~\ref{sec-ave-xsc-afb} and Table~\ref{tab:lep2-ff-hfresults}.

 \item
 $\LEPII$ heavy flavour averages $\Abb$ and $\Acc$ for all $\LEPII$ energies
 from Table~\ref{tab:lep2-ff-hfresults}.

\end{list}

\begin{figure}[tp]
 \begin{center}
  \begin{tabular}{c}
    \includegraphics[width=0.48\linewidth]{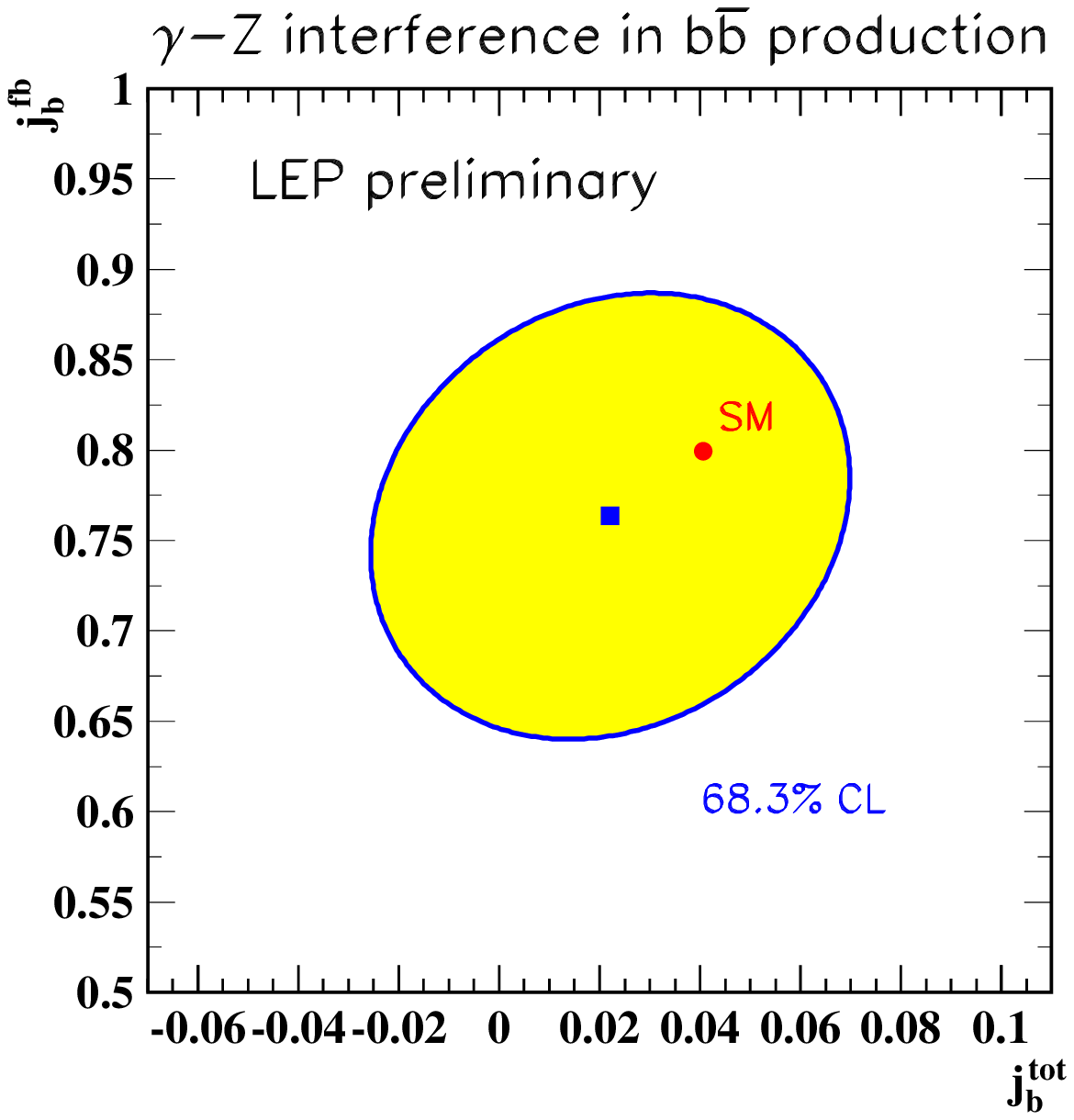}
    \includegraphics[width=0.48\linewidth]{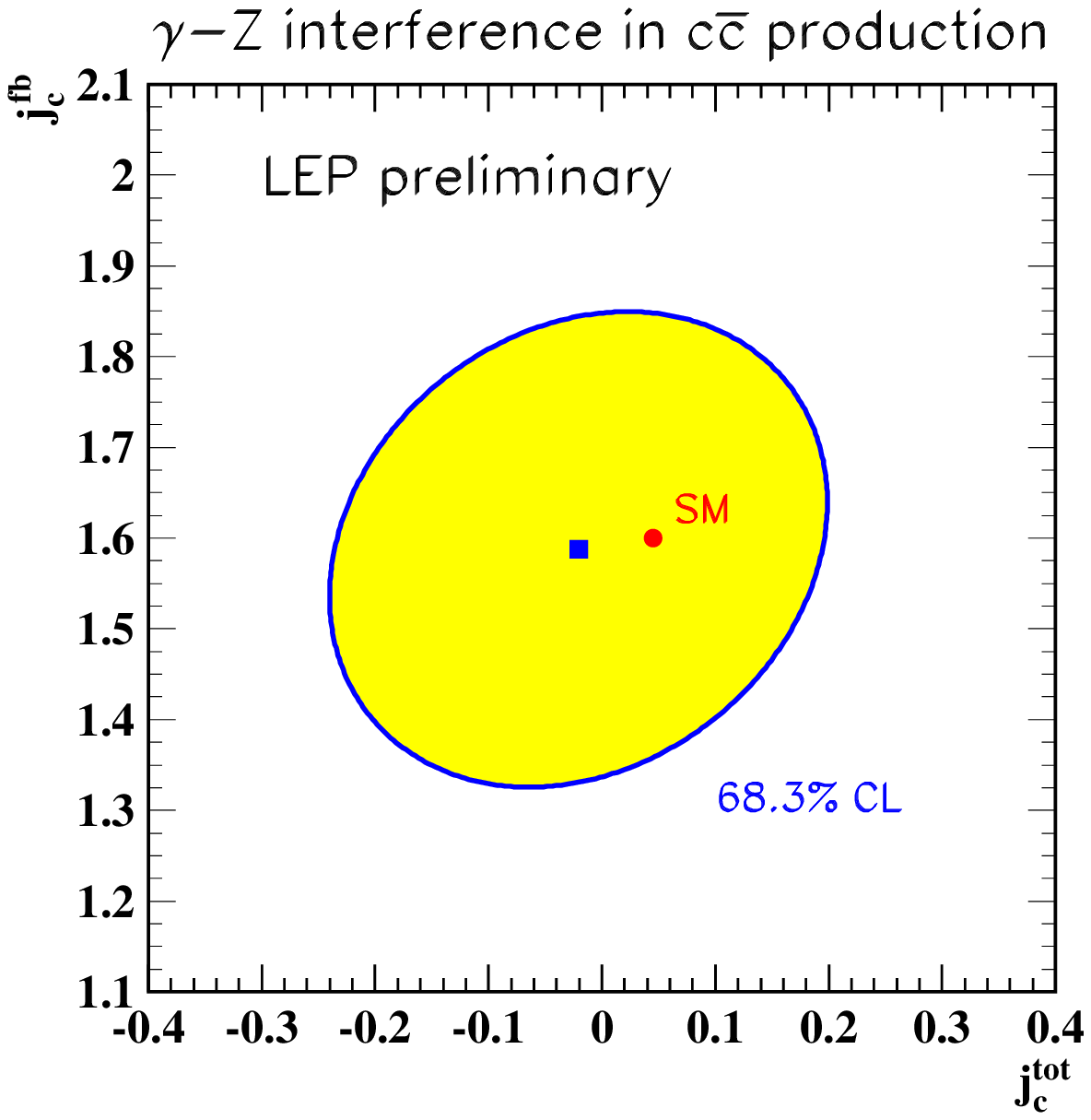}
  \end{tabular}
 \end{center}
 \caption{Preliminary combined LEP results on the S-Matrix parameters
           describing $\gamma$-$\Zzero$ interference in 
           $\bb$ and $\cc$ production.
           The expectations of the SM are also shown.}
 \label{fig-hf_smat}
\end{figure}

The results for the S-Matrix parameters $\jbctot$ and $\jbcfb$ are
shown in Figure~\ref{fig-hf_smat} for both bottom and charm quark production.
Good agreement is observed with the Standard Model 
prediction~\cite{ref:SMATASY} for  $\gamma$-$\Zzero$  interference in 
heavy quark production.

\clearpage
\boldmath
\section{Measurement of W Boson Properties at $\LEPII$}
\unboldmath
\label{sec-MW}
\updates{There are new results on the WW cross section and on the W mass.  
The W decay branching
  ratios are updated.
}

\noindent
In the year 2000 LEP ran at \CoM\  energies larger than 200 \GeV,
with a maximum of 209 \GeV\ (see Table~\ref{tab:lep2-ff-ecms}). 
The collected data are divided
in two ranges of \roots, below and above 205.5 \GeV.
The two data sets have mean \CoM\  energies of 204.9 and 206.7 \GeV~ 
and the respective integrated luminosities used for the analyses
considered in this note  are 60 and 30 \pb. The same
\CoM\  energy binning is also applied for the measurement of the
Z--pair cross section.
All data recorded 
from 1996 to 1999 over a range of \CoM\ energies, $\roots=161-202$~\GeV\ 
are  used to extract the W boson mass and width.

\subsection{W--pair Production Cross Section}
\label{sec:WWxsec}

All experiments have published final results on the W--pair (CC03) production 
cross section for \CoM\  energies up to 189 \GeV~\cite{bib:aleww189,bib:delww189,
bib:ltrww189a,bib:opaww189}\footnote{In the combination the preliminary 
results of the L3  collaboration~\cite{bib:ltrww189b} are still used,
as the final results only became available shortly before the ICHEP conference.}.
ALEPH~\cite{bib:aleww1999}, DELPHI~\cite{bib:delww1999} and L3
\cite{bib:ltrww1999} preliminary results at $\roots=192$--202 \GeV\ are
unchanged with respect to combination of results prepared for the
year 2000 winter conferences.
OPAL  has updated its preliminary 
results at $\roots=200$ and 202 \GeV~\cite{bib:opaww2000} to include the
full luminosity collected in 1999, whereas the results at $\roots=192$ and 
196 \GeV~\cite{bib:opaww1999} are unchanged since the winter conferences.
All experiments have contributed preliminary results~\cite{bib:aleww2000,
bib:delww2000,bib:ltrww2000,bib:opaww2000} based on the analysis of 
year 2000 data.

Results from different experiments are combined using the method described in
\cite{bib:lyons} and taking into account, when relevant, the correlations between 
the systematic uncertainties, which arise mainly from the use of the same \MC\  
programs to predict background cross sections and to simulate the hadronisation processes.

The results from each experiments for the W--pair production cross section, 
assuming Standard Model values for the W decay branching ratios, are shown 
in Table~\ref{tab:wwxsec}, together with the LEP combination. In the averaging
procedure the QCD component of the systematic errors
\footnote{The QCD component of the systematic error, includes for example uncertainties 
on the 4--jet rate, 
on the fragmentation modelling and estimates of possible effects 
of colour reconnection and Bose-Einstein correlations.}
 from each individual
measurement is taken to be fully correlated between experiments.
This common error ranges between 0.07 and 0.10 pb. These results
supersede the ones presented in~\cite{bib:wwmor00} for $\roots=189$,
200 and 202 \GeV.

\begin{figure}[htbp]
\centering
\begin{center}
  \mbox{\includegraphics[width=\linewidth]{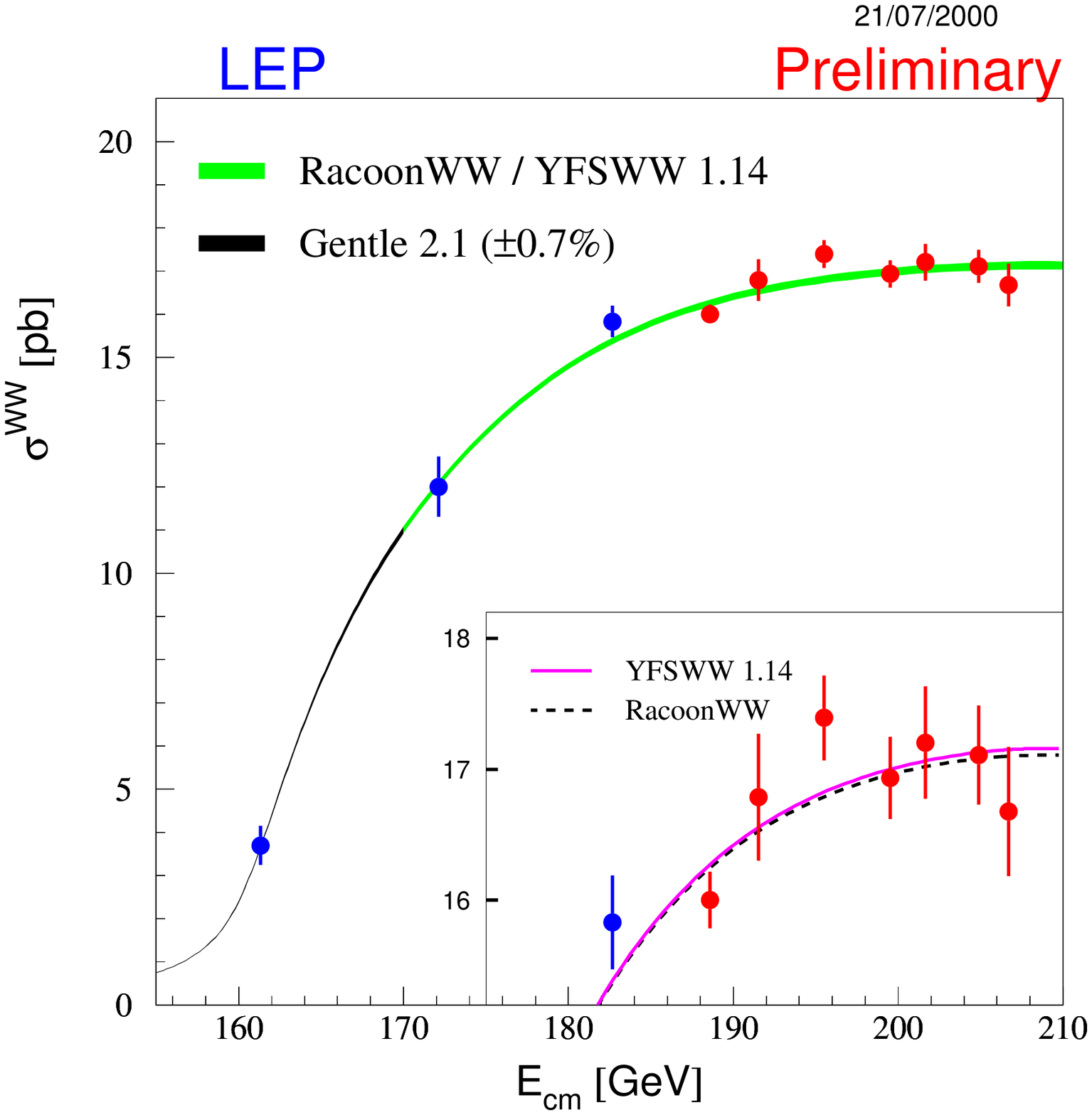}}
\end{center}
\vspace*{-0.6cm}
\caption{Measurements of the W--pair production cross section 
compared to the predictions of \RacoonWW, and \YFSWW\  and \Gentle\
for $\Mw=80.35$ \GeV. 
The
shaded area represent the uncertainty on the prediction 
which is $\pm0.7$\% for $\roots<170$ \GeV\ and 
varies between $\pm0.4$ and $\pm0.7$\% for $\roots>170$ \GeV.}
\label{fig:sww_vs_sqrts}
\end{figure}

Figure~\ref{fig:sww_vs_sqrts} shows the total W--pair cross section 
measured as a function of the LEP \CoM\  energy.
The experimental
points are compared with new calculations (\RacoonWW~\cite{bib:racoonww}
and \YFSWW~\cite{bib:yfsww}) based on the double pole approximation 
(DPA)~\cite{bib:dpa}
for $\Mw=80.35$ \GeV. 
These two codes have been extensively compared
and agree at a level better than 0.2\% at the \LEPII\ energies. The
theoretical uncertainty of the DPA decreases from 0.7\% at 170~\GeV\
to a level of 0.4\% at \CoM\  energies larger 
than 200~\GeV\footnote{The theoretical uncertainty $\Delta\sigma/\sigma$ on the 
                        W--pair production cross section calculated in the DPA 
                        can be parametrised as
                        $\Delta\sigma/\sigma=0.4\oplus0.072\cdot t_1\cdot t_2$,
                        where $t_1=(200-2\Mw)/(\roots-2\Mw)$ and
                        $t_2=(1-(\frac{2\cdot \Mw}{200})^2)/
                        (1-(\frac{2\cdot \Mw}{\sqrt{s}})^2)$\cite{bib:dpaerr}.}.
This theoretical uncertainty is represented by
the grey band in Figure~\ref{fig:sww_vs_sqrts}. An error
of 50 \MeV\ on the W mass translates into a 0.1\% error
on the cross section predictions at 200 \GeV.
The DPA is only valid 
away from the production threshold, therefore below 170 \GeV\ the experimental
results are compared with the \Gentle~\cite{bib:gentle} prediction,
which is adjusted to reproduce the DPA results at higher energies. All
results, up to the highest \CoM\  energies, are in agreement with
the theoretical predictions.

\begin{table}[hbtp]
\centering
\begin{tabular}{|c|c|c|c|c|c|c|} 
\hline
\roots & \multicolumn{5}{|c|}{Cross section  (pb)} & $\chi^2/\textrm{d.o.f.}$ \\
\cline{2-6} 
(\GeV)          & ALEPH          & DELPHI         &
                 L3           & OPAL           &
                 LEP              &                  \\
\hline
161.33         & $\pz4.23\pm0.75^*$    & $\pz3.67\pm0.93^*$ &
                  $\pz2.89\pm0.77^*$   & $\pz3.62\pm0.89^*$ &
                  $\pz3.69\pm0.45$     & 1.3/3           \\
172.12         & $11.7\pz\pm1.3^*\pz$     & $11.58\pm1.4^*\pz$ &
                  $12.27\pm1.4^*\pz$   & $12.3\pz\pm1.3^*\pz$ &
                 $12.0\pz\pm0.7\pz$ & 0.22/3           \\
182.67         & $15.57\pm0.68^*$   & $15.86\pm0.74^*$ &
                  $16.53\pm0.72^*$  & $15.43\pm0.66^*$ &
                  $15.83\pm0.36$    & 1.49/3           \\
188.63         & $15.71\pm0.38^*$   & $15.83\pm0.43^*$ &
                 $16.20\pm0.46\pz$  & $16.30\pm0.38^*$ &
                 $16.00\pm0.21$     & 1.57/3           \\
191.6\phz      & $17.23\pm0.91\pz$  & $16.90\pm1.02\pz$  &
                 $16.39\pm0.93\pz$  & $16.60\pm0.97\pz$  &
                 $16.78\pm0.48$     & 0.47/3           \\
195.5\phz      & $17.00\pm0.57\pz$  & $17.86\pm0.63\pz$  &
                 $16.67\pm0.60\pz$  & $18.59\pm0.74\pz$  &
                 $17.39\pm0.32$     & 5.26/3           \\
199.5\phz      & $16.98\pm0.56\pz$  & $17.35\pm0.60\pz$  &
                 $16.94\pm0.62\pz$  & $16.32\pm0.66\pz$  &
                 $16.93\pm0.31$     & 1.39/3           \\
201.6\phz      & $16.16\pm0.76\pz$  & $17.67\pm0.84\pz$  &
                 $16.95\pm0.88\pz$  & $18.48\pm0.91\pz$  &
                 $17.20\pm0.43$     & 4.27/3           \\
204.9\phz      & $16.70\pm0.64\pz$  & $18.81\pm0.80\pz$  &
                 $17.70\pm0.86\pz$  & $15.84\pm0.71\pz$  &
                 $17.11\pm0.38$     & 8.79/3           \\
206.7\phz      & $17.01\pm0.88\pz$  & $16.50\pm1.05\pz$  &
                 $17.20\pm1.03\pz$  & $15.96\pm0.96\pz$  &
                 $16.68\pm0.49$     & 1.01/3           \\
\hline
\end{tabular}
\caption{W--pair production cross sections from the four LEP
experiments and combined values for the seven of the recorded \CoM\  
energies. All results are preliminary with the exception of those indicated
by $^*$. A common systematic error of (0.07--0.10) pb is taken
into account in the averaging procedure. Final results for the
combined W--pair production cross section at lower \CoM\  energies
can be found in \cite{bib-EWPPE97-154,bib-EWEP-18,bib-EWEP-99} .}
\label{tab:wwxsec}
\end{table}

\subsection{W Decay Branching Fractions}
\label{sec:Wbranching}

From the cross sections for the individual W decay channels measured by the
four experiments at all energies larger than 161 \GeV, the W decay branching 
ratios ($\mathcal{B}(\mathrm{W}\rightarrow\textrm{f}\overline{\textrm{f}}')$) 
are determined, with and without the assumption of lepton universality. 
ALEPH~\cite{bib:aleww2000} and OPAL~\cite{bib:opaww2000} have updated their 
results to include the data collected in the year 2000 at \CoM\  energies up 
to 208 \GeV, whereas DELPHI~\cite{bib:delww1999} and L3~\cite{bib:ltrww1999} 
have not updated their preliminary results prepared for the year 2000 winter conferences. 
The results from the single experiments are given in Table~\ref{tab:wwbra} 
together with the result of the LEP combination. Correlated errors between the 
different channels of individual experiments are taken into account in the 
averaging procedure, together with the QCD component of the systematic error 
which is assumed to be fully correlated among experiments. 

The results of the fit which does not make use of the lepton universality
assumption show a negative correlation of 21.0\% (18.7\%) between the 
\Wtotnu\  and \Wtoenu\  (\Wtomnu)  branching ratios, while between the
electron and muon decay channels there is a positive correlation of 4.4\%.
These branching ratios are all consistent within the errors and allow to
test lepton universality in the decay of on--shell W bosons at the level 
of 3.3\%. Assuming lepton universality, the measured hadronic branching 
ratio is $(67.78\pm0.32)\%$ and the leptonic one is $(10.74\pm0.10)\%$. 
These results are consistent with the Standard Model expectations. 
Statistical and systematic uncertainties give approximately equal 
contributions to the total errors on the branching ratios. The systematic 
error is divided equally in the two components which are correlated and 
uncorrelated between experiments.

\begin{table}[hbtp]
\centering
\begin{tabular}{|c|c|c|c|c|} 
\cline{2-5}
\multicolumn{1}{c|}{$\quad$} & \multicolumn{3}{|c|}{Lepton} & Lepton \\
\multicolumn{1}{c|}{$\quad$} & \multicolumn{3}{|c|}{non--universality} & universality \\
\hline
Experiment & \Wtoenu\  & \Wtomnu\  & \Wtotnu\  & {\mbox{$\mathrm{W}\rightarrow\mathrm{hadrons}$}} \\
           & (\%) & (\%) & (\%) & (\%)  \\
\hline
ALEPH    & $11.19\pm0.34$ & $11.05\pm0.32$ & $10.53\pm0.42$ & $67.22\pm0.53$ \\
DELPHI   & $10.33\pm0.45$ & $10.68\pm0.34$ & $11.28\pm0.56$ & $67.81\pm0.61$ \\
L3     & $10.22\pm0.36$ & $\phz9.87\pm0.38$ & $11.64\pm0.51$ & $68.47\pm0.59$ \\
OPAL     & $10.52\pm0.37$ & $10.56\pm0.35$ & $11.15\pm0.49$ & $67.86\pm0.62$ \\
\hline
LEP        & $10.62\pm0.20$ & $10.60\pm0.18$ & $11.07\pm0.25$ & $67.78\pm0.32$ \\
\hline
$\chi^2/\textrm{d.o.f.}$ & \multicolumn{3}{|c|}{11.1/9} & 13.2/9 \\
\hline
\end{tabular}
\caption{Summary of W branching ratios derived from W--pair production 
cross sections measurements up to 208 \GeV\ \CoM\  energy (DELPHI  and
L3   results are based only on data collected up to $\roots=202 $ \GeV).
A common systematic error of (0.04--0.13)\% is taken into account in the 
averaging procedure.} 
\label{tab:wwbra} 
\end{table}

Within the Standard Model, the branching fractions of the W boson depend
on the six matrix elements $|\mathrm{V}_{\mathrm{qq'}}|$ of the 
Cabibbo-Kobayashi-Maskawa (CKM) quark mixing matrix not involving the top
quark. In terms of these $|\mathrm{V}_{\mathrm{qq'}}|$  the leptonic
branching ratio of the W boson, $\mathcal{B}(\Wtolnu)$, is given by

\begin{equation}
1/\mathcal{B}(\Wtolnu)\quad=\quad 3 \bigl\{1+[1+\alpha_s(\Mw^2)/\pi] \sum_{i=\mathrm{u,c},\;j=\mathrm{d,s,b}} |\mathrm{V}_{ij}|^2\bigr\},
\end{equation} 

\noindent
where $\alpha_s(\Mw^2)$ is the strong coupling
constant. Taking $\alpha_s(\Mw^2)=(0.121\pm0.002)$,
the measured leptonic branching ratio of the W yields

\begin{equation}
\sum_{i=\mathrm{u,c},\;j=\mathrm{d,s,b}} |\mathrm{V}_{ij}|^2 \quad = \quad
2.026\,\pm\,0.030\,(\mathrm{Br.})\,\pm\,0.001\,(\alpha_s),
\end{equation} 

\noindent
where the first error is due to the uncertainty on
the branching ratio measurement and the second to the uncertainty on
$\alpha_s$. Using the experimental knowledge of the sum
$|\mathrm{V}_{\mathrm{ud}}|^2+|\mathrm{V}_{\mathrm{us}}|^2+|\mathrm{V}_{\mathrm{ub}}|^2+
 |\mathrm{V}_{\mathrm{cd}}|^2+|\mathrm{V}_{\mathrm{cb}}|^2=(1.0477\pm0.0074)$~\cite{bib-pdg2000}, 
the above result can be interpreted as a measurement of $|\mathrm{V}_{\mathrm{cs}}|$ 
which is the least well determined of these matrix elements:

\begin{equation}
|\mathrm{V}_{\mathrm{cs}}|\quad=\quad0.989\,\pm\,0.016.
\end{equation}

The error includes the uncertainties on $\alpha_s$ and on the other
$|\mathrm{V}_{\mathrm{qq'}}|$ matrix elements but is dominated
by the experimental error on the W branching fractions.
The uncertainty in the sum of the other CKM matrix elements, which
is dominated by the uncertainty on $|\mathrm{V}_{cd}|$, contributes 
a negligible uncertainty of 0.004 to this determination of 
$|\mathrm{V}_{cs}|$.

\subsection{W Mass Measurement}

The W boson mass results presented here are obtained from
data recorded over a range of \CoM\ energies, $\roots=161-202$~\GeV\ 
during the 1996-1999 operation of the LEP collider. These data correspond to
an integrated luminosity of approximately 460~pb$^{-1}$ per experiment. 
No results from the year 2000 LEP running are included.
The results on the W mass quoted below correspond to a mass
definition based on a Breit-Wigner denominator with an s-dependent width,
$|s-\Mw^2 + i s \Gw/\Mw|$.

Since 1996 the LEP $\epem$ collider is operating above the
threshold for $\WW$ pair production. Initially 10~pb$^{-1}$ of data
were recorded close to $\WW$ pair production threshold. At this
energy the $\WW$ cross section is sensitive to the W boson mass, $\Mw$.
Table~\ref{tab:wmass_threshold} summarises the W mass results from the four 
LEP collaborations based on these data~\cite{bib:A-mw161,bib:D-mw161,bib:L-mw161,bib:O-mw161}.
\begin{table}[htbp]
 \begin{center}
  \begin{tabular}{|r|c|}\hline
     \multicolumn{2}{|c|}{Threshold  Cross Section } \\
Experiment &   \Mw /\GeV     \\ \hline
   ALEPH~\cite{bib:A-mw161}    & $80.14\pm0.35$  \\ 
   DELPHI~\cite{bib:D-mw161}   & $80.40\pm0.45$  \\
   L3~\cite{bib:L-mw161}       & $80.80^{+0.48}_{-0.42}$  \\
   OPAL~\cite{bib:O-mw161}     & $80.40^{+0.46}_{-0.43}$  \\ \hline
\end{tabular}
 \caption{W mass measurements from the $\WW$ threshold cross section 
          at $\roots=161$~\GeV. The errors
          include statistical and systematic contributions.}
 \label{tab:wmass_threshold}
\end{center}
\end{table} 

Subsequently LEP has operated well above the $\WW$ threshold,
delivering 450~pb$^{-1}$ per experiment (excluding the operation in 2000).
Above threshold, the $\epem\rightarrow\WW$ cross section has little 
sensitivity to $\Mw$. For these data the W boson mass is measured using the
method of direct reconstruction where the measured four-momenta
of the reconstructed jets and leptons are used to reconstruct the 
invariant mass on an event-by-event basis. 
Table~\ref{tab:wmass_experiments} summarises the W mass results from the four
LEP experiments using the method of direct reconstruction. 
In addition to
the combined numbers, each experiment presents mass measurements from
$\WWqqln$ and $\WWqqqq$ channels separately. 
The DELPHI, L3 and OPAL collaborations
obtain results for the $\qqln$  and $\qqqq$ channels  using independent
fits to data from each channel, taking into account correlated
systematic errors. The $\qqln$ and $\qqqq$ results quoted by the
ALEPH collaboration are obtained from a simultaneous fit to all data 
which, in addition to other correlations, takes into account the 
correlated systematic uncertainties between the two channels.
The large variation in the systematic uncertainties in the $\WWqqqq$ 
channel are caused by differing
estimates of the possible effects of colour reconnection (CR) and
Bose-Einstein correlations (BEC), discussed below. The systematic
errors in the $\WWqqln$ channel are dominated by uncertainties from
hadronisation, with estimates ranging from 20-40~\MeV.
\begin{table}[htbp]
 \begin{center}
  \begin{tabular}{|r|c|c||c|}\hline
    \multicolumn{1}{|c|}{ } & \multicolumn{3}{c|}{DIRECT RECONSTRUCTION } \\
           & \WWqqln         & \WWqqqq         & Combined        \\   
Experiment & \Mw /\GeV        & \Mw /\GeV        & \Mw /\GeV        \\ \hline
   ALEPH~\cite{bib:A-mw183,bib:A-mw189,bib:A-mw19x} 
  & $80.435\pm0.063\pm0.048$ & $80.467\pm0.064\pm0.057$ & 
                             $80.449\pm0.045\pm0.047$ \\ 
   DELPHI~\cite{bib:D-mw172,bib:D-mw183,bib:D-mw189,bib:D-mw19x}
  & $80.381\pm0.088\pm0.048$ & $80.372\pm0.064\pm0.063$ & 
                             $80.380\pm0.053\pm0.047$ \\
   L3~\cite{bib:L-mw172,bib:L-mw183,bib:L-mw189,bib:L-mw19x}]
  & $80.273\pm0.089\pm0.046$ & $80.461\pm0.077\pm0.069$ & 
                             $80.362\pm0.058\pm0.052$ \\
   OPAL~\cite{bib:O-mw172,bib:O-mw183,bib:O-mw189,bib:O-mw19x,bib:O-mwlvlv}    
 & $80.510\pm0.067\pm0.031$ & $80.408\pm0.066\pm0.100$ & 
                             $80.486\pm0.053\pm0.039$ \\ \hline
\end{tabular}
 \caption{Preliminary W mass measurements from direct reconstruction
         ($\roots=172-202$~\GeV).
         Results are given for the
         semileptonic, fully hadronic channels and the combined value.
         The errors are statistical and 
         systematic respectively. The combined values of $\Mw$ from
         each collaboration take into account the correlated systematic
         uncertainties between the two channels and between the different
         years of data taking.  The $\WWqqln$ results from the 
         ALEPH and OPAL collaborations include mass information from 
         the $\WWlnln$ channel.
       }
 \label{tab:wmass_experiments}
\end{center}
\end{table}

\subsubsection{Combination Procedure}
 
A LEP combined W mass measurement is obtained from the results 
of the four experiments. To combine the measurements, a more detailed input 
than that given in Table~\ref{tab:wmass_experiments} is required.
Each experiment provides a W mass measurement
for both $\WWqqln$ and $\WWqqqq$ channels for each of the four
years data taking (1996-1999), a total of 30 measurements 
(the L3 Collaboration provides results from the 1996 and 1997 data
already combined).  
The subdivision into years enables a proper treatment of the 
correlated systematic uncertainty from the LEP beam energy and allows
for the possibility that the systematic uncertainties depend on
the \CoM\ energy and/or data-taking period. 
For each result a detailed breakdown of the sources of systematic uncertainty 
is provided and the correlations are specified. 
In the combination these inter-year, inter-channel and inter-experiment 
correlations are included. The main sources of correlated systematic errors 
are: colour reconnection, Bose-Einstein correlations, hadronisation,
the LEP beam energy, and uncertainties from initial and final state 
radiation. The full correlation matrix  for the LEP beam energy is
employed\cite{bib:energy}. 

The four LEP collaborations give different estimates of the
systematic errors arising from final state interactions,
ranging from 30-66~\MeV\ for colour reconnection and from 
20-67~\MeV\ for Bose-Einstein correlations. These spreads could be 
due to different experimental sensitivities to these effects or,
alternatively,  they could be a reflection of the different 
models used to assess the uncertainties. This question is 
addressed using common Monte Carlo samples with and without CR effects. 
These are passed through the full detector simulations of each of the 
four experiments. Studies of these 
samples demonstrate that the four experiments are equally sensitive to 
colour reconnection effects, {\em i.e.} when looking at the same CR model 
similar biases are seen by all experiments.
For this reason a common value of the CR systematic 
uncertainty is used in the combination. For Bose-Einstein 
Correlations, no similar test is made  
of the respective experimental sensitivities.
However, in the absence of evidence that the experiments have
different sensitivities to the effect, a common value of the 
systematic uncertainty from BEC is assumed. In the combination a common 
colour reconnection error of 50~\MeV\ and a common Bose-Einstein systematic 
uncertainty of 25~\MeV\ are used. These uncertainties are representative
of the numbers estimated by the four LEP collaborations.

\subsubsection{LEP Combined W Boson Mass }

The combined W mass from direct reconstruction, taking into account 
all correlations including those between the $\WWqqln$ and $\WWqqqq$
channels, years and experiments gives 
\begin{eqnarray*}
        \Mw(direct) = 80.428\pm0.030(\mathrm{stat.})\pm0.036(\mathrm{syst.})~\GeV,
\end{eqnarray*}
with a $\chi^2$/d.o.f. of 27.1/29, corresponding to a $\chi^2$ probability 
of 57\%.
The weight of the fully hadronic channel in the combined fit is
0.27. The reduced weight is a consequence of the relatively large
size of the current estimates of the systematic errors from 
CR and BEC. Table~\ref{tab:errors} gives a breakdown of the contribution
to the total error of the systematic errors from different sources
and the contribution to total error from statistics. The largest
contribution to the systematic error comes from hadronisation uncertainties,
which are treated as correlated between the two channels. 
In the absence of systematic effects the current
LEP statistical precision on $\Mw$ would be $25$~\MeV. The statistical
error contribution to the total error from the LEP combination is larger
than this (30~\MeV) due to the significantly reduced weight of the 
fully hadronic channel.
Table~\ref{tab:errors} also shows the error breakdown for W mass measurements 
from the two channels separately (described in the following section). 
\begin{table}[htbp]
 \begin{center}
  \begin{tabular}{|l|c|c||c|}\hline
       Source  &  \multicolumn{3}{|c|}{Systematic Error on \Mw\ (\MeV)}  \\  
                             &  \qqln & \qqqq  & Combined  \\ \hline   
 ISR/FSR                     &  8 & 10 &  8 \\
 Hadronisation               & 26 & 23 & 24 \\
 Detector Systematics        & 11 &  7 & 10 \\
 LEP Beam Energy             & 17 & 17 & 17 \\
 Colour Reconnection         & $-$& 50 & 13 \\
 Bose-Einstein Correlations  & $-$& 25 &  7 \\
 Other                       &  5 &  5 &  4 \\ \hline
 Total Systematic            & 35 & 64 & 36 \\ \hline
 Statistical                 & 38 & 34 & 30 \\ \hline\hline
 Total                       & 51 & 73 & 47 \\ \hline
\end{tabular}
 \caption{Error decomposition for the combined W mass results. 
          Detector systematics include uncertainties
          in the jet and lepton energy scales and resolution. The `Other'
          category refers to errors, all of which are uncorrelated
          between experiments, arising from: Monte Carlo statistics,
          background, four-fermion treatment, fitting and event 
          selection. The error decomposition in the $\qqln$ and $\qqqq$
          channels refers to the independent fits to the results from 
          the two channels separately.}
 \label{tab:errors}
\end{center}
\end{table}

In addition to the above results, the W boson mass is measured at
LEP from the
data recorded at threshold (10~pb$^{-1}$/experiment) for W pair production:
\begin{eqnarray*}
      {\Mw(threshold) = 
  80.40\pm0.20(\mathrm{stat.})\pm
          0.07(\mathrm{syst.})\pm0.03(\mathrm{LEP})~\GeV}.
\end{eqnarray*}
When this is combined with the much more precise results obtained 
from direct reconstruction one obtains a W mass measurement of 
\begin{eqnarray*}
           \Mw = 80.427\pm0.046~\GeV.
\end{eqnarray*}
In combining the threshold and direct reconstruction measurements the 
only correlated systematic uncertainty is that from the LEP beam energy. 

\subsubsection{Consistency Checks}

In addition to fitting for the combined W mass from the $\WWqqln$ and
$\WWqqqq$ channels, the difference between the W boson mass measurements
obtained from the fully hadronic and semileptonic channels, 
$\Delta\Mw(\qqqq-\qqln)$, can be determined.
A significant non-zero value for $\Delta\Mw$ could indicate that 
FSI effects are biasing the value of \Mw\ determined from \WWqqqq\ events.
Since $\Delta\Mw$ is primarily of interest as a check of the possible 
effects of final state interactions, the errors from CR and BEC are set to 
zero in its determination, giving:
\begin{eqnarray*}
 \Delta\Mw(\qqqq-\qqln) =  +5\pm51~{\mathrm{\MeV}}.    
\end{eqnarray*}
This result is obtained from a fit to the 
results of the four experiments taking into account correlations between: 
the experiments, the two channels and different years. 
This result is almost unchanged if the systematic part of the error on \Mw\
from fragmentation effects is considered as uncorrelated between experiments 
and channels, 
although the uncertainty increases by about 17\%.
The masses from the two channels obtained from this fit are
\begin{eqnarray*}
 \Mw(\WWqqln) = 80.427\pm0.038(\mathrm{stat.})\pm0.035(\mathrm{syst.})~\GeV, \\
 \Mw(\WWqqqq) = 80.432\pm0.034(\mathrm{stat.})\pm0.064(\mathrm{syst.})~\GeV.  
\end{eqnarray*}
These two results are correlated having a correlation coefficient of 
0.29. The value of $\chi^2$/d.o.f is 27.1/28, corresponding to a
$\chi^2$ probability of 52$\%$.  
These results and the correlation between them
can be used to combine the two measurements or to form the mass 
difference. 

Experimentally,  
separate $\Mw$ measurements are obtained from the
$\WWqqln$ and $\WWqqqq$ channels. 
The combination using only the $\qqlv$ measurements yields
\begin{eqnarray*}
 m_{\mathrm{W}}^{indep}(\WWqqln) = 80.429\pm0.038(\mathrm{stat.})\pm0.035(\mathrm{syst.})~\GeV. 
\end{eqnarray*}
This result is independent of the measurements in the $\WWqqqq$ channel.
The  systematic error is dominated by
fragmentation uncertainties  ($\pm26$~\MeV) and the 
uncertainty in the LEP \CoM\ energy  ($\pm17$~\MeV).
The independent average over all $\qqqq$ measurements only gives
\begin{eqnarray*}
  m_{\mathrm{W}}^{indep}(\WWqqqq) = 80.424\pm0.034(\mathrm{stat.})\pm0.064(\mathrm{syst.})~\GeV.  
\end{eqnarray*}
where the dominant contributions to the systematic error arise from 
BE/CR ($\pm56$~\MeV), fragmentation ($\pm23$~\MeV)
and from the uncertainty in the LEP \CoM\ energy ($\pm17$~\MeV).

\subsubsection{LEP Combined W Boson Width}

The method of direct reconstruction is also used for the
direct measurement of the width of the W boson. The results of the four 
LEP experiments are shown in Table~\ref{tab:wwidth_experiments}.
\begin{table}[htbp]
 \begin{center}
  \begin{tabular}{|l|c|}\hline
  Experiment & \Gw /\GeV        \\ \hline
   ALEPH    & $2.17\pm0.16\pm0.12$ \\ 
   DELPHI   & $2.09\pm0.12\pm0.09$ \\
   L3       & $2.19\pm0.14\pm0.16$ \\
   OPAL     & $2.04\pm0.16\pm0.09$ \\ \hline
   Combined & $2.12\pm0.08\pm0.07$ \\ \hline
\end{tabular}
 \caption{Preliminary W width measurements ($\roots=172-202$~\GeV) 
         from the individual experiments. The first error is statistical
         and the second systematic. The ALEPH results at 192-202~\GeV\
         do not use the $\WWqqqq$ channel and the OPAL results do not include
         data recorded at \CoM\ energies greater than 189~\GeV.
         Only OPAL and L3 include data from 172~\GeV.}
 \label{tab:wwidth_experiments}
\end{center}
\end{table}

A simultaneous fit to the results of the four LEP collaborations is
performed in the same way as for the $\Mw$ measurement. Correlated
systematic uncertainties are taken into account to give: 
\begin{eqnarray*}
      \Gw = 2.12\pm0.08(\mathrm{stat.})\pm0.07
                                     (\mathrm{syst.})~\GeV,
\end{eqnarray*}
with a $\chi^2$/d.o.f. of 13.7/17.

Currently no LEP combined value for $\Gw$ is obtained from the
$\WWqqln$ and $\WWqqqq$ channels separately.

\subsubsection{Summary}

The results of the four LEP experiments on the mass and width of the
W boson are combined taking into account correlated systematic
uncertainties, giving:
\begin{eqnarray*}
       \Mw & = & 80.427\pm0.046~\GeV, \\
       \Gw & = &  2.12\pm0.11~\GeV.
\end{eqnarray*}


\clearpage
\boldmath
\section{Single--W Production Cross Section}
\label{sec-SGW}
\updates{This is a new section. }
\unboldmath

Preliminary single--W results for data collected at $\roots=183$--202 \GeV\
are available.
No preliminary
analysis of the data collected in the year 2000 at $\roots>200$ \GeV\ was
prepared in time for the summer conferences.

A common definition of the single--W production cross section is adopted,
 to allow direct comparisons and the combination of the results 
obtained by the 
four experiments. Single--W production is defined by the complete $t$-channel 
subset of 
Feynman diagrams contributing to the $\mathrm{e} \nu f f'$ final states, 
with additional cuts on 
kinematic variables to exclude the regions of the phase space dominated by 
multiperipheral diagrams, where the cross section calculation is affected 
by large uncertainties.
The advantage of this definition is that a gauge invariant set of diagrams 
is taken, 
which has a small interference with the $s$-channel one. In addition, 
no angular cuts due 
to the specific detector acceptances are used.

The kinematic cuts used in the cross section definition are 
(charge conjugation is  implied in the definition):

\begin{itemize}
\item $M_{\mathrm{qq'}} > 45 $ \GeV ~for the $\mathrm{e} \nu \mathrm{q q'}$ final states,
\item $E_{\ell^{+}} > 20$ \GeV ~for the $\mathrm{e} \nu\ell \nu$ final states 
(with $\ell=\mu,\tau$),
\item $|cos(\theta_{\mathrm{e}^{-}})|>0.95, |cos(\theta_{\mathrm{e}^{+}})|<0.95, E_{\mathrm{e}^{+}} > 20$ 
\GeV ~for the $\mathrm{e} \nu \mathrm{e} \nu$ final state.
\end{itemize}

The results of the single experiments and of the LEP average are
summarised for the different \CoM\  energies in Table~\ref{tab:swxsec}.
Given the statistical 
precision of the single--W measurements, results are combined assuming 
uncorrelated systematic errors between experiments and 
using expected statistical errors.

\begin{figure}[p]
\begin{center}
  \mbox{\includegraphics[width=\linewidth]{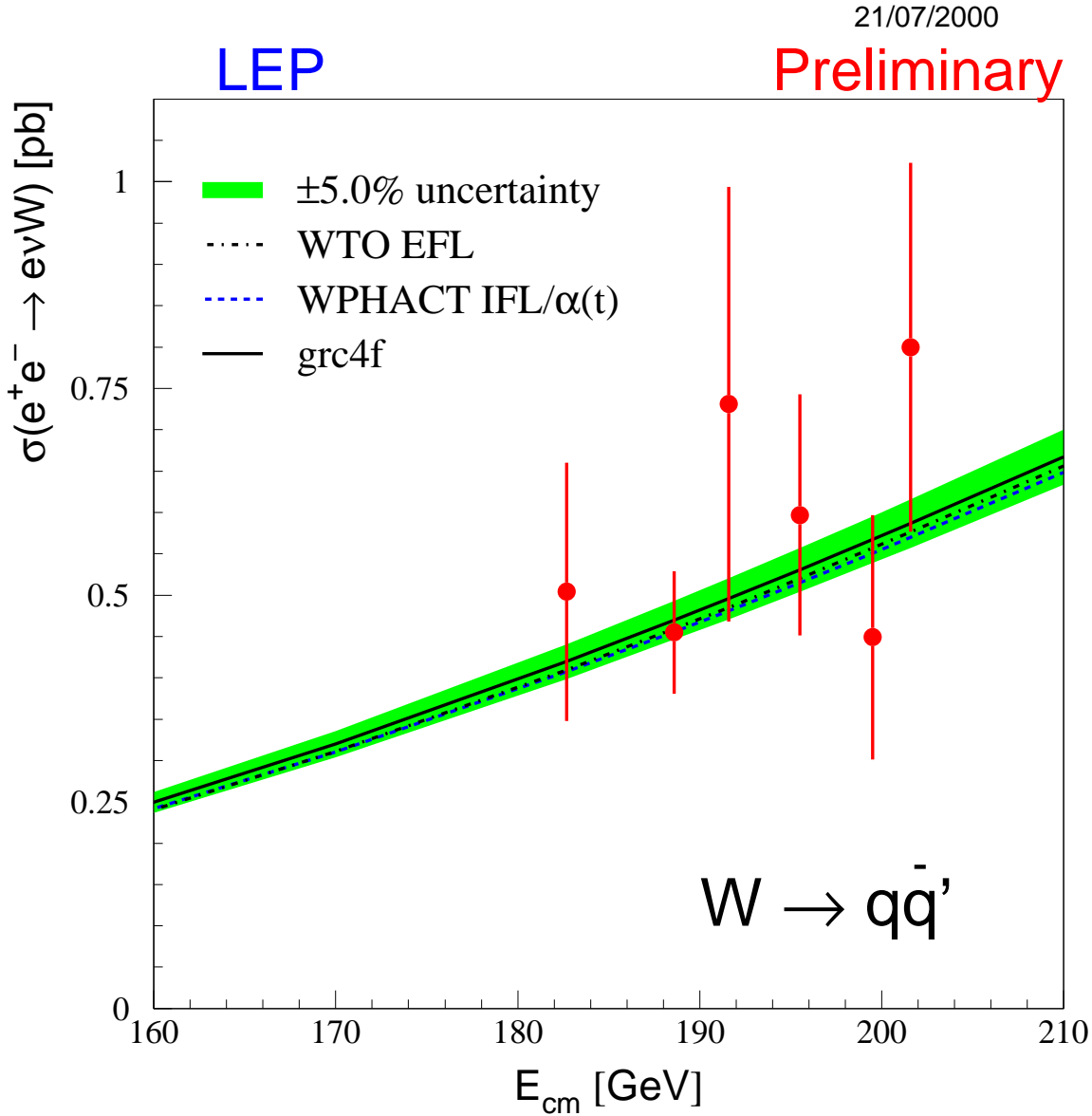}}
\end{center}
\vspace*{-0.6cm}
\caption{Measurements of the single--W production cross section in the
hadronic decay channel of the W boson compared to the predictions of 
\WTO, \WPHACT\  and \Grace. The shaded area represents a $\pm5$\% uncertainty 
on the predictions.}
\label{fig:swen_had}
\end{figure}

Since winter 2000 conferences~\cite{bib:swmor00},
the only update concerns the addition of one theoretical prediction
\cite{bib:wto} to the single--W production cross section curve as a function
of the \CoM\  energy for the hadronic final state, based on the exact 
treatment of the evolution of the energy scale for 
$\alpha_{em}$~\cite{bib:efloop} in the 4f matrix element. 
This evolution is only approximated
(at the level of 1--2\%) by one of the codes previously used~\cite{bib:wphact},
and introduced by rescaling the cross section in one other code
\cite{bib:grace}. The data at $\roots=183$--202 \GeV\ are compared with 
the theoretical predictions in Figure~\ref{fig:swen_had}. 

It should be remarked that all the theoretical predictions are scaled 
upward by 4\% , since all codes implement QED radiative corrections (mainly
due to initial state radiation) using the wrong energy scale {\it s} in
the ISR structure functions which is not adequate for this process dominated
by $t$-channel exchange diagrams. 
The correction factor of 4\% is derived in
\cite{bib:fourfrep} from a comparison of \Grace\  \cite{bib:gracet} and 
\SWAP\  \cite{bib:swap} for different scales in the structure functions.
Since this is only an approximation of the complete $\mathcal{O}(\alpha)$
correction for the single W process, a 100\% uncertainty is conservatively
assigned to this 4\% correction. This constitutes the major part of the
theoretical uncertainty currently assigned to the predictions (5\%, as discussed
in~\cite{bib:fourfrep}), which is becoming a limiting factor for the extraction 
of triple gauge couplings from the single W events~\cite{bib:tgc2krep}.

\begin{table}[hbtp]
\centering
\begin{tabular}{|c|c|c|c|c|c|c|} 
\hline
\roots & \multicolumn{5}{|c|}{Cross section  (pb)} & $\chi^2/\textrm{d.o.f.}$ \\
\cline{2-6} 
(\GeV)          & ALEPH                  & DELPHI                 &  
                  L3                     & OPAL                   &
                 LEP                     &                          \\ 
\hline
182.67         & $0.40\pm0.24$  & &
                 $0.58^{+0.23}_{-0.20}$  & &
                 $0.50\pm0.16$          & 0.33/1                   \\
188.63         & $0.31\pm0.14$  & $0.44^{+0.28}_{-0.25}$          &
                 $0.52^{+0.14}_{-0.13}$  & $0.53^{+0.13}_{-0.13}$  &
                 $0.46\pm0.07$          & 1.71/3                   \\
191.6\phz      & $0.94\pm0.44$   & $0.01^{+0.19}_{-0.07}$            &
                 $0.85^{+0.45}_{-0.37}$   & &
                 $0.73\pm0.26$            & 1.70/2                   \\
195.5\phz      & $0.45\pm0.23$   & $0.78^{+0.38}_{-0.34}$            &
                 $0.66^{+0.25}_{-0.23}$  &   &
                 $0.60\pm0.15$            & 0.79/2                   \\
199.5\phz      & $0.82\pm0.26$   & $0.16^{+0.29}_{-0.17}$            &
                 $0.34^{+0.23}_{-0.20}$   & &
                 $0.45\pm0.15$            & 3.26/2                   \\
201.6\phz      & $0.68\pm0.35$   & $0.55^{+0.47}_{-0.40}$            &
                 $1.09^{+0.42}_{-0.37}$   & &
                 $0.80\pm0.22$            & 1.03/2                   \\
\hline
\end{tabular}
\caption{
Single--W cross section for the hadronic decay of the W, from the four experiments and 
LEP combined value for the five \CoM\ energies. All numbers are 
preliminary.
}
\label{tab:swxsec}
\end{table}
 
\clearpage
\boldmath
\section{Measurement of the ZZ Production Cross Section}
\label{sec-ZZ}
\unboldmath

\updates{ All experiments have published results up to 189 \GeV,
new results at higher energies are available.  }

\noindent
At  \CoM\ energies above twice the Z boson mass, the
production of pairs of Z bosons is possible.
All experiments have published final results~\cite{bib:alezz189,bib:delzz189,
bib:ltrzz189,bib:opazz189} on the Z--pair (NC02) production cross section at
$\roots=183$ and 189 \GeV. L3~\cite{bib:ltrzz1999} has updated its
preliminary results for $\roots=192$--202 \GeV, whereas ALEPH~\cite{bib:alezz1999},
DELPHI~\cite{bib:delzz1999} and OPAL~\cite{bib:opazz1999} preliminary results
are unchanged since the winter conferences. All experiments have
provided preliminary results~\cite{bib:aleww2000,bib:delww2000,bib:ltrww2000,
bib:opaww2000} based on an integrated luminosity of 
approximately 90 \pb\  collected in the year 2000 at \CoM\  energies between
200 and 208 \GeV.

The combination of results is performed using the expected statistical
error of each analysis, to avoid biases due to the limited number of
events selected. The following components of the systematic errors are
considered correlated between experiments:
\begin{itemize}
\item $\pm5\%$ uncertainty on the 2--fermion background rate.
\item $\pm2\%$ uncertainty on the WW background rate.
\item $\pm10\%$ uncertainty on the $\mathrm{Z^0e^+e^-}$ background rate.
\item the uncertainty on the b quark modelling.
\end{itemize}
The common error ranges from 0.01 to 0.03 pb. 

\begin{figure}[p]
\begin{center}
  \mbox{\includegraphics[width=\linewidth]{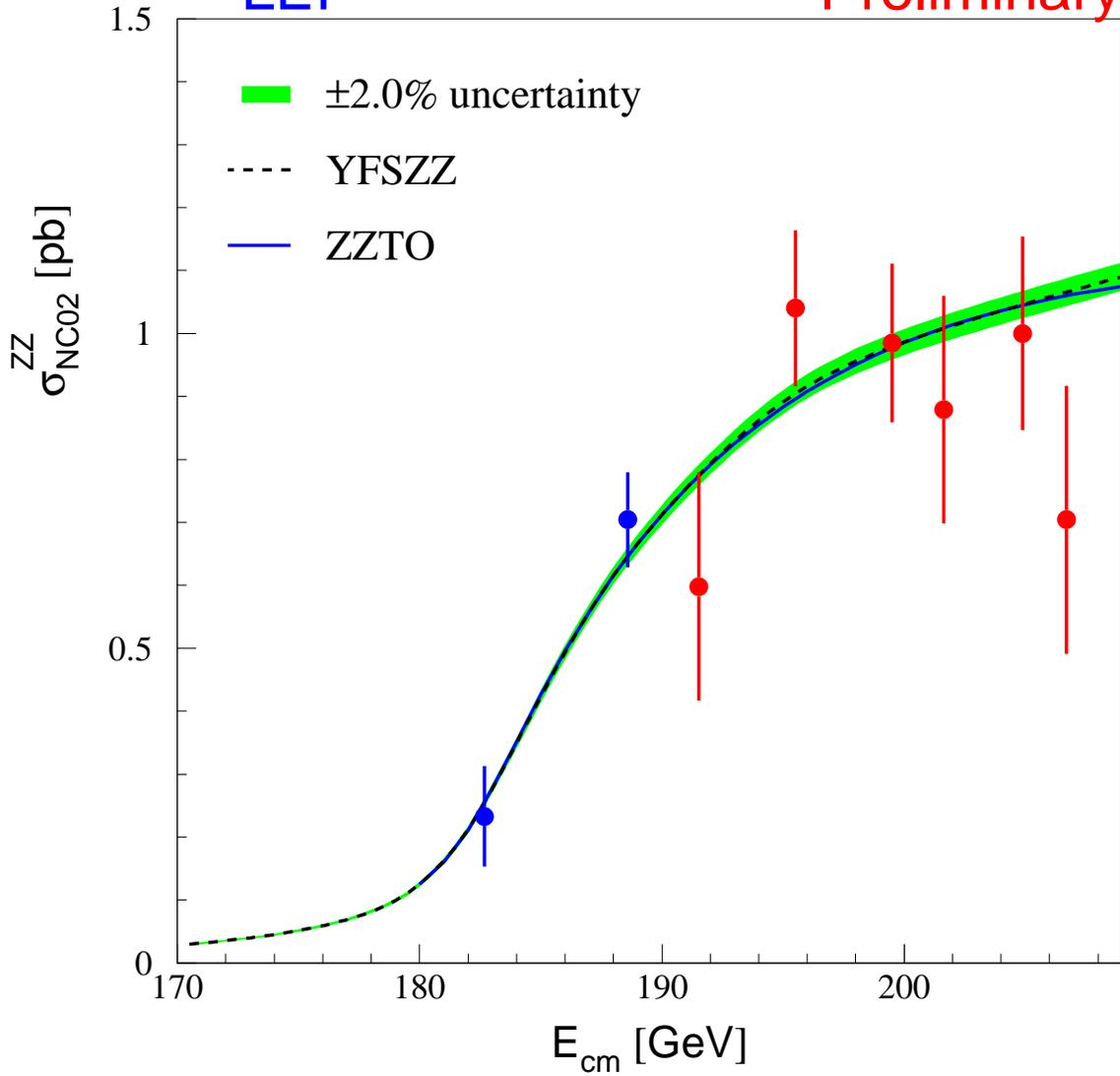}}
\end{center}
\vspace*{-0.6cm}
\caption{Measurements of the Z--pair production cross section
compared to the predictions of \YFSZZ\  and \ZZTO. The shaded area 
represent a $\pm2$\% uncertainty on the predictions.}
\label{fig:szz_vs_sqrts}
\end{figure}

The results of the single experiments and of the LEP average are
summarised for the different \CoM\  energies in Table~\ref{tab:zzxsec}.
All results are preliminary, with the exception of those at $\roots=183$ 
and 189 \GeV. The results are shown in Figure~\ref{fig:szz_vs_sqrts} as a function 
of the LEP \CoM\  energy and compared to the YFSZZ~\cite{bib:yfszz} and
ZZTO~\cite{bib:zzto} predictions, which have an estimated 
uncertainty of $\pm2\%$~\cite{bib:fourfrep}. These results supersede
those presented in~\cite{bib:zzmor00} for $\roots=183$--202 \GeV. The 
data do not show any significant deviation from the theoretical
expectations.
 
\begin{table}[hbtp]
\centering
\begin{tabular}{|c|c|c|c|c|c|c|} 
\hline
\roots & \multicolumn{5}{|c|}{Cross section  (pb)} & $\chi^2/\textrm{d.o.f.}$ \\
\cline{2-6} 
(\GeV)          & ALEPH                  & DELPHI                 &  
                 L3                   & OPAL                   &
                 LEP                      &                          \\ 
\hline
182.67         & $0.11^{+0.16*}_{-0.12}$  & $0.38\pm0.18^*$          &
                 $0.31^{+0.17*}_{-0.15}$  & $0.12^{+0.20*}_{-0.18}$  &
                 $0.23\pm0.08^*$          & 2.28/3                   \\
188.63         & $0.67^{+0.14*}_{-0.13}$  & $0.60\pm0.15^*$          &
                 $0.73^{+0.15*}_{-0.14}$  & $0.80^{+0.15*}_{-0.14}$  &
                 $0.70\pm0.08^*$          & 0.92/3                   \\
191.6\phz      & $0.53^{+0.34}_{-0.27}$   & $0.55\pm0.34$            &
                 $0.27^{+0.20}_{-0.23}$   & $1.13^{+0.47}_{-0.41}$   &
                 $0.60\pm0.18$            & 2.99/3                   \\
195.5\phz      & $0.69^{+0.23}_{-0.20}$   & $1.17\pm0.28$            &
                 $1.11\pm0.23$            & $1.28^{+0.29}_{-0.27}$   &
                 $1.04\pm0.12$            & 3.43/3                   \\
199.5\phz      & $0.70^{+0.22}_{-0.20}$   & $1.08\pm0.26$            &
                 $1.19\pm0.24$            & $1.01^{+0.26}_{-0.23}$   &
                 $0.99\pm0.13$            & 2.31/3                   \\
201.6\phz      & $0.70^{+0.33}_{-0.28}$   & $0.87\pm0.33$            &
                 $0.89^{+0.36}_{-0.37}$   & $1.09^{+0.40}_{-0.35}$   &
                 $0.88\pm0.18$            & 0.57/3                   \\
204.9\phz      & $0.86^{+0.29}_{-0.26}$   & $1.28\pm0.34$            &
                 $0.56\pm0.28$            & $1.41^{+0.35}_{-0.32}$   &
                 $1.00\pm0.15$            & 4.74/3                   \\
206.7\phz      & $0.51^{+0.35}_{-0.28}$   & $1.11\pm0.42$            &
                 $0.84\pm0.39$            & $0.30^{+0.37}_{-0.29}$   &
                 $0.70\pm0.21$            & 2.13/3                   \\
\hline
\end{tabular}
\caption{Z--pair production cross section from the four LEP
experiments and combined values for the eight \CoM\  energies.
All results are preliminary with the exception of those indicated
by~$^*$. A common systematic error of (0.01--0.03) pb is taken
into account in the averaging procedure.}
\label{tab:zzxsec}
\end{table}

\clearpage
\boldmath
\section{Electroweak Gauge Boson Couplings}
\label{sec-GC}
\unboldmath

\updates{ This is a new section }

\noindent
The W-pair production process, $\mathrm{e^+e^-\rightarrow\WW}$,
involves the triple gauge boson vertices between the $\WW$ and the Z
or the photon.  Up-to the end of 1999, more than 7000 W-pair events
were collected by each experiment.  Single W ($\enw$) and single photon
($\nng$) production at LEP are sensitive to the $\WWg$ vertex.  The
measurement of these charged triple gauge boson couplings and the
search for possible anomalous contributions due to the effects of new
physics beyond the Standard Model is one of the principal physics
goals at \LepII~\cite{LEP2YR}.

At \CoM\ energies exceeding twice the Z boson mass, pair
production of Z bosons is kinematically allowed. Here, one searches
for the possible existence of triple vertices involving only neutral
electroweak gauge bosons. Such vertices could also contribute to
Z$\gamma$ production.  In contrast to triple gauge boson vertices with
two charged gauge bosons, purely neutral gauge boson vertices do not
occur in the Standard Model of electroweak interactions.

Within the Standard Model, quartic electroweak gauge boson vertices
with at least two charged gauge bosons exist. In $\ee$ collisions at
\LepII\ \CoM\ energies, the $\WWZg$ and $\WWgg$ vertices
contribute to $\WWg$ and $\nngg$ production in $s$-channel and
$t$-channel, respectively.  The effect of the Standard Model quartic
electroweak vertices is below the sensitivity of \LepII.  Thus,
anomalous quartic vertices are searched for in the production of
$\WWg$ and $\nngg$ but also $\Zgg$ final states.

In this note we present the combination of results on electroweak
gauge boson couplings based on the data collected by the four LEP
experiments at \LepII.  Charged
TGCs~\cite{ALEPH-cTGC,DELPHI-cTGC,L3-cTGC,OPAL-cTGC}, neutral
TGCs~\cite{DELPHI-hTGC,L3-hTGC,OPAL-hTGC,DELPHI-fTGC,L3-fTGC,OPAL-fTGC}
as well as QGCs~\cite{ALEPH-QGC,L3-QGC,OPAL-QGC} are combined.

\subsection{Charged Triple Gauge Boson Couplings }
\label{sec:cTGCs}

The parametrisation of the charged triple gauge boson vertices is
described in References~\cite{GAEMERS,n-hagiwara,HAGIWARA,BILENKY,
  KUSS,PAPADOPOULOSCP,LEP2YR}.  The most general Lorentz invariant
Lagrangian which describes the charged triple gauge boson interaction 
contains fourteen independent complex couplings, seven describing the
WW$\gamma$ vertex and seven describing the WWZ vertex.  Assuming
electromagnetic gauge invariance as well as CP conservation, the
number of independent TGCs reduces to six.  One common set is \{$\gz$,
$\kg$, $\kz$, $\lg$, $\lz$, $\gvz$\} where $\gz = \kg = \kz = 1$ and
$\lg = \lz = \gvz = 0$ in the Standard Model.  Except $\gvz$, all
these couplings also conserve C and P separately.  

The parameters proposed in~\cite{LEP2YR} and used here are:
\begin{eqnarray}
\dgz & \equiv & \gz - 1\,,\\
\dkg & \equiv & \kg - 1\,,\\
\lg  & and  &  \gvz 
\end{eqnarray}                               
with the gauge constraints:
\begin{eqnarray}
\dkz & \equiv & \kz - 1 ~ = ~ \dgz - \dkg \twsq  \\
\lz  & = & \lg \,,
\end{eqnarray}                               
where $\Delta$ indicates the deviation of the respective coupling from
its Standard Model value of unity, and $\theta_W$ is the weak mixing
angle.  The couplings are considered as real, with the imaginary parts
fixed to zero.

\subsection{Neutral Triple Gauge Boson Couplings}
\label{sec:nTGCs}

There are two possible anomalous triple vertices in the neutral
sector. Each is parametrised by the most general Lorentz and
$U(1)_{\mathrm{EM}}$ invariant Lagrangian plus Bose symmetry, as described in
References~\cite{n-gounaris,n-hagiwara}.

The first vertex refers to anomalous Z$\gamma\gamma$ and ZZ$\gamma$
couplings which are accessible at LEP in the process $\mathrm{e^{+}
  e^{-}} \rightarrow {\rm Z} \gamma$. The parametrisation contains
eight couplings: $h_i^{V}$ with $i=1,...,4$ and $V=\gamma$,Z.  The
superscript $\gamma$ refers to Z$\gamma\gamma$ couplings and
superscript Z refers to ZZ$\gamma$ couplings.  The photon and the Z
boson in the final state are considered as on-shell particles, while
the third boson at the triple vertex is off shell.  The couplings
$h_{1}^{V}$ and $h_{2}^{V}$ are CP-odd while $h_{3}^{V}$ and
$h_{4}^{V}$ are CP-even.

The second vertex refers to anomalous ZZ$\gamma$ and ZZZ couplings
which are accessible at LEP in the process $\mathrm{e^{+} e^{-}}
\rightarrow$ ZZ.  This anomalous vertex is parametrised in terms of
four couplings: $f_{i}^{V}$ with $i=4,5$ and $V=\gamma$,Z.  The
superscript $\gamma$ refers to ZZ$\gamma$ couplings and the
superscript Z refers to ZZZ couplings.  Both Z bosons in the final
state are assumed to be on-shell, while the third boson at the triple
vertex is off-shell.  The couplings $f_{4}^{V}$ are CP-odd whereas
$f_{5}^{V}$ are CP-even.

Note that the $h_i^{V}$ and $f_{i}^{V}$ couplings are independent of
each other.  They are considered as real, with the imaginary parts
fixed to zero.  They vanish in the Standard Model at tree level.

\subsection{Quartic Gauge Boson Couplings}
\label{sec:QGCs}

Anomalous contributions to electroweak quartic vertices are considered
in the framework of References~\cite{QGC-old,QGC-VVV,QGC-nngg}.  Three
lowest-dimensional operators leading to quartic vertices not causing
anomalous TGCs are considered.  The corresponding couplings are
$\azl$, $\acl$ and $\anl$, where $\Lambda$ represents the energy scale
of new physics.  The couplings $\azl$ and $\acl$ parametrise
variations in the $\WWgg$ and $\ZZgg$ vertices, while $\anl$ affects
purely the $\WWZg$ vertex.  The couplings $\azl$ and $\acl$ conserve C
and P, while the coupling $\anl$ is CP violating.  The production of
$\WWg$ depends on all three couplings.  The production of $\nngg$ and
$\Zgg$ depends only on $\azl$ and $\acl$, the former through the
$\WWgg$ vertex, the latter through the $\ZZgg$ vertex.  The couplings
are considered as real, with the imaginary parts fixed to zero.  They
vanish in the Standard Model at tree level.

A more detailed description of QGCs has recently
appeared~\cite{QGC-new}, in which it is suggested that the coupling
$\azl$ and $\acl$ describing the anomalous $\WWgg$ vertex should not
be assumed to be the same couplings as those describing the anomalous
$\ZZgg$ vertex.  The experimental results on these vertices are
therefore also given separately.

\subsection{Measurements}

The results presented comprise measurements of all electroweak gauge
boson couplings discussed above.  In most cases, the results are based
on the data collected at \LepII{} until the end of 1999 at
\CoM\ energies up to 202~\GeV.  The experiments already
combine different processes and final states for each coupling.  In
each case, the individual references should be consulted for the
details on the data samples used.

For charged TGCs, the experimental analyses study different channels,
typically the semileptonic, fully hadronic and fully leptonic W-pair
decays~\cite{ALEPH-cTGC,DELPHI-cTGC,L3-cTGC,OPAL-cTGC}.  Anomalous
TGCs affect both the total production cross section and the shape of
the differential cross section as a function of the polar W$^-$
production angle.  The relative contributions of each helicity state
of the W bosons are also changed, which in turn affects the
distributions of their decay products.  The analyses presented by each
experiment make use of different combinations of each of these
quantities.  In general, however, all analyses use at least the
expected variations of the total production cross section and the
W$^-$ production angle.  Results from $\enw$ and $\nng$ production are
included by some experiments.  Single W production is particularly
sensitive to \kg, thus providing information complementary to that
from W-pair production.

For neutral TGCs, the analyses mainly use  measurements of the total
cross sections of Z$\gamma$ and ZZ production, though the use of
differential distributions increases the sensitivity.  The individual
references should be consulted concerning the details of the analyses
for the $h_i^V$ couplings~\cite{DELPHI-hTGC,L3-hTGC,OPAL-hTGC} and the
$f_i^V$ couplings~\cite{DELPHI-fTGC,L3-fTGC,OPAL-fTGC}.\footnote{For
  quoting results on the $f_i^V$ couplings, DELPHI~\cite{DELPHI-fTGC}
  uses a convention for the sign of $f_i^V$ opposite to that used by
  L3~\cite{L3-fTGC} and OPAL~\cite{OPAL-fTGC}.  For DELPHI and
  combined LEP results presented in this note, the DELPHI $f_i^V$
  measurements are modified to conform to the same sign
  convention as used by L3 and OPAL.}

For QGCs, the experiments investigate $\WWg$, $\nngg$ and $\Zgg$
production.  Besides the total cross section, the sensitive variables
analysed are photon energies ($\WWg$ and $\Zgg$) and recoil masses
($\nngg$).  Again, the individual references should be consulted for
the details in the determination of the QGCs $\azl$, $\acl$ and
$\anl$~\cite{ALEPH-QGC,L3-QGC,OPAL-QGC}.

\subsection{Combination Procedure}
\label{sec-combination}

The combination procedure is modified compared to previous LEP
combinations of electroweak gauge boson
couplings~\cite{bib-EWEP-99,moriond00} in order to treat systematic
uncertainties correlated between the experiments.

Each experiment provides the negative log likelihood, $\LL$, as a
function of the coupling parameters (one, two, or three) to be
combined.  The single parameter analyses are performed fixing the
values of all other parameters to their Standard Model value.  The
two- and three-parameter analyses are performed setting the remaining
parameters to their Standard Model value.  For the charged TGCs, the
gauge constraints listed in Section~\ref{sec:cTGCs} are always
enforced.

The $\LL$ functions from each experiment include statistical as well
as all systematic uncertainties considered as uncorrelated between
experiments.  For both single- and multi-parameter combinations, the
individual $\LL$ functions are added.  It is necessary to use the
$\LL$ functions directly for the combination as in some cases they are
not parabolic, so that it is not possible to combine the results
correctly by simply taking weighted averages of the measurements.

The main contributions to the systematic uncertainties uncorrelated
between experiments arise from detector effects, backgrounds in the
selected samples, limited Monte Carlo statistics and the fitting
methods.  Their importance varies for each experiment and the
individual references should be consulted for details.

The systematic uncertainties arising from the following sources are
treated as correlated between the experiments: the uncertainty on the
theoretical cross section prediction ($\WW$, $\enw$ and Z$\gamma$
production), fragmentation effects ($\WW$ production), and
Bose-Einstein correlations and colour-reconnection effects (fully
hadronic W-pair decays).  While the correlated systematic
uncertainties are small in Z$\gamma$ production, they are sizeable for
charged TGCs.  In particular, the
uncertainty on the theoretical cross section prediction for the $\enw$
process causes a large reduction in the sensitivity of this process to
charged TGCs.  For ZZ production, the uncertainty on the theoretical
cross section prediction is small compared to the statistical accuracy
and is neglected.  For other sources of correlated systematic
uncertainties, such as those arising from the LEP beam energy or the W
mass, the effect of their correlation on the combination result is
negligible.

The correlated systematic uncertainties are applied by scaling the
likelihood functions by the squared ratio of the statistical and
uncorrelated systematic uncertainty over the total uncertainty
including all correlated uncertainties.  This procedure to treat
correlations is an approximation in the case where the log-likelihood
function is not parabolic.

The one standard deviation uncertainties (68\% confidence level) are
obtained by taking the coupling values where $\LL=+0.5$ above
the minimum.  The 95\% confidence level (C.L.)  limits are given by
the coupling values where $\LL=+1.92$ above the minimum.  These
cut-off values are used for obtaining the results of both single- and
multi-parameter analyses reported here.  Note that in the case of the
neutral TGCs and the QGCs, double minima structures appear in the
negative log-likelihood curves.  For multi-parameter analyses, the
68\%~C.L.  contour curves for any pair of couplings are obtained by
requiring $\LL=+1.15$, while for the 95\% C.L.  contour curves
$\LL=+3.0$ is required.

\subsection{Results}

We present below results on combination of the four LEP experiments 
on the various
electroweak gauge boson couplings.
The
results quoted individually for each experiment, summarised in 
Appendix~\ref{app-GC} 
are calculated using
the method described in Section~\ref{sec-combination}.  Thus, they 
sometimes differ slightly from those reported in the individual references.


\subsubsection{Charged Triple Gauge Boson Couplings}

\begin{figure}[htbp]
\begin{center}
\includegraphics[width=\linewidth]{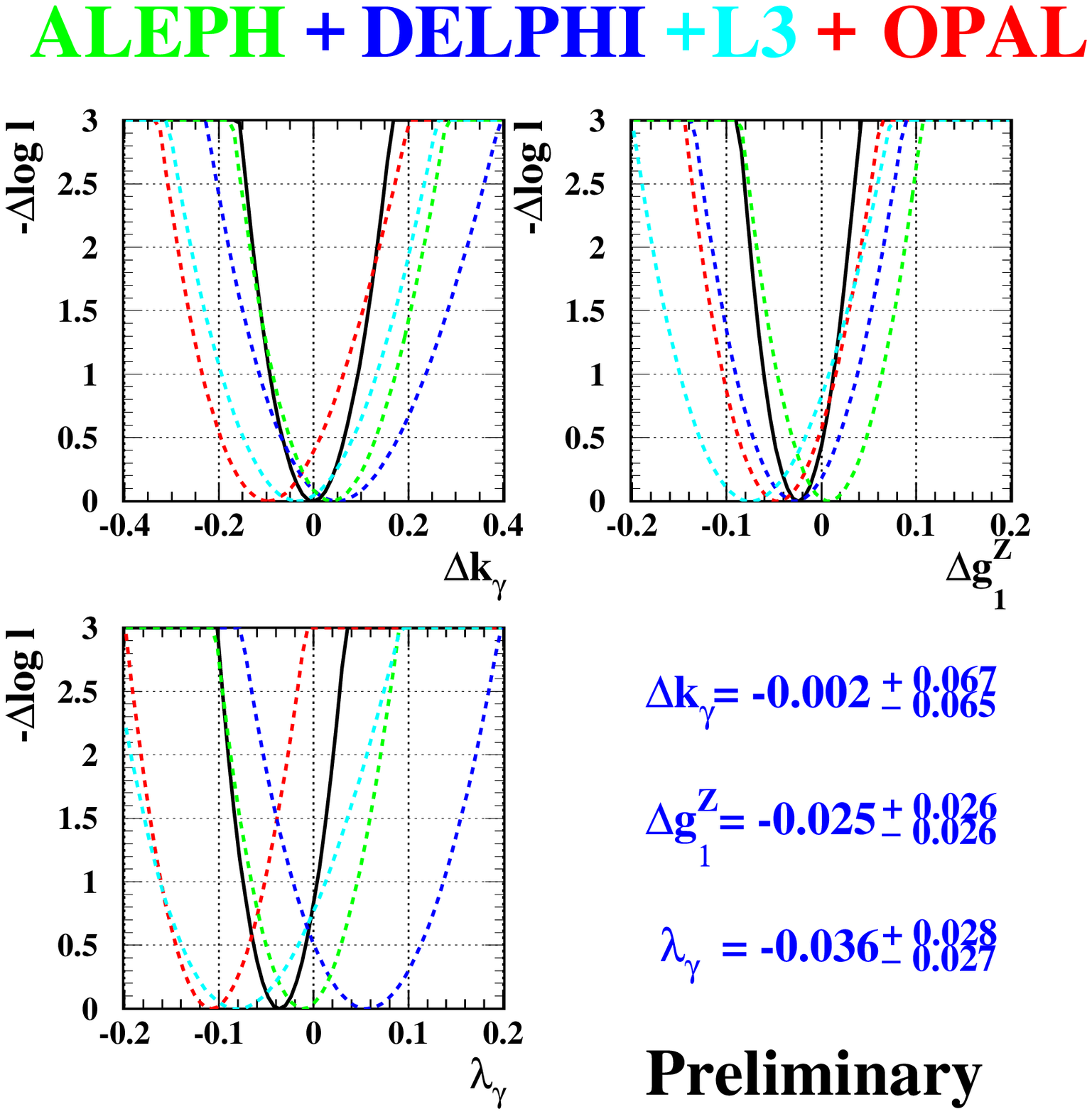}
\caption[]{
  The $\LL$ curves of the four experiments (dashed lines) and the LEP
  combined curve (solid line) for the three charged TGCs $\dgz$,
  $\dkg$ and $\lg$.  In each case, the minimal value is subtracted.  }
\label{fig:cTGC-1}
\end{center}
\end{figure}

\begin{figure}[htbp]
  \begin{center}
    \leavevmode
    \mbox{
    \includegraphics[width=0.49\linewidth]{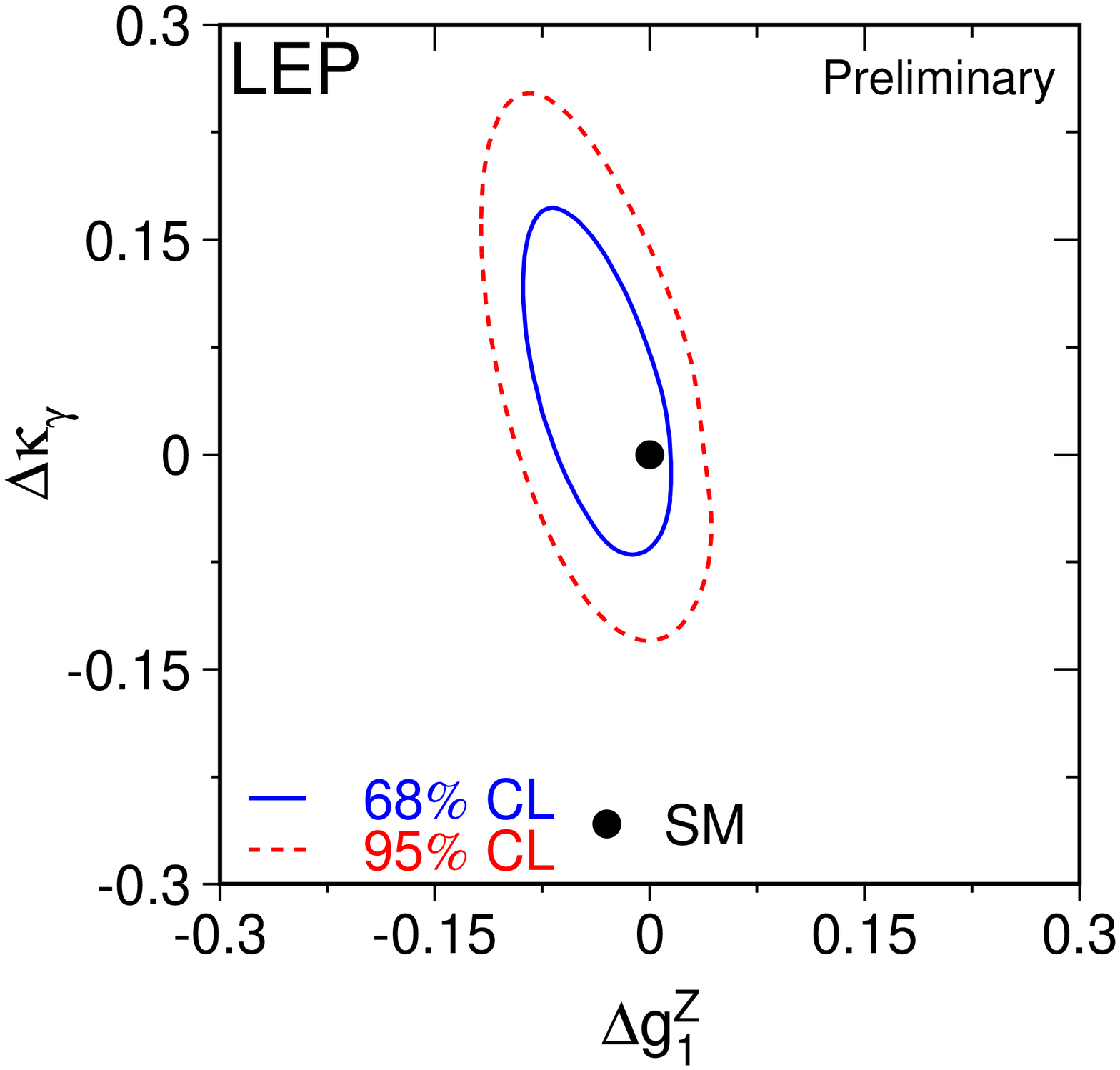}
    \includegraphics[width=0.49\linewidth]{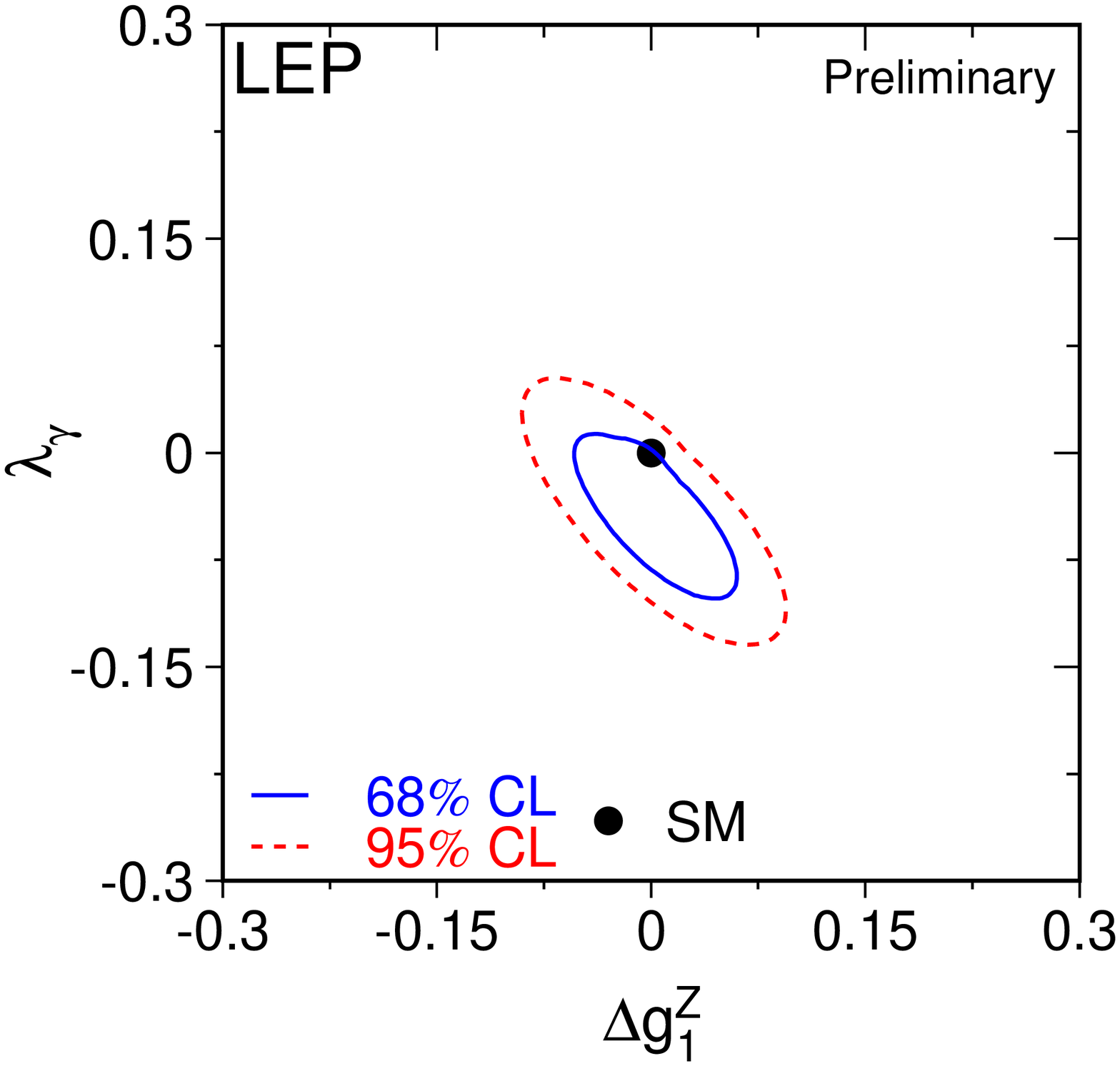}
    }
    \begin{flushleft}
    \mbox{
    \includegraphics[width=0.49\linewidth]{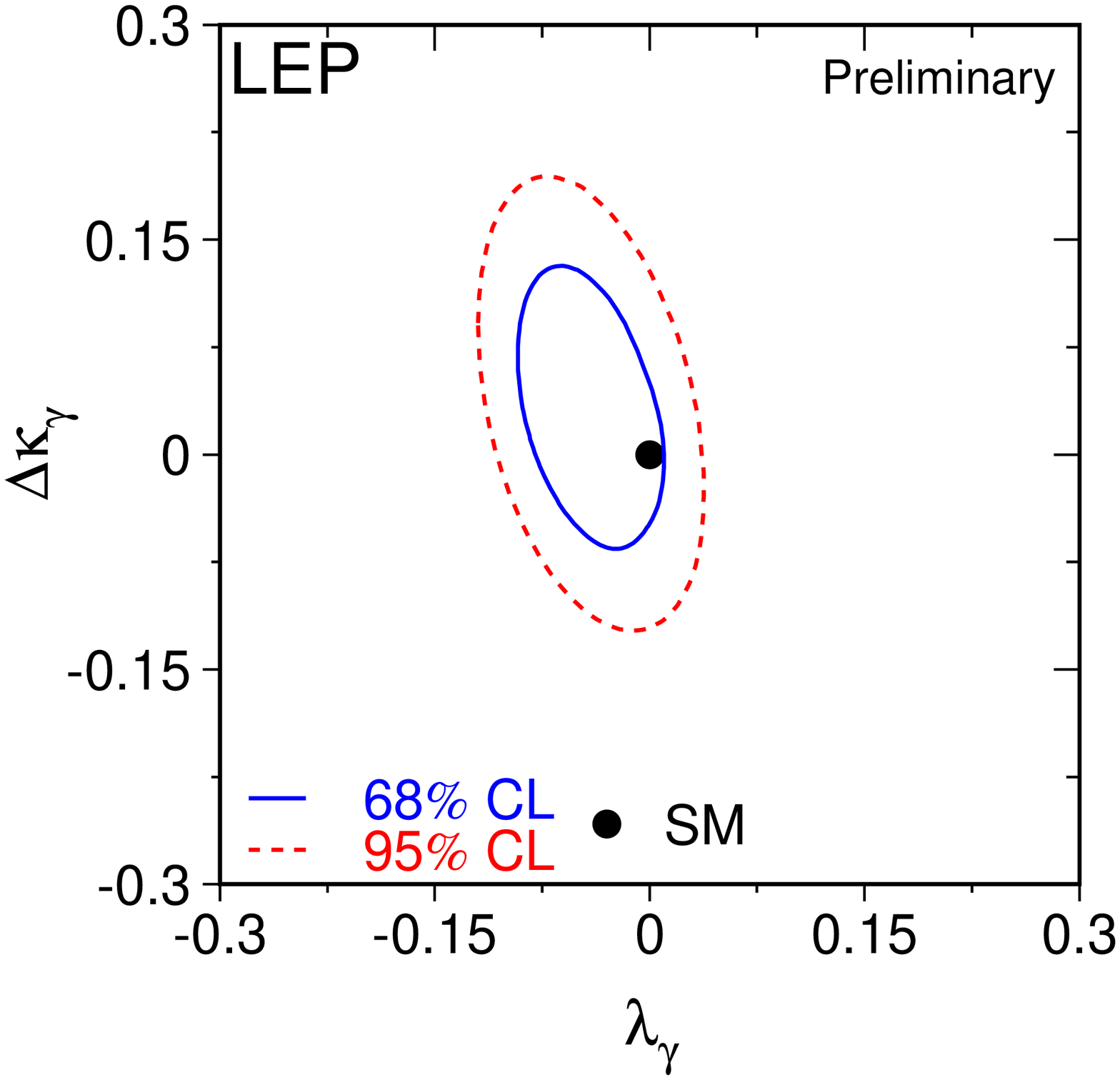}
    }
    \end{flushleft}
    \caption[]{
   Contour curves of 68\% C.L. and 95\% C.L. in the planes $(\dgz,\dkg)$,
   $(\dgz,\lg)$, $(\lg,\dkg)$
   showing the LEP combined results.
      }
    \label{fig:cTGC-2D}
  \end{center}
\end{figure}

Results of the combination of charged triple gauge boson couplings are
given in Tables~\ref{tab:cTGC-1-LEP}, \ref{tab:cTGC-2-LEP} and 
\ref{tab:cTGC-3-LEP} for the single, two and three-parameter analyses
respectively.
Individual $\LL$ curves and their sum are
shown in Figure~\ref{fig:cTGC-1} for the single-parameter analysis.
The 68\% C.L. and 95\% C.L. contours curves
resulting from the combinations of the two-dimensional likelihood
curves are shown in
Figure~\ref{fig:cTGC-2D}.

\begin{table}[htbp]
\begin{center}
\renewcommand{\arraystretch}{1.3}
\begin{tabular}{|l||r|c|} 
\hline
Parameter  & 68\% C.L.   & 95\% C.L.      \\
\hline
\hline
$\gvz$     & $+0.05\apm{0.16}{0.17}$     & [$-0.28,~~+0.36$]  \\ 
\hline
\hline
$\dgz$     & $-0.025\apm{0.026}{0.026}$  & [$-0.074,~~+0.028$]  \\ 
\hline
$\dkg$     & $-0.002\apm{0.067}{0.065}$  & [$-0.13,~~+0.13$]  \\ 
\hline
$\lg$      & $-0.036\apm{0.028}{0.027}$  & [$-0.089,~~+0.020$]  \\ 
\hline
\end{tabular}
\caption[]{Single-parameter analyses : 
  the central values, 68\% C.L. uncertainties and 95\%
  C.L. intervals ($\LL=0.5,~1.92$) obtained combining the
  results from the four LEP experiments.  In each case the parameter
  listed is varied while the remaining ones are fixed to their
  Standard Model value.  Both statistical and systematic uncertainties
  are included.  }
 \label{tab:cTGC-1-LEP}
\end{center}
\end{table}

\begin{table}[htbp]
\begin{center}
\renewcommand{\arraystretch}{1.3}
\begin{tabular}{|l||r|c|rr|} 
\hline
Parameter  & 68\% C.L.   & 95\% C.L.    & \multicolumn{2}{|c|}{Correlations} \\
\hline
\hline
$\dgz$        &$-0.038\apm{0.041}{0.034}$  
              & [$-0.10,~~+0.03$] & $ 1.00$ & $-0.43$ \\ 
$\dkg$        &$+0.053\apm{0.074}{0.089}$  
              & [$-0.10,~~+0.21$] & $-0.43$ & $ 1.00$ \\ 
\hline
$\dgz$        &$+0.006\apm{0.035}{0.039}$ 
              & [$-0.072,~~+0.076$] & $ 1.00$ & $-0.70$ \\ 
$\lg$         &$-0.048\apm{0.043}{0.036}$
              & [$-0.12,~~+0.03$] & $-0.70$ & $ 1.00$ \\ 
\hline
$\lg$         &$-0.046\apm{0.041}{0.031}$  
              & [$-0.11,~~+0.02$] & $ 1.00$ & $-0.19$ \\ 
$\dkg$        &$+0.037\apm{0.060}{0.069}$  
              & [$-0.09,~~+0.16$] & $-0.19$ & $ 1.00$ \\ 
\hline
\end{tabular}
\caption{Two-parameter analyses : 
  the central values, 68\% C.L. uncertainties and 95\% C.L. intervals 
  ($\LL=0.5,~1.92$) obtained combining the results from the four
  LEP experiments.  In each case the two parameters listed are varied
  while the remaining one is fixed to its Standard Model value. Both
  statistical and systematic uncertainties are included.  Since the
  shape of the log-likelihood is not parabolic, there is some
  ambiguity in the definition of the correlation coefficients and the
  values quoted here are approximate.  }
\label{tab:cTGC-2-LEP}
\end{center}
\end{table}

\begin{table}[htbp]
\begin{center}
\renewcommand{\arraystretch}{1.3}
\begin{tabular}{|l||r|c|rrr|} 
\hline
Parameter  & 68\% C.L.   & 95\% C.L.   
           & \multicolumn{3}{|c|}{Correlations}    \\
\hline
\hline
$\dgz$     & $+0.023\apm{0.039}{0.041}$  & [$-0.06,~~+0.10$] 
           & $ 1.00$ & $-0.20$ & $-0.66$ \\
$\dkg$     & $+0.027\apm{0.073}{0.072}$  & [$-0.11,~~+0.18$] 
           & $-0.20$ & $ 1.00$ & $-0.15$ \\
$\lg$      & $-0.067\apm{0.044}{0.037}$  & [$-0.14,~~+0.02$] 
           & $-0.66$ & $-0.15$ & $ 1.00$ \\
\hline
\end{tabular}
\caption{Three-parameter analyses : 
  the central values, 68\% C.L. uncertainties and 95\% C.L. 
  intervals ($\LL=0.5,~1.92$) obtained combining the results
  from ALEPH, L3 and OPAL.  All three parameters listed are varied.
  Both statistical and systematic uncertainties are included. Since
  the shape of the log-likelihood is not parabolic, there is some
  ambiguity in the definition of the correlation coefficients and the
  values quoted here are approximate.  }
\label{tab:cTGC-3-LEP}
\end{center}
\end{table}

\clearpage

\subsubsection{Neutral Triple Gauge Boson Couplings in Z\boldmath$\gamma$ 
Production}

\begin{figure}[htbp]
  \begin{center}
    \leavevmode
    \mbox{
    \includegraphics[width=0.49\linewidth]{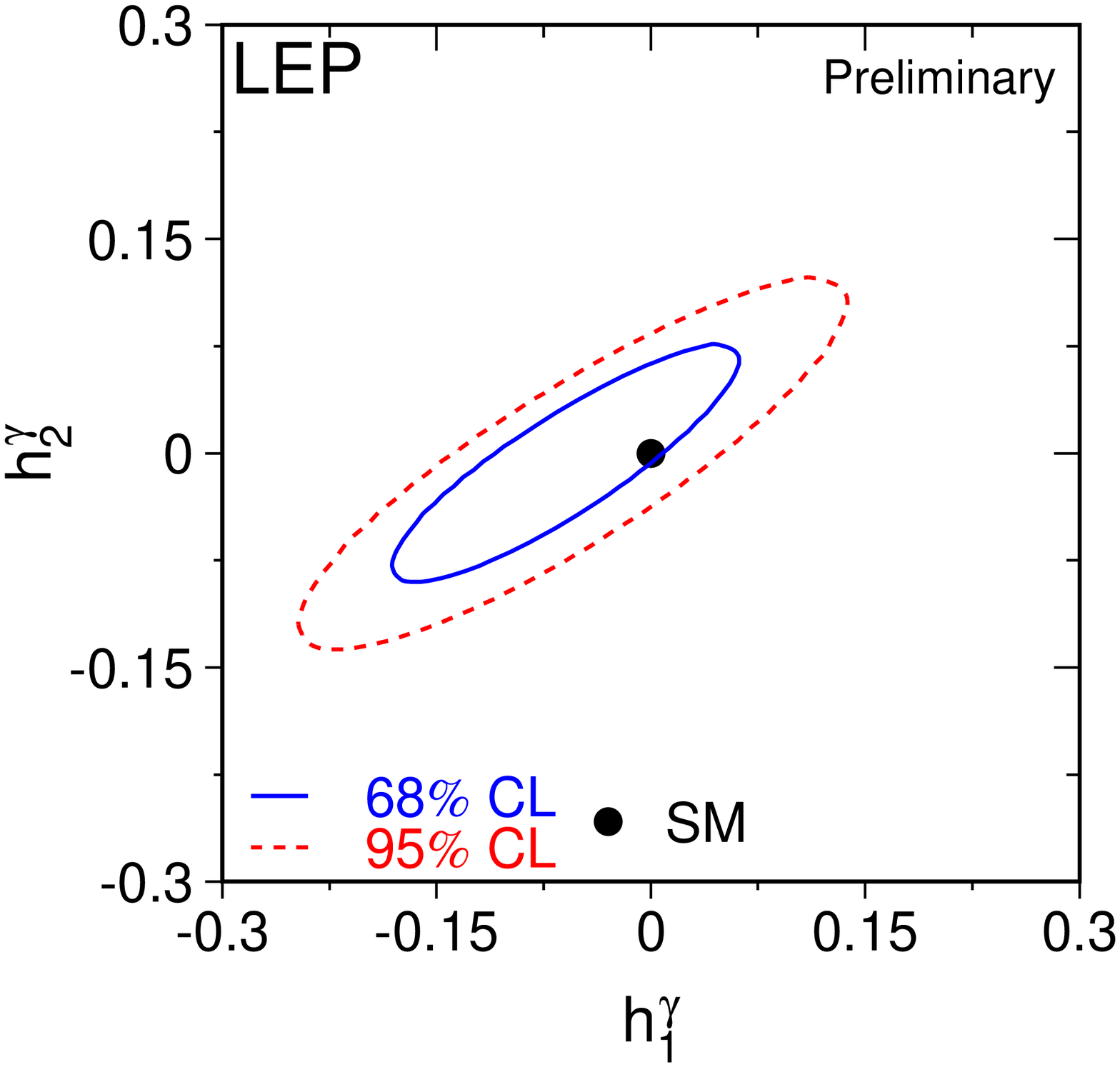}
    \includegraphics[width=0.49\linewidth]{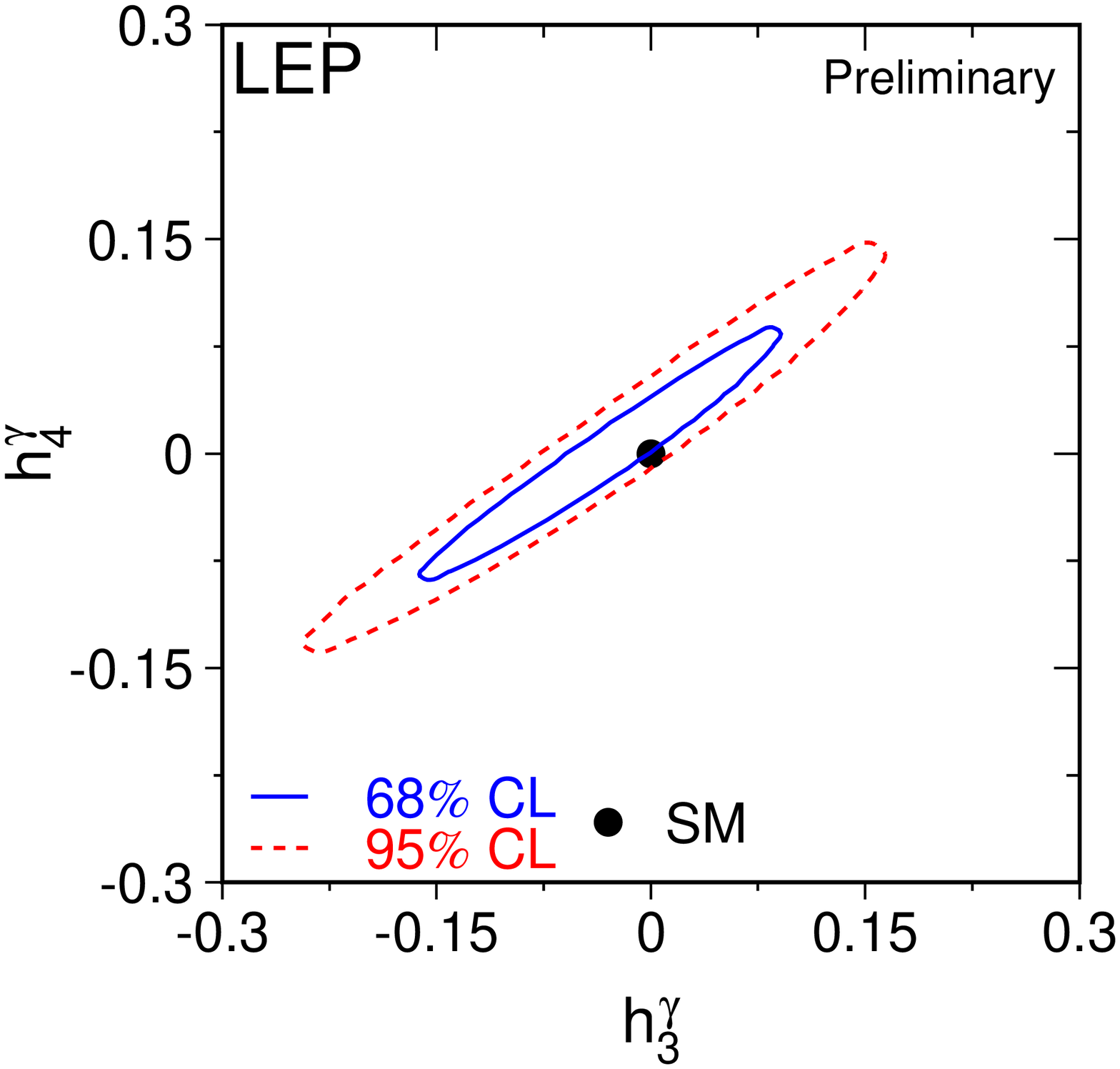}
    }
    \mbox{
    \includegraphics[width=0.49\linewidth]{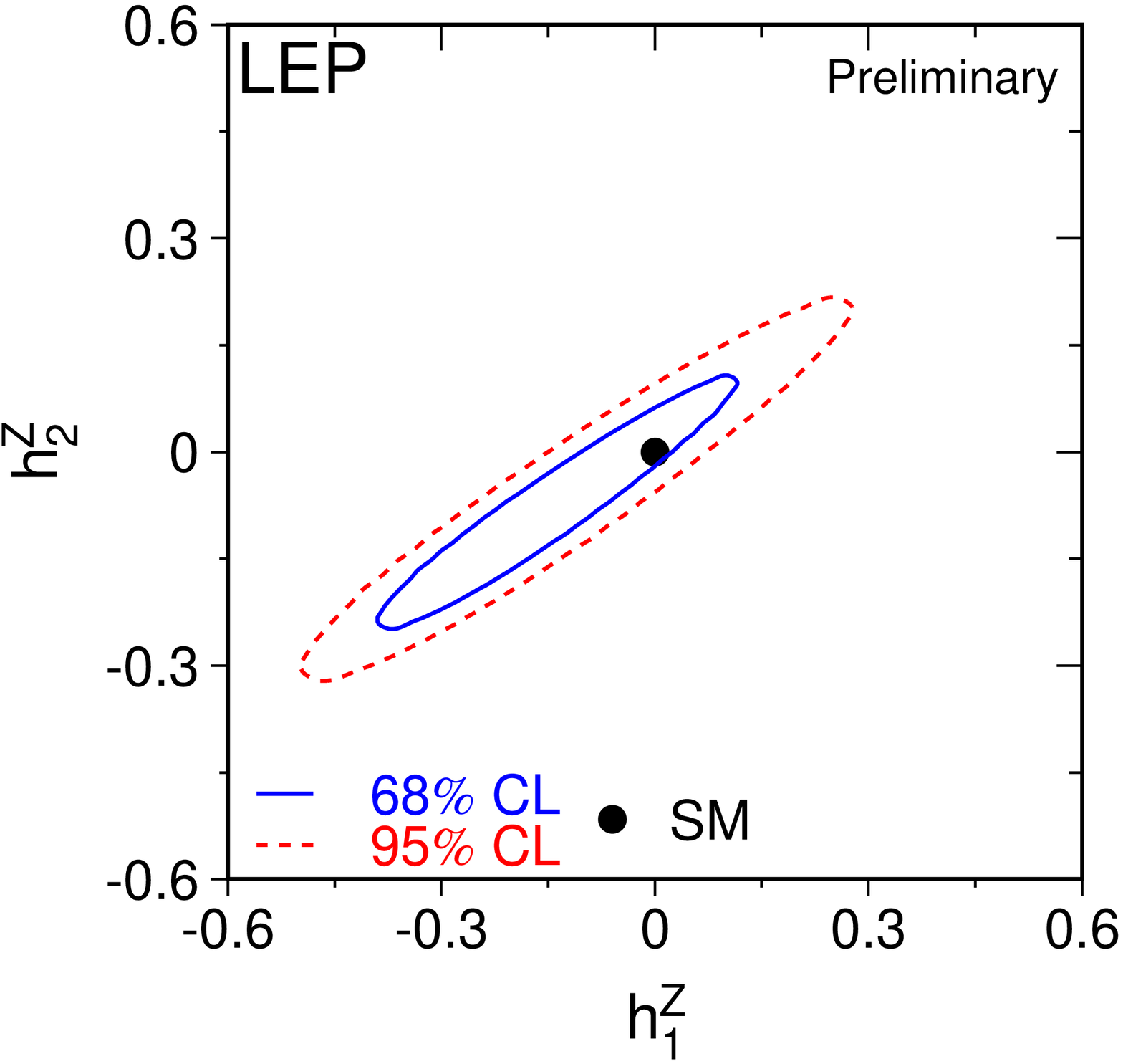}
    \includegraphics[width=0.49\linewidth]{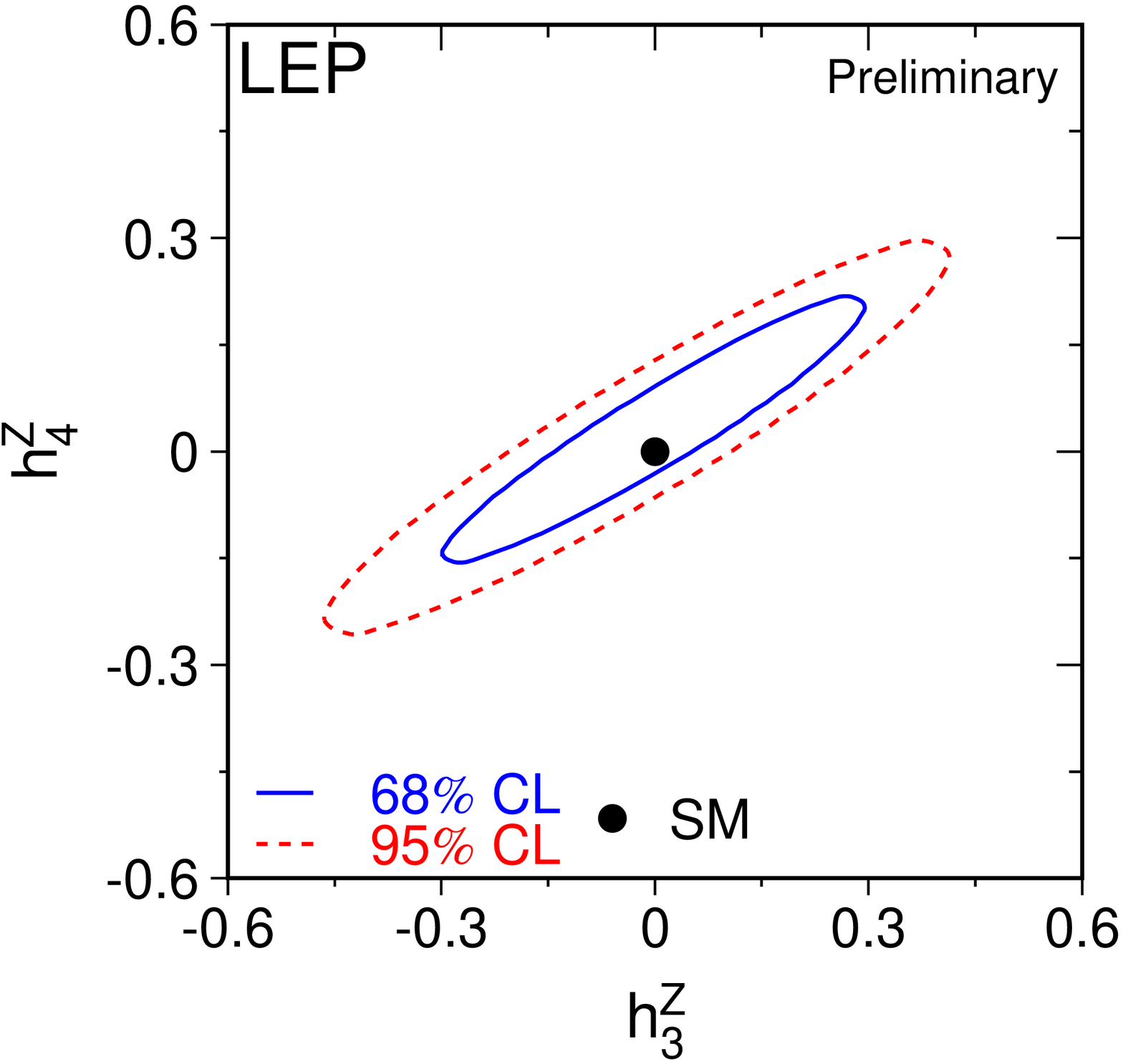}
    }
    \caption[]{
   Contour curves of 68\% C.L. and 95\% C.L. in the planes 
   $(h_1^\gamma,h_2^\gamma)$,
   $(h_3^\gamma,h_4^\gamma)$,
   $(h_1^Z,h_2^Z)$ and
   $(h_3^Z,h_4^Z)$
   showing the LEP combined results.
      }
    \label{fig:hzg-2D}
  \end{center}
\end{figure}

Results of the combination of 
neutral triple gauge boson couplings in Z$\gamma$ 
production
are
given in Tables~\ref{tab:hTGC-1-LEP} and \ref{tab:hTGC-2-LEP} 
for the single, two-parameter analyses
respectively.
The 68\% C.L. and 95\% C.L. contours curves
resulting from the combinations of the two-dimensional likelihood
curves are shown in
Figure~\ref{fig:hzg-2D}.

\begin{table}[htbp]
\begin{center}
\renewcommand{\arraystretch}{1.3}
\begin{tabular}{|l||c|} 
\hline
Parameter     & 95\% C.L.      \\
\hline
\hline
$h_1^\gamma$  & [$-0.10,~~+0.03$]  \\ 
\hline
$h_2^\gamma$  & [$-0.036,~~+0.061$]  \\ 
\hline
$h_3^\gamma$  & [$-0.075,~~-0.004$]  \\ 
\hline
$h_4^\gamma$  & [$+0.005,~~+0.056$]  \\ 
\hline
$h_1^Z$       & [$-0.13,~~+0.04$]  \\ 
\hline
$h_2^Z$       & [$-0.041,~~+0.080$]  \\ 
\hline
$h_3^Z$       & [$-0.16,~~+0.07$]  \\ 
\hline
$h_4^Z$       & [$-0.04,~~+0.10$]  \\ 
\hline
\end{tabular}
\caption[]{Single-parameter analyses : 
 The 95\% C.L. intervals ($\LL=1.92$) obtained
  combining the results from DELPHI, L3 and OPAL.  In each case the
  parameter listed is varied while the remaining ones are fixed to
  their Standard Model value.  Both statistical and systematic
  uncertainties are included.  }
 \label{tab:hTGC-1-LEP}
\end{center}
\end{table}

\begin{table}[htbp]
\begin{center}
\renewcommand{\arraystretch}{1.3}
\begin{tabular}{|l||c|rr|} 
\hline
Parameter  & 95\% C.L. & \multicolumn{2}{|c|}{Correlations} \\
\hline
\hline
$h_1^\gamma$  & [$-0.21,~~+0.10$]    & $ 1.00$ & $+0.88$ \\ 
$h_2^\gamma$  & [$-0.11,~~+0.10$]    & $+0.88$ & $ 1.00$ \\ 
\hline
$h_3^\gamma$  & [$-0.20,~~+0.13$]    & $ 1.00$ & $+0.98$ \\ 
$h_4^\gamma$  & [$-0.11,~~+0.12$]    & $+0.98$ & $ 1.00$ \\ 
\hline
$h_1^Z$       & [$-0.44,~~+0.20$]    & $ 1.00$ & $+0.96$ \\ 
$h_2^Z$       & [$-0.28,~~+0.16$]    & $+0.96$ & $ 1.00$ \\ 
\hline
$h_3^Z$       & [$-0.38,~~+0.35$]    & $ 1.00$ & $+0.95$ \\ 
$h_4^Z$       & [$-0.21,~~+0.26$]    & $+0.95$ & $ 1.00$ \\ 
\hline
\end{tabular}
\caption[]{Two-parameter analyses : 
  The 95\% C.L. intervals ($\LL=1.92$) obtained
  combining the results from DELPHI and L3.  In each case the two
  parameters listed are varied while the remaining ones are fixed to
  their Standard Model value.  Both statistical and systematic
  uncertainties are included.  Since the shape of the log-likelihood
  is not parabolic, there is some ambiguity in the definition of the
  correlation coefficients and the values quoted here are approximate.
  }
 \label{tab:hTGC-2-LEP}
\end{center}
\end{table}

\clearpage

\subsubsection{Neutral Triple Gauge Boson Couplings in ZZ Production}

\begin{figure}[htbp]
  \begin{center}
    \leavevmode
    \mbox{
    \includegraphics[width=0.49\linewidth]{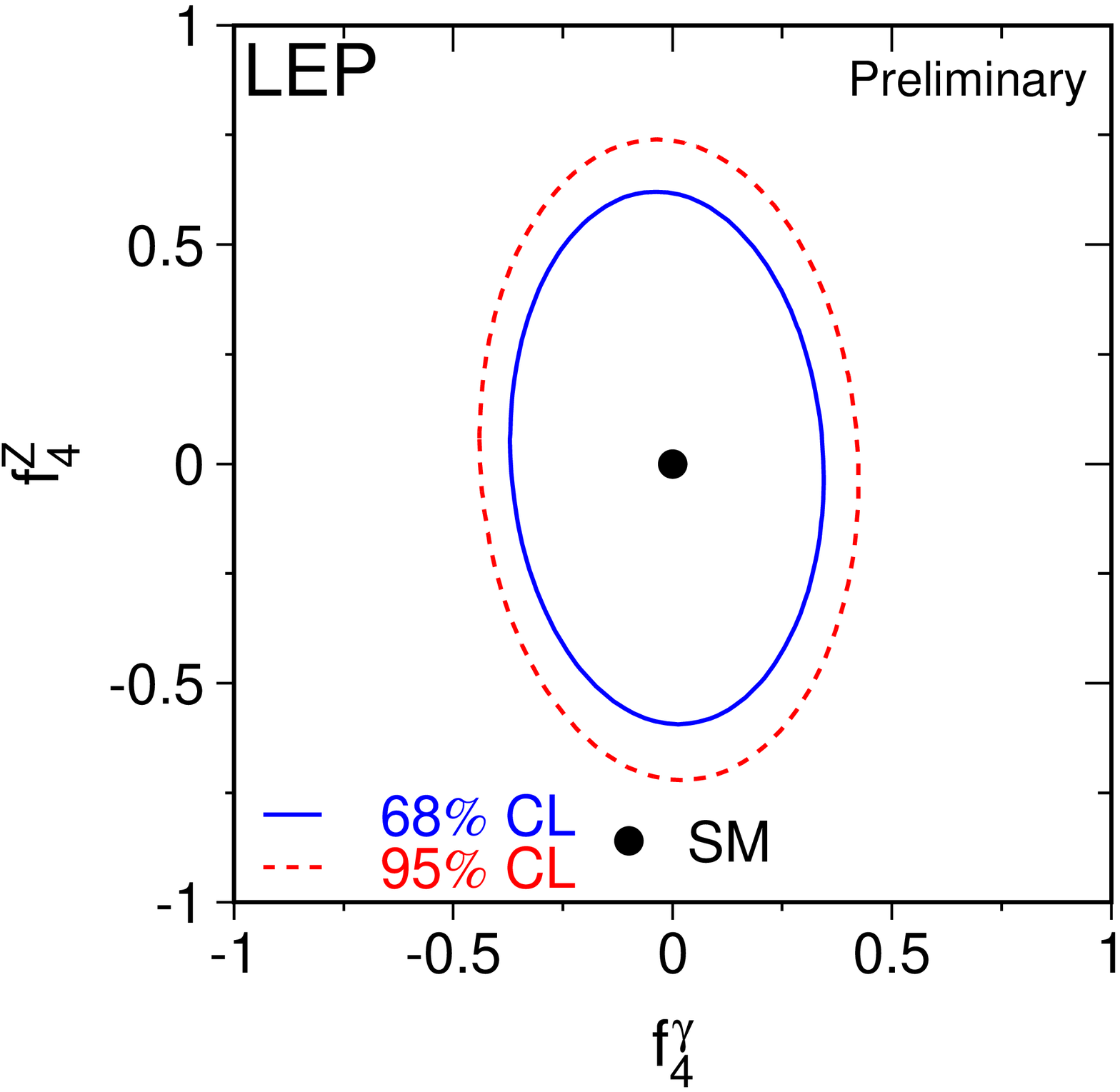}
    \includegraphics[width=0.49\linewidth]{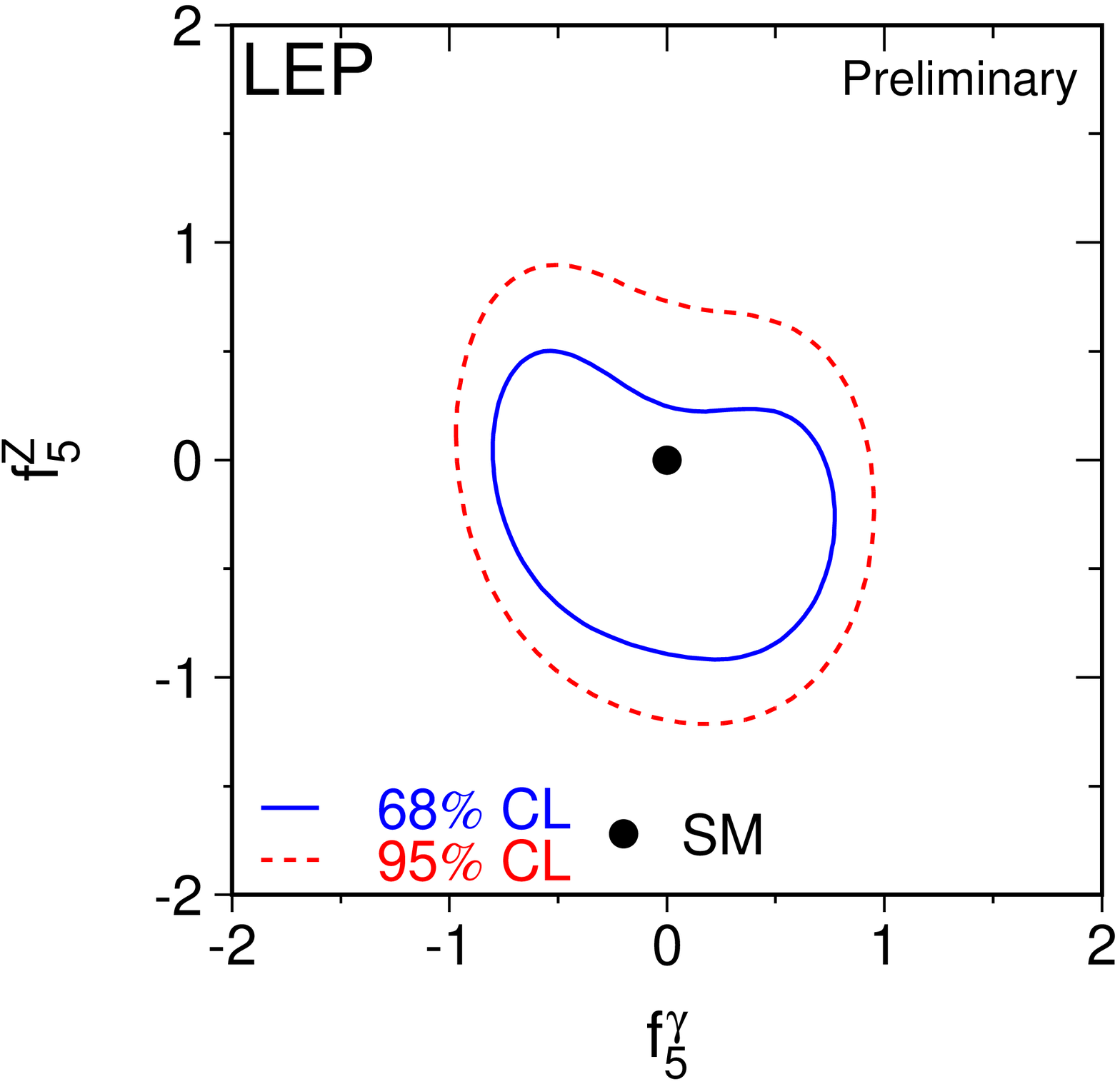}
    }
    \caption[]{
   Contour curves of 68\% C.L. and 95\% C.L. in the planes 
   $(f_4^\gamma,f_4^Z)$ and
   $(f_5^\gamma,f_5^Z)$
   showing the LEP combined results.
      }
    \label{fig:fTGC-2D}
  \end{center}
\end{figure}

Results of the combination of 
neutral triple gauge boson couplings in ZZ production
are
given in Tables~\ref{tab:fTGC-1-LEP} and \ref{tab:fTGC-2-LEP} 
for the single, two-parameter analyses
respectively.
The 68\% C.L. and 95\% C.L. contours curves
resulting from the combinations of the two-dimensional likelihood
curves are shown in
Figure~\ref{fig:fTGC-2D}.

\begin{table}[htbp]
\begin{center}
\renewcommand{\arraystretch}{1.3}
\begin{tabular}{|l||c|} 
\hline
Parameter     & 95\% C.L.     \\
\hline
\hline
$f_4^\gamma$  & [$-0.41,~~+0.39$]  \\ 
\hline
$f_4^Z$       & [$-0.66,~~+0.68$]  \\ 
\hline
$f_5^\gamma$  & [$-0.89,~~+0.84$]  \\ 
\hline
$f_5^Z$       & [$-1.1,~~+0.5$]  \\ 
\hline
\end{tabular}
\caption[]{Single-parameter analyses : 
  The 95\% C.L. intervals ($\LL=1.92$) obtained
  combining the results from DELPHI, L3 and OPAL.  In each case the
  parameter listed is varied while the remaining ones are fixed to
  their Standard Model value.  Both statistical and systematic
  uncertainties are included.  }
 \label{tab:fTGC-1-LEP}
\end{center}
\end{table}

\begin{table}[htbp]
\begin{center}
\renewcommand{\arraystretch}{1.3}
\begin{tabular}{|l||c|rr|} 
\hline
Parameter     & 95\% C.L. & \multicolumn{2}{|c|}{Correlations} \\
\hline
\hline
$f_4^\gamma$  &[$-0.40,~~+0.38$] & $ 1.00$ & $+0.88$\\ 
$f_4^Z$       &[$-0.66,~~+0.68$] & $+0.88$ & $ 1.00$\\ 
\hline
$f_5^\gamma$  &[$-0.88,~~+0.86$] & $ 1.00$ & $-0.52$\\ 
$f_5^Z$       &[$-1.1,~~+0.7$]   & $-0.52$ & $ 1.00$\\ 
\hline
\end{tabular}
\caption[]{Two-parameter analyses : 
  The 95\% C.L. intervals ($\LL=1.92$) obtained
  combining the results from DELPHI, L3 and OPAL.  In each case the
  two parameters listed are varied while the remaining ones are fixed
  to their Standard Model value.  Both statistical and systematic
  uncertainties are included. Since the shape of the log-likelihood is
  not parabolic, there is some ambiguity in the definition of the
  correlation coefficients and the values quoted here are approximate.
  }
 \label{tab:fTGC-2-LEP}
\end{center}
\end{table}

\clearpage

\subsubsection{Quartic Gauge Boson Couplings}

\begin{figure}[htbp]
\begin{center}
\includegraphics[width=\linewidth]{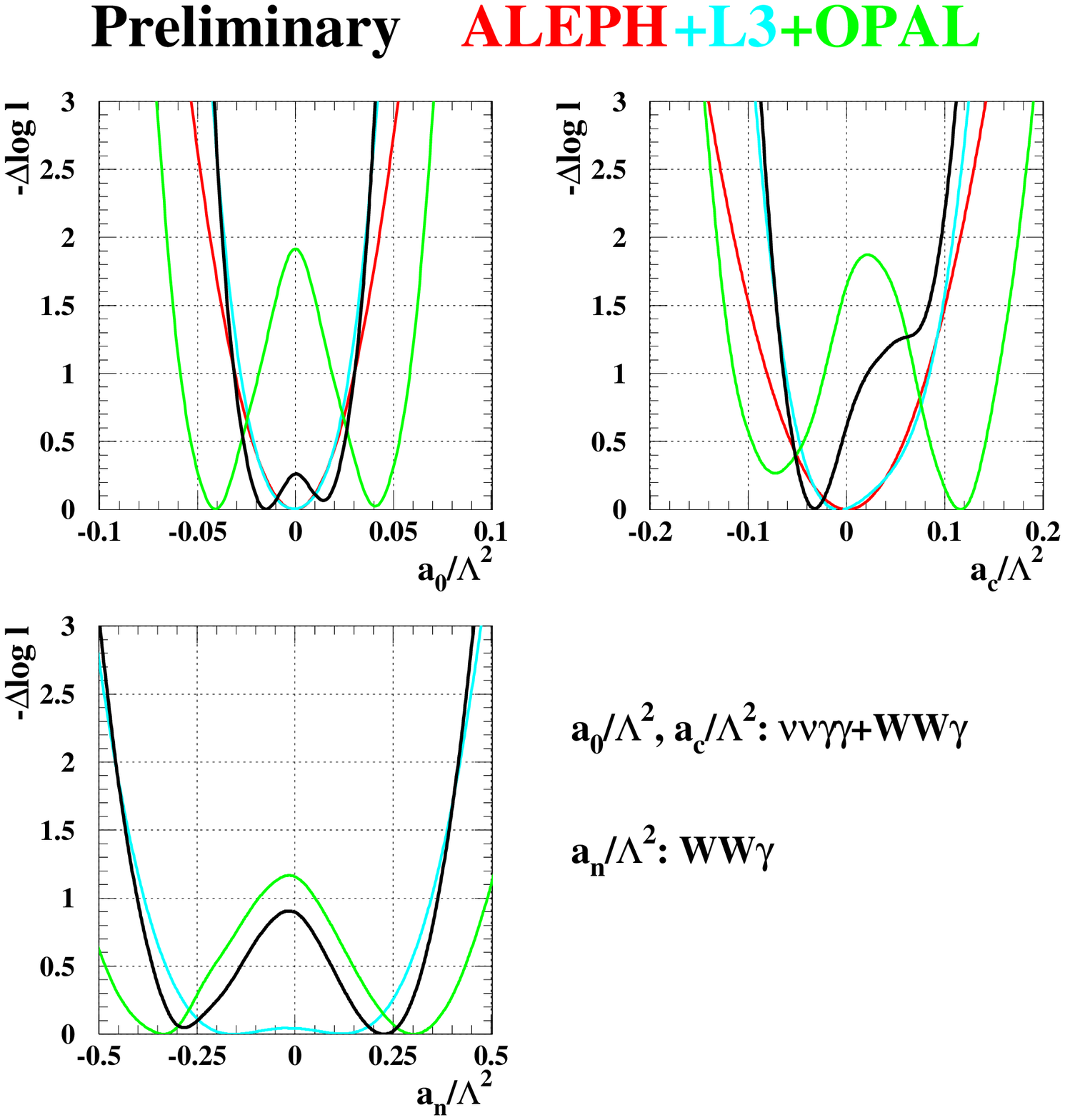}
\caption[]{
  The $\LL$ curves of ALEPH, L3 and OPAL, and the LEP combined curve
  for the QGCs derived from $\WWg$ and $\nngg$ production.  In each
  case, the minimal value is subtracted.  }
\label{fig:wQGC-1}
\end{center}
\end{figure}

The individual $\LL$ curves and their sum are
shown in Figure~\ref{fig:wQGC-1} for the $\WWg$ and $\nngg$ channels.
The results
of the combination are given in Table~\ref{tab:QGC-1-LEP}.

\begin{table}[htbp]
\begin{center}
\renewcommand{\arraystretch}{1.3}
\begin{tabular}{|l||c|} 
\hline
Parameter  & 95\% C.L.      \\
$[\GeV^{-2}]$ & \\
\hline
\hline
$\azl$     & [$-0.037,~~+0.036$]  \\ 
\hline
$\acl$     & [$-0.077,~~+0.095$]  \\ 
\hline
$\anl$     & [$-0.45,~~+0.41$]  \\ 
\hline
\hline
$\azl$     & [$-0.0048,~~+0.0056$]  \\ 
\hline
$\acl$     & [$-0.0052,~~+0.0099$]  \\ 
\hline
\end{tabular}
\caption[]{ The 95\% C.L. intervals ($\LL=1.92$) obtained
  combining the results from ALEPH, L3 and OPAL.  In each case the
  parameter listed is varied while the remaining ones are fixed to
  their Standard Model value.  Both statistical and systematic
  uncertainties are included.  Top: $\WWg$ and $\nngg$. Bottom:
  $\Zgg$.  }
 \label{tab:QGC-1-LEP}
\end{center}
\end{table}

\subsection{Conclusions}

The existence of triple gauge boson couplings among the electroweak
gauge bosons is experimentally verified.  No significant deviation
from the Standard Model predictions is seen for any of the electroweak
gauge boson couplings investigated.  As an example, these data allow
the Kaluza-Klein theory~\cite{klein}, in which $\kg = -2$, to be
excluded~\cite{maiani}.

\clearpage

\clearpage

\clearpage
\boldmath
\section{Interpretation of Results}
\label{sec-IofR}
\unboldmath

\updates{
Results on $\swsqeffl$ from asymmetries are averaged for
leptonic and hadronic channels separately.  
A new fit for the top mass is performed where
all data except the direct measurement of $\Mt$ is included.
}

\boldmath
\subsection{Number of Neutrino Species}
\label{sec-Nnu}
\unboldmath

An important aspect of our measurement concerns the information
related to $\Zzero$ decays into invisible channels. Using the results
of Table~\ref{tab-zparavg}, the ratio of
the $\Zzero$ decay width into invisible particles and the leptonic
decay width is determined:
\begin{eqnarray}
\Ginv / \Gll & = & 5.942\pm 0.016\,.
\end{eqnarray}
The Standard Model value for the ratio of the partial widths to
neutrinos and charged leptons is:
\begin{eqnarray}
(\Gnu / \Gll)_{\mathrm{SM}} & = & 1.9912\pm 0.0012\,.
\end{eqnarray}
The central value is evaluated for $\MZ=91.1875$~\GeV{} 
and the error quoted accounts for
a variation of $\Mt$ in the range $\Mt=174.3\pm5.1~\GeV$ and a
variation of $\MH$ in the range $\HI$.  The number of light neutrino
species is given by the ratio of the two expressions listed above:
\begin{eqnarray}
\Nnu & = & 2.9841\pm 0.0083,
\end{eqnarray}
which is 2 standard deviations below the expected value of 3.

Alternatively, one can assume 3 neutrino species and determine the
width from additional invisible decays of the Z.  This yields
\begin{eqnarray}
  \Delta\Ginv & = & -2.7 \pm 1.6\ \MeV.
\end{eqnarray}
The measured total width
is below the Standard Model expectation.  
If
a conservative approach is taken to limit the result to only positive
values of $\Delta\Ginv$, then the 95\% CL upper limit on additional
invisible decays of the Z is
\begin{eqnarray}
  \Delta\Ginv & < & 2.0\ \MeV.
\end{eqnarray}

The uncertainties on $\Nnu$ and  $\Delta\Ginv$ are
dominated by the theoretical error on the luminosity.  These results
have therefore improved due to the improved theoretical calculations on
Bhabha scattering\cite{bib-lumth99}.

\boldmath
\subsection{The Coupling Parameters $\cAf$}
\label{sec-AF}
\unboldmath

The coupling parameters $\cAf$ are defined in terms of the effective
vector and axial-vector neutral current couplings of fermions
(Equation~(\ref{eqn-cAf})).  The LEP measurements of the
forward-backward asymmetries of charged leptons (Section~\ref{sec-LS})
and b and c quarks (Section~\ref{sec-HF}) determine the products
$\Afbzf=\frac{3}{4}\cAe\cAf$ (Equation~(\ref{eqn-apol})). The LEP
measurements of the $\tau$ polarisation (Section~\ref{sec-TP}),
$\ptau(\cos\theta)$, determine $\cAt$ and $\cAe$ separately
(Equation~(\ref{eqn-taupol})).

Table~\ref{tab-AF-L} shows the results for the leptonic coupling
parameter $\cAl$ from the LEP and SLD measurements, assuming lepton universality.

\begin{table}[tbp]
\begin{center}
\renewcommand{\arraystretch}{1.15}
\begin{tabular}{|c||c|c|r|}
\hline
                   & $\cAl$            & Cumulative Average & $\chi^2$/d.o.f.\\
\hline
\hline
$\Afbzl$           & $0.1512\pm0.0042$ &                    &                \\
$\ptau(\cos\theta)$& $0.1464\pm0.0032$ & $0.1482\pm 0.0025$ &  0.8/1         \\
\hline
$\cAl$ (SLD)& $0.1513\pm0.0021$ & $0.1500\pm 0.0016$ &  1.7/2          \\
\hline
\end{tabular}
\end{center}
\caption[]{
  Determination of the leptonic coupling parameter $\cAl$ assuming
  lepton universality. The second column lists the $\cAl$ values
  derived from the quantities listed in the first column. The third
  column contains the cumulative averages of these $\cAl$ results. The
  averages are derived assuming no correlations between the
  measurements. The $\chi^2$ per degree of freedom for the cumulative
  averages is given in the last column.  }
\label{tab-AF-L}
\end{table}

\begin{table}[tbp]
\begin{center}
\renewcommand{\arraystretch}{1.15}
\begin{tabular}{|c||c|c|c|c|}
\hline
         &    LEP                  & SLD  & LEP+SLD  & \mcc{Standard} \\
         & ($\cAl=0.1482\pm0.0025$)&      & ($\cAl=0.1500\pm0.0016$) & \mcc{Model fit}\\
\hline
\hline
$\cAb$   & $0.890\pm0.024$    & $0.922 \pm 0.023$  & $0.898\pm0.015$ & 0.935 \\
$\cAc$   & $0.619\pm0.035$    & $0.631 \pm 0.026$  & $0.623\pm0.020$ & 0.668 \\
\hline
\end{tabular}
\end{center}
\caption[]{
  Determination of the quark coupling parameters $\cAb$ and $\cAc$
  from LEP data alone (using the LEP average for $\cAl$), from SLD
  data alone, and from LEP+SLD data (using the LEP+SLD average for
  $\cAl$) assuming lepton universality.  }
\label{tab-AF-Q}
\end{table}

\begin{figure}[htbp]
  \begin{center}
    \leavevmode
   \mbox{
    \includegraphics[width=0.48\textwidth]{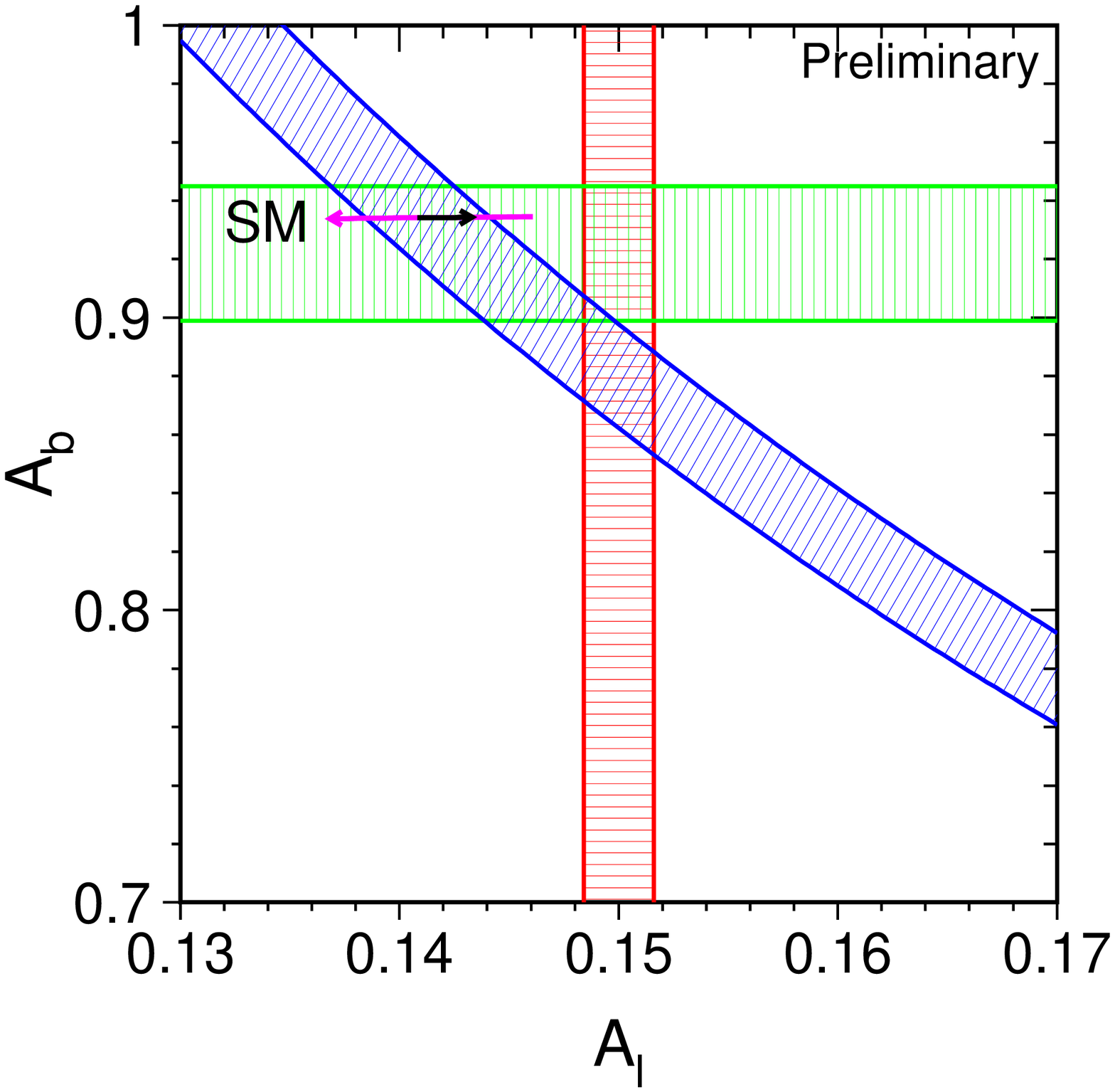}
    \includegraphics[width=0.48\textwidth]{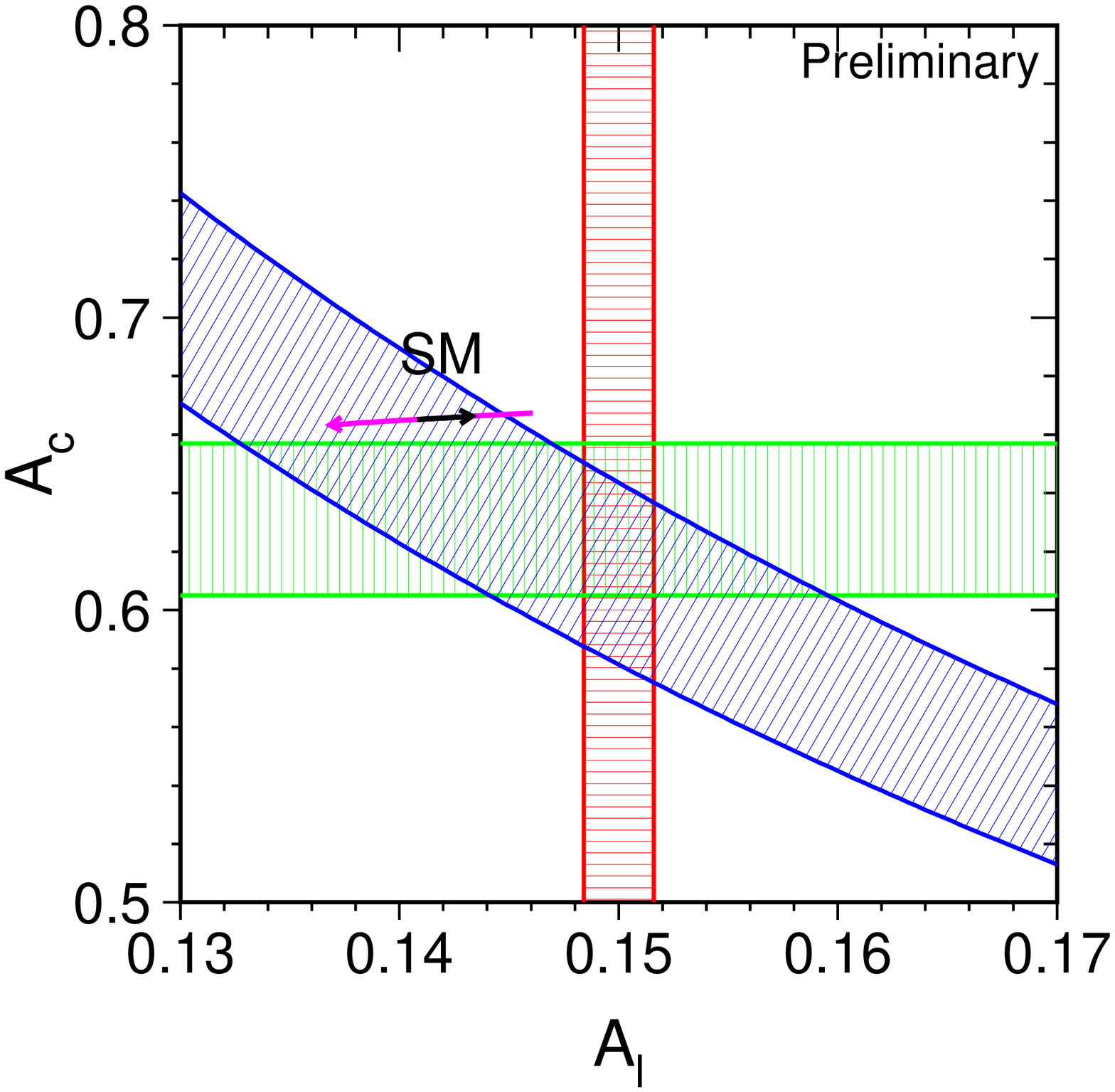}
    }
    \caption{The measurements of the combined LEP+SLD $\cAl$ (vertical
    band), SLD $\cAb$,$\cAc$ (horizontal bands) and LEP $\Afbzb$,$\Afbzc$ (diagonal
    bands), compared to the Standard Model expectations (arrows).  The
    arrow pointing to the left shows the variation in the SM
    prediction for $\MH$ in the range $300^{+700}_{-187}$ \GeV, and the arrow
    pointing to the right for $\Mt$ in the range $174.3 \pm 5.1$
    \GeV.  
    Although 
    the $\Afbzb$ measurements
    prefer a high Higgs mass, the Standard Model fit to the full set
    of measurements prefers a low Higgs mass, for
    example because of the influence
    of $\cAl$.
    }
    \label{fig-ae_ab}
  \end{center}
\end{figure}

Using the measurements of $\cAl$ one can extract $\cAb$ and $\cAc$
from the LEP measurements of the b and c quark asymmetries.  The SLD
measurements of the left-right forward-backward asymmetries for b and
c quarks are direct determinations of $\cAb$ and
$\cAc$.  Table~\ref{tab-AF-Q} shows the results on the quark coupling
parameters $\cAb$ and $\cAc$ derived from LEP or SLD measurements
separately (Equations~\ref{eqn-hf4} and~\ref{eqn-hf6}) and from the
combination of LEP+SLD measurements (Equation~\ref{eqn-hf6}).  The
LEP extracted values of $\cAb$ and $\cAc$ in excellent agreement
with the SLD measurements, and in reasonable agreement with the
Standard Model predictions (0.935 and 0.668, respectively, 
essentially
independent of $\Mt$ and $\MH$).  
The combination of LEP and
SLD of $\cAb$ and $\cAc$ 
are  2.5 and 2.3 sigma below the Standard Model respectively.  
This is mainly because the 
$\cAb$ value deduced from the measured $\Afbzb$ 
and the combined $\cAl$ is low 
compared to both the Standard Model and 
the direct measurement of $\cAb$,
this can also be seen in Figure~\ref{fig-ae_ab}.

\boldmath
\subsection{The Effective Vector and Axial-Vector Coupling Constants}
\label{sec-GAGV}
\unboldmath

The partial widths of the $\Zzero$ into leptons and the lepton
forward-backward asymmetries (Section~\ref{sec-LS}), the $\tau$
polarisation and the $\tau$ polarisation asymmetry
(Section~\ref{sec-TP}) are combined to determine the effective
vector and axial-vector couplings for $\rm e$, $\mu$ and $\tau$. The
asymmetries (Equations~(\ref{eqn-apol}) and~(\ref{eqn-taupol}))
determine the ratio $\gvhatl/\gahatl$ (Equation~(\ref{eqn-cAf})),
while the leptonic partial widths determine the sum of the squares of
the couplings:
\begin{eqnarray}
\label{eqn-Gll}
\Gll & = &
{{\GF\MZ^3}\over{6\pi\sqrt 2}}
(\gvhatl^{2}+\gahatl^{2})(1+\delta^{QED}_\ell)\,,
\end{eqnarray}
where $\delta^{QED}_\ell=3q^2_\ell\alpha(\MZ^2)/(4\pi)$ accounts for
final state photonic corrections. Corrections due to lepton masses,
neglected in Equation~\ref{eqn-Gll}, are taken into account for the
results presented below.

\begin{figure}[htbp]
\begin{center}
  \mbox{\includegraphics[width=\linewidth]{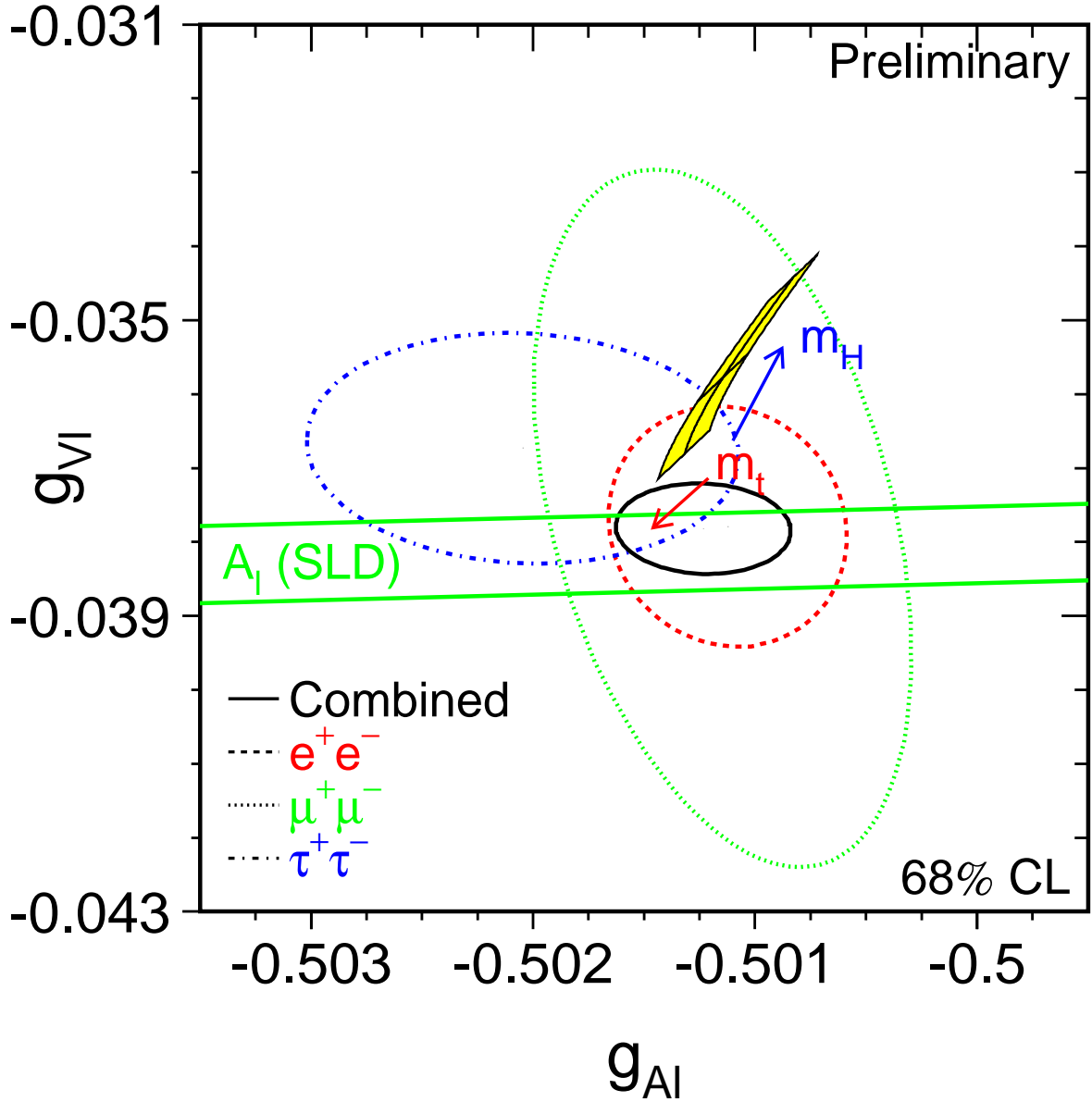}}
\end{center}
\caption[]{
  Contours of 68\% probability in the $\gvhatl$-$\gahatl$ plane from
  LEP measurements.  Also shown is the one standard deviation band
  resulting from the $\cAl$ measurement of SLD\@. The solid contour
  results from a fit to the LEP and SLD results assuming
  lepton universality. The shaded region
  corresponds to the Standard Model prediction for $\Mt = 174.3 \pm
  5.1$~\GeV{} and $\MH=300^{+700}_{-187}~\GeV$. The arrows point in
  the direction of increasing values of $\Mt$ and $\MH$.  }
\label{fig-gagv}
\end{figure}

The averaged results for the effective lepton couplings are given in
Table~\ref{tab-coup} for both the LEP data alone as well as for the
LEP and SLD measurements.  
Figure~\ref{fig-gagv} shows the 68\% probability
contours in the $\gahatl$-$\gvhatl$ plane for the individual lepton
species from the LEP data. The signs of $\gahatl$ and
$\gvhatl$ are based on the convention $\gahate < 0$. With this
convention the signs of the couplings of all charged leptons follow
from LEP data alone. For comparison, the $\gvhatl$-$\gahatl$ relation
following from the measurement of $\cAl$ from SLD\cite{ref:sld-s99} is
indicated as a band in the $\gahatl$-$\gvhatl$-plane of
Figure~\ref{fig-gagv}.  
The measured ratios of the $\rm e$, $\mu$ and $\tau$ couplings provide a
test of lepton universality and are shown in
Table~\ref{tab-coup}.  All values are consistent with lepton
universality.  The combined results assuming universality are also
given in the table and are shown as a solid contour in
Figure~\ref{fig-gagv}. 

The neutrino couplings to the $\Zzero$ can be derived from the
measured value of the invisible width of the $\Zzero$, $\Ginv$ (see
Table~\ref{tab-widths}), attributing it exclusively to the decay into
three identical neutrino generations ($\Ginv=3\Gnu$) and assuming
$\gahatn\equiv\gvhatn\equiv\ghatn$.  The relative sign of $\ghatn$ is
chosen to be in agreement with neutrino scattering
data\cite{ref:CHARMIIgn}, resulting in $\ghatn = +0.50068\pm 0.00075$.
\begin{table}[htbp]
\renewcommand{\arraystretch}{1.15}
\begin{center}
\begin{tabular}{|l||c|c|}
\hline
&\multicolumn{2}{|c|}{Without Lepton Universality:} \\
 & LEP & LEP+SLD\\
\hline
\hline
$\gvhate$    & $-0.0378  \pm 0.0011 $ & $-0.03809 \pm 0.00047 $ \\
$\gvhatmu$   & $-0.0376  \pm 0.0031 $ & $-0.0360  \pm 0.0024 $ \\
$\gvhattau$  & $-0.0367  \pm 0.0010 $ & $-0.0364  \pm 0.0010 $ \\
$\gahate$    & $-0.50112 \pm 0.00035$ & $-0.50111 \pm 0.00035$ \\
$\gahatmu$   & $-0.50115 \pm 0.00056$ & $-0.50120 \pm 0.00054$ \\
$\gahattau$  & $-0.50205 \pm 0.00064$ & $-0.50204 \pm 0.00064$ \\
\hline
\hline
&\multicolumn{2}{|c|}{Ratios of couplings:} \\
 & LEP & LEP+SLD\\
\hline
\hline
$\gvhatmu/\gvhate$ & $0.995\pm0.093$    &$0.962\pm0.062$\\
$\gvhattau/\gvhate$& $0.972\pm0.039$    &$0.958\pm0.029$\\
$\gahatmu/\gahate$ & $1.0001\pm0.0014$  &$1.0002\pm0.0014$\\
$\gahattau/\gahate$& $1.0018\pm0.0015$  &$1.0019\pm0.0015$\\
\hline
\hline
&\multicolumn{2}{|c|}{With Lepton Universality:   } \\
 & LEP & LEP+SLD\\
\hline
\hline
$\gvhatl$    & $-0.03734 \pm 0.00065$& $-0.03782 \pm 0.00040$\\
$\gahatl$    & $-0.50126 \pm 0.00026$& $-0.50123 \pm 0.00026$\\
\hline
$\ghatn$     & $+0.50068 \pm 0.00075$& $+0.50068 \pm 0.00075$\\
\hline
\end{tabular}
\end{center}
\caption[]{%
  Results for the effective vector and axial-vector couplings derived
  from the combined LEP data without and with the assumption of lepton
  universality. For the right column the SLD measurements of $\ALRz$,
  $\cAe$, $\cAm$ and $\cAt$ are also included.}
\label{tab-coup}
\end{table}

\boldmath
\subsection{The Effective Electroweak Mixing Angle $\swsqeffl$}
\label{sec-SW}
\unboldmath

The asymmetry measurements from LEP 
and SLD can be combined into a single
observable, the effective electroweak mixing angle, $\swsqeffl$,
defined as:
\begin{eqnarray}
\label{eqn-sw}
\swsqeffl & \equiv &
\frac{1}{4}\left(1-\frac{\gvhatl}{\gahatl}\right)\,,
\end{eqnarray}
without making strong model-specific assumptions.

\begin{figure}[p]
  \begin{center}
    \leavevmode
    \includegraphics[width=0.9\linewidth]{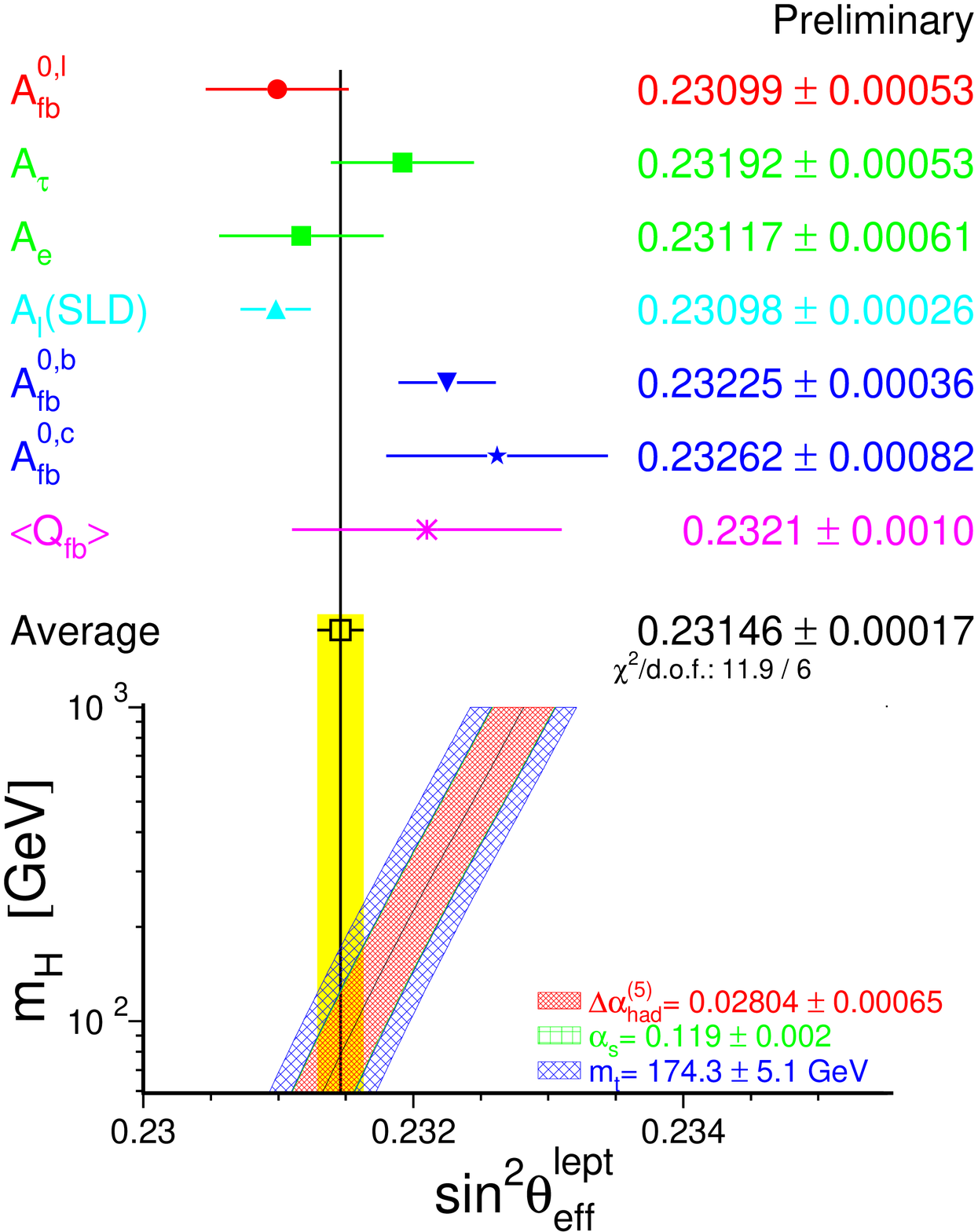}
  \end{center}
  \caption[]{
    Comparison of several determinations of $\swsqeffl$ from 
    asymmetries.  In the average, the small correlation between
    $\Afbzb$ and $\Afbzc$ is included.
    Also shown is the prediction of the Standard Model
    as a function of $\MH$.  The width of the Standard Model band is
    due to the uncertainties in
    $\Delta\alpha_{\mathrm{had}}^{(5)}(\MZ^2)$ (see Section~\ref{sec-SM}),
    $\alfmz$ and $\Mt$.
    The total width of the band is the linear sum of these effects.}
  \label{fig-swsq}
\end{figure}

For a combined average of $\swsqeffl$ from $\Afbpol$, $\cAt$ and
$\cAe$ only the assumption of lepton universality, already inherent in
the definition of $\swsqeffl$, is needed.
Also
the value derived from the measurements of $\cAl$ from SLD is given.  
We can also include the hadronic forward-backward asymmetries if we
assume the quark couplings to be given by the Standard Model.  This
is justified within the Standard Model as the hadronic asymmetries
$\Afbzb$ and $\Afbzc$ have a reduced sensitivity to corrections
particular to the quark vertex.  The  
results of these determinations of $\swsqeffl$ and their combination
are shown in Table~\ref{tab-swsq} and in Figure~\ref{fig-swsq}.  
The combinations based on 
leptonic and hadronic results show a difference
of 3.0 sigma, mainly caused by the
two most precise measurements of $\swsqeffl$, $\ALRz$ (SLD) and
$\Afbzb$ (LEP). This is the same effect as discussed
in section~\ref{sec-AF} and shown in Figure~\ref{fig-ae_ab}.

\begin{table}[htbp]
\renewcommand{\arraystretch}{1.25}
\begin{center}
\begin{tabular}{|l||c|c|c|c|}
\hline
     & $\swsqeffl$&\mco{Average by Group}&Cumulative &       \\
     &            &\mco{of Observations} &Average    &$\chi^2$/d.o.f.\\
\hline
\hline
$\Afbpol$ & $0.23099\pm 0.00053$ &&& \\
$\cAt$    & $0.23192\pm 0.00053$ &&& \\
$\cAe$    & $0.23117\pm 0.00061$ &$0.23138\pm0.00032$
                                 &$0.23138\pm0.00032$&1.7/2\\
\hline
$\cAl$ (SLD)&$0.23098\pm0.00026$ &$0.23098\pm0.00026$
                                 &$0.23114\pm0.00020$&2.6/3\\
\hline
$\Afbzb$  & $0.23225\pm 0.00036$ &&& \\
$\Afbzc$  & $0.23262\pm 0.00082$ &&& \\
$\avQfb$  & $0.2321 \pm 0.0010 $ &$0.23228\pm0.00032$
                                 &$0.23146\pm0.00017$&11.9/6\\
\hline
\end{tabular}\end{center}
\caption[]{
  Determinations of $\swsqeffl$ from asymmetries.  
  The second
  column lists the $\swsqeffl$ values derived from the quantities
  listed in the first column. The third column contains the averages
  of these numbers by groups of observations, where the groups are
  separated by the horizontal lines. The fourth column shows the
  cumulative averages. The $\chi^2$ per degree of freedom for the
  cumulative averages is also given. The averages have
  been performed including the small correlation between $\Afbzb$ and 
  $\Afbzc$. }
\label{tab-swsq}
\end{table}

\clearpage

\boldmath
\subsection{Constraints on the Standard Model}
\label{sec-SM}
\unboldmath

The precise electroweak measurements performed at LEP and SLC and elsewhere
can be used to 
check the validity of the Standard Model and, within its framework, to
infer valuable information about its fundamental parameters. The
accuracy of the measurements makes them sensitive to the mass of the top quark
$\Mt$, and to the mass of the Higgs boson $\MH$ through loop
corrections. While the leading $\Mt$ dependence is quadratic, the
leading $\MH$ dependence is logarithmic.  Therefore, the inferred
constraints on $\MH$ are not very strong.

The LEP and SLD\cite{ref:sld-s99} measurements used are summarised in Table~\ref{tab-SMIN}
together with the results of the Standard Model fit.  Also shown are the
results of
measurements of $\MW$ from UA2\cite{bib-UA2MW},
CDF\cite{bib-CDFMW1,bib-CDFMW2}, and D\O\cite{bib-D0MW}\footnote{See
  Reference~\citen{bib-MWAVE-00} for a combination of these $\MW$
  measurements.}, measurements of the top quark mass by
CDF\cite{bib-topCDF} and D\O\cite{bib-topD0}\footnote{
  See Reference~\citen{bib-Tevatop} for a combination of these
  $\Mt$ measurements.}, and  measurements of the neutrino-nucleon
 neutral to charged
current ratios from CCFR\cite{bib-CCFRnn} and NuTeV\cite{bib-NuTeV}.
Although 
these latter
results are quoted in terms of $\sin^2\theta_W=\swsq$, 
radiative corrections result
in small $\Mt$ and $\MH$ dependences\footnote{The formula used is 
$\delta\sin^2\theta_W = -0.00142 \frac{\Mt^2 -
  (175\GeV)^2}{(100\GeV)^2} + 0.00048 \ln(\frac{\MH}{150\GeV}).$ See
Reference \citen{bib-NuTeV} for details.}
that are included in the fit.   
In addition, the value
of the electromagnetic coupling constant $\alpha(\MZ^2)$, which is
used in the fits, is shown.  An additional input parameter, not shown
in the table, is the Fermi constant $G_F$, determined from the $\mu$
lifetime, $G_F = (1.16637 \pm 0.00001) \times 10^{-5}
\GeV^{-2}$\cite{bib-Gmu}.  The relative error of $G_F$ is
comparable to that of $\MZ$;  both errors have negligible effects in the fit
results.

\begin{table}[p]
\begin{center}
\renewcommand{\arraystretch}{1.10}
\begin{tabular}{|ll||r|r|r|r|}
\hline
 && \mcc{Measurement with}  &\mcc{Systematic} & \mcc{Standard} & \mcc{Pull} \\
 && \mcc{Total Error}       &\mcc{Error}      & \mcc{Model fit}&            \\
\hline
\hline
&&&&& \\[-3mm]
& $\Delta\alpha^{(5)}_{\mathrm{had}}(\MZ^2)$\cite{bib-JEG2,bib-alphalept}
                & $0.02804 \pm 0.00065$ & 0.00064 &0.02804& $ 0.0$ \\
&&&&& \\[-3mm]
\hline
a) & \underline{LEP}     &&&& \\
   & line-shape and      &&&& \\
   & lepton asymmetries: &&&& \\
&$\MZ$ [\GeV{}] & $91.1875\pm0.0021\pz$
                               &${}^{(a)}$0.0017$\pz$ &91.1874$\pz$ & $ 0.0$ \\
&$\GZ$ [\GeV{}] & $2.4952 \pm0.0023\pz$
                               &${}^{(a)}$0.0012$\pz$ & 2.4962$\pz$ & $-0.4$ \\
&$\shad$ [nb]   & $41.540 \pm0.037\pzz$ & ${}^{(b)}$0.028$\pzz$ &41.480$\pzz$ & $ 1.6$ \\
&$\RZ$          & $20.767 \pm0.025\pzz$ & ${}^{(b)}$0.007$\pzz$ &20.740$\pzz$ & $ 1.1$ \\
&$\Afbpol$      & $0.0171 \pm0.0010\pz$ & ${}^{(b)}$0.0003\pz & 0.0164\pz     & $ 0.8$ \\
&+ correlation matrix Table~\ref{tab-zparavg} &&&& \\
&                                                &&&& \\[-3mm]
&$\tau$ polarisation:                            &&&& \\
&$\cAt$         & $0.1439\pm 0.0042\pz$ & 0.0026$\pz$ & 0.1480$\pz$ & $-1.0$ \\
&$\cAe$         & $0.1498\pm 0.0048\pz$ & 0.0009$\pz$ & 0.1480$\pz$ & $ 0.4$ \\
&                       &&&& \\[-3mm]
&$\qq$ charge asymmetry:                      &&&& \\
&$\swsqeffl$
($\avQfb$)     & $0.2321\pm0.0010\pz$ & 0.0008$\pz$ & 0.23140     & $ 0.7$ \\
&                                             &&&& \\[-3mm]
&$\MW$ [\GeV{}]
& $80.427 \pm 0.046 \pzz$& 0.035$\pzz$ &80.402$\pzz$ & $ 0.5$ \\
&&&&& \\[-3mm]

\hline
b) & \underline{SLD}\cite{ref:sld-s99} &&&& \\
&$\swsqeffl$ ($\cAl$)
& $0.23098\pm0.00026 $ & 0.00018     & 0.23140     & $-1.6$ \\
&&&&& \\[-3mm]
\hline
c) & \underline{LEP and SLD Heavy Flavour} &&&& \\
&$\Rbz{}$        & $0.21653\pm0.00069$ &  0.00053     & 0.21578     & $ 1.1$ \\
&$\Rcz{}$        & $0.1709\pm0.0034\pz$ & 0.0022$\pz$ & 0.1723$\pz$ & $-0.4$ \\
&$\Afbzb{}$      & $0.0990\pm0.0020\pz$ & 0.0009$\pz$ & 0.1038$\pz$ & $-2.4$ \\
&$\Afbzc{}$      & $0.0689\pm0.0035\pz$ & 0.0017$\pz$ & 0.0742$\pz$ & $-1.5$ \\
&$\cAb$          & $0.922\pm 0.023\pzz$ & 0.016$\pzz$ & 0.935$\pzz$ & $-0.6$ \\
&$\cAc$          & $0.631\pm 0.026\pzz$ & 0.016$\pzz$ & 0.668$\pzz$ & $-1.4$ \\
&+ correlation matrix Table~\ref{tab:14parcor} &&&& \\
&                                             &&&& \\[-3mm]
\hline
d) & \underline{$\pp$ and $\nu$N} &&&& \\
&$\MW$ [\GeV{}] ($\pp$\cite{bib-MWAVE-00})
& $80.452 \pm 0.062\pzz$& 0.050$\pzz$ &80.402$\pzz$ & $ 0.8$ \\
&$\swsq$ ($\nu$N\cite{bib-CCFRnn,bib-NuTeV})
& $0.2255\pm0.0021\pz$ & 0.0010$\pz$ & 0.2226$\pz$ &$ 1.2$ \\
&$\Mt$ [\GeV{}] ($\pp$\cite{bib-Tevatop})
& $174.3\pm 5.1\pzz\pzz$
                                        & 4.0$\pzz\pzz$
                                                      & 174.3$\pzz\pzz$
                                                                    & $ 0.0$ \\
\hline
\end{tabular}\end{center}
\caption[]{
  Summary of measurements included in the combined analysis of
  Standard Model parameters. Section~a) summarises LEP averages,
  Section~b) SLD results ($\swsqeffl$ includes $\ALR$ and the
  polarised lepton asymmetries), Section~c) the LEP and SLD heavy
  flavour results and Section~d) electroweak measurements from $\pp$
  colliders and $\nu$N scattering.  The total errors in column 2
  include the systematic errors listed in column 3.  Although the systematic
  errors include both correlated and uncorrelated sources, the determination 
  of the systematic part of each error is approximate.  The $\SM$
  results in column~4 and the pulls (difference between measurement
  and fit in units of the total measurement error) in column~5 are
  derived from the Standard Model fit including all data
  (Table~\ref{tab-BIGFIT}, column~5) with the Higgs mass treated as a
  free parameter.\\
  $^{(a)}$\small{The systematic errors on $\MZ$ and $\GZ$ contain the
    errors arising from the uncertainties in the LEP energy only.}\\
  $^{(b)}$\small{Only common systematic errors are indicated.}\\
}
\label{tab-SMIN}
\end{table}

Detailed studies of the theoretical uncertainties in the Standard
Model predictions due to missing higher-order electroweak corrections
and their interplay with QCD corrections have been carried out in the
working group on `Precision calculations for the $\Zzero$
resonance'\cite{bib-PCLI}. Theoretical uncertainties are evaluated by
comparing different but, within our present knowledge, equivalent
treatments of aspects such as resummation techniques, momentum
transfer scales for vertex corrections and factorisation schemes. 
The effects of these theoretical uncertainties have been reduced by
the inclusion of higher-order corrections\cite{bib-twoloop,bib-QCDEW} in the 
electroweak libraries\cite{bib-SMNEW}.
The use of the new QCD corrections\cite{bib-QCDEW} increases the value
of $\alfmz$ by 0.001, as expected.
The effects of missing higher-order QCD corrections on $\alfmz$ covers missing higher-order
electroweak corrections and uncertainties in the interplay of
electroweak and QCD corrections 
and is estimated to be about
0.002~\cite{bib-SMALFAS}.  A discussion of theoretical uncertainties
in the determination of $\alfas$ can be found in
References~\citen{bib-PCLI} and~\citen{bib-SMALFAS}.  
For the moment, the determination of the size of remaining theoretical 
uncertainties is still under study.
All theoretical errors discussed in this paragraph are neglected
for the results presented in Table~\ref{tab-BIGFIT}.

At present the impact of theoretical uncertainties on the
determination of SM parameters from the 
precise electroweak measurements is small
compared with the error due to the uncertainty in the value of
$\alpha(\MZ^2)$. The uncertainty in $\alpha(\MZ^2)$ arises from the
contribution of light quarks to the photon vacuum polarisation
($\Delta\alpha_{\mathrm{had}}^{(5)}(\MZ^2)$):
\begin{equation}
\alpha(\MZ^2) = \frac{\alpha(0)}%
   {1 - \Delta\alpha_\ell(\MZ^2) -
   \Delta\alpha_{\mathrm{had}}^{(5)}(\MZ^2) -
   \Delta\alpha_{\mathrm{top}}(\MZ^2)}.
\end{equation}

The top contribution depends on the mass of the top quark, and is
therefore determined inside the electroweak libraries\cite{bib-SMNEW}.
The leptonic contribution is calculated to third
order\cite{bib-alphalept} to be $0.031498$.  For the hadronic
contribution, we use the value  $0.02804 \pm 0.00065$\cite{bib-JEG2},
which results in $1/\alpha^{(5)}(\MZ^2) = 128.878 \pm 0.090$.
This
uncertainty causes an error of 0.00023 on the $\SM$ prediction of
$\swsqeffl$, an error of 1~\GeV{} on $\Mt$, and 0.2 on $\log(\MH)$,
which are included in the results.  The effect on the $\SM$ prediction
for $\Gll$ is negligible. The $\alfmz$ values for the $\SM$ fits
presented in this Section are stable against a variation of
$\alpha(\MZ^2)$ in the interval quoted.


At the ICHEP 2000 conference in Osaka, the BES collaboration reported
on new preliminary measurements of the hadronic cross section in
electron-positron collisions at 2 to 5~$\GeV$ \CoM\
energy~\cite{BES_Osaka_00}.  With these data, the hadronic vacuum
polarisation is determined with improved precision:
$\Delta\alpha^{(5)}_{\mathrm{had}}(\MZ^2)= 0.02755 \pm 0.00046$
\cite{Bolek_Osaka_00}.  To show the effects of the uncertainty of
$\alpha(\MZ^2)$, we also use this evaluation of the hadronic vacuum
polarisation.  There are also several evaluations of
$\Delta\alpha^{(5)}_{\mathrm{had}}(\MZ^2)$%
\cite{bib-Swartz,bib-Zeppe,bib-JEG2,bib-Burk,bib-Alemany,bib-Davier,bib-alphaKuhn,bib-Erler,bib-ADMartin,bib-jeger99}
which are more theory driven.  The most recent of these (Reference
\citen{bib-ADMartin}) also includes the new preliminary results
from BES.  All these evaluations obtain values for
$\Delta\alpha^{(5)}_{\mathrm{had}}(\MZ^2)$ consistently lower than 
the old value of $0.02804 \pm 0.00065$.


\renewcommand{\arraystretch}{1.0}

\begin{figure}[p]
\begin{center}
  \mbox{\includegraphics[width=\linewidth]{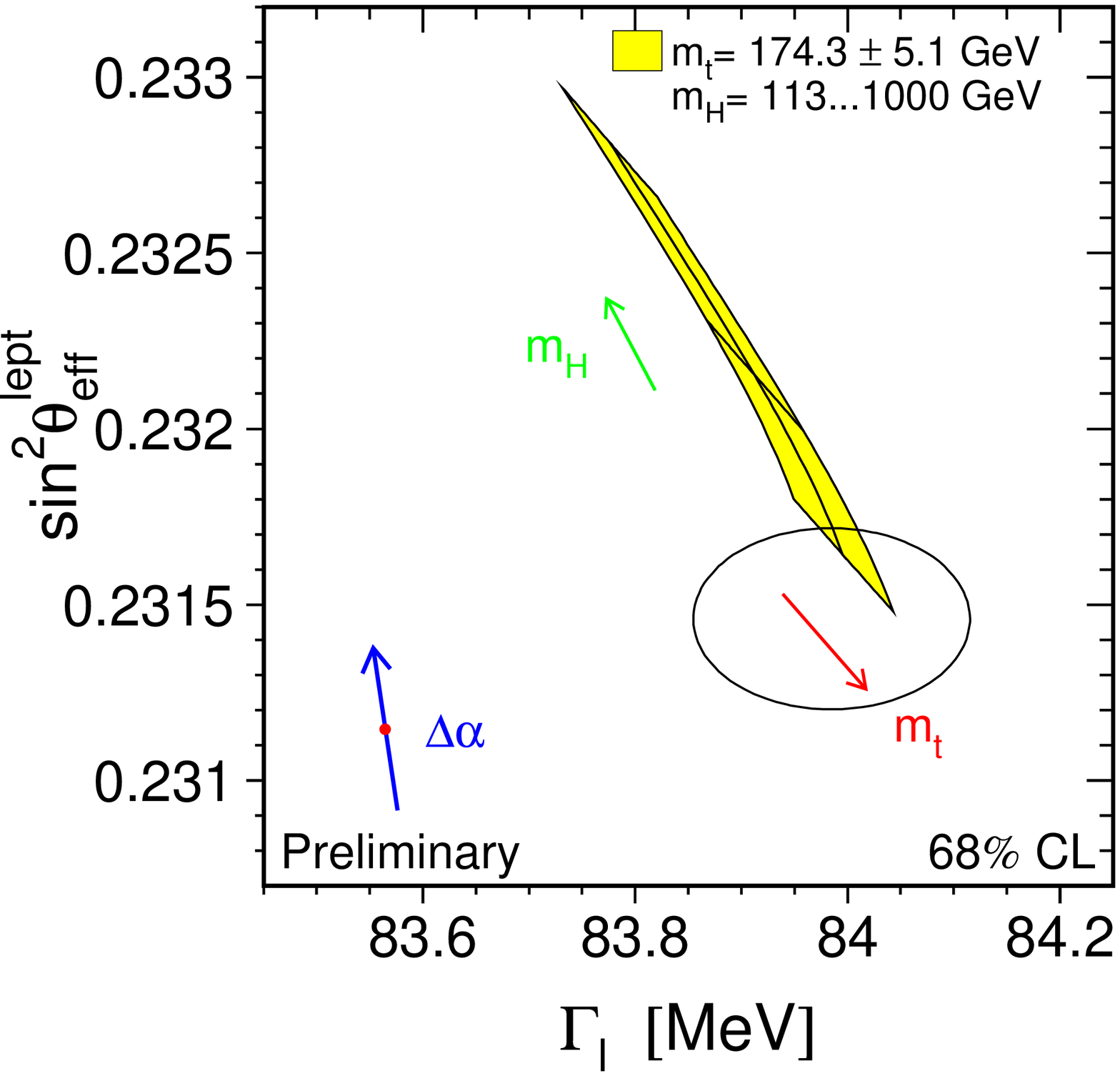}}
\end{center}
\caption[]{%
  $\LEPI$+SLD measurements of $\swsqeffl$ (Table~\ref{tab-swsq}) and
  $\Gll$ (Table~\ref{tab-widths}) and the Standard Model prediction.
  The point shows the predictions if among the electroweak radiative
  corrections only the photon vacuum polarisation is included. The
  corresponding arrow shows variation of this prediction if
  $\alpha(\MZ^2)$ is changed by one standard deviation. This variation
  gives an additional uncertainty to the Standard Model prediction
  shown in the figure.  }
\label{fig-gllsef}
\end{figure}

Figure~\ref{fig-gllsef} shows a comparison of the leptonic partial
width from LEP (Table~\ref{tab-widths}) and the effective electroweak
mixing angle from asymmetries measured at LEP and SLD
(Table~\ref{tab-swsq}), with the Standard Model. Good agreement with
the $\SM$ prediction is observed. 
The point with the arrow shows the prediction if
among the electroweak radiative corrections only the photon vacuum
polarisation is included, which shows an example of evidence that
LEP+SLD data are 
sensitive to electroweak corrections.  
Note that the error due to the
uncertainty on $\alpha(\MZ^2)$ (shown as the length of the arrow) 
is larger than the experimental error on
$\swsqeffl$ from LEP and SLD. This underlines the growing importance
of a precise measurement of $\sigma(\mathrm{e^+e^-\rightarrow
  hadrons})$ at low \CoM\ energies.

Of the measurements given in Table~\ref{tab-SMIN}, $\RZ$ is 
one of the most
sensitive to QCD corrections.  
For $\MZ=91.1875$~\GeV{}, and imposing
$\Mt=174.3\pm5.1$~\GeV{} as a constraint,
$\alfas=0.123\pm0.004$ is obtained.
Alternatively, $\sll$ (see Table~\ref{tab-widths}) which
has higher sensitivity to QCD corrections and less dependence on
$\MH$ yields : $\alfas=0.118\pm0.003$. Typical
errors arising from variation $\MH$  are of the order of $0.001$, somewhat smaller
for $\sll$.
These
results are in very good agreement with the world average ($\alfmz=0.119
\pm 0.002$\cite{bib-pdg2000}).

\begin{figure}[htbp]
\begin{center}
  \mbox{\includegraphics[width=\linewidth]{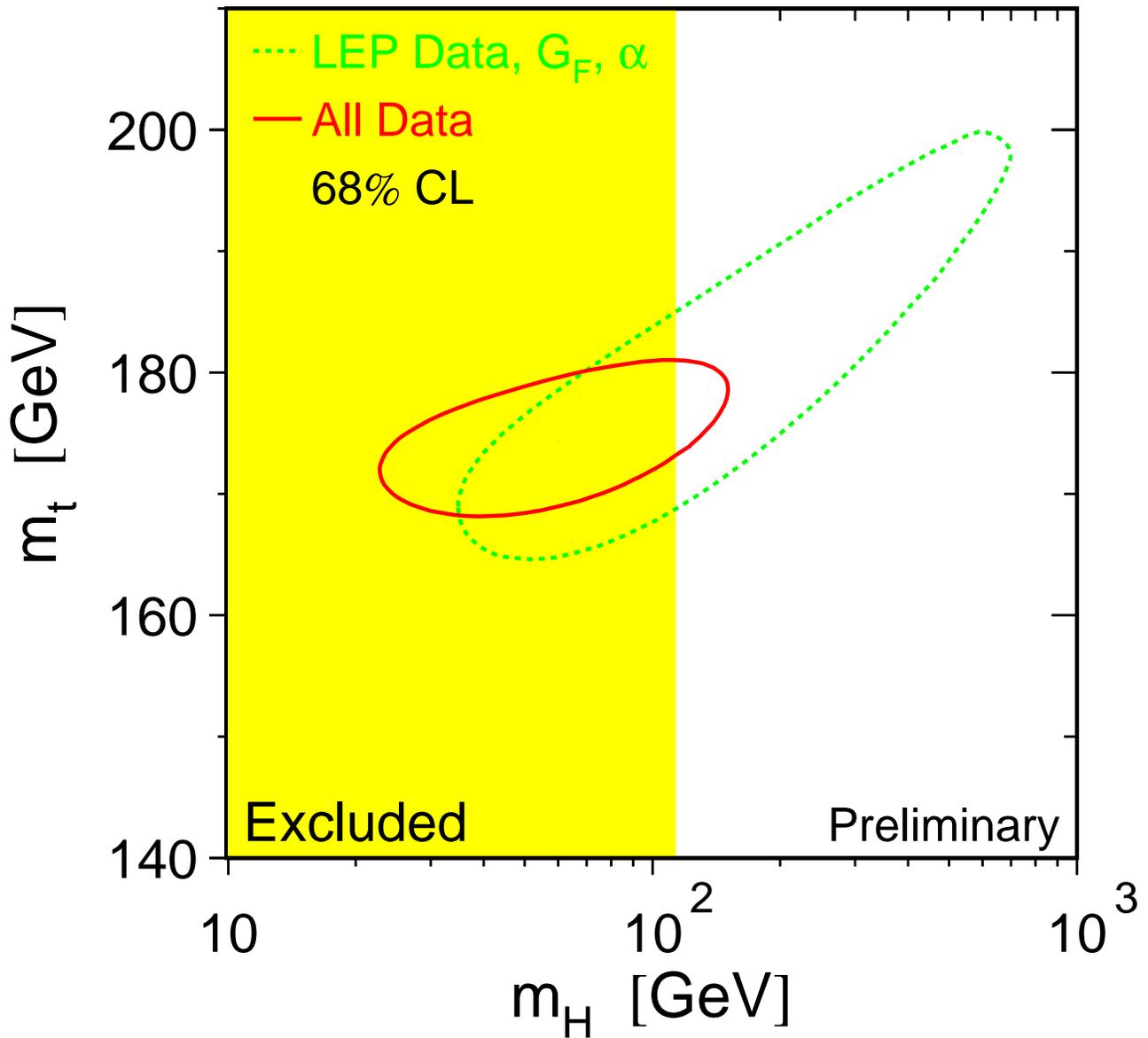}}
\end{center}
\vspace*{-0.6cm}
\caption[]{
  The 68\% confidence level contours in $\Mt$ and $\MH$ for the fits
  to LEP data only (dashed curve) and to all data including the
  CDF/D\O{} $\Mt$ measurement (solid curve).  The vertical band shows
  the 95\% CL exclusion limit on $\MH$ from the direct search.  }
\label{fig-mtmh}
\end{figure}

To test the agreement between the LEP data and the Standard Model,
a fit to the data (including the $\LEPII$ $\MW$
determination) leaving the top quark mass and the Higgs mass as free
parameters is performed.  
The result is shown in Table~\ref{tab-BIGFIT}, column~1.
This fit shows that the LEP data prefer a light top quark and a light
Higgs boson, albeit with very large errors.  The strongly asymmetric errors
on $\MH$ are due to the fact that to first order, the radiative
corrections in the Standard Model are proportional to $\log(\MH)$.
The correlation between the top quark mass and the Higgs mass is 0.63
(see Figure~\ref{fig-mtmh}).

\begin{figure}[htbp]
\begin{center}
  \leavevmode \includegraphics[width=\linewidth]{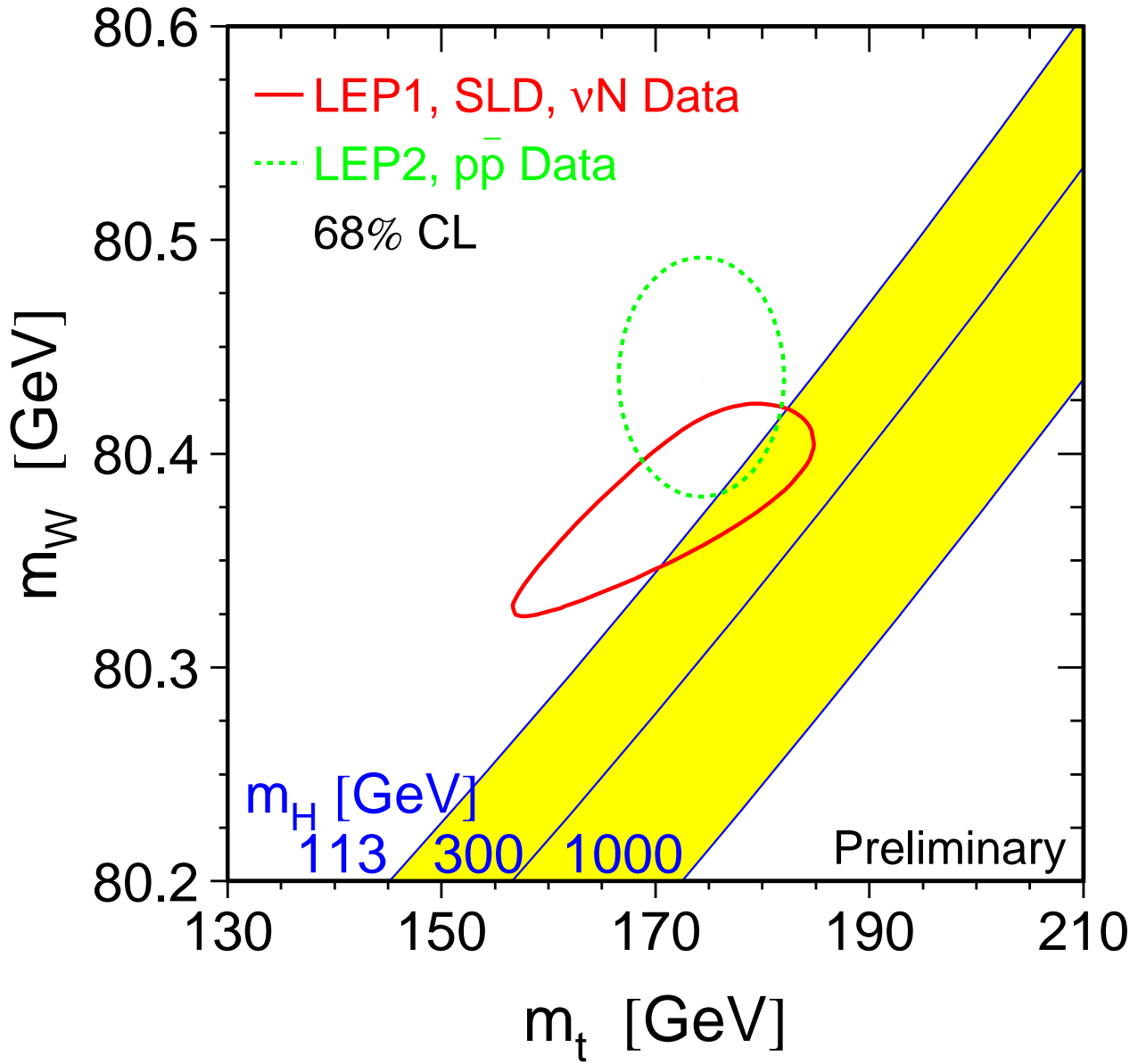}
\caption[]{
  The comparison of the indirect measurements of $\MW$ and $\Mt$
  ($\LEPI$+SLD+$\nu$N data) (solid contour) and the direct
  measurements ($\pp$ colliders and $\LEPII$ data) (dashed contour).  In both
  cases the 68\% CL contours are plotted.  Also shown is the Standard
  Model relationship for the masses as a function of the Higgs mass.}
\label{fig:mtmW}
\end{center}
\end{figure}

The data can also be used within the Standard Model to determine the
top quark and W masses indirectly, which can be compared  to the direct
measurements performed at the $\pp$ colliders and LEP.  
In the second fit, 
all the results in Table~\ref{tab-SMIN}, except the $\LEPII$ and
$\pp$ colliders $\MW$ and $\Mt$ results are used.  
The results are shown in
column~2 of Table~\ref{tab-BIGFIT}.  The indirect measurements of
$\MW$ and $\Mt$ from this data sample are shown in
Figure~\ref{fig:mtmW}, compared with the direct measurements. Also
shown is the Standard Model predictions for Higgs masses between 113
and 1000~\GeV.  As can be seen in the figure, the indirect and direct
measurements of $\MW$ and $\Mt$ are in good agreement, and both sets 
prefer a low Higgs mass.  

For the third fit,
the direct $\Mt$ measurements is used to obtain the best
indirect determination of $\MW$.  The result is shown in
column~3 of Table~\ref{tab-BIGFIT}.  
Also here,  the indirect determination of W boson mass 
$80.386\pm0.025$ \GeV\
is in excellent agreement with    
the combination of
direct measurements from LEP and $\pp$ colliders~\cite{bib-MWAVE-00}
of $\MW= 80.436\pm0.037$ \GeV.

For the next fit, 
(column~4 of Table~\ref{tab-BIGFIT}), 
the direct $\MW$ measurements from LEP 
and $\pp$ colliders are included
to obtain $\Mt= 174.4^{+9.3}_{-6.7}$ \GeV, in good
agreement with the direct 
measurement of $\Mt = 174.3\pm5.1$ \GeV. 

\begin{figure}[p]
\vspace*{-2.0cm}
\begin{center}
  \mbox{\includegraphics[height=21cm]{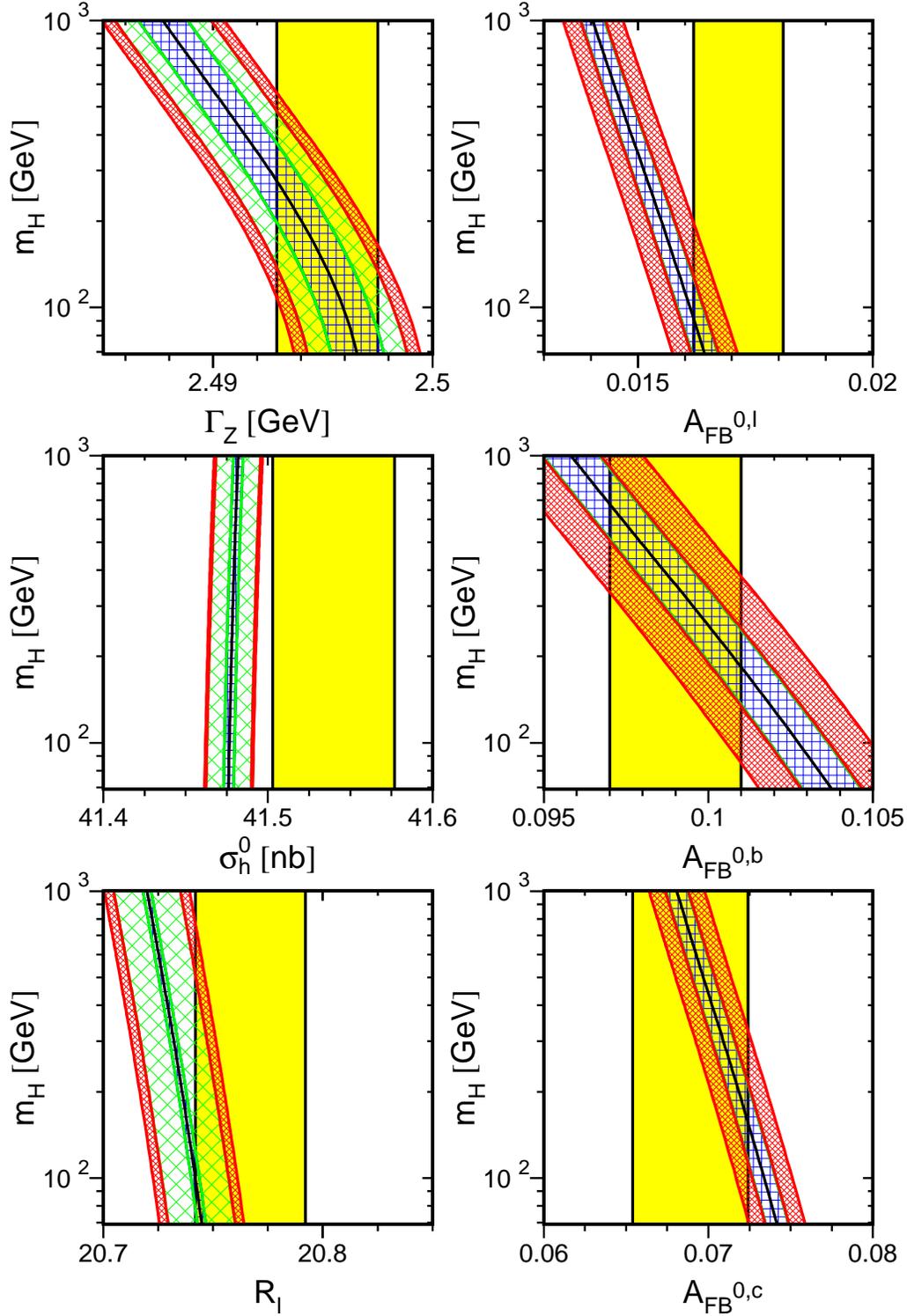}}
\end{center}
\vspace*{-0.6cm}
\caption[]{%
  Comparison of $\LEPI$ measurements with the Standard Model
  prediction as a function of $\MH$.  The measurement with its error
  is shown as the vertical band.  The width of the Standard Model band
  is due to the uncertainties in
  $\Delta\alpha^{(5)}_{\mathrm{had}}(\MZ^2)$, $\alfmz$ and $\Mt$. 
  The total width of the band is the linear sum of these effects.  See
  Figure~\protect\ref{fig-higgs2} for the definition of these
  uncertainties.  }
\label{fig-higgs1}
\end{figure}

\begin{figure}[p]
\vspace*{-2.0cm}
\begin{center}
  \mbox{\includegraphics[height=21cm]{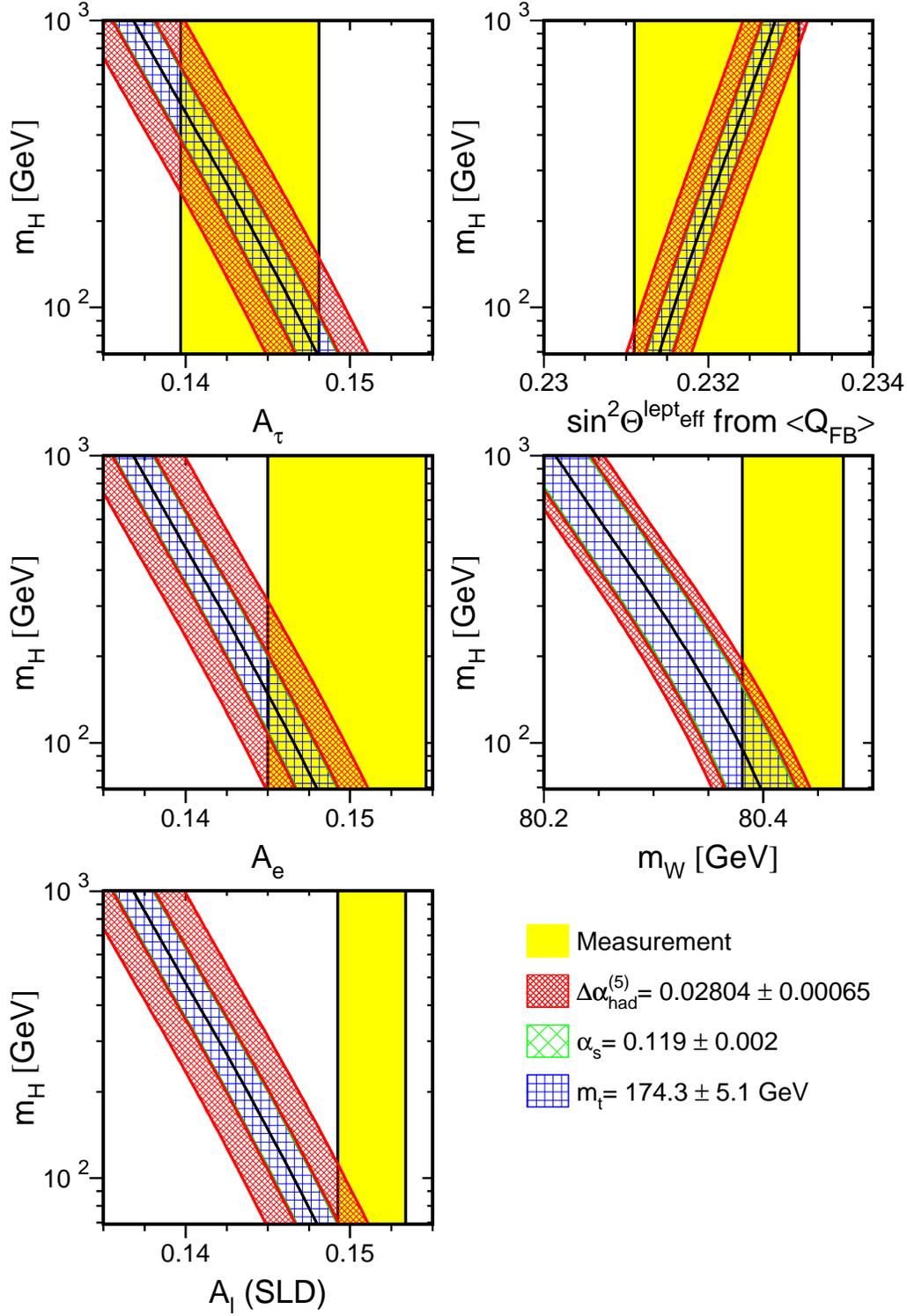}}
\end{center}
\vspace*{-0.6cm}
\caption[]{%
  Comparison of $\LEPI$ measurements with the Standard Model
  prediction as a function of $\MH$ ({\it c.f.\/}
  Figure~\protect\ref{fig-higgs1}).  Also shown is the comparison of
  the SLD measurement of $\ALRz$ with the Standard Model.  }
\label{fig-higgs2}
\end{figure}

\begin{figure}[htbp]
\begin{center}
  \mbox{\includegraphics[width=\linewidth]{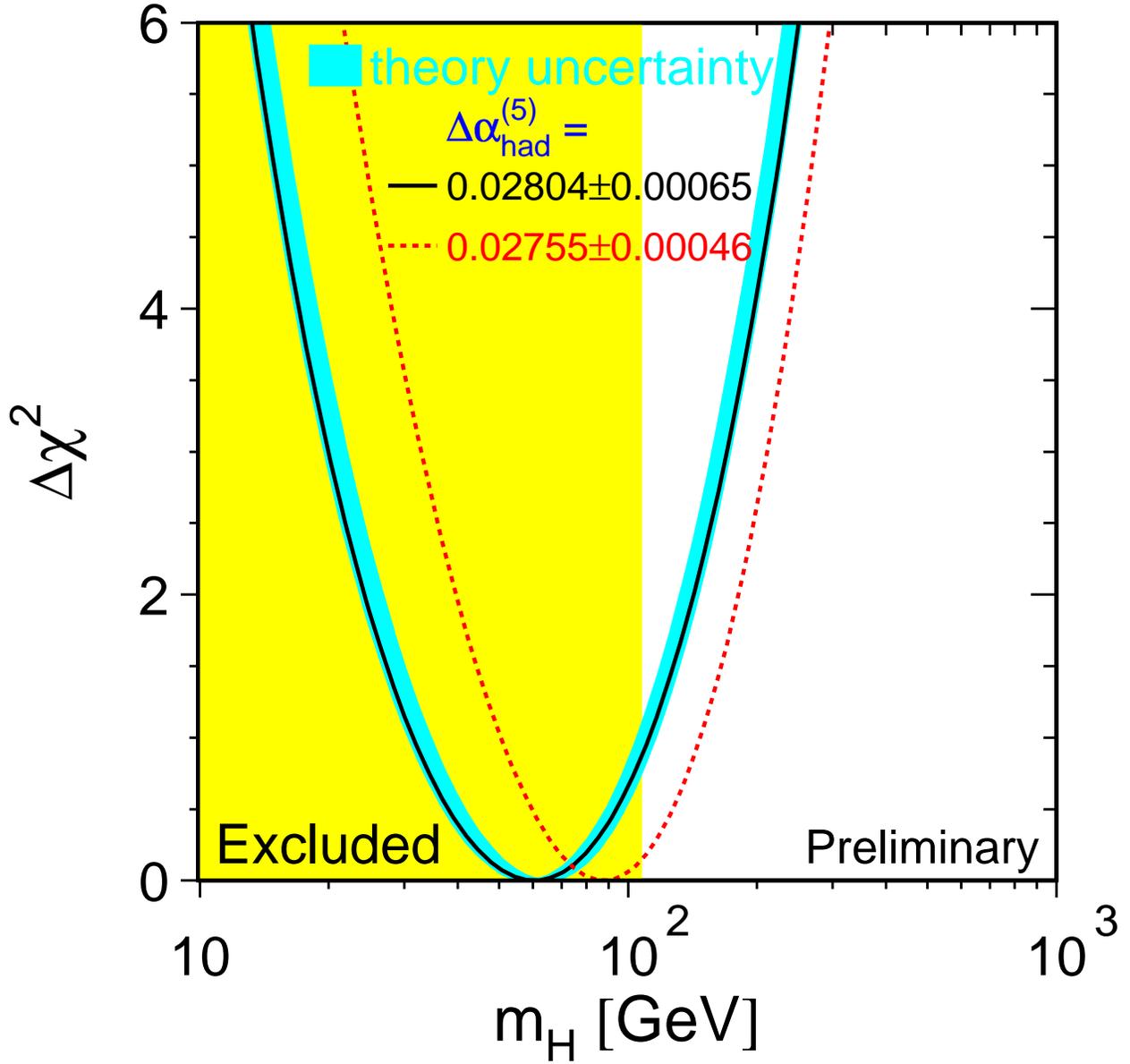}}
\end{center}
\vspace*{-0.6cm}
\caption[]{%
  $\Delta\chi^{2}=\chi^2-\chi^2_{min}$ {\it vs.} $\MH$ curve.  The
  line is the result of the fit using all data (last column of
  Table~\protect\ref{tab-BIGFIT}); the band represents an estimate of
  the theoretical error due to missing higher order corrections.  The
  vertical band shows the 95\% CL exclusion limit on $\MH$ from the
  direct search.  The dashed curve is the result obtained using the
  evaluation of $\Delta\alpha^{(5)}_{\mathrm{had}}(\MZ^2)$ from
  Reference~\citen{Bolek_Osaka_00}. }
\label{fig-chiex}
\end{figure}

Finally, the best constraints on $\MH$ are obtained when all data 
are used in the fit.  The results of
this fit are shown in column~5 of Table~\ref{tab-BIGFIT} and in
Figure~\ref{fig-mtmh}.  In Figures~\ref{fig-higgs1}
and~\ref{fig-higgs2} the sensitivity of the LEP and SLD measurements
to the Higgs mass is shown.  As can be seen, the most sensitive
measurements are the asymmetries.  A reduced uncertainty for the value
of $\alpha(\MZ^2)$ would therefore result in an improved constraint on
$\MH$, as shown in Figures~\ref{fig-gllsef} and \ref{fig-chiex}.

In Figure~\ref{fig-chiex} the observed value of $\Delta\chi^2 \equiv
\chi^2 - \chi^2_{\mathrm{min}}$ as a function of $\MH$ is plotted for
the fit including all data.  The solid curve is the result using
ZFITTER, and corresponds to the last column of Table~\ref{tab-BIGFIT}.
The shaded band represents the uncertainty due to uncalculated
higher-order corrections, as estimated by ZFITTER and TOPAZ0.  
The 95\% confidence level upper limit on
$\MH$ (taking the band into account) is 165 \GeV.  
The lower limit on $\MH$ of approximately 113
\GeV{} obtained from direct searches\cite{ref:HiggsOsaka} has not been
used in this limit determination.
Also
shown is the result (dashed curve) obtained 
when using  $\alpha^{(5)}(\MZ^2)$ of Reference \citen{Bolek_Osaka_00}.  
The fit results in  $\log(\MH/\GeV) = 1.94^{+0.22}_{-0.24}$ 
corresponding to $\MH= 88^{+60}_{-37}$ \GeV~ and an upper limit on 
$\MH$ of approximately 206 \GeV.


\vfill

\begin{table}[htbp]
\renewcommand{\arraystretch}{1.5}
  \begin{center}
    \leavevmode
    \begin{tabular}{|c||c|c|c|c|c|}
\hline
  &    - 1 -         &     - 2 -          &      - 3 -           &    - 4 -             &    - 5 -    \\
  & LEP including    & all data except    & all data except      & all data except      &
  all data \\[-3mm]
  & $\LEPII$ $\MW$   & $\MW$ and $\Mt$    &           $\MW$      &           $\Mt$      &           \\
\hline
\hline
$\Mt$\hfill[\GeV] & $179^{+13}_{-10}$ & $169^{+10}_{-8} $ & $173.0^{+4.7}_{-4.5}   $ & $174.4^{+9.3}_{-6.7}   $ & $174.3^{+4.4}_{-4.1}$     \\
$\MH$\hfill[\GeV] &$135^{+262}_{-83}$ & $56^{+75}_{-27}$ & $74^{+68}_{-37}$  &$61^{+94}_{-32}$  & $ 60^{+52}_{-29}$\\
$\log(\MH/\GeV)$  & $2.13^{+0.47}_{-0.41}$ & $1.75^{+0.37}_{-0.29} $ & $1.87^{+0.28}_{-0.30}$ &$1.78^{+0.41}_{-0.32}$ & $1.78^{+0.27}_{-0.28}$\\
$\alfmz$          & $0.120 \pm 0.003$ & $0.119 \pm 0.003$& $0.119\pm0.003$   &$0.118\pm0.003$   & $0.118 \pm 0.003$ \\
\hline
$\chi^2$/d.o.f.{} & $13/9$            & $19/12 $         & $20/13$           &$21/14$           & $21/15$           \\
\hline
\hline
$\swsqeffl$       & $\pz0.23167$& $\pz0.23147$& $\pz0.23147$& $\pz0.23140$& $\pz0.23140$ \\
                  & $\pm0.00020$& $\pm0.00017$& $\pm0.00017$& $\pm0.00016$& $\pm0.00016$ \\
$\swsq$           & $\pz0.2229$& $\pz0.2231$& $\pz0.2229$& $\pz0.2226$& $\pz0.2226$     \\
                  & $\pm0.0006$& $\pm0.0007$& $\pm0.0005$& $\pm0.0005$& $\pm0.0004$     \\
$\MW$\hfill[\GeV]   & $80.383\pm0.029$ & $80.374\pm0.034$   & $80.386\pm0.025$   &$80.402\pm0.025$   & $80.402\pm0.020$        \\
\hline
    \end{tabular}
    \caption[]{
      Results of the fits to LEP data alone, to all data except the
      direct determinations of $\Mt$ and $\MW$ ($\pp$ collider and
      $\LEPII$), to all data except direct $\MW$ determinations,
      and to all data.  As
      the sensitivity to $\MH$ is logarithmic, both $\MH$ as well as
      $\log(\MH/\GeV)$ are quoted.  The bottom part of the table lists
      derived results for $\swsqeffl$, $\swsq$ and $\MW$.  See text
      for a discussion of theoretical errors not included in the
      errors above.  }
    \label{tab-BIGFIT}
  \end{center}
\renewcommand{\arraystretch}{1.0}
\end{table}
\vfill


\clearpage

\boldmath
\section{Prospects for the Future}
\label{sec-Future}
\unboldmath

Most of the measurements from data taken at or near the Z resonance, both
at LEP as well as at SLC, that
are presented in this report are either final or
are being finalised.  The major improvements will therefore take place
in the high energy data. 
With more than 600~pb$^{-1}$ per experiment,  $\LEPII$
 will lead to substantially improved
measurements of certain electroweak parameters.  As a result, the
measurements of $\MW$ are likely to
match the error obtained via the radiative corrections of the \Zzero{}
data, providing a further important test of the Standard Model.  
In
the measurement of the WW$\gamma$ and WWZ triple-gauge-boson couplings
the increase in $\LEPII$ statistics, together with the increased
sensitivity at higher beam energies, will lead to an
improvement in the current precision.

\boldmath
\section{Conclusions}
\label{sec-Conc}
\unboldmath

The combination of the many precise electroweak results yields
stringent constraints on the Standard Model. All measurements
agree with the predictions.  In addition, the results
are sensitive to the Higgs mass.

The experiments wish to stress that this report reflects a
preliminary status at the time of the 2000 summer conferences. A
definitive statement on these results must wait for publication by
each collaboration.

\section*{Acknowledgements}

We would like to thank the CERN accelerator divisions for the
efficient operation of the LEP accelerator, the precise information on
the absolute energy scale and their close cooperation with the four
experiments.  
The SLD collaboration would like to thank 
the SLAC accelerator
department for the efficient operation of the SLC accelerator. 
We would also like to thank members of the CDF, D\O{} and NuTeV
Collaborations for making results available to us in advance
of the conferences and for useful discussions concerning their
combination.  Finally, the results of the section on Standard Model
constraints would not have been possible without the close
collaboration of many theorists.  

\clearpage

\clearpage

\begin{appendix}
\noindent{\Huge{\bf Appendix}}
\section{Heavy-Flavour Fit including Off-Peak Asymmetries}\label{app-HF}
The full 18 parameter fit to the LEP and SLD data gave the following results:
\begin{eqnarray*}
  \Rbz    &=& 0.21652   \pm  0.00069\\
  \Rcz    &=& 0.1702    \pm  0.0034 \\
  \Abl    &=& 0.0572    \pm  0.0075 \\
  \Acl    &=&-0.027     \pm  0.016  \\
  \Abp    &=& 0.0973    \pm  0.0021 \\
  \Acp    &=& 0.0621    \pm  0.0036 \\
  \Abh    &=& 0.1106    \pm  0.0065 \\
  \Ach    &=& 0.131     \pm  0.013  \\
  \cAb    &=& 0.922     \pm  0.022  \\
  \cAc    &=& 0.631     \pm  0.026  \\
  \Brbl   &=& 0.1056    \pm  0.0019 \\
  \Brbclp &=& 0.0801    \pm  0.0026 \\
  \Brcl   &=& 0.0984    \pm  0.0032 \\
  \chiM   &=& 0.1194    \pm  0.0043 \\
  \fDp    &=& 0.237     \pm  0.016  \\
  \fDs    &=& 0.121     \pm  0.025  \\
  \fcb    &=& 0.090     \pm  0.022  \\
  \PcDst  &=& 0.1631    \pm  0.0050 \,
\end{eqnarray*}
with a $\chi^2/$d.o.f.{} of  $52/(98-18)$. The corresponding correlation
matrix is given in Table~\ref{tab:18parcor}.
The energy for the  peak$-$2, peak and peak+2 results are respectively
89.55 \GeV{}, 91.26 \GeV{} and 92.94 \GeV.
Note that the asymmetry results shown here are not the pole
asymmetries shown in Section~\ref{sec-HFSUM-LEP-SLD}.
The non-electroweak parameters do not depend on the treatment of the 
asymmetries.

\begin{table}[p]
\begin{center}
\begin{sideways}
\begin{minipage}[b]{\textheight}
\begin{center}
\footnotesize
\begin{tabular}{|l||rrrrrrrrrrrrrrrrrr|}
\hline
&\makebox[0.45cm]{$1)$}
&\makebox[0.45cm]{$2)$}
&\makebox[0.45cm]{$3)$}
&\makebox[0.45cm]{$4)$}
&\makebox[0.45cm]{$5)$}
&\makebox[0.45cm]{$6)$}
&\makebox[0.45cm]{$7)$}
&\makebox[0.45cm]{$8)$}
&\makebox[0.45cm]{$9)$}
&\makebox[0.45cm]{$10)$}
&\makebox[0.45cm]{$11)$}
&\makebox[0.45cm]{$12)$}
&\makebox[0.45cm]{$13)$}
&\makebox[0.45cm]{$14)$}
&\makebox[0.45cm]{$15)$}
&\makebox[0.45cm]{$16)$}
&\makebox[0.45cm]{$17)$}
&\makebox[0.45cm]{$18)$}\\
&\makebox[0.45cm]{\Rb}
&\makebox[0.45cm]{\Rc}
&\makebox[0.45cm]{$\Abb$}
&\makebox[0.45cm]{$\Acc$}
&\makebox[0.45cm]{$\Abb$}
&\makebox[0.45cm]{$\Acc$}
&\makebox[0.45cm]{$\Abb$}
&\makebox[0.45cm]{$\Acc$}
&\makebox[0.45cm]{\cAb}
&\makebox[0.45cm]{\cAc}
&\makebox[0.45cm]{BR}
&\makebox[0.45cm]{BR}
&\makebox[0.45cm]{BR}
&\makebox[0.45cm]{\chiM}
&\makebox[0.45cm]{$\fDp$}
&\makebox[0.45cm]{$\fDs$}
&\makebox[0.45cm]{$f(c_{bar.})$}
&\makebox[0.55cm]{PcDst}\\
&
&
&\makebox[0.45cm]{$(-2)$}
&\makebox[0.45cm]{$(-2)$}
&\makebox[0.45cm]{(pk)}
&\makebox[0.45cm]{(pk)}
&\makebox[0.45cm]{$(+2)$}
&\makebox[0.45cm]{$(+2)$}
&
&
&\makebox[0.45cm]{$(1)$}
&\makebox[0.45cm]{$(2)$}
&\makebox[0.45cm]{$(3)$}
&
&
&
&
&\\
\hline\hline
1)  &$  1.00$&$ -0.13$&$  0.00$&$ -0.01$&$ -0.02$&$  0.01$&$ -0.01$&$  0.00$&$ -0.04$&
$  0.02$&$ -0.09$&$ -0.02$&$ -0.02$&$ -0.02$&$ -0.16$&$ -0.04$&$  0.12$&$  0.11$\\
2)  &$ -0.13$&$  1.00$&$  0.01$&$  0.01$&$  0.05$&$ -0.01$&$  0.02$&$ -0.01$&$  0.02$&
$ -0.02$&$  0.05$&$ -0.01$&$ -0.32$&$  0.04$&$ -0.13$&$  0.19$&$  0.18$&$ -0.49$\\
3)  &$  0.00$&$  0.01$&$  1.00$&$  0.14$&$  0.04$&$  0.01$&$  0.02$&$  0.00$&$  0.01$&
$  0.00$&$  0.02$&$ -0.03$&$  0.01$&$  0.06$&$  0.00$&$  0.00$&$  0.00$&$ -0.01$\\
4)  &$ -0.01$&$  0.01$&$  0.14$&$  1.00$&$  0.01$&$  0.02$&$  0.00$&$  0.00$&$  0.00$&
$  0.00$&$  0.01$&$ -0.01$&$  0.02$&$  0.01$&$  0.00$&$  0.00$&$  0.00$&$  0.01$\\
5)  &$ -0.02$&$  0.05$&$  0.04$&$  0.01$&$  1.00$&$  0.10$&$  0.10$&$  0.00$&$  0.02$&
$  0.00$&$  0.01$&$ -0.09$&$  0.02$&$  0.16$&$  0.01$&$  0.03$&$  0.00$&$ -0.03$\\
6)  &$  0.01$&$ -0.01$&$  0.01$&$  0.02$&$  0.10$&$  1.00$&$  0.00$&$  0.09$&$  0.00$&
$  0.01$&$  0.14$&$ -0.17$&$ -0.05$&$  0.16$&$  0.00$&$  0.00$&$ -0.01$&$  0.00$\\
7)  &$ -0.01$&$  0.02$&$  0.02$&$  0.00$&$  0.10$&$  0.00$&$  1.00$&$  0.13$&$  0.01$&
$  0.00$&$ -0.01$&$ -0.03$&$  0.01$&$  0.07$&$  0.01$&$  0.01$&$  0.00$&$ -0.01$\\
8)  &$  0.00$&$ -0.01$&$  0.00$&$  0.00$&$  0.00$&$  0.09$&$  0.13$&$  1.00$&$  0.00$&
$  0.00$&$  0.02$&$ -0.04$&$ -0.03$&$  0.02$&$ -0.01$&$ -0.01$&$  0.00$&$  0.00$\\
9)  &$ -0.04$&$  0.02$&$  0.01$&$  0.00$&$  0.02$&$  0.00$&$  0.01$&$  0.00$&$  1.00$&
$  0.14$&$ -0.02$&$  0.00$&$  0.05$&$  0.09$&$ -0.01$&$  0.00$&$  0.00$&$ -0.01$\\
10) &$  0.02$&$ -0.02$&$  0.00$&$  0.00$&$  0.00$&$  0.01$&$  0.00$&$  0.00$&$  0.14$&
$  1.00$&$  0.02$&$ -0.03$&$ -0.03$&$  0.01$&$ -0.01$&$  0.00$&$  0.01$&$  0.01$\\
11) &$ -0.09$&$  0.05$&$  0.02$&$  0.01$&$  0.01$&$  0.14$&$ -0.01$&$  0.02$&$ -0.02$&
$  0.02$&$  1.00$&$ -0.33$&$  0.14$&$  0.43$&$  0.04$&$  0.01$&$ -0.02$&$ -0.02$\\
12) &$ -0.02$&$ -0.01$&$ -0.03$&$ -0.01$&$ -0.09$&$ -0.17$&$ -0.03$&$ -0.04$&$  0.00$&
$ -0.03$&$ -0.33$&$  1.00$&$ -0.03$&$ -0.40$&$  0.01$&$ -0.01$&$  0.00$&$  0.01$\\
13) &$ -0.02$&$ -0.32$&$  0.01$&$  0.02$&$  0.02$&$ -0.05$&$  0.01$&$ -0.03$&$  0.05$&
$ -0.03$&$  0.14$&$ -0.03$&$  1.00$&$  0.20$&$  0.02$&$ -0.04$&$ -0.04$&$  0.16$\\
14) &$ -0.02$&$  0.04$&$  0.06$&$  0.01$&$  0.16$&$  0.16$&$  0.07$&$  0.02$&$  0.09$&
$  0.01$&$  0.43$&$ -0.40$&$  0.20$&$  1.00$&$  0.01$&$  0.01$&$ -0.01$&$ -0.03$\\
15) &$ -0.16$&$ -0.13$&$  0.00$&$  0.00$&$  0.01$&$  0.00$&$  0.01$&$ -0.01$&$ -0.01$&
$ -0.01$&$  0.04$&$  0.01$&$  0.02$&$  0.01$&$  1.00$&$ -0.38$&$ -0.27$&$  0.11$\\
16) &$ -0.04$&$  0.19$&$  0.00$&$  0.00$&$  0.03$&$  0.00$&$  0.01$&$ -0.01$&$  0.00$&
$  0.00$&$  0.01$&$ -0.01$&$ -0.04$&$  0.01$&$ -0.38$&$  1.00$&$ -0.46$&$ -0.11$\\
17) &$  0.12$&$  0.18$&$  0.00$&$  0.00$&$  0.00$&$ -0.01$&$  0.00$&$  0.00$&$  0.00$&
$  0.01$&$ -0.02$&$  0.00$&$ -0.04$&$ -0.01$&$ -0.27$&$ -0.46$&$  1.00$&$ -0.17$\\
18) &$  0.11$&$ -0.49$&$ -0.01$&$  0.01$&$ -0.03$&$  0.00$&$ -0.01$&$  0.00$&$ -0.01$&
$  0.01$&$ -0.02$&$  0.01$&$  0.16$&$ -0.03$&$  0.11$&$ -0.11$&$ -0.17$&$  1.00$\\
\hline
\end{tabular}
\normalsize
\end{center}
\caption[]{
  The correlation matrix for the set of the 18 heavy flavour
  parameters. BR(1), BR(2) and BR(3) denote $\Brbl$, $\Brbclp$ and $\Brcl$
  respectively, PcDst denotes $\PcDst$.  }
\label{tab:18parcor}
\end{minipage}
\end{sideways}
\end{center}
\end{table}

\clearpage
\section*{The Measurements used in the Heavy Flavour Averages}

In the following 20 tables the results used in the combination are listed.
In each case an indication of the dataset used and the type of analysis is
given.
Preliminary results are indicated by the symbol ``\dag''. 
The values of centre-of-mass energy are given where
relevant.  In each table, the result used as input to the average
procedure is given followed by the statistical error, the correlated
and uncorrelated systematic errors, the total systematic error, and
any dependence on other electroweak parameters.  In the case of the
asymmetries, the measurement moved to a common energy (89.55 \GeV{},
91.26 \GeV{} and 92.94 \GeV{}, respectively, for peak$-$2, peak and
peak+2 results) is quoted as {\it corrected\/} asymmetry.

Contributions to the correlated systematic error quoted here are from
any sources of error shared with one or more other results from
different experiments in the same table, and the uncorrelated errors
from the remaining sources. In the case of \cAc{} and \cAb{} from SLD
the quoted correlated systematic error has contributions from any
source shared with one or more other measurements from LEP experiments.
Constants such as $a(x)$ denote the dependence on the assumed value of
$x^{\rm{used}}$, which is also given.
 \begin{table}[htb]
 \begin{center}
 \begin{tabular}{|l||c|c|c|c|c|}
 \hline
           & \mca{1}{ALEPH} & \mca{1}{DELPHI} & \mca{1}{L3} & \mca{1}{OPAL} & \mca{1}{SLD} \\
 \hline
           &92-95 &92-95 &94-95 &92-95 &93-98\dag\\
           &\tmcite{ref:alife} &\tmcite{ref:drb} &\tmcite{ref:lrbmixed} &\tmcite{ref:omixed} &\tmcite{ref:SLD_RB_RC}\\
 \hline\hline
 \Rb      &    0.2157 &   0.2163 &   0.2174 &   0.2174 &   0.2167\\
 \hline
 Statistical          &    0.0009 &   0.0007 &   0.0015 &   0.0011 &   0.0009\\
 \hline
 Uncorrelated         &    0.0007 &   0.0004 &   0.0015 &   0.0009 &   0.0008\\
 Correlated           &    0.0007 &   0.0004 &   0.0018 &   0.0008 &   0.0005\\
 \hline
 Total Systematic     &    0.0009 &   0.0006 &   0.0023 &   0.0012 &   0.0010\\
 \hline
 \hline
 $a( \Rc      )$ &   -0.0033 &  -0.0041 &  -0.0376 &  -0.0122 &  -0.0057\\
 $\Rc     ^{\mathrm{used}}$ &    0.1720 &   0.1720 &   0.1734 &   0.1720 &   0.1710\\
 \hline
 $a( \Brcl    )$ &           &          &  -0.0133 &  -0.0067 &         \\
 $\Brcl   ^{\mathrm{used}}$ &           &          &     9.80 &     9.80 &         \\
 \hline
 $a( \fDp     )$ &   -0.0010 &  -0.0010 &  -0.0086 &  -0.0029 &  -0.0008\\
 $\fDp    ^{\mathrm{used}}$ &    0.2330 &   0.2330 &   0.2330 &   0.2380 &   0.2370\\
 \hline
 $a( \fDs     )$ &   -0.0001 &   0.0001 &  -0.0005 &  -0.0001 &  -0.0003\\
 $\fDs    ^{\mathrm{used}}$ &    0.1020 &   0.1030 &   0.1030 &   0.1020 &   0.1140\\
 \hline
 $a( \fLc     )$ &    0.0002 &   0.0003 &   0.0008 &   0.0003 &  -0.0003\\
 $\fLc    ^{\mathrm{used}}$ &    0.0650 &   0.0630 &   0.0630 &   0.0650 &   0.0730\\
 \hline
 \end{tabular}
 \caption{The measurements of 
 \Rb     .
 All measurements use a lifetime tag enhanced by other features like
 invariant mass cuts or high $p_T$ leptons.
 }
 \label{tab:Rbinp}
 \end{center}
 \end{table}
 \begin{table}[htb]
 \begin{center}
 \begin{tabular}{|l||c|c|c|c|c|c|c|c| }
 \hline
           & \mca{3}{ALEPH} & \mca{2}{DELPHI} & \mca{2}{OPAL} & \mca{1}{SLD} \\
 \hline
           &91-95   &91-95 &92-95 &92-95 & 92-95 &91-94 & 90-95 &93-97\dag\\
           &c-count &D meson &lepton &c-count &D meson& c-count & D meson &vtx-mass\\
           &\tmcite{ref:arcc} &\tmcite{ref:arcd} &\tmcite{ref:arcd} &\tmcite{ref:drcc} 
           &\tmcite{ref:drcc} &\tmcite{ref:orcc} &\tmcite{ref:orcd}&\tmcite{ref:SLD_RB_RC}\\
 \hline\hline
 \Rc               &    0.1734 &   0.1679 &  0.1668 & 0.1692& 0.1610 & 0.164 & 0.1760 &  0.1732\\
 \hline                                    
 Statistical       &    0.0049 &   0.0082 &  0.0062 & 0.0047 & 0.0104 & 0.011 & 0.0095  &  0.0041\\
 \hline                                    
 Uncorrelated      &    0.0057 &   0.0078 &  0.0059 & 0.0050 & 0.0064 & 0.012 & 0.0102 & 0.0025\\
 Correlated        &    0.0101 &   0.0026 &  0.0010 & 0.0083 & 0.0060 & 0.010 & 0.0062 & 0.0004\\
 \hline                                    
 Total Systematic  &    0.0116 &   0.0082 &  0.0059 & 0.0097 & 0.0088 & 0.016 & 0.0120& 0.0025\\
 \hline
 \hline
 $a( \Rb      )$ & &   -0.0050 &          & & & & & -0.0239\\
 $\Rb     ^{\mathrm{used}}$ & &    0.2159 & & &   & &      &   0.2175\\
 \hline
 $a( \Brcl    )$ & &           &  -0.1646 & & & & &       \\
 $\Brcl   ^{\mathrm{used}}$ & &           &     9.80 && & & &         \\
 \hline
 \end{tabular}
 \end{center}
 \caption{The measurements of 
 $\Rcz$. 
``c-count'' denotes the determination of \Rcz{} from the sum of production rates
of weakly decaying charmed hadrons. ``D meson'' denotes any single/double tag
analysis using exclusive and/or inclusive  D meson reconstruction.
}
 \label{tab:Rcinp}
 \end{table}
 \begin{table}[htb]
 \begin{center}
 \begin{sideways}
 \begin{minipage}[b]{\textheight}
 \begin{center}
 \begin{tabular}{|l||c|c|c|c|c|c|c|c|c|c|c|}
 \hline
           & \mca{4}{ALEPH} & \mca{3}{DELPHI} & \mca{1}{L3} & \mca{3}{OPAL} \\
 \hline
           &90-95 &90-95 &90-95 &91-95 &93-95\dag &92-95 &92-95 &90-95 &91-95 &90-95\dag &90-95\\
           &lepton &lepton &lepton &jet charge &lepton &$D$-meson &jet charge &lepton &jet charge &lepton &$D$-meson\\
           &\tmcite{ref:alasy} &\tmcite{ref:alasy} &\tmcite{ref:alasy} &\tmcite{ref:ajet} &\tmcite{ref:dlasy} &\tmcite{ref:ddasy} &\tmcite{ref:djasy} &\tmcite{ref:llasy} &\tmcite{ref:ojet} &\tmcite{ref:olasy} &\tmcite{ref:odsac}\\
 \hline\hline
 \roots\ (\GeV)       &  88.380   & 89.380   & 90.210   & 89.430   & 89.433   & 89.434   & 89.550   & 89.500   & 89.440   & 89.490   & 89.490  \\
 \hline
 \Abl     &     -3.53 &     5.47 &     9.11 &     7.46 &     5.90 &     5.65 &     6.80 &     6.14 &     4.10 &     3.56 &    -9.30\\
 \hline
 \hline\hline
 \Abl     Corrected &   \mca{3}{5.87}
&     7.75 &     6.18 &     5.93 &     6.80 &     6.26 &     4.36 &     3.70 &    -9.16\\
 \hline
 Statistical          &    \mca{3}{1.90}
 &     1.78 &     2.20 &     7.59 &     1.80 &     2.93 &     2.10 &     1.73 &    10.80\\
 \hline
 Uncorrelated         &   \mca{3}{0.39}   
 &     0.19 &     0.08 &     0.91 &     0.12 &     0.37 &     0.25 &     0.16 &     2.51\\
 Correlated           &    \mca{3}{0.70} 
 &     0.15 &     0.08 &     0.08 &     0.01 &     0.19 &     0.02 &     0.04 &     1.41\\
 \hline
 Total Systematic     &     \mca{3}{0.80} 
 &     0.24 &     0.12 &     0.91 &     0.13 &     0.41 &     0.25 &     0.16 &     2.88\\
 \hline
 \hline
 $a( \Rb      )$ &     \mca{3}{-0.3069}
 &  -0.2430 &  -1.1543 &          &  -0.1962 &  -1.4467 &  -0.7300 &  -0.1000 &         \\
 $\Rb     ^{\mathrm{used}}$ &   \mca{3}{0.2192} 
 &   0.2155 &   0.2164 &          &   0.2158 &   0.2170 &   0.2150 &   0.2155 &         \\
 \hline
 $a( \Rc      )$ &   \mca{3}{0.0362} 
&   1.4800 &   1.0444 &          &   0.3200 &   0.3612 &   0.0700 &   0.1000 &         \\
 $\Rc     ^{\mathrm{used}}$ &  \mca{3}{0.1710}
 &   0.1726 &   0.1671 &          &   0.1720 &   0.1734 &   0.1730 &   0.1720 &         \\
 \hline
 $a( \Acl     )$ &   \mca{3}{-0.2244}
 &  -0.2501 &          &          &          &  -0.1000 &  -0.3156 &          &         \\
 $\Acl    ^{\mathrm{used}}$ &      \mca{3}{-2.34}
 &    -2.70 &          &          &          &    -2.50 &    -2.81 &          &         \\
 \hline
 $a( \Brbl    )$ &    \mca{3}{-0.2486}
 &          &  -1.0154 &          &          &  -1.0290 &          &   0.3406 &         \\
 $\Brbl   ^{\mathrm{used}}$ &        \mca{3}{11.34}
 &          &    10.56 &          &          &    10.50 &          &    10.90 &         \\
 \hline
 $a( \Brbclp  )$ &    \mca{3}{-0.1074}
&          &  -0.1424 &          &          &  -0.1440 &          &  -0.5298 &         \\
 $\Brbclp ^{\mathrm{used}}$ &     \mca{3}{7.86}  
 &          &     8.07 &          &          &     8.00 &          &     8.30 &         \\
 \hline
 $a( \Brcl    )$ &   \mca{3}{-0.0474} 
 &          &   0.7224 &          &          &   0.5096 &          &   0.1960 &         \\
 $\Brcl   ^{\mathrm{used}}$ &        \mca{3}{9.80}
 &          &     9.90 &          &          &     9.80 &          &     9.80 &         \\
 \hline
 $a( \chiM    )$ &    \mca{3}{5.259}
&          &   1.3054 &          &          &          &          &          &         \\
 $\chiM   ^{\mathrm{used}}$ &  \mca{3}{ 0.12460}
 &          &  0.11770 &          &          &          &          &          &         \\
 \hline
 $a( \fDp     )$ &   \mca{3}{}      
     &          &          &   0.5083 &   0.0949 &          &          &          &         \\
 $\fDp    ^{\mathrm{used}}$ &   \mca{3}{}       
          &          &          &   0.2210 &   0.2330 &          &          &          &         \\
 \hline
 $a( \fDs     )$ &           \mca{3}{}
         &          &          &   0.1742 &   0.0035 &          &          &          &         \\
 $\fDs    ^{\mathrm{used}}$ &          \mca{3}{}
           &          &          &   0.1120 &   0.1020 &          &          &          &         \\
 \hline
 $a( \fLc     )$ &       \mca{3}{}
   &          &          &  -0.0191 &  -0.0225 &          &          &          &         \\
 $\fLc    ^{\mathrm{used}}$ &     \mca{3}{} 
      &          &          &   0.0840 &   0.0630 &          &          &          &         \\
 \hline
 \end{tabular}
 \end{center}
 \caption{The measurements of 
 \Abl    . 
  All numbers are given in \%.
}
 \label{tab:Ablinp}
 \end{minipage}
 \end{sideways}
 \end{center}
 \end{table}
 \begin{table}[htb]
 \begin{center}
 \begin{tabular}{|l||c|c|c|c|c|}
 \hline
           & \mca{1}{ALEPH} & \mca{2}{DELPHI} & \mca{2}{OPAL} \\
 \hline
           &91-95 &93-95\dag &92-95 &90-95\dag &90-95\\
           &$D$-meson &lepton &$D$-meson &lepton &$D$-meson\\
           &\tmcite{ref:adsac} &\tmcite{ref:dlasy} &\tmcite{ref:ddasy} &\tmcite{ref:olasy} &\tmcite{ref:odsac}\\
 \hline\hline
 \roots\ (\GeV)       &  89.370   & 89.433   & 89.434   & 89.490   & 89.490  \\
 \hline
 \Acl     &     -1.10 &     1.11 &    -5.04 &    -6.92 &     3.90\\
 \hline
 \hline\hline
 \Acl     Corrected &     -0.02 &     1.81 &    -4.35 &    -6.56 &     4.26\\
 \hline
 Statistical          &      4.30 &     3.60 &     3.69 &     2.44 &     5.10\\
 \hline
 Uncorrelated         &      1.00 &     0.53 &     0.40 &     0.38 &     0.80\\
 Correlated           &      0.09 &     0.16 &     0.09 &     0.23 &     0.30\\
 \hline
 Total Systematic     &      1.00 &     0.55 &     0.41 &     0.44 &     0.86\\
 \hline
 \hline
 $a( \Rb      )$ &           &  -0.2886 &          &  -3.4000 &         \\
 $\Rb     ^{\mathrm{used}}$ &           &   0.2164 &          &   0.2155 &         \\
 \hline
 $a( \Rc      )$ &           &   1.0096 &          &   3.2000 &         \\
 $\Rc     ^{\mathrm{used}}$ &           &   0.1671 &          &   0.1720 &         \\
 \hline
 $a( \Abl     )$ &   -1.3365 &          &          &          &         \\
 $\Abl    ^{\mathrm{used}}$ &      6.13 &          &          &          &         \\
 \hline
 $a( \Brbl    )$ &           &  -1.0966 &          &  -1.7031 &         \\
 $\Brbl   ^{\mathrm{used}}$ &           &    10.56 &          &    10.90 &         \\
 \hline
 $a( \Brbclp  )$ &           &   1.1156 &          &  -1.4128 &         \\
 $\Brbclp ^{\mathrm{used}}$ &           &     8.07 &          &     8.30 &         \\
 \hline
 $a( \Brcl    )$ &           &   1.0703 &          &   3.3320 &         \\
 $\Brcl   ^{\mathrm{used}}$ &           &     9.90 &          &     9.80 &         \\
 \hline
 $a( \chiM    )$ &           &  -0.0856 &          &          &         \\
 $\chiM   ^{\mathrm{used}}$ &           &  0.11770 &          &          &         \\
 \hline
 $a( \fDp     )$ &           &          &  -0.3868 &          &         \\
 $\fDp    ^{\mathrm{used}}$ &           &          &   0.2210 &          &         \\
 \hline
 $a( \fDs     )$ &           &          &  -0.1742 &          &         \\
 $\fDs    ^{\mathrm{used}}$ &           &          &   0.1120 &          &         \\
 \hline
 $a( \fLc     )$ &           &          &  -0.0878 &          &         \\
 $\fLc    ^{\mathrm{used}}$ &           &          &   0.0840 &          &         \\
 \hline
 \end{tabular}
 \end{center}
 \caption{The measurements of 
 \Acl    .  All numbers are given in \%.
 }
 \label{tab:Aclinp}
 \end{table}
 \begin{table}[htb]
 \begin{center}
 \begin{sideways}
 \begin{minipage}[b]{\textheight}
 \begin{center}
 \begin{tabular}{|l||c|c|c|c|c|c|c|c|c|c|c|}
 \hline
           & \mca{2}{ALEPH} & \mca{4}{DELPHI} & \mca{2}{L3} & \mca{3}{OPAL} \\
 \hline
           &91-95\dag &91-95 &91-92 &93-95\dag &92-95 &92-95 &91-95 &90-95 &91-95 &90-95\dag &90-95\\
           &lepton &jet charge &lepton &lepton &$D$-meson &jet charge &jet charge &lepton &jet charge &lepton &$D$-meson\\
           &\tmcite{ref:alasy} &\tmcite{ref:ajet} &\tmcite{ref:dlasy} &\tmcite{ref:dlasy} &\tmcite{ref:ddasy} &\tmcite{ref:djasy} &\tmcite{ref:ljet} &\tmcite{ref:llasy} &\tmcite{ref:ojet} &\tmcite{ref:olasy} &\tmcite{ref:odsac}\\
 \hline\hline
 \roots\ (\GeV)       &  91.210   & 91.250   & 91.270   & 91.223   & 91.235   & 91.260   & 91.240   & 91.260   & 91.210   & 91.240   & 91.240  \\
 \hline
 \Abp     &      9.71 &    10.40 &    10.89 &     9.86 &     7.59 &     9.83 &     9.31 &     9.85 &    10.06 &     9.14 &     8.90\\
 \hline
 \hline\hline
 \Abp     Corrected &      9.81 &    10.42 &    10.87 &     9.93 &     7.63 &     9.83 &     9.35 &     9.85 &    10.15 &     9.18 &     8.94\\
 \hline
 Statistical          &      0.40 &     0.40 &     1.30 &     0.64 &     1.97 &     0.47 &     1.01 &     0.67 &     0.52 &     0.44 &     2.70\\
 \hline
 Uncorrelated         &      0.16 &     0.23 &     0.33 &     0.15 &     0.77 &     0.14 &     0.51 &     0.27 &     0.41 &     0.14 &     2.15\\
 Correlated           &      0.12 &     0.22 &     0.27 &     0.14 &     0.07 &     0.04 &     0.21 &     0.14 &     0.20 &     0.15 &     0.45\\
 \hline
 Total Systematic     &      0.20 &     0.32 &     0.43 &     0.20 &     0.77 &     0.14 &     0.55 &     0.31 &     0.46 &     0.20 &     2.20\\
 \hline
 \hline
 $a( \Rb      )$ &   -0.9545 &  -0.2430 &  -2.8933 &  -2.0201 &          &  -0.1962 &  -9.1622 &  -2.1700 &  -7.6300 &  -0.7000 &         \\
 $\Rb     ^{\mathrm{used}}$ &    0.2172 &   0.2155 &   0.2170 &   0.2164 &          &   0.2158 &   0.2170 &   0.2170 &   0.2150 &   0.2155 &         \\
 \hline
 $a( \Rc      )$ &    0.6450 &   1.4900 &   1.0993 &   1.1488 &          &   0.8400 &   1.0831 &   1.3005 &   0.4600 &   0.6000 &         \\
 $\Rc     ^{\mathrm{used}}$ &    0.1720 &   0.1726 &   0.1710 &   0.1671 &          &   0.1720 &   0.1733 &   0.1734 &   0.1730 &   0.1720 &         \\
 \hline
 $a( \Acp     )$ &           &   0.6345 &          &          &          &          &   1.1603 &   0.9262 &   0.6870 &          &         \\
 $\Acp    ^{\mathrm{used}}$ &           &     6.85 &          &          &          &          &     6.91 &     7.41 &     6.19 &          &         \\
 \hline
 $a( \Brbl    )$ &   -1.8480 &          &  -3.8824 &  -2.0308 &          &          &          &  -2.0160 &          &  -0.3406 &         \\
 $\Brbl   ^{\mathrm{used}}$ &     10.78 &          &    11.00 &    10.56 &          &          &          &    10.50 &          &    10.90 &         \\
 \hline
 $a( \Brbclp  )$ &    0.4233 &          &   0.4740 &  -0.3798 &          &          &          &  -0.1280 &          &  -0.3532 &         \\
 $\Brbclp ^{\mathrm{used}}$ &      8.14 &          &     7.90 &     8.07 &          &          &          &     8.00 &          &     8.30 &         \\
 \hline
 $a( \Brcl    )$ &    0.5096 &          &   0.7840 &   1.0703 &          &          &          &   1.5288 &          &   0.5880 &         \\
 $\Brcl   ^{\mathrm{used}}$ &      9.80 &          &     9.80 &     9.90 &          &          &          &     9.80 &          &     9.80 &         \\
 \hline
 $a( \chiM    )$ &    2.9904 &          &   3.4467 &   1.6692 &          &          &          &          &          &          &         \\
 $\chiM   ^{\mathrm{used}}$ &   0.12460 &          &  0.12100 &  0.11770 &          &          &          &          &          &          &         \\
 \hline
 $a( \fDp     )$ &           &          &          &          &   0.0442 &   0.2761 &          &          &          &          &         \\
 $\fDp    ^{\mathrm{used}}$ &           &          &          &          &   0.2210 &   0.2330 &          &          &          &          &         \\
 \hline
 $a( \fDs     )$ &           &          &          &          &  -0.0788 &   0.0106 &          &          &          &          &         \\
 $\fDs    ^{\mathrm{used}}$ &           &          &          &          &   0.1120 &   0.1020 &          &          &          &          &         \\
 \hline
 $a( \fLc     )$ &           &          &          &          &  -0.0115 &  -0.0495 &          &          &          &          &         \\
 $\fLc    ^{\mathrm{used}}$ &           &          &          &          &   0.0840 &   0.0630 &          &          &          &          &         \\
 \hline
 \end{tabular}
 \end{center}
 \caption{The measurements of 
 \Abp    .   All numbers are given in \%.
}
 \label{tab:Abpinp}
 \end{minipage}
 \end{sideways}
 \end{center}
 \end{table}
 \begin{table}[htb]
 \begin{center}
 \begin{tabular}{|l||c|c|c|c|c|c|c|c|}
 \hline
           & \mca{2}{ALEPH} & \mca{3}{DELPHI} & \mca{1}{L3} & \mca{2}{OPAL} \\
 \hline
           &91-95\dag &91-95 &91-92 &93-95\dag &92-95 &90-95 &90-95\dag &90-95\\
           &lepton &$D$-meson &lepton &lepton &$D$-meson &lepton &lepton &$D$-meson\\
           &\tmcite{ref:alasy} &\tmcite{ref:adsac} &\tmcite{ref:dlasy} &\tmcite{ref:dlasy} &\tmcite{ref:ddasy} &\tmcite{ref:llasy} &\tmcite{ref:olasy} &\tmcite{ref:odsac}\\
 \hline\hline
 \roots\ (\GeV)       &  91.210   & 91.220   & 91.270   & 91.223   & 91.235   & 91.240   & 91.240   & 91.240  \\
 \hline
 \Acp     &      5.69 &     6.20 &     8.05 &     6.28 &     6.58 &     7.94 &     5.97 &     6.60\\
 \hline
 \hline\hline
 \Acp     Corrected &      5.94 &     6.39 &     8.00 &     6.46 &     6.70 &     8.04 &     6.07 &     6.70\\
 \hline
 Statistical          &      0.53 &     0.90 &     2.26 &     1.00 &     0.97 &     3.70 &     0.59 &     1.20\\
 \hline
 Uncorrelated         &      0.24 &     0.23 &     1.25 &     0.53 &     0.25 &     2.40 &     0.37 &     0.49\\
 Correlated           &      0.36 &     0.17 &     0.49 &     0.27 &     0.04 &     0.49 &     0.32 &     0.24\\
 \hline
 Total Systematic     &      0.44 &     0.28 &     1.35 &     0.60 &     0.25 &     2.45 &     0.49 &     0.54\\
 \hline
 \hline
 $a( \Rb      )$ &    1.4318 &          &   2.8933 &  -2.3087 &          &   4.3200 &   4.1000 &         \\
 $\Rb     ^{\mathrm{used}}$ &    0.2172 &          &   0.2170 &   0.2164 &          &   0.2160 &   0.2155 &         \\
 \hline
 $a( \Rc      )$ &   -2.9383 &          &  -6.4736 &   5.4307 &          &  -6.7600 &  -3.8000 &         \\
 $\Rc     ^{\mathrm{used}}$ &    0.1720 &          &   0.1710 &   0.1671 &          &   0.1690 &   0.1720 &         \\
 \hline
 $a( \Abp     )$ &           &  -2.1333 &          &          &          &   6.4274 &          &         \\
 $\Abp    ^{\mathrm{used}}$ &           &     9.79 &          &          &          &     8.84 &          &         \\
 \hline
 $a( \Brbl    )$ &    1.8993 &          &   4.8529 &  -2.7618 &          &   3.5007 &   5.1094 &         \\
 $\Brbl   ^{\mathrm{used}}$ &     10.78 &          &    11.00 &    10.56 &          &    10.50 &    10.90 &         \\
 \hline
 $a( \Brbclp  )$ &   -1.0745 &          &  -3.9500 &   2.2786 &          &  -3.2917 &  -1.7660 &         \\
 $\Brbclp ^{\mathrm{used}}$ &      8.14 &          &     7.90 &     8.07 &          &     7.90 &     8.30 &         \\
 \hline
 $a( \Brcl    )$ &   -3.2732 &          &  -7.2520 &   4.8965 &          &  -6.5327 &  -3.9200 &         \\
 $\Brcl   ^{\mathrm{used}}$ &      9.80 &          &     9.80 &     9.90 &          &     9.80 &     9.80 &         \\
 \hline
 $a( \chiM    )$ &    0.0453 &          &          &   0.3852 &          &          &          &         \\
 $\chiM   ^{\mathrm{used}}$ &   0.12460 &          &          &  0.11770 &          &          &          &         \\
 \hline
 $a( \fDp     )$ &           &          &          &          &  -0.0221 &          &          &         \\
 $\fDp    ^{\mathrm{used}}$ &           &          &          &          &   0.2210 &          &          &         \\
 \hline
 $a( \fDs     )$ &           &          &          &          &   0.0788 &          &          &         \\
 $\fDs    ^{\mathrm{used}}$ &           &          &          &          &   0.1120 &          &          &         \\
 \hline
 $a( \fLc     )$ &           &          &          &          &   0.0115 &          &          &         \\
 $\fLc    ^{\mathrm{used}}$ &           &          &          &          &   0.0840 &          &          &         \\
 \hline
 \end{tabular}
 \end{center}
 \caption{The measurements of 
 \Acp    .   All numbers are given in \%.
}
 \label{tab:Acpinp}
 \end{table}
 \begin{table}[htb]
 \begin{center}
 \begin{sideways}
 \begin{minipage}[b]{\textheight}
 \begin{center}
 \begin{tabular}{|l||c|c|c|c|c|c|c|c|c|c|c|}
 \hline
           & \mca{4}{ALEPH} & \mca{3}{DELPHI} & \mca{1}{L3} & \mca{3}{OPAL} \\
 \hline
           &90-95 &90-95 &90-95 &91-95 &93-95\dag &92-95 &92-95 &90-95 &91-95 &90-95\dag &90-95\\
           &lepton &lepton &lepton &jet charge &lepton &$D$-meson &jet charge &lepton &jet charge &lepton &$D$-meson\\
           &\tmcite{ref:alasy} &\tmcite{ref:alasy} &\tmcite{ref:alasy} &\tmcite{ref:ajet} &\tmcite{ref:dlasy} &\tmcite{ref:ddasy} &\tmcite{ref:djasy} &\tmcite{ref:llasy} &\tmcite{ref:ojet} &\tmcite{ref:olasy} &\tmcite{ref:odsac}\\
 \hline\hline
 \roots\ (\GeV)       &  92.050   & 92.940   & 93.900   & 92.970   & 92.990   & 92.990   & 92.940   & 93.100   & 92.910   & 92.950   & 92.950  \\
 \hline
 \Abh     &      3.93 &    10.60 &     9.03 &     9.24 &    10.10 &     8.78 &    12.30 &    13.78 &    14.60 &    10.75 &    -3.40\\
 \hline
 \hline\hline
 \Abh     Corrected &       \mca{3}{10.03}
 &     9.21 &    10.05 &     8.73 &    12.30 &    13.62 &    14.63 &    10.74 &    -3.41\\
 \hline
 Statistical          &       \mca{3}{1.51}
 &     1.79 &     1.80 &     6.37 &     1.60 &     2.40 &     1.70 &     1.43 &     9.00\\
 \hline
 Uncorrelated         &      \mca{3}{0.14}
 &     0.45 &     0.14 &     0.97 &     0.25 &     0.34 &     0.64 &     0.25 &     2.03\\
 Correlated           &       \mca{3}{0.24}
 &     0.26 &     0.16 &     0.13 &     0.05 &     0.20 &     0.34 &     0.28 &     1.74\\
 \hline
 Total Systematic     &        \mca{3}{0.28}
 &     0.52 &     0.21 &     0.98 &     0.26 &     0.40 &     0.73 &     0.37 &     2.68\\
 \hline
 \hline
 $a( \Rb      )$ &   \mca{3}{-1.964} 
&  -0.2430 &  -2.8859 &          &  -0.1962 &  -3.3756 & -12.9000 &  -0.8000 &         \\
 $\Rb     ^{\mathrm{used}}$ &      \mca{3}{0.2192} 
 &   0.2155 &   0.2164 &          &   0.2158 &   0.2170 &   0.2150 &   0.2155 &         \\
 \hline
 $a( \Rc      )$ &     \mca{3}{1.575}
&   1.4900 &   1.3577 &          &   1.2000 &   1.9869 &   0.6900 &   0.8000 &         \\
 $\Rc     ^{\mathrm{used}}$ &   \mca{3}{0.1710}
 &   0.1726 &   0.1671 &          &   0.1720 &   0.1734 &   0.1730 &   0.1720 &         \\
 \hline
 $a( \Ach     )$ &     \mca{3}{1.081}
&   1.2018 &          &          &          &   0.5206 &   1.3287 &          &         \\
 $\Ach    ^{\mathrm{used}}$ &     \mca{3}{12.51}
 &    12.96 &          &          &          &    12.39 &    12.08 &          &         \\
 \hline
 $a( \Brbl    )$ &   \mca{3}{-1.762}
 &          &  -2.3557 &          &          &  -2.0790 &          &  -1.3625 &         \\
 $\Brbl   ^{\mathrm{used}}$ &    \mca{3}{11.34}
&          &    10.56 &          &          &    10.50 &          &    10.90 &         \\
 \hline
 $a( \Brbclp  )$ &   \mca{3}{-0.2478}
 &          &  -0.7595 &          &          &  -1.1200 &          &   0.7064 &         \\
 $\Brbclp ^{\mathrm{used}}$ &    \mca{3}{7.86} 
&          &     8.07 &          &          &     8.00 &          &     8.30 &         \\
 \hline
 $a( \Brcl    )$ &    \mca{3}{1.524}
&          &   1.0703 &          &          &   1.9796 &          &   0.7840 &         \\
 $\Brcl   ^{\mathrm{used}}$ &    \mca{3}{9.80}
&          &     9.90 &          &          &     9.80 &          &     9.80 &         \\
 \hline
 $a( \chiM    )$ &    \mca{3}{6.584}
 &          &   1.6050 &          &          &          &          &          &         \\
 $\chiM   ^{\mathrm{used}}$ &    \mca{3}{0.12460}
 &          &  0.11770 &          &          &          &          &          &         \\
 \hline
 $a( \fDp     )$ &       \mca{3}{}     
  &          &          &   0.3978 &   0.4229 &          &          &          &         \\
 $\fDp    ^{\mathrm{used}}$ &   \mca{3}{}       
         &          &          &   0.2210 &   0.2330 &          &          &          &         \\
 \hline
 $a( \fDs     )$ &      \mca{3}{}      
         &          &          &  -0.0788 &   0.0211 &          &          &          &         \\
 $\fDs    ^{\mathrm{used}}$ &      \mca{3}{} 
         &          &          &   0.1120 &   0.1020 &          &          &          &         \\
 \hline
 $a( \fLc     )$ &        \mca{3}{} 
        &          &          &   0.0573 &  -0.0855 &          &          &          &         \\
 $\fLc    ^{\mathrm{used}}$ &    \mca{3}{}       
          &          &          &   0.0840 &   0.0630 &          &          &          &         \\
 \hline
 \end{tabular}
 \end{center}
 \caption{The measurements of 
 \Abh    .   All numbers are given in \%.
}
 \label{tab:Abhinp}
 \end{minipage}
 \end{sideways}
 \end{center}
 \end{table}
 \begin{table}[htb]
 \begin{center}
 \begin{tabular}{|l||c|c|c|c|c|}
 \hline
           & \mca{1}{ALEPH} & \mca{2}{DELPHI} & \mca{2}{OPAL} \\
 \hline
           &91-95 &93-95\dag &92-95 &90-95\dag &90-95\\
           &$D$-meson &lepton &$D$-meson &lepton &$D$-meson\\
           &\tmcite{ref:adsac} &\tmcite{ref:dlasy} &\tmcite{ref:ddasy} &\tmcite{ref:olasy} &\tmcite{ref:odsac}\\
 \hline\hline
 \roots\ (\GeV)       &  92.960   & 92.990   & 92.990   & 92.950   & 92.950  \\
 \hline
 \Ach     &     10.94 &    10.50 &    11.78 &    15.65 &    16.70\\
 \hline
 \hline\hline
 \Ach     Corrected &     10.89 &    10.37 &    11.65 &    15.62 &    16.67\\
 \hline
 Statistical          &      3.30 &     2.90 &     3.20 &     2.02 &     4.10\\
 \hline
 Uncorrelated         &      0.79 &     0.41 &     0.52 &     0.57 &     0.92\\
 Correlated           &      0.18 &     0.29 &     0.07 &     0.62 &     0.51\\
 \hline
 Total Systematic     &      0.81 &     0.50 &     0.52 &     0.84 &     1.05\\
 \hline
 \hline
 $a( \Rb      )$ &           &  -4.0402 &          &   9.6000 &         \\
 $\Rb     ^{\mathrm{used}}$ &           &   0.2164 &          &   0.2155 &         \\
 \hline
 $a( \Rc      )$ &           &   7.5891 &          &  -8.9000 &         \\
 $\Rc     ^{\mathrm{used}}$ &           &   0.1671 &          &   0.1720 &         \\
 \hline
 $a( \Abh     )$ &   -2.6333 &          &          &          &         \\
 $\Abh    ^{\mathrm{used}}$ &     12.08 &          &          &          &         \\
 \hline
 $a( \Brbl    )$ &           &  -3.2492 &          &   9.5375 &         \\
 $\Brbl   ^{\mathrm{used}}$ &           &    10.56 &          &    10.90 &         \\
 \hline
 $a( \Brbclp  )$ &           &   1.5191 &          &  -1.5894 &         \\
 $\Brbclp ^{\mathrm{used}}$ &           &     8.07 &          &     8.30 &         \\
 \hline
 $a( \Brcl    )$ &           &   8.1341 &          &  -9.2120 &         \\
 $\Brcl   ^{\mathrm{used}}$ &           &     9.90 &          &     9.80 &         \\
 \hline
 $a( \chiM    )$ &           &  -0.2140 &          &          &         \\
 $\chiM   ^{\mathrm{used}}$ &           &  0.11770 &          &          &         \\
 \hline
 $a( \fDp     )$ &           &          &  -0.2984 &          &         \\
 $\fDp    ^{\mathrm{used}}$ &           &          &   0.2210 &          &         \\
 \hline
 $a( \fDs     )$ &           &          &   0.0539 &          &         \\
 $\fDs    ^{\mathrm{used}}$ &           &          &   0.1120 &          &         \\
 \hline
 $a( \fLc     )$ &           &          &   0.0764 &          &         \\
 $\fLc    ^{\mathrm{used}}$ &           &          &   0.0840 &          &         \\
 \hline
 \end{tabular}
 \end{center}
 \caption{The measurements of 
 \Ach    .   All numbers are given in \%.
}
 \label{tab:Achinp}
 \end{table}
 \begin{table}[htb]
 \begin{center}
 \begin{tabular}{|l||c|c|c|c|}
 \hline
           & \mca{4}{SLD} \\
 \hline
           &93-98\dag &93-98\dag &94-98\dag &97-98\dag\\
           &lepton &jet charge &$K^{\pm}$ &multi\\
           &\tmcite{ref:SLD_ABL} &\tmcite{ref:SLD_ABJ} &\tmcite{ref:SLD_ABK} &\tmcite{ref:SLD_vtxasy}\\
 \hline\hline
 \roots\ (\GeV)       &  91.280   & 91.280   & 91.280   & 91.280  \\
 \hline
 \cAb     &     0.922 &    0.882 &    0.960 &    0.926\\
 \hline
 Statistical          &     0.029 &    0.020 &    0.040 &    0.019\\
 \hline
 Uncorrelated         &     0.019 &    0.029 &    0.056 &    0.027\\
 Correlated           &     0.008 &    0.001 &    0.002 &    0.001\\
 \hline
 Total Systematic     &     0.021 &    0.029 &    0.056 &    0.027\\
 \hline
 \hline
 $a( \Rb      )$ &   -0.0542 &          &          &         \\
 $\Rb     ^{\mathrm{used}}$ &    0.2168 &          &          &         \\
 \hline
 $a( \Rc      )$ &    0.0424 &          &          &         \\
 $\Rc     ^{\mathrm{used}}$ &    0.1697 &          &          &         \\
 \hline
 $a( \cAc     )$ &    0.0449 &   0.0134 &  -0.0112 &   0.0133\\
 $\cAc    ^{\mathrm{used}}$ &     0.667 &    0.670 &    0.666 &    0.667\\
 \hline
 $a( \Brbl    )$ &   -0.2160 &          &          &         \\
 $\Brbl   ^{\mathrm{used}}$ &     10.80 &          &          &         \\
 \hline
 $a( \Brbclp  )$ &    0.0888 &          &          &         \\
 $\Brbclp ^{\mathrm{used}}$ &      8.05 &          &          &         \\
 \hline
 $a( \Brcl    )$ &    0.0479 &          &          &         \\
 $\Brcl   ^{\mathrm{used}}$ &      9.83 &          &          &         \\
 \hline
 $a( \chiM    )$ &    0.3052 &          &          &         \\
 $\chiM   ^{\mathrm{used}}$ &   0.11990 &          &          &         \\
 \hline
 \end{tabular}
 \end{center}
 \caption{The measurements of 
 \cAb    . }
 \label{tab:cAbinp}
 \end{table}
 \begin{table}[htb]
 \begin{center}
 \begin{tabular}{|l||c|c|c|}
 \hline
           & \mca{3}{SLD} \\
 \hline
           &93-98\dag &93-97\dag &93-97\dag\\
           &lepton &$D$-meson &K+vertex\\
           &\tmcite{ref:SLD_ACL} &\tmcite{ref:SLD_ACD} &\tmcite{ref:SLD_ACV}\\
 \hline\hline
 \roots\ (\GeV)       &  91.280   & 91.280   & 91.280  \\
 \hline
 \cAc     &     0.567 &    0.688 &    0.603\\
 \hline
 Statistical          &     0.051 &    0.035 &    0.028\\
 \hline
 Uncorrelated         &     0.056 &    0.020 &    0.023\\
 Correlated           &     0.018 &    0.003 &    0.001\\
 \hline
 Total Systematic     &     0.059 &    0.021 &    0.023\\
 \hline
 \hline
 $a( \Rb      )$ &    0.2173 &          &         \\
 $\Rb     ^{\mathrm{used}}$ &    0.2173 &          &         \\
 \hline
 $a( \Rc      )$ &   -0.4089 &          &         \\
 $\Rc     ^{\mathrm{used}}$ &    0.1730 &          &         \\
 \hline
 $a( \cAb     )$ &    0.2151 &  -0.0673 &  -0.0306\\
 $\cAb    ^{\mathrm{used}}$ &     0.935 &    0.935 &    0.900\\
 \hline
 $a( \Brbl    )$ &    0.2328 &          &         \\
 $\Brbl   ^{\mathrm{used}}$ &     11.06 &          &         \\
 \hline
 $a( \Brbclp  )$ &   -0.1178 &          &         \\
 $\Brbclp ^{\mathrm{used}}$ &      8.02 &          &         \\
 \hline
 $a( \Brcl    )$ &   -0.4077 &          &         \\
 $\Brcl   ^{\mathrm{used}}$ &      9.80 &          &         \\
 \hline
 $a( \chiM    )$ &    0.1138 &          &         \\
 $\chiM   ^{\mathrm{used}}$ &   0.12170 &          &         \\
 \hline
 $a( \fDp     )$ &           &          &  -0.0140\\
 $\fDp    ^{\mathrm{used}}$ &           &          &   0.2300\\
 \hline
 $a( \fDs     )$ &           &          &  -0.0028\\
 $\fDs    ^{\mathrm{used}}$ &           &          &   0.1150\\
 \hline
 $a( \fLc     )$ &           &          &   0.0005\\
 $\fLc    ^{\mathrm{used}}$ &           &          &   0.0740\\
 \hline
 \end{tabular}
 \end{center}
 \caption{The measurements of 
 \cAc    . }
 \label{tab:cAcinp}
 \end{table}
 \begin{table}[htb]
 \begin{center}
 \begin{tabular}{|l||c|c|c|c|c|}
 \hline
           & \mca{1}{ALEPH} & \mca{1}{DELPHI} & \mca{2}{L3} & \mca{1}{OPAL} \\
 \hline
           &91-95\dag & 94-95\dag & 92    &94-95\dag &92-95\\
           &multi     & multi     &lepton &multi     &multi \\
           &\tmcite{ref:abl} &\tmcite{ref:dbl} &\tmcite{ref:lbl} &\tmcite{ref:lrbmixed} &\tmcite{ref:obl}\\
 \hline\hline
 \Brbl                &     11.55 &    10.65 &    10.68 &    10.21 &    10.83\\
 \hline                                                            
 Statistical          &      0.09 &     0.07 &     0.11 &     0.13 &     0.10\\
 \hline                                                            
 Uncorrelated         &      0.17 &     0.23 &     0.36 &     0.20 &     0.20\\
 Correlated           &      0.22 &     0.42 &     0.22 &     0.31 &     0.21\\
 \hline
 Total Systematic     &      0.27 &     0.48 &     0.42 &     0.36 &     0.29\\
 \hline
 \hline
 $a( \Rb      )$      &           &          &  -9.2571 &          &  -0.1808\\
 $\Rb     ^{\mathrm{used}}$                                        
                      &           &          &   0.2160 &          &   0.2169\\
 \hline                                                            
 $a( \Rc      )$      &           &          &          &   1.4450 &   0.4867\\
 $\Rc     ^{\mathrm{used}}$                                        
                      &           &          &          &   0.1734 &   0.1770\\
 \hline                                                            
 $a( \Brbclp  )$      &           &          &  -1.1700 &   0.1618 &         \\
 $\Brbclp ^{\mathrm{used}}$                                        
                      &           &          &     9.00 &     8.09 &         \\
 \hline                                                            
 $a( \Brcl    )$      &    -0.3078 &  -0.1960 &  -2.5480 &   0.9212 &         \\
 $\Brcl   ^{\mathrm{used}}$                                        
                      &      9.85 &     9.80 &     9.80 &     9.80 &         \\
 \hline                                                            
 $a( \chiM    )$      &    0.7683 &          &          &          &         \\
 $\chiM   ^{\mathrm{used}}$                                        
                      &    0.1178 &          &          &          &         \\
 \hline                                                            
 $a( \fDp     )$      &           &          &          &   0.5523 &   0.1445\\
 $\fDp    ^{\mathrm{used}}$                                        
                      &           &          &          &   0.2330 &   0.2380\\
 \hline                                                            
 $a( \fDs     )$      &           &          &          &   0.0213 &   0.0055\\
 $\fDs    ^{\mathrm{used}}$                                        
                      &           &          &          &   0.1030 &   0.1020\\
 \hline                                                            
 $a( \fLc     )$      &           &          &          &  -0.0427 &  -0.0157\\
 $\fLc    ^{\mathrm{used}}$                                        
                      &           &          &          &   0.0630 &   0.0650\\
 \hline
 \end{tabular}
 \end{center}
 \caption{The measurements of 
 \Brbl   .   All numbers are given in \%.
}
 \label{tab:Brblinp}
 \end{table}
 \begin{table}[htb]
 \begin{center}
 \begin{tabular}{|l||c|c|c|}
 \hline
           & \mca{1}{ALEPH} & \mca{1}{DELPHI} & \mca{1}{OPAL} \\
 \hline
           &91-95\dag &94-95\dag &92-95 \\
           &multi &multi &multi \\
           &\tmcite{ref:abl} &\tmcite{ref:dbl} &\tmcite{ref:obl} \\
 \hline\hline
 \Brbclp  &      8.04 &     7.88 &     8.40\\
 \hline
 Statistical          &      0.14 &     0.13 &     0.16\\
 \hline
 Uncorrelated         &      0.20 &     0.26 &     0.19\\
 Correlated           &      0.14 &     0.36 &     0.34\\
 \hline
 Total Systematic     &      0.25 &     0.45 &     0.39\\
 \hline
 \hline
 $a( \Rb      )$ &           &          &  -0.1808 \\
 $\Rb     ^{\mathrm{used}}$ &           &          &   0.2169 \\
 \hline
 $a( \Rc      )$ &    0.8916 &          &   0.3761 \\
 $\Rc     ^{\mathrm{used}}$ &    0.1694 &          &   0.1770 \\
 \hline
 $a( \Brcl    )$ &    0.3078 &  -0.1960 &                  \\
 $\Brcl   ^{\mathrm{used}}$ &      9.85 &     9.80 &                   \\
 \hline
 $a( \chiM    )$ &   -1.2804 &          &                  \\
 $\chiM   ^{\mathrm{used}}$ &   0.11780 &                    &         \\
 \hline
 $a( \fDp     )$ &           &          &   0.1190 \\
 $\fDp    ^{\mathrm{used}}$ &           &         &   0.2380\\
 \hline
 $a( \fDs     )$ &           &          &      0.0028\\
 $\fDs    ^{\mathrm{used}}$ &           &   &   0.1020\\
 \hline
 $a( \fLc     )$ &           &          &   -0.0110\\
 $\fLc    ^{\mathrm{used}}$ &           &    &   0.0660\\
 \hline
 \end{tabular}
 \end{center}
 \caption{The measurements of 
 \Brbclp .   All numbers are given in \%.
}
 \label{tab:Brbclpinp}
 \end{table}
 \begin{table}[htb]
 \begin{center}
 \begin{tabular}{|l||c|c|}
 \hline
           & \mca{1}{DELPHI} & \mca{1}{OPAL} \\
 \hline
           &92-95 &90-95\\
           &$D$+lepton &$D$+lepton\\
           &\tmcite{ref:drcd} &\tmcite{ref:ocl}\\
 \hline\hline
 $\Brcl$              &      9.59 &     9.60\\
 \hline
 Statistical          &      0.42 &     0.60\\
 \hline
 Uncorrelated         &      0.24 &     0.49\\
 Correlated           &      0.14 &     0.43\\
 \hline
 Total Systematic     &      0.27 &     0.65\\
 \hline
 \hline
 $a( \Brbl    )$ &   -0.5600 &  -1.4335\\
 $\Brbl   ^{\mathrm{used}}$ &     11.20 &    10.99\\
 \hline
 $a( \Brbclp  )$ &   -0.4100 &  -0.7800\\
 $\Brbclp ^{\mathrm{used}}$ &      8.20 &     7.80\\
 \hline
 \end{tabular}
 \end{center}
 \caption{The measurements of 
 $\Brcl$.   All numbers are given in \%.
}
 \label{tab:Brclinp}
 \end{table}
 \begin{table}[htb]
 \begin{center}
 \begin{tabular}{|l||c|c|c|c|}
 \hline
           & \mca{1}{ALEPH} & \mca{1}{DELPHI} & \mca{1}{L3} & \mca{1}{OPAL} \\
 \hline
           &90-95 &94-95\dag &90-95 &90-95\dag\\
           &multi &multi &lepton &lepton\\
           &\tmcite{ref:alasy} &\tmcite{ref:dbl} &\tmcite{ref:llasy} &\tmcite{ref:olasy}\\
 \hline\hline
 \chiM    &   0.12461 &  0.12700 &  0.11920 &  0.11390\\
 \hline
 Statistical          &   0.00515 &  0.01300 &  0.00680 &  0.00540\\
 \hline
 Uncorrelated         &   0.00252 &  0.00566 &  0.00214 &  0.00306\\
 Correlated           &   0.00397 &  0.00554 &  0.00252 &  0.00324\\
 \hline
 Total Systematic     &   0.00470 &  0.00792 &  0.00330 &  0.00446\\
 \hline
 \hline
 $a( \Rb      )$ &    0.0341 &          &   0.0000 &         \\
 $\Rb     ^{\mathrm{used}}$ &    0.2192 &          &   0.2170 &         \\
 \hline
 $a( \Rc      )$ &    0.0009 &          &   0.0004 &         \\
 $\Rc     ^{\mathrm{used}}$ &    0.1710 &          &   0.1734 &         \\
 \hline
 $a( \Brbl    )$ &    0.0524 &          &   0.0550 &   0.0170\\
 $\Brbl   ^{\mathrm{used}}$ &     11.34 &          &    10.50 &    10.90\\
 \hline
 $a( \Brbclp  )$ &   -0.0440 &          &  -0.0466 &  -0.0318\\
 $\Brbclp ^{\mathrm{used}}$ &      7.86 &          &     8.00 &     8.30\\
 \hline
 $a( \Brcl    )$ &    0.0035 &  -0.0020 &   0.0006 &   0.0039\\
 $\Brcl   ^{\mathrm{used}}$ &      9.80 &     9.80 &     9.80 &     9.80\\
 \hline
 \end{tabular}
 \end{center}
 \caption{The measurements of 
 \chiM   . }
 \label{tab:chiMinp}
 \end{table}
 \begin{table}[htb]
 \begin{center}
 \begin{tabular}{|l||c|c|}
 \hline
           & \mca{1}{DELPHI} & \mca{1}{OPAL} \\
 \hline
           &92-95 &90-95\\
           &$D$-meson &$D$-meson\\
           &\tmcite{ref:drcd} &\tmcite{ref:orcd}\\
 \hline\hline
 \PcDst   &    0.1740 &   0.1513\\
 \hline
 Statistical          &    0.0100 &   0.0096\\
 \hline
 Uncorrelated         &    0.0040 &   0.0088\\
 Correlated           &    0.0007 &   0.0011\\
 \hline
 Total Systematic     &    0.0041 &   0.0089\\
 \hline
 \hline
 $a( \Rb      )$ &    0.0293 &         \\
 $\Rb     ^{\mathrm{used}}$ &    0.2166 &         \\
 \hline
 $a( \Rc      )$ &   -0.0158 &         \\
 $\Rc     ^{\mathrm{used}}$ &    0.1735 &         \\
 \hline
 \end{tabular}
 \end{center}
 \caption{The measurements of 
 \PcDst  . }
 \label{tab:PcDstinp}
 \end{table}
 \clearpage
 \begin{table}[htb]
 \begin{center}
 \begin{tabular}{|l||c|c|c|}
 \hline
           & \mca{1}{ALEPH} & \mca{1}{DELPHI} & \mca{1}{OPAL} \\
 \hline
           &91-95 &92-95 &91-94\\
           &$D$ meson &$D$ meson &$D$ meson\\
           &\tmcite{ref:arcc} &\tmcite{ref:drcc} &\tmcite{ref:orcc}\\
 \hline\hline
 \RcfDp   &    0.0406 &   0.0384 &   0.0390\\
 \hline
 Statistical          &    0.0013 &   0.0013 &   0.0050\\
 \hline
 Uncorrelated         &    0.0014 &   0.0015 &   0.0042\\
 Correlated           &    0.0032 &   0.0025 &   0.0031\\
 \hline
 Total Systematic     &    0.0035 &   0.0030 &   0.0052\\
 \hline
 \hline
 $a( \fDp     )$ &           &   0.0008 &         \\
 $\fDp    ^{\mathrm{used}}$ &           &   0.2210 &         \\
 \hline
 $a( \fDs     )$ &           &  -0.0002 &         \\
 $\fDs    ^{\mathrm{used}}$ &           &   0.1120 &         \\
 \hline
 $a( \fLc     )$ &           &   0.0000 &         \\
 $\fLc    ^{\mathrm{used}}$ &           &   0.0840 &         \\
 \hline
 \end{tabular}
 \end{center}
 \caption{The measurements of 
 \RcfDp  . }
 \label{tab:RcfDpinp}
 \end{table}
 \begin{table}[htb]
 \begin{center}
 \begin{tabular}{|l||c|c|c|}
 \hline
           & \mca{1}{ALEPH} & \mca{1}{DELPHI} & \mca{1}{OPAL} \\
 \hline
           &91-95 &92-95 &91-94\\
           &$D$ meson &$D$ meson &$D$ meson\\
           &\tmcite{ref:arcc} &\tmcite{ref:drcc} &\tmcite{ref:orcc}\\
 \hline\hline
 \RcfDs   &    0.0207 &   0.0213 &   0.0160\\
 \hline
 Statistical          &    0.0033 &   0.0017 &   0.0042\\
 \hline
 Uncorrelated         &    0.0011 &   0.0010 &   0.0016\\
 Correlated           &    0.0053 &   0.0054 &   0.0043\\
 \hline
 Total Systematic     &    0.0054 &   0.0055 &   0.0046\\
 \hline
 \hline
 $a( \fDp     )$ &           &   0.0007 &         \\
 $\fDp    ^{\mathrm{used}}$ &           &   0.2210 &         \\
 \hline
 $a( \fDs     )$ &           &  -0.0009 &         \\
 $\fDs    ^{\mathrm{used}}$ &           &   0.1120 &         \\
 \hline
 $a( \fLc     )$ &           &  -0.0001 &         \\
 $\fLc    ^{\mathrm{used}}$ &           &   0.0840 &         \\
 \hline
 \end{tabular}
 \end{center}
 \caption{The measurements of 
 \RcfDs  . }
 \label{tab:RcfDsinp}
 \end{table}
 \begin{table}[htb]
 \begin{center}
 \begin{tabular}{|l||c|c|c|}
 \hline
           & \mca{1}{ALEPH} & \mca{1}{DELPHI} & \mca{1}{OPAL} \\
 \hline
           &91-95 &92-95 &91-94\\
           &$D$ meson &$D$ meson &$D$ meson\\
           &\tmcite{ref:arcc} &\tmcite{ref:drcc} &\tmcite{ref:orcc}\\
 \hline\hline
 \RcfLc   &    0.0157 &   0.0169 &   0.0091\\
 \hline
 Statistical          &    0.0016 &   0.0035 &   0.0050\\
 \hline
 Uncorrelated         &    0.0005 &   0.0016 &   0.0015\\
 Correlated           &    0.0044 &   0.0045 &   0.0035\\
 \hline
 Total Systematic     &    0.0045 &   0.0048 &   0.0038\\
 \hline
 \hline
 $a( \fDp     )$ &           &   0.0002 &         \\
 $\fDp    ^{\mathrm{used}}$ &           &   0.2210 &         \\
 \hline
 $a( \fDs     )$ &           &  -0.0001 &         \\
 $\fDs    ^{\mathrm{used}}$ &           &   0.1120 &         \\
 \hline
 $a( \fLc     )$ &           &  -0.0002 &         \\
 $\fLc    ^{\mathrm{used}}$ &           &   0.0840 &         \\
 \hline
 \end{tabular}
 \end{center}
 \caption{The measurements of 
 \RcfLc  . }
 \label{tab:RcfLcinp}
 \end{table}
 \begin{table}[htb]
 \begin{center}
 \begin{tabular}{|l||c|c|c|}
 \hline
           & \mca{1}{ALEPH} & \mca{1}{DELPHI} & \mca{1}{OPAL} \\
 \hline
           &91-95 &92-95 &91-94\\
           &$D$ meson &$D$ meson &$D$ meson\\
           &\tmcite{ref:arcc} &\tmcite{ref:drcc} &\tmcite{ref:orcc}\\
 \hline\hline
 \RcfDz   &    0.0964 &   0.0926 &   0.0997\\
 \hline
 Statistical          &    0.0029 &   0.0026 &   0.0070\\
 \hline
 Uncorrelated         &    0.0040 &   0.0038 &   0.0057\\
 Correlated           &    0.0045 &   0.0023 &   0.0041\\
 \hline
 Total Systematic     &    0.0060 &   0.0044 &   0.0070\\
 \hline
 \hline
 $a( \fDp     )$ &           &   0.0020 &         \\
 $\fDp    ^{\mathrm{used}}$ &           &   0.2210 &         \\
 \hline
 $a( \fDs     )$ &           &  -0.0004 &         \\
 $\fDs    ^{\mathrm{used}}$ &           &   0.1120 &         \\
 \hline
 $a( \fLc     )$ &           &  -0.0004 &         \\
 $\fLc    ^{\mathrm{used}}$ &           &   0.0840 &         \\
 \hline
 \end{tabular}
 \end{center}
 \caption{The measurements of 
 \RcfDz  . }
 \label{tab:RcfDzinp}
 \end{table}
 \begin{table}[htb]
 \begin{center}
 \begin{tabular}{|l||c|c|}
 \hline
           & \mca{1}{DELPHI} & \mca{1}{OPAL} \\
 \hline
           &92-95 &90-95\\
           &$D$ meson &$D$-meson\\
           &\tmcite{ref:drcc} &\tmcite{ref:orcd}\\
 \hline\hline
 \RcPcDst &    0.0282 &   0.0266\\
 \hline
 Statistical          &    0.0007 &   0.0005\\
 \hline
 Uncorrelated         &    0.0010 &   0.0010\\
 Correlated           &    0.0007 &   0.0009\\
 \hline
 Total Systematic     &    0.0012 &   0.0014\\
 \hline
 \hline
 $a( \fDp     )$ &    0.0006 &         \\
 $\fDp    ^{\mathrm{used}}$ &    0.2210 &         \\
 \hline
 $a( \fDs     )$ &   -0.0001 &         \\
 $\fDs    ^{\mathrm{used}}$ &    0.1120 &         \\
 \hline
 $a( \fLc     )$ &   -0.0004 &         \\
 $\fLc    ^{\mathrm{used}}$ &    0.0840 &         \\
 \hline
 \end{tabular}
 \end{center}
 \caption{The measurements of 
 \RcPcDst. }
 \label{tab:RcPcDstinp}
 \end{table}


\clearpage
\section{The Measurements used in the Electroweak Gauge Boson Couplings}\label{app-GC}
In the following, results from individual experiments used in the 
electroweak gauge boson couplings averages are summarised.

\subsection*{Charged Triple Gauge Boson Couplings}

Results from each experiment are shown in
Tables~\ref{tab:cTGC-1-ADLO}, 
\ref{tab:cTGC-2-ADLO} and ~\ref{tab:cTGC-3-ADLO} for
the single, two and three-parameter analyses respectively.
Uncertainties include both statistical
and systematic effects.  

\begin{table}[htbp]
\begin{center}
\renewcommand{\arraystretch}{1.3}
\begin{tabular}{|l||r|r|r|r|} 
\hline
Parameter  &ALEPH~\cite{ALEPH-cTGC} 
           &DELPHI~\cite{DELPHI-cTGC}
           &L3~\cite{L3-cTGC}
           &OPAL~\cite{OPAL-cTGC} \\
\hline
\hline
\gvz       & $+0.13\apm{0.21}{0.21}$ & ---  
           & $-0.04\apm{0.24}{0.25}$ & --- \\
\hline
\hline
\dgz       & $+0.008\apm{0.040}{0.039}$ & $-0.028\apm{0.046}{0.045}$ 
           & $-0.075\apm{0.058}{0.054}$ & $-0.046\apm{0.043}{0.041}$ \\ 
\hline
\dkg       & $+0.04\apm{0.09}{0.09}$ & $+0.05\apm{0.13}{0.12}$  
           & $-0.03\apm{0.12}{0.11}$ & $-0.10\apm{0.11}{0.10}$ \\  
\hline
\lg        & $-0.012\apm{0.041}{0.039}$ & $+0.056\apm{0.057}{0.056}$ 
           & $-0.081\apm{0.065}{0.058}$ & $-0.108\apm{0.040}{0.038}$ \\ 
\hline
\end{tabular}
\caption[]{The measured central values and one standard deviation
  uncertainties ($\Delta\LL=0.5$) obtained by the four LEP
  experiments.  In each case the parameter listed is varied while the
  remaining ones are fixed to their Standard Model value.  Both
  statistical and systematic uncertainties are included.  }
\label{tab:cTGC-1-ADLO}
\end{center}
\end{table}

\begin{table}[htbp]
\begin{center}
\renewcommand{\arraystretch}{1.3}
\begin{tabular}{|l||r|r|r|r|r|} 
\hline
Parameter  
           &ALEPH~\cite{ALEPH-cTGC}
           &DELPHI~\cite{DELPHI-cTGC}
           &L3~\cite{L3-cTGC}
           &OPAL~\cite{OPAL-cTGC} \\
\hline
\hline
\dgz       & $-0.001\apm{0.043}{0.045}$ & $-0.048\apm{0.049}{0.046}$
           & $-0.088\apm{0.095}{0.064}$ & $-0.017\apm{0.049}{0.061}$  \\ 
\dkg       & $+0.02\apm{0.09}{0.09}$    & $+0.14\apm{0.11}{0.11}$ 
           & $+0.04\apm{0.13}{0.17}$    & $-0.11\apm{0.17}{0.10}$  \\ 
\hline
\dgz       & $+0.019\apm{0.052}{0.051}$ & $-0.103\apm{0.063}{0.058}$ 
           & $-0.02\apm{0.10}{0.16}$    & $+0.075\apm{0.059}{0.059}$  \\ 
\lg        & $-0.026\apm{0.053}{0.055}$ & $+0.111\apm{0.069}{0.072}$ 
           & $-0.06\apm{0.19}{0.10}$    & $-0.158\apm{0.054}{0.051}$  \\ 
\hline
\lg        & $-0.019\apm{0.044}{0.044}$ & $+0.032\apm{0.060}{0.058}$ 
           & $-0.088\apm{0.093}{0.067}$ & $-0.103\apm{0.044}{0.044}$  \\ 
\dkg       & $+0.03\apm{0.09}{0.09}$    & $+0.07\apm{0.11}{0.10}$ 
           & $+0.03\apm{0.12}{0.14}$    & $-0.03\apm{0.14}{0.11}$  \\ 
\hline
\end{tabular}
\caption[]{ The measured central values and one standard deviation 
  uncertainties ($\Delta\LL=0.5$) obtained by the four LEP
  experiments.  In each case the two parameters listed are varied
  while the remaining one is fixed to its Standard Model value.  Both
  statistical and systematic uncertainties are included.  }
\label{tab:cTGC-2-ADLO}
\end{center}
\end{table}

\begin{table}
\begin{center}
\renewcommand{\arraystretch}{1.3}
\begin{tabular}{|l||r|r|r|r|} 
\hline
Parameter  
           &ALEPH~\cite{ALEPH-cTGC}
           &L3~\cite{L3-cTGC}
           &OPAL~\cite{OPAL-cTGC} \\
\hline
\hline
\dgz       & $+0.018\apm{0.043}{0.043}$
           & $-0.02\apm{0.11}{0.18}$ & $+0.081\apm{0.054}{0.058}$  \\ 
\dkg       & $+0.013\apm{0.071}{0.071}$ 
           & $-0.02\apm{0.14}{0.15}$ & $-0.07\apm{0.11}{0.09}$     \\ 
\lg        & $-0.028\apm{0.044}{0.047}$ 
           & $-0.07\apm{0.21}{0.11}$ & $-0.149\apm{0.055}{0.053}$  \\ 
\hline
\end{tabular}
\caption[]{ The measured central values and one standard deviation 
  uncertainties ($\Delta\LL=0.5$) obtained by ALEPH, L3 and OPAL.
  All three parameters listed are varied.  Both statistical and
  systematic uncertainties are included.  }
\label{tab:cTGC-3-ADLO}
\end{center}
\end{table}

\clearpage
\subsection*{Neutral Triple Gauge Boson Couplings in Z\boldmath$\gamma$ 
Production}

Results from each experiment are shown in
Tables~\ref{tab:hTGC-1-ADLO} and 
\ref{tab:hTGC-2-ADLO} for
the single and two-parameter analyses respectively.
Uncertainties include both statistical
and systematic effects.

\begin{table}[htbp]
\begin{center}
\renewcommand{\arraystretch}{1.3}
\begin{tabular}{|l||r|r|r|} 
\hline
Parameter  & DELPHI~\cite{DELPHI-hTGC}  &  L3~\cite{L3-hTGC}    & OPAL~\cite{OPAL-hTGC} \\
\hline
\hline
$h_1^\gamma$ & [$-0.17,~~+0.17$] & [$-0.14,~~+0.03$] & [$-0.11,~~+0.11$] \\
\hline
$h_2^\gamma$ & [$-0.11,~~+0.11$] & [$-0.039,~~+0.079$] & [$-0.077,~~+0.077$] \\
\hline
$h_3^\gamma$ & [$-0.058,~~+0.051$] & [$-0.095,~~-0.001$] & [$-0.17,~~-0.01$] \\
\hline
$h_4^\gamma$ & [$-0.036,~~+0.041$] & [$+0.005,~~+0.072$] & [$+0.01,~~+0.14$] \\
\hline
$h_1^Z$      & [$-0.28,~~+0.29$] & [$-0.16,~~+0.04$] & [$-0.19,~~+0.19$] \\
\hline
$h_2^Z$      & [$-0.18,~~+0.18$] & [$-0.042,~~+0.093$] & [$-0.13,~~+0.13$] \\
\hline
$h_3^Z$      & [$-0.37,~~+0.21$] & [$-0.16,~~+0.11$] & [$-0.27,~~+0.12$] \\
\hline
$h_4^Z$      & [$-0.14,~~+0.22$] & [$-0.05,~~+0.11$] & [$-0.08,~~+0.18$] \\
\hline
\end{tabular}
\caption[]{The 95\% C.L. intervals ($\Delta\LL=1.92$) measured by
  DELPHI, L3 and OPAL.  In each case the parameter listed is varied
  while the remaining ones are fixed to their Standard Model value.
  Both statistical and systematic uncertainties are included.  }
\label{tab:hTGC-1-ADLO}
\end{center}
\end{table}

\begin{table}[htbp]
\begin{center}
\renewcommand{\arraystretch}{1.3}
\begin{tabular}{|l||r|r|} 
\hline
Parameter  & DELPHI~\cite{DELPHI-hTGC}  &  L3~\cite{L3-hTGC}  \\
\hline
\hline
$h_1^\gamma$ & [$-0.35,~~+0.35$] & [$-0.23,~~+0.10$] \\
$h_2^\gamma$ & [$-0.23,~~+0.23$] & [$-0.12,~~+0.10$] \\
\hline
$h_3^\gamma$ & [$-0.26,~~+0.38$] & [$-0.20,~~+0.12$] \\
$h_4^\gamma$ & [$-0.32,~~+0.43$] & [$-0.10,~~+0.12$] \\
\hline
$h_1^Z$      & [$-0.58,~~+0.58$] & [$-0.49,~~+0.19$] \\
$h_2^Z$      & [$-0.38,~~+0.38$] & [$-0.31,~~+0.16$] \\
\hline
$h_3^Z$      & [$-0.70,~~+0.50$] & [$-0.37,~~+0.40$] \\
$h_4^Z$      & [$-0.41,~~+0.37$] & [$-0.20,~~+0.28$] \\
\hline
\end{tabular}
\caption[]{The 95\% C.L. intervals ($\Delta\LL=1.92$) measured  by
  DELPHI and L3.  In each case the two parameters listed are varied
  while the remaining ones are fixed to their Standard Model value.
  Both statistical and systematic uncertainties are included.  }
\label{tab:hTGC-2-ADLO}
\end{center}
\end{table}

\clearpage
\subsection*{Neutral Triple Gauge Boson Couplings in ZZ Production}

Results from each experiment are shown in
Tables~\ref{tab:fTGC-1-ADLO} and 
\ref{tab:fTGC-2-ADLO} for
the single and two-parameter analyses respectively.
Uncertainties include both statistical
and systematic effects.  

\begin{table}[htbp]
\begin{center}
\renewcommand{\arraystretch}{1.3}
\begin{tabular}{|l||r|r|r|} 
\hline
Parameter  & DELPHI~\cite{DELPHI-fTGC}  &  L3~\cite{L3-fTGC}   & OPAL~\cite{OPAL-fTGC}  \\
\hline
\hline
$f_4^\gamma$ & [$-0.49,~~+0.49$] & [$-0.59,~~+0.58$] & [$-0.45,~~+0.43$] \\ 
\hline
$f_4^Z$      & [$-0.82,~~+0.83$] & [$-0.97,~~+0.99$] & [$-0.74,~~+0.75$] \\ 
\hline
$f_5^\gamma$ & [$-1.2,~~+1.2$] & [$-1.4,~~+1.4$] & [$-0.9,~~+0.8$] \\ 
\hline
$f_5^Z$      & [$-1.9,~~+2.1$] & [$-2.2,~~+2.5$] & [$-1.0,~~+0.5$] \\ 
\hline
\end{tabular}
\caption[]{The 95\% C.L. intervals ($\Delta\LL=1.92$) measured by
  DELPHI, L3 and OPAL.  In each case the parameter listed is varied
  while the remaining ones are fixed to their Standard Model value.
  Both statistical and systematic uncertainties are included.  }
\label{tab:fTGC-1-ADLO}
\end{center}
\end{table}

\begin{table}[htbp]
\begin{center}
\renewcommand{\arraystretch}{1.3}
\begin{tabular}{|l||r|r|r|} 
\hline
Parameter  & DELPHI~\cite{DELPHI-fTGC}  &  L3~\cite{L3-fTGC}   & OPAL~\cite{OPAL-fTGC}  \\
\hline
\hline
$f_4^\gamma$ & [$-0.49,~~+0.49$] & [$-0.58,~~+0.58$] & [$-0.45,~~+0.43$] \\ 
$f_4^Z$      & [$-0.82,~~+0.83$] & [$-0.97,~~+0.99$] & [$-0.73,~~+0.75$] \\ 
\hline
$f_5^\gamma$ & [$-1.2,~~+1.2$] & [$-1.4,~~+1.4$] & [$-0.9,~~+0.9$] \\ 
$f_5^Z$      & [$-1.9,~~+2.1$] & [$-2.2,~~+2.5$] & [$-1.0,~~+0.7$] \\ 
\hline
\end{tabular}
\caption[]{The 95\% C.L. intervals ($\Delta\LL=1.92$) measured by
  DELPHI, L3 and OPAL.  In each case the two parameters listed are
  varied while the remaining ones are fixed to their Standard Model
  value.  Both statistical and systematic uncertainties are included.
  }
\label{tab:fTGC-2-ADLO}
\end{center}
\end{table}

\clearpage
\subsection*{Quartic Gauge Boson Couplings}

Results from each experiment are shown in
Table~\ref{tab:QGC-1-ADLO}, where the uncertainties include both statistical
and systematic effects.

\begin{table}[htbp]
\begin{center}
\renewcommand{\arraystretch}{1.3}
\begin{tabular}{|l||r|r|r|} 
\hline
Parameter     & ALEPH~\cite{ALEPH-QGC}   &  L3~\cite{L3-QGC}   & OPAL~\cite{OPAL-QGC}  \\
$[\GeV^{-2}]$ &
              &
              & \\
\hline
\hline
\azl   & [$-0.043,~~+0.042$] & [$-0.036,~~+0.035$] & [$-0.065,~~+0.065$] \\ 
\hline
\acl   &  [$-0.11,~~+0.11$]  &  [$-0.08,~~+0.11$]  &  [$-0.13,~~+0.17$] \\ 
\hline
\anl   &              ---    &  [$-0.45,~~+0.42$]  &  [$-0.61,~~+0.57$] \\  
\hline
\hline
\azl   &              ---    & [$-0.006,~~+0.006$] & [$-0.006,~~+0.008$] \\
\hline
\acl   &              ---    & [$-0.006,~~+0.010$] &  [$-0.008,~~+0.012$] \\ 
\hline
\end{tabular}
\caption[]{The 95\% C.L. intervals ($\Delta\LL=1.92$) measured by
  ALEPH, L3 and OPAL.  In each case the parameter listed is varied
  while the remaining ones are fixed to their Standard Model value.
  Both statistical and systematic uncertainties are included. Top:
  $\WWg$ and $\nngg$.  Bottom: $\Zgg$. }
\label{tab:QGC-1-ADLO}
\end{center}
\end{table}


\clearpage
\end{appendix}

\clearpage
\bibliographystyle{unsrt}
\bibliography{lepew00,lepew00_ffbar,lepew00_hf,lepew00_w,lepew00_singlew,lepew00_zz,lepew00_tgc}
\end{document}
